\documentclass[11pt]{article}

\usepackage{amsfonts,amssymb,chet,dsfont,graphicx,mathrsfs,pgfplots,tikz}

\newcommand{\1}{\mathds{1}}
\newcommand{\spin}[2]{\genfrac{}{}{0pt}{1}{#1}{#2}}
\newcommand{\Op}[2]{\mathcal{O}_{#1}(\eta_{#2})}

\newcommand{\ee}[3]{(\eta_{#1}\cdot\eta_{#2})^{#3}}
\newcommand{\e}[3]{\eta_{#1}^{#2_{#3}}}
\newcommand{\D}{\mathcal{D}}
\newcommand{\A}{\mathcal{A}}
\newcommand{\cOPE}[4]{{}_{#1}c_{#2#3}^{\phantom{#2#3}#4}}
\newcommand{\DOPE}[4]{{}_{#1}\D_{#2#3}^{\phantom{#2#3}#4}}
\newcommand{\tOPE}[6]{{}_{#1}t_{#2#3}^{#5#6#4}}
\newcommand{\cCF}[4]{{}_{#1}c_{#2#3#4}}
\newcommand{\tCF}[6]{{}_{#1}t_{#2#3#4}^{#5#6}}
\newcommand{\vev}[1]{\langle{#1}\rangle}
\newcommand{\Vev}[1]{\left\langle{#1}\right\rangle}

\usetikzlibrary{calc}
\newcommand\ellipsebyfoci[4]{% options, focus pt1, focus pt2, cste
\path[#1] let \p1=(#2), \p2=(#3), \p3=($(\p1)!.5!(\p2)$)
in \pgfextra{
\pgfmathsetmacro{\angle}{atan2(\y2-\y1,\x2-\x1)}
\pgfmathsetmacro{\focal}{veclen(\x2-\x1,\y2-\y1)/2/1cm}
\pgfmathsetmacro{\lentotcm}{\focal*2*#4}
\pgfmathsetmacro{\axeone}{(\lentotcm - 2 * \focal)/2+\focal}
\pgfmathsetmacro{\axetwo}{sqrt((\lentotcm/2)*(\lentotcm/2)-\focal*\focal}
}
(\p3) ellipse[x radius=\axeone cm,y radius=\axetwo cm, rotate=\angle];
}

\title{New Methods for Conformal Correlation Functions}

\author{Jean-Fran\c{c}ois Fortin$^{\ast,}$\email{jean-francois.fortin@phy.ulaval.ca} and Witold Skiba$^{\dagger,}$\email{witold.skiba@yale.edu}}

\affiliation{
$^\ast$D\'epartement de Physique, de G\'enie Physique et d'Optique\\Universit\'e Laval, Qu\'ebec, QC G1V 0A6, Canada\\
$^\dagger$Department of Physics, Yale University, New Haven, CT 06520, USA
}%Choices for affiliations $^{\ast,\dagger,\$,\S,\ddag,}$

\abstract{The most general operator product expansion in conformal field theory is obtained using the embedding space formalism and a new uplift for general quasi-primary operators.  The uplift introduced here, based on quasi-primary operators with spinor indices only and standard projection operators, allows a unified treatment of all quasi-primary operators irrespective of their Lorentz group irreducible representations.  This unified treatment works at the level of the operator product expansion and hence applies to all correlation functions.  A very useful differential operator appearing in the operator product expansion is established and its action on appropriate products of embedding space coordinates is explicitly computed.  This computation leads to tensorial generalizations of the usual Exton function for all correlation functions.  Several important identities and contiguous relations are also demonstrated for these new tensorial functions.  From the operator product expansion all correlation functions for all quasi-primary operators, irrespective of their Lorentz group irreducible representations, can be computed recursively in a systematic way.  The resulting answer can be expressed in terms of tensor structures that carry all the Lorentz group information and linear combinations of the new tensorial functions.  Finally, a summary of the well-defined rules allowing the computation of all correlation functions constructively is presented.}

\date{May 2019} %Uncomment this line for month to be fixed

\begin{document}

\maketitle

\toc

%%%%%%%%%%%%%%%%%%%%%%%%%%%%%%%%%%%%%%%%%%%%%%%%%%
%%%%%%%%%%%%%%%%%%%%%%%%%%%%%%%%%%%%%%%%%%%%%%%%%%

\section{Introduction}\label{SecIntro}

Conformal field theories (CFTs) are special quantum field theories (QFTs) with extended spacetime symmetry groups.  They correspond to fixed points in the renormalization group flows of usual QFTs.  Hence, barring the existence of scale invariant but not conformal invariant field theories \cite{Zamolodchikov:1986gt,Polchinski:1987dy,Cardy:1988cwa,Jack:1990eb,Osborn:1991gm,Komargodski:2011vj,Luty:2012ww,Fortin:2012hn}, usual QFTs could be understood as relevant deformations of CFTs.  Because of the connection between QFT at imaginary time and statistical physics, CFTs also describe second-order phase transitions in condensed matter systems.  It is thus clear that a thorough understanding of CFTs cannot be overstated.  In particular, the study and classification of all CFTs in any spacetime dimension is of utmost importance.

The extended spacetime symmetry group of a CFT includes the usual $SO(1,d-1)$ Poincar\'e generators plus a dilatation generator, responsible for re-scalings, and a vector of generators, responsible for special conformal transformations.  Together, they generate the conformal group $SO(2,d)$ that strongly constrains properties of CFTs.  Indeed, the existence of the conformal algebra leads to a natural separation between quasi-primary operators on the one side and their descendants on the other.  The quasi-primary operators are the special operators that are annihilated by the special conformal generators at the origin of the coordinate system.  The descendants are obtained from the quasi-primary operators by acting with the spacetime translation generators.  Therefore, the special conformal generators and the spacetime translation generators can be thought of as raising and lowering operators, respectively, while the quasi-primary operators can be understood as highest weight states.  Hence, the behavior of descendants is completely fixed by their quasi-primary operators and the conformal algebra.  It is thus possible to focus on correlation functions of quasi-primary operators only.

Moreover, symmetry considerations alone completely fix the form of the two-point and three-point correlation functions up to a finite number of constant coefficients.  These coefficients are related to the coefficients appearing in the operator product expansion (OPE) \cite{Ferrara:1971vh,Ferrara:1971zy,Ferrara:1972cq,Ferrara:1973eg,Dobrev:1975ru,Mack:1976pa}.  In CFT, the OPE is a well-defined quantity.  It re-expresses the product of two quasi-primary operators at two different spacetime points in terms of an infinite sum of quasi-primary operators weighted by their OPE coefficients.  Hence, the OPE is a fundamental object that can be used to compute $M$-point correlation functions from $(M-1)$-point correlation functions.  Since two-point correlation functions are completely determined by the conformal algebra, once the OPE is known technically all correlation functions in a CFT are known up to the OPE coefficients.

At this point, associativity of the correlation functions, or crossing symmetry, can be used to express the same correlation functions in terms of different combinations of OPE coefficients times well-behaved conformal objects.  This idea, introduced in \cite{Ferrara:1973yt,Polyakov:1974gs}, is behind the conformal bootstrap approach.  In principle, the conformal bootstrap approach leads to constraints on the OPE coefficients which were not directly fixed by symmetry considerations alone.  A partial list of nice analytical and numerical results from the conformal bootstrap approach can be found in \cite{Dobrev:1977qv,Cornalba:2007fs,Cornalba:2009ax,Pappadopulo:2012jk,Costa:2012cb,Hogervorst:2013sma,Hartman:2015lfa,Kim:2015oca,Li:2015itl,Hartman:2016dxc,Simmons-Duffin:2016gjk,Hofman:2016awc,Hartman:2016lgu,Afkhami-Jeddi:2016ntf,Gadde:2017sjg,Hogervorst:2017sfd,Caron-Huot:2017vep,Hogervorst:2017kbj,Kulaxizi:2017ixa,Li:2017lmh,Cuomo:2017wme,Dey:2017oim,Simmons-Duffin:2017nub,Elkhidir:2017iov,Kravchuk:2018htv,Karateev:2018oml,Liendo:2019jpu,Albayrak:2019gnz} and \cite{Rattazzi:2008pe,Rychkov:2009ij,Caracciolo:2009bx,Poland:2010wg,Rattazzi:2010gj,Poland:2011ey,Rychkov:2011et,ElShowk:2012ht,Liendo:2012hy,ElShowk:2012hu,Gliozzi:2013ysa,Alday:2013opa,Gaiotto:2013nva,El-Showk:2014dwa,Chester:2014fya,Kos:2014bka,Caracciolo:2014cxa,Paulos:2014vya,Beem:2014zpa,Simmons-Duffin:2015qma,Bobev:2015jxa,Beem:2015aoa,Iliesiu:2015qra,Poland:2015mta,Lemos:2015awa,Lin:2015wcg,Chester:2016wrc,Rychkov:2016iqz,Behan:2016dtz,El-Showk:2016mxr,Lin:2016gcl,Lemos:2016xke,Beem:2016wfs,Li:2017ddj,Collier:2017shs,Cornagliotto:2017dup,Rychkov:2017tpc,Nakayama:2017vdd,Chang:2017xmr,Dymarsky:2017yzx,Poland:2018epd,Karateev:2019pvw} respectively.\footnote{For work on the conformal bootstrap with global symmetry, see \cite{Rattazzi:2010yc,Vichi:2011ux,Kos:2013tga,Berkooz:2014yda,Nakayama:2014lva,Nakayama:2014yia,Nakayama:2014sba,Bae:2014hia,Chester:2014gqa,Kos:2015mba,Chester:2015qca,Chester:2015lej,Dey:2016zbg,Nakayama:2016knq,Li:2016wdp,Pang:2016xno,Dymarsky:2017xzb,Stergiou:2018gjj,Kousvos:2018rhl,Stergiou:2019dcv}, for work on the conformal bootstrap with higher spins, see \cite{Alday:2015eya,Alday:2015ota,Alday:2015ewa,Alday:2016mxe,Alday:2016njk,Alday:2016jfr}.}

The use of the OPE inside four-point correlation functions leads to the conformal blocks \cite{Ferrara:1973vz,Ferrara:1974nf,Dobrev:1975ru,Dolan:2000ut,Dolan:2003hv,Dolan:2011dv}, which are the well-behaved conformal objects that account for the exchange of specific quasi-primary operators as well as their infinite towers of descendants.  As mentioned above, the conformal blocks are completely prescribed by conformal invariance.  However, the technical details are complicated and very few analytical results exist on conformal blocks for quasi-primary operators in general irreducible representations (see \cite{Dolan:2000ut,Dolan:2003hv,Giombi:2011rz,Costa:2011mg,Dolan:2011dv,Costa:2011dw,SimmonsDuffin:2012uy,Costa:2014rya,Elkhidir:2014woa,Echeverri:2015rwa,Hijano:2015zsa,Rejon-Barrera:2015bpa,Penedones:2015aga,Iliesiu:2015akf,Echeverri:2016dun,Isachenkov:2016gim,Costa:2016hju,Costa:2016xah,Chen:2016bxc,Nishida:2016vds,Cordova:2016emh,Schomerus:2016epl,Kravchuk:2016qvl,Gliozzi:2017hni,Castro:2017hpx,Dyer:2017zef,Sleight:2017fpc,Chen:2017yia,Pasterski:2017kqt,Cardoso:2017qmj,Karateev:2017jgd,Kravchuk:2017dzd,Dey:2017fab,Hollands:2017chb,Schomerus:2017eny,Isachenkov:2017qgn,Faller:2017hyt,Rong:2017cow,Chen:2017xdz,Sleight:2018epi,Costa:2018mcg,Kobayashi:2018okw,Bhatta:2018gjb,Lauria:2018klo,Liu:2018jhs,Gromov:2018hut,Zhou:2018sfz,Kazakov:2018gcy,Rosenhaus:2018zqn} for different research avenues on the computations of conformal blocks).

The embedding space, introduced in \cite{Dirac:1936fq}, is the natural arena where CFTs are defined.  Indeed, the embedding space is a hypercone on which quasi-primary operators naturally live.  More importantly, the action of the conformal algebra on quasi-primary operators in embedding space is homogeneous.  Although natural, focus on CFTs from the embedding space formalism has not completely blossomed as of now (see \textit{e.g.} \cite{Mack:1969rr,Ferrara:1971vh,Ferrara:1971zy,Ferrara:1972cq,Ferrara:1973eg,Ferrara:1973yt,Weinberg:2010fx,Weinberg:2012mz,Fortin:2016lmf,Fortin:2016dlj,Comeau:2019xco} for interesting results using the embedding space formalism).  For example, most analytic results for conformal blocks have not been computed using the embedding space formalism.

Since all correlation functions can be computed and the conformal bootstrap can be undertaken once the OPE is known, it is clear that the most fundamental building block of a CFT is its OPE.  Recently \cite{Fortin:2016lmf,Fortin:2016dlj,Comeau:2019xco}, the authors proposed to use the embedding space formalism to derive the OPE (partial results on the embedding space OPE can be found in \cite{Ferrara:1971vh,Ferrara:1971zy,Ferrara:1972cq,Ferrara:1973eg,Ferrara:1973yt}) and to employ the OPE directly in embedding space to compute the conformal blocks appearing in $M$-point correlation functions of quasi-primary operators in general irreducible representations.  The purpose of this paper is to complete this goal by defining the embedding space OPE in the most useful way possible and investigate some of its consequences for the computations of $M$-point correlation functions.  In the process, a new uplift of quasi-primary operators in position space to quasi-primary operators in embedding space is introduced.  With the knowledge of the OPE, it is then relatively straightforward to use it recursively to compute all correlation functions and then implement crossing symmetry.

As already mentioned, the OPE is reminiscent of an algebra which encodes how the product of two quasi-primary operators at different points can be expanded in a series in quasi-primary operators and their descendants.  More explicitly, the embedding space OPE can be expressed as
\eqn{\Op{i}{1}\Op{j}{2}=\sum_k\sum_{a=1}^{N_{ijk}}\cOPE{a}{i}{j}{k}\DOPE{a}{i}{j}{k}(\eta_1,\eta_2)\Op{k}{2},}[EqSOPE]
where the sum over quasi-primary operators $\Op{k}{2}$ is infinite but the sum over the $N_{ijk}$ OPE coefficients $\cOPE{a}{i}{j}{k}$, with appropriate differential operators $\DOPE{a}{i}{j}{k}(\eta_1,\eta_2)$, is finite.  All linearly-independent quasi-primary operators appear in the sum in \eqref{EqSOPE} and the differential operator $\DOPE{a}{i}{j}{k}(\eta_1,\eta_2)$ takes care of the infinite towers of descendants.

To build the embedding space OPE \eqref{EqSOPE}, several consistency conditions are used.  For example, unlike in position space, to remain on the embedding space hypercone the quasi-primary operators on the right-hand side of \eqref{EqSOPE} must either be located at $\eta_1^A$ or $\eta_2^A$.  The correlation functions computed from the OPE must not depend on this particular choice.  Other consistency conditions originating from homogeneity, transversality, and the Lorentz symmetry as well as simplicity-of-use arguments determine the OPE explicitly.  Therefore, the main goal of this paper is to obtain the most useful definitions of embedding space quasi-primary operators $\Op{}{}$ and the differential operators $\DOPE{a}{i}{j}{k}(\eta_1,\eta_2)$ for quasi-primary operators in general irreducible representations of the Lorentz group.  The embedding space OPE \eqref{EqSOPE} needs to be practical for computations of $M$-point correlation functions.

This paper is organized as follows: Section \ref{SecPS} gives a quick review of the conformal algebra and its action on quasi-primary operators when these operators are chosen to carry spinor indices only.  To this end, a simple introduction to Clifford algebras in odd and even dimensions is presented.  The treatment in the even-dimensional case is somewhat different than usual due to the emphasis on irreducible representations of the Lorentz group.

Section \ref{SecES} discusses the usual embedding space formalism where the action of the conformal algebra on quasi-primary operators is homogeneous.  To uplift quasi-primary operators in defining irreducible representations of the Lorentz group, the proper Clifford algebras in embedding space in odd and even dimensions are presented and the usual consistency conditions of homogeneity and transversality are introduced.  It is shown how homogeneity restricts quasi-primary operators to be functions of spacetime coordinates only as in the usual embedding space formalism.  It is also shown how transversality restricts the quasi-primary operators in embedding space with spinor indices only to the proper quasi-primary operators in position space, removing the extra degrees of freedom to reach the appropriate number in position space.  This is all done at the level of quasi-primary operators in defining irreducible representations of the Lorentz group since the latter uplift to quasi-primary operators in defining irreducible representations of the conformal group.

In Section \ref{SecDO}, the most useful differential operator appearing in the OPE is obtained after a careful treatment of all first-order differential operators which are well defined on the light-cone.  The differential operator is a tensorial object that is fully symmetric and traceless with respect to the embedding space metric.  Its action on appropriate products of embedding space coordinates is then computed for any number of embedding space coordinates, leading to tensorial functions of the uncontracted embedding space coordinates and the conformal cross-ratios.  These new tensorial functions are the proper generalizations to all quasi-primary operators of the Exton function of scalar exchange in four-point correlation functions of four scalars.  Several important properties (identities and contiguous relations) of the new tensorial functions are then shown.

Next, Section \ref{SecOPE} constructs the most general OPE using consistency conditions coming from homogeneity, transversality and the Lorentz group.  A new doubly-transverse embedding space metric, introduced in the context of the OPE differential operator, allows the uplift of quasi-primary operators in general irreducible representations of the Lorentz group to embedding space at the level of the OPE.  The OPE is always given in terms of tensor structures that relate the different irreducible representations of the quasi-primary operators and the tensorial differential operator.  The former can be seen as the sole quantities containing all non-trivial group-theoretical information.  The latter implies that the action of the OPE in correlation functions is known.  It is then shown from the OPE that the number of independent tensor structures corresponds to the expected number of OPE coefficients.  The tensor structures can be constructed systematically from the Lorentz group information only and do not rely on any conformal information---they are purely group-theoretical objects.

After several consistency checks and useful identities, in Section \ref{SecnptCF} the OPE is finally exploited in the computation of $M$-point correlation functions.  The OPE is used recursively to obtain $M$-point correlation functions from $(M-1)$-point correlation functions.  Two-point correlation functions are given explicitly and three-point correlation functions are given up to simple use of the differential operator.  Higher-point correlation functions are computed schematically.  It is shown that the new tensorial functions appear directly in four-point correlation functions but some re-summations are still necessary for $(M>4)$-point correlation functions.  A recurrence relation for the seed conformal block of higher-point correlation functions in terms of the action of the OPE differential operator, which is known, is obtained.  A short discussion of the conformal bootstrap with the usual argument showing that associativity of all four-point correlation functions is sufficient to constrain all OPE coefficients is presented.

Section \ref{SecSum} summarizes all the results of this paper by explicitly giving a set of systematic rules allowing the computation of any correlation function.  Finally, Section \ref{SecConc} concludes, pointing out what can be done with the formalism introduced here.  For completeness and reference, Appendix \ref{SecIrrep} provides a lengthy introduction to irreducible representations of $SO(p,q)$ useful for the construction of the several Lorentz group objects appearing throughout the paper.  Appendix \ref{SecPtoE} lists several properties of the embedding space quantities.

%%%%%%%%%%%%%%%%%%%%%%%%%%%%%%%%%%%%%%%%%%%%%%%%%%
%%%%%%%%%%%%%%%%%%%%%%%%%%%%%%%%%%%%%%%%%%%%%%%%%%

\section{Conformal Field Theories in Position Space}\label{SecPS}

This section summarizes some simple properties of CFTs in any dimension $d$ that will be useful when introducing the uplift to embedding space in the next section.  The metric is chosen to be mostly-negative, \textit{i.e.} $g_{\mu\nu}=\text{diag}(+1,-1,\ldots,-1)$, with Lorentz group $SO(d)\equiv SO(1,d-1)$.\footnote{Although the focus is on $SO(1,d-1)$, all equations apply to $SO(p,q)$ as per Appendix \ref{SecIrrep}.}  The notation used for quasi-primary operators is free of indices.

%%%%%%%%%%%%%%%%%%%%%%%%%%%%%%%%%%%%%%%%%%%%%%%%%%

\subsection{Conformal Algebra in Position Space}\label{SSecAlgPS}

The non-trivial part of the conformal algebra in position space is
\eqna{
[M_{\mu\nu},M_{\lambda\rho}]&=-(s_{\mu\nu})_\lambda^{\phantom{\lambda}\delta}M_{\delta\rho}-(s_{\mu\nu})_\rho^{\phantom{\rho}\delta}M_{\lambda\delta},\\
[M_{\mu\nu},P_\lambda]&=-(s_{\mu\nu})_\lambda^{\phantom{\lambda}\rho}P_\rho,\\
[M_{\mu\nu},K_\lambda]&=-(s_{\mu\nu})_\lambda^{\phantom{\lambda}\rho}K_\rho,\\
[P_\mu,K_\nu]&=2i(g_{\mu\nu}D-M_{\mu\nu}),\\
[P_\mu,D]&=iP_\mu,\\
[K_\mu,D]&=-iK_\mu,
}[EqAlgPS]
where $M_{\mu\nu}$ are the Lorentz generators, $P_\mu$ are the translation generators, $K_\mu$ are the special conformal generators, and $D$ is the dilatation generator.  The matrices $s_{\mu\nu}$ are the Lorentz generators in the fundamental vector representation and are given by
\eqn{(s_{\mu\nu})^{\lambda\rho}=i(\delta_\mu^{\phantom{\mu}\lambda}\delta_\nu^{\phantom{\nu}\rho}-\delta_\mu^{\phantom{\mu}\rho}\delta_\nu^{\phantom{\nu}\lambda}).}
They satisfy the Lorentz algebra by construction, \textit{i.e.}
\eqn{[s_{\mu\nu},s_{\lambda\rho}]=-(s_{\mu\nu})_{\lambda}^{\phantom{\lambda}\lambda'}s_{\lambda'\rho}-(s_{\mu\nu})_{\rho}^{\phantom{\rho}\rho'}s_{\lambda\rho'}.}
%

%%%%%%%%%%%%%%%%%%%%%%%%%%%%%%%%%%%%%%%%%%%%%%%%%%

\subsection{Spinors in Position Space}\label{SSecSPS}

To completely determine the action of the conformal algebra on all quasi-primary operators, it is convenient to first discuss the fundamental spinor representations of the Lorentz algebra.  This is usually accomplished by studying the appropriate Clifford algebras.  Since irreducible Clifford representations in even dimensions are reducible Lorentz representations it is convenient to investigate odd and even dimensions separately.

Note that, since the metric has Lorentzian signature, for all dimensions $d=1,2,3\text{ mod }8$ it is possible to impose an independent reality condition on the operators, the Majorana condition, which reduces the number of real degrees of freedom of the fundamental spinor by half.  Indeed for such dimensions there is a purely real or a purely imaginary representation of the matrices introduced below.

\subsubsection{Odd Dimensions}\label{SSecSPSodd}

In odd dimensions the Clifford representations are irreducible spinor representations of the Lorentz algebra.  By introducing $2^{(d-1)/2}$-dimensional square $\gamma$-matrices which satisfy the Clifford algebra
\eqn{\gamma^\mu\gamma^\nu+\gamma^\nu\gamma^\mu=2g^{\mu\nu}\1,}
it is easy to define Lorentz generators in the fundamental spinor representation by
\eqn{\sigma_{\mu\nu}=\frac{i}{4}(\gamma_\mu\gamma_\nu-\gamma_\nu\gamma_\mu).}
These Lorentz generators satisfy the following identities
\eqn{
\begin{gathered}
\sigma_{\mu\nu}\gamma_\lambda-\gamma_\lambda\sigma_{\mu\nu}=-(s_{\mu\nu})_{\lambda}^{\phantom{\lambda}\lambda'}\gamma_{\lambda'},\\
[\sigma_{\mu\nu},\sigma_{\lambda\rho}]=-(s_{\mu\nu})_{\lambda}^{\phantom{\lambda}\lambda'}\sigma_{\lambda'\rho}-(s_{\mu\nu})_{\rho}^{\phantom{\rho}\rho'}\sigma_{\lambda\rho'},
\end{gathered}
}
and thus verify the Lorentz algebra.  Moreover, there exists matrices $A$, $C$, $B$ which relate the adjoint, contragredient and conjugate representations respectively to the original representation (see Section \ref{SSSecOddSpinor}).  They satisfy the appropriate relations \eqref{EqSymgammaOdd} for $p=1$ and $q=d-1$.

\subsubsection{Even Dimensions}\label{SSecSPSeven}

In even dimensions the Clifford representations are reducible spinor representations of the Lorentz algebra.  To study directly the two irreducible (Weyl) representations obtained, it is necessary to introduce two types of $2^{(d-2)/2}$-dimensional square matrices, $\gamma$ and $\tilde{\gamma}$, which satisfy
\eqna{
\gamma^\mu\tilde{\gamma}^\nu+\gamma^\nu\tilde{\gamma}^\mu&=2g^{\mu\nu}\1,\\
\tilde{\gamma}^\mu\gamma^\nu+\tilde{\gamma}^\nu\gamma^\mu&=2g^{\mu\nu}\1.
}
By defining the Lorentz generators in the two fundamental spinor representations as
\eqna{
\sigma_{\mu\nu}=\frac{i}{4}(\gamma_\mu\tilde{\gamma}_\nu-\gamma_\nu\tilde{\gamma}_\mu),\\
\tilde{\sigma}_{\mu\nu}=\frac{i}{4}(\tilde{\gamma}_\mu\gamma_\nu-\tilde{\gamma}_\nu\gamma_\mu),
}
it is a simple matter to verify that
\eqn{
\begin{gathered}
\sigma_{\mu\nu}\gamma_\lambda-\gamma_\lambda\tilde{\sigma}_{\mu\nu}=-(s_{\mu\nu})_{\lambda}^{\phantom{\lambda}\lambda'}\gamma_{\lambda'},\\
\tilde{\sigma}_{\mu\nu}\tilde{\gamma}_\lambda-\tilde{\gamma}_\lambda\sigma_{\mu\nu}=-(s_{\mu\nu})_{\lambda}^{\phantom{\lambda}\lambda'}\tilde{\gamma}_{\lambda'},\\
[\sigma_{\mu\nu},\sigma_{\lambda\rho}]=-(s_{\mu\nu})_{\lambda}^{\phantom{\lambda}\lambda'}\sigma_{\lambda'\rho}-(s_{\mu\nu})_{\rho}^{\phantom{\rho}\rho'}\sigma_{\lambda\rho'},\\
[\tilde{\sigma}_{\mu\nu},\tilde{\sigma}_{\lambda\rho}]=-(s_{\mu\nu})_{\lambda}^{\phantom{\lambda}\lambda'}\tilde{\sigma}_{\lambda'\rho}-(s_{\mu\nu})_{\rho}^{\phantom{\rho}\rho'}\tilde{\sigma}_{\lambda\rho'}.
\end{gathered}
}
As expected, the two Lorentz generators satisfy the Lorentz algebra.  Finally, there also exists matrices $A$, $C$, $B$ and $\tilde{A}$, $\tilde{C}$, $\tilde{B}$ which relate the adjoint, contragredient and conjugate representations respectively to the two original representations (see Section \ref{SSSecEvenSpinor}).  They satisfy the appropriate relations \eqref{EqSymgammaEven} for $p=1$ and $q=d-1$.

%%%%%%%%%%%%%%%%%%%%%%%%%%%%%%%%%%%%%%%%%%%%%%%%%%

\subsection{Quasi-Primary Operators in Position Space}\label{SSecOPS}

Quasi-primary operators are labeled by their irreducible (finite-dimensional and non-unitary) representations of the Lorentz group as well as their conformal dimensions.  More importantly though, they are operators which transform under the conformal algebra in specific ways.  Indeed, the action of the conformal algebra on a quasi-primary operator $\mathcal{O}^{(x)}(x)$ in position space is
\eqna{
[M_{\mu\nu},\mathcal{O}^{(x)}(x)]&=-i\left(x_\mu\frac{\partial}{\partial x^\nu}-x_\nu\frac{\partial}{\partial x^\mu}\right)\mathcal{O}^{(x)}(x)-(\sigma_{\mu\nu}\mathcal{O}^{(x)})(x),\\
[P_\mu,\mathcal{O}^{(x)}(x)]&=-i\frac{\partial}{\partial x^\mu}\mathcal{O}^{(x)}(x),\\
[K_\mu,\mathcal{O}^{(x)}(x)]&=-i\left(2x_\mu x^\nu\frac{\partial}{\partial x^\nu}-x^2\frac{\partial}{\partial x^\mu}+2\Delta_\mathcal{O}x_\mu\right)\mathcal{O}^{(x)}(x)-2x^\nu(\sigma_{\mu\nu}\mathcal{O}^{(x)})(x),\\
[D,\mathcal{O}^{(x)}(x)]&=-i\left(x^\mu\frac{\partial}{\partial x^\mu}+\Delta_\mathcal{O}\right)\mathcal{O}^{(x)}(x).
}[EqAlgOPS]
Irreducible representations of the Lorentz group, which are denoted by a set of non-negative integers $\boldsymbol{N}^\mathcal{O}=\{N_1^\mathcal{O},\ldots,N_r^\mathcal{O}\}$ with $r$ the rank of the Lorentz group, can be solely written in terms of fundamental spinor indices with specific symmetry properties.  The Lorentz generators for quasi-primary operators are therefore appropriate sums of the generators in the fundamental spinor representation (with the Majorana condition imposed if possible) as above.  Thus, when all $n_\mathcal{O}$ quasi-primary operator indices are chosen to be fundamental spinor indices, where
\eqna{
\text{$d$ odd:}&\qquad n_\mathcal{O}\equiv2S_\mathcal{O}=2\sum_{i=1}^{r-1}N_i^\mathcal{O}+N_r^\mathcal{O},\\
\text{$d$ even:}&\qquad n_\mathcal{O}\equiv2S_\mathcal{O}=2\sum_{i=1}^{r-2}N_i^\mathcal{O}+N_{r-1}^\mathcal{O}+N_r^\mathcal{O},
}
and $S_\mathcal{O}$ is the ``spin'' of the quasi-primary operator, an index-free notation can be used.  In this notation, spinor index contraction is written without parenthesis when the contraction occurs on one particular spinor index only and with parenthesis when the contraction occurs on all spinor indices of the quasi-primary operator $\mathcal{O}_{\alpha_1\cdots\alpha_{n_\mathcal{O}}}^{(x)}(x)$, \textit{e.g.}
\eqna{
\sigma_{\mu\nu}^{(i)}\mathcal{O}^{(x)}(x)&\equiv(\sigma_{\mu\nu})_{\alpha_i}^{\phantom{\alpha_i}\beta_i}\mathcal{O}_{\alpha_1\cdots\beta_i\cdots\alpha_{n_\mathcal{O}}}^{(x)}(x),\\
(\sigma_{\mu\nu}\mathcal{O}^{(x)})(x)&\equiv\sum_{i=1}^{n_\mathcal{O}}(\sigma_{\mu\nu})_{\alpha_i}^{\phantom{\alpha_i}\beta_i}\mathcal{O}_{\alpha_1\cdots\beta_i\cdots\alpha_{n_\mathcal{O}}}^{(x)}(x).
}
When no sums are involved, the superscript $(i)$ indicates which spinor index is contracted.  Although different types of spinor indices can exist (\textit{e.g.} up and down or tilde and untidle), in explicit computations spinor indices of quasi-primary operators are assumed down and untilde for simplicity.  The reader is referred to Appendix \ref{SecIrrep} for more detail on the irreducible representations of $SO(p,q)$.

Finally, note that $\Delta_\mathcal{O}$ is the conformal dimension of the quasi-primary operator and that quasi-primary operators located at the origin of position space are annihilated by the special conformal generators, \textit{i.e.} $[K_\mu,\mathcal{O}^{(x)}(0)]=0$.

%%%%%%%%%%%%%%%%%%%%%%%%%%%%%%%%%%%%%%%%%%%%%%%%%%
%%%%%%%%%%%%%%%%%%%%%%%%%%%%%%%%%%%%%%%%%%%%%%%%%%

\section{Conformal Field Theories in Embedding Space}\label{SecES}

The action of the conformal algebra on quasi-primary operators \eqref{EqAlgOPS} is rather intricate.  Fortunately, it is possible to render it manifestly covariant by going to the embedding space.  This section reviews the embedding space formalism \cite{Dirac:1936fq,Mack:1969rr,Ferrara:1971vh,Ferrara:1971zy,Ferrara:1972cq,Ferrara:1973eg,Ferrara:1973yt,Weinberg:2010fx,Weinberg:2012mz,Fortin:2016lmf,Fortin:2016dlj,Comeau:2019xco}, establishes the notation, and introduces the particular uplift of quasi-primary operators in position space to embedding space.  Again, the index-free notation is used for quasi-primary operators.

%%%%%%%%%%%%%%%%%%%%%%%%%%%%%%%%%%%%%%%%%%%%%%%%%%

\subsection{Conformal Algebra in Embedding Space}\label{SSecAlgES}

By defining the conformal generators in embedding space as
\eqn{L_{AB}=\left(\begin{array}{ccc}M_{\mu\nu}&-\frac{1}{2}(P_\mu-K_\mu)&-\frac{1}{2}(P_\mu+K_\mu)\\\frac{1}{2}(P_\nu-K_\nu)&0&-D\\\frac{1}{2}(P_\nu+K_\nu)&D&0\end{array}\right),}[EqLtoM]
the conformal algebra \eqref{EqAlgPS} in embedding space is simply $SO(d_E)\equiv SO(2,d)$, \textit{i.e.}
\eqn{[L_{AB},L_{CD}]=-(S_{AB})_C^{\phantom{C}C'}L_{C'D}-(S_{AB})_D^{\phantom{D}D'}L_{CD'},}[EqAlgES]
where the metric is $g_{AB}=\text{diag}(+1,-1,\ldots,-1;-1,+1)$.  The matrices $S_{AB}$ are the fundamental vector generators of $SO(d_E)$ and are explicitly given by
\eqn{(S_{AB})^{CD}=i(g_A^{\phantom{A}C}g_B^{\phantom{B}D}-g_A^{\phantom{A}D}g_B^{\phantom{B}C}).}

The embedding space is a $d_E=d+2$ dimensional projective space which can be described by the hypercone
\eqn{\eta^2\equiv g_{AB}\eta^A\eta^B=0,}[EqHC]
where $\eta^A=(\eta^\mu,\eta^{d+1},\eta^{d+2})$ are the coordinates on the hypercone.  The connection with position space is given by
\eqn{x^\mu=\frac{\eta^\mu}{-\eta^{d+1}+\eta^{d+2}},}
due to the projective nature of the cone where $\lambda\eta^A$ is identified with $\eta^A$ for $\lambda>0$, while the conformal generators are
\eqn{M_{\mu\nu}=L_{\mu\nu},\quad\quad P_\mu=-(L_{\mu,d+1}+L_{\mu,d+2}),\quad\quad K_\mu=L_{\mu,d+1}-L_{\mu,d+2},\quad\quad D=-L_{d+1,d+2}.}

The Casimirs of the conformal algebra are straightforwardly obtained in embedding space from the $SO(d_E)$ algebra.  Indeed $SO(d_E)$ has $\lfloor d_E/2\rfloor$ Casimirs given by
\eqna{
\text{$d_E$ odd:}&\qquad\{|L^{2i}| : i\in\mathbb{N}, i\leq\lfloor d_E/2\rfloor\},\\
\text{$d_E$ even:}&\qquad\{|L^{2i}| : i\in\mathbb{N}, i<\lfloor d_E/2\rfloor\}\cup\{\epsilon^{A_1\cdots A_{d_E}}L_{A_1A_2}\cdots L_{A_{d_E-1}A_{d_E}}\},
}[EqCasimirs]
where $|L^n|\equiv(L^n)_A^{\phantom{A}A}$ (for $d_E=2$, the Casimir can be chosen as $|L^2|$ since $\epsilon^{AB}L_{AB}\propto L_{01}$).  This can be seen from the following identities, which can all be shown by induction,
\eqna{
[L_{AB},(L^n)_{CD}]&=-(S_{AB})_C^{\phantom{C}E}(L^n)_{ED}-(S_{AB})_D^{\phantom{D}E}(L^n)_{CE},\\
[L_{AC},(L^n)^C_{\phantom{C}B}]&=i(d_E-1)(L^n)_{AB}+i(L^n)_{BA}-ig_{AB}|L^n|,\\
[L_{AB},|L^n|]&=0.
}[EqAlgDO]
The second and third identities are obvious consequences of the first identity.  The last identity shows that all $|L^n|$ are Casimirs of the $SO(d_E)$ algebra.  However, by using $L_{BA}=-L_{AB}$ one can also show that all $|L^{2n+1}|$ can effectively be written in terms of (powers of) the $|L^{2i}|$ with $i\leq n$.  Thus the odd $|L^n|$ are not independent Casimirs.  Since $d_E\times d_E$ matrices have characteristic polynomials of degree $d_E$, for $d_E$ even and $d_E>2$, $|L^{d_E}|$ is always expressible in terms of the remaining Casimirs.  Since the epsilon tensor is an invariant tensor of the conformal algebra, it is then a trivial matter to show that $\epsilon^{A_1\cdots A_{d_E}}L_{A_1A_2}\cdots L_{A_{d_E-1}A_{d_E}}$ is the last Casimir needed.  For future convenience, conformal Casimirs will be denoted by $C_{2n}=|L^{2n}|$ and $C_\epsilon=\epsilon^{A_1\cdots A_{d_E}}L_{A_1A_2}\cdots L_{A_{d_E-1}A_{d_E}}$.  Note that going from odd $d_E$ to even $d_E+1$ the missing Casimir is simply $C_\epsilon$ which thus distinguishes between the two irreducible (Weyl) spinor representations of even $d_E+1$.

%%%%%%%%%%%%%%%%%%%%%%%%%%%%%%%%%%%%%%%%%%%%%%%%%%

\subsection{Spinors in Embedding Space}\label{SSecSES}

There are several ways to embed the fundamental spinor representations in embedding space.  The following way will be particularly convenient to investigate the uplift of generic operators to embedding space, particularly with respect to Lorentz covariance and the Majorana condition.  As in position space, it is natural to treat odd and even dimensions separately.

Note that in embedding space, where the metric is $g_{AB}=\text{diag}(+1,-1,\ldots,-1;-1,+1)$, it is for all dimensions $d_E=3,4,5\text{ mod }8$ that the independent Majorana condition, which halves the number of real degrees of freedom of the fundamental spinor, can be imposed.  Since $d_E=d+2$, it implies that the independent Majorana condition can be imposed in embedding space if and only if it can be imposed in position space.  Therefore the lift of quasi-primary operators from position space to embedding space is consistent.  The matrices introduced here will straightforwardly have a purely real or purely imaginary representation if the $\gamma$-matrices in position space do.

%%%%%%%%%%%%%%%%%%%%%%%%%%%%%%%%%%%%%%%%%%%%%%%%%%

\subsubsection{Odd Dimensions}\label{SSecSESodd}

Define the $2^{(d_E-1)/2}$-dimensional square matrices $\Gamma^A$ in odd dimensions as
\eqn{\Gamma^\mu=\left(\begin{array}{cc}\gamma^\mu&0\\0&-\gamma^\mu\end{array}\right),\quad\quad\Gamma^{d+1}=\alpha\left(\begin{array}{cc}0&\1\\-\alpha^2\1&0\end{array}\right),\quad\quad\Gamma^{d+2}=\alpha\left(\begin{array}{cc}0&\1\\\alpha^2\1&0\end{array}\right).}
They verify the appropriate algebra
\eqn{\Gamma^A\Gamma^B+\Gamma^B\Gamma^A=2g^{AB}\1,}
for $\alpha=\pm1$ ($\alpha=\pm i$) and are purely real (imaginary) if the $\gamma$-matrices are purely real (imaginary).  By defining the $SO(d_E)$ generators in the fundamental spinor representation as
\eqn{\Sigma_{AB}=\frac{i}{4}(\Gamma_A\Gamma_B-\Gamma_B\Gamma_A),}
one shows that
\eqn{
\begin{gathered}
\Sigma_{AB}\Gamma_C-\Gamma_C\Sigma_{AB}=-(S_{AB})_{C}^{\phantom{C}C'}\Gamma_{C'},\\
[\Sigma_{AB},\Sigma_{CD}]=-(S_{AB})_{C}^{\phantom{C}C'}\Sigma_{C'D}-(S_{AB})_{D}^{\phantom{D}D'}\Sigma_{CD'}.
\end{gathered}
}
Thus, the $SO(d_E)$ generators satisfy the appropriate algebra.  The $SO(d_E)$ generators written in terms of the position space matrices are given by
\eqn{
\begin{gathered}
\Sigma_{\mu\nu}=\left(\begin{array}{cc}\sigma_{\mu\nu}&0\\0&\sigma_{\mu\nu}\end{array}\right),\quad\quad\Sigma_{\mu,d+1}=-\frac{\alpha i}{2}\left(\begin{array}{cc}0&\gamma_\mu\\\alpha^2\gamma_\mu&0\end{array}\right),\\
\Sigma_{\mu,d+2}=-\frac{\alpha i}{2}\left(\begin{array}{cc}0&-\gamma_\mu\\\alpha^2\gamma_\mu&0\end{array}\right),\quad\quad\Sigma_{d+1,d+2}=-\frac{i}{2}\left(\begin{array}{cc}\1&0\\0&-\1\end{array}\right).
\end{gathered}
}
Moreover, as the matrices $\Gamma$ in embedding space can be written in terms of the matrices $\gamma$ in position space as shown above, the same can be done for the matrices $A_\Gamma$, $C_\Gamma$, $B_\Gamma$ in embedding space.  These matrices relate the adjoint, contragredient, and conjugate representations, respectively, to the original representation (see Section \ref{SSSecOddSpinor}).  They are simply written in terms of the corresponding ones in position space (with $p=1$ and $q=d-1$, see Section \ref{SSSecOddSpinor}), and are given by
\begingroup\makeatletter\def\f@size{9}\check@mathfonts\def\maketag@@@#1{\hbox{\m@th\large\normalfont#1}}%
\eqn{
\begin{gathered}
A_\Gamma=\alpha\left(\begin{array}{cc}0&A\\(-1)^d\alpha^2A&0\end{array}\right),\quad\quad C_\Gamma=\left(\begin{array}{cc}0&C\\(-1)^{r+1}C&0\end{array}\right),\quad\quad B_\Gamma=\alpha\left(\begin{array}{cc}(-1)^d\alpha^2B&0\\0&(-1)^{r+1}B\end{array}\right),
\end{gathered}
}
\endgroup
with $p_E=p+1=2$, $q_E=q+1=d$ and $r_E=r+1$.  They satisfy the appropriate relations \eqref{EqSymgammaOdd} for $r_E$ and $q_E$ since $A$, $C$, and $B$ satisfy the appropriate relations \eqref{EqSymgammaOdd} for $r$ and $q$.

\subsubsection{Even Dimensions}\label{SSecSESeven}

In even dimensions the two irreducible representations in embedding space are obtained with the help of the $2^{(d_E-2)/2}$-dimensional square matrices $\Gamma^A$ and $\tilde{\Gamma}^A$, defined as
\eqn{
\begin{gathered}
\Gamma^\mu=\left(\begin{array}{cc}\gamma^\mu&0\\0&-\tilde{\gamma}^\mu\end{array}\right),\quad\quad\Gamma^{d+1}=\alpha\left(\begin{array}{cc}0&\1\\-\alpha^2\1&0\end{array}\right),\quad\quad\Gamma^{d+2}=\alpha\left(\begin{array}{cc}0&\1\\\alpha^2\1&0\end{array}\right),\\
\tilde{\Gamma}^\mu=\left(\begin{array}{cc}\tilde{\gamma}^\mu&0\\0&-\gamma^\mu\end{array}\right),\quad\quad\tilde{\Gamma}^{d+1}=\alpha\left(\begin{array}{cc}0&\1\\-\alpha^2\1&0\end{array}\right),\quad\quad\tilde{\Gamma}^{d+2}=\alpha\left(\begin{array}{cc}0&\1\\\alpha^2\1&0\end{array}\right).
\end{gathered}
}
For $\alpha=\pm1$ ($\alpha=\pm i$), they are purely real (imaginary) if the position space matrices are purely real (imaginary) and they verify the appropriate algebra
\eqna{
\Gamma^A\tilde{\Gamma}^B+\Gamma^B\tilde{\Gamma}^A&=2g^{AB}\1,\\
\tilde{\Gamma}^A\Gamma^B+\tilde{\Gamma}^B\Gamma^A&=2g^{AB}\1.
}
By defining the $SO(d_E)$ generators in the two fundamental spinor representations as
\eqna{
\Sigma_{AB}=\frac{i}{4}(\Gamma_A\tilde{\Gamma}_B-\Gamma_B\tilde{\Gamma}_A),\\
\tilde{\Sigma}_{AB}=\frac{i}{4}(\tilde{\Gamma}_A\Gamma_B-\tilde{\Gamma}_B\Gamma_A),
}
it is a simple matter to verify that
\eqn{
\begin{gathered}
\Sigma_{AB}\Gamma_C-\Gamma_C\tilde{\Sigma}_{AB}=-(S_{AB})_{C}^{\phantom{C}C'}\Gamma_{C'},\\
\tilde{\Sigma}_{AB}\tilde{\Gamma}_C-\tilde{\Gamma}_C\Sigma_{AB}=-(S_{AB})_{C}^{\phantom{C}C'}\tilde{\Gamma}_{C'},\\
[\Sigma_{AB},\Sigma_{CD}]=-(S_{AB})_{C}^{\phantom{C}C'}\Sigma_{C'D}-(S_{AB})_{D}^{\phantom{D}D'}\Sigma_{CD'},\\
[\tilde{\Sigma}_{AB},\tilde{\Sigma}_{CD}]=-(S_{AB})_{C}^{\phantom{C}C'}\tilde{\Sigma}_{C'D}-(S_{AB})_{D}^{\phantom{D}D'}\tilde{\Sigma}_{CD'}.
\end{gathered}
}
As expected, the two $SO(d_E)$ generators satisfy the $SO(d_E)$ algebra.  In terms of the position space matrices the generators are
\eqn{
\begin{gathered}
\Sigma_{\mu\nu}=\left(\begin{array}{cc}\sigma_{\mu\nu}&0\\0&\tilde{\sigma}_{\mu\nu}\end{array}\right),\quad\quad\Sigma_{\mu,d+1}=-\frac{\alpha i}{2}\left(\begin{array}{cc}0&\gamma_\mu\\\alpha^2\tilde{\gamma}_\mu&0\end{array}\right),\\
\Sigma_{\mu,d+2}=-\frac{\alpha i}{2}\left(\begin{array}{cc}0&-\gamma_\mu\\\alpha^2\tilde{\gamma}_\mu&0\end{array}\right),\quad\quad\Sigma_{d+1,d+2}=-\frac{i}{2}\left(\begin{array}{cc}\1&0\\0&-\1\end{array}\right),\\
\tilde{\Sigma}_{\mu\nu}=\left(\begin{array}{cc}\tilde{\sigma}_{\mu\nu}&0\\0&\sigma_{\mu\nu}\end{array}\right),\quad\quad\tilde{\Sigma}_{\mu,d+1}=-\frac{\alpha i}{2}\left(\begin{array}{cc}0&\tilde{\gamma}_\mu\\\alpha^2\gamma_\mu&0\end{array}\right),\\
\tilde{\Sigma}_{\mu,d+2}=-\frac{\alpha i}{2}\left(\begin{array}{cc}0&-\tilde{\gamma}_\mu\\\alpha^2\gamma_\mu&0\end{array}\right),\quad\quad\tilde{\Sigma}_{d+1,d+2}=-\frac{i}{2}\left(\begin{array}{cc}\1&0\\0&-\1\end{array}\right).
\end{gathered}
}
Finally, just as the matrices $\Gamma$ and $\tilde{\Gamma}$ in embedding space can be written in terms of the matrices $\gamma$ and $\tilde{\gamma}$ in position space, so can the matrices $A_\Gamma$, $C_\Gamma$, $B_\Gamma$ and $\tilde{A}_\Gamma$, $\tilde{C}_\Gamma$, $\tilde{B}_\Gamma$ in embedding space  (see Section \ref{SSSecEvenSpinor}) be written in terms of the equivalent matrices in position space (with $p=1$ and $q=d-1$, see Section \ref{SSSecEvenSpinor}) as
\begingroup\makeatletter\def\f@size{9}\check@mathfonts\def\maketag@@@#1{\hbox{\m@th\large\normalfont#1}}%
\eqn{
\begin{gathered}
A_\Gamma=\alpha\left(\begin{array}{cc}0&\tilde{A}\\(-1)^d\alpha^2A&0\end{array}\right),\quad\quad C_\Gamma=\left(\begin{array}{cc}0&\tilde{C}\\(-1)^{r+1}C&0\end{array}\right),\quad\quad B_\Gamma=\alpha\left(\begin{array}{cc}(-1)^d\alpha^2B&0\\0&(-1)^{r+1}\tilde{B}\end{array}\right),\\
\tilde{A}_\Gamma=\alpha\left(\begin{array}{cc}0&A\\(-1)^d\alpha^2\tilde{A}&0\end{array}\right),\quad\quad\tilde{C}_\Gamma=\left(\begin{array}{cc}0&C\\(-1)^{r+1}\tilde{C}&0\end{array}\right),\quad\quad\tilde{B}_\Gamma=\alpha\left(\begin{array}{cc}(-1)^d\alpha^2\tilde{B}&0\\0&(-1)^{r+1}B\end{array}\right),\nonumber
\end{gathered}
}
\endgroup
with $p_E=p+1=2$, $q_E=q+1=d$ and $r_E=r+1$.  They satisfy the appropriate relations \eqref{EqSymgammaEven} for $r_E$ and $q_E$ since $A$, $C$, $B$ and $\tilde{A}$, $\tilde{C}$, $\tilde{B}$ satisfy the appropriate relations \eqref{EqSymgammaEven} for $r$ and $q$.  Again, these relate the adjoint, contragredient, and conjugate representations respectively to the two original representations.

%%%%%%%%%%%%%%%%%%%%%%%%%%%%%%%%%%%%%%%%%%%%%%%%%%

\subsection{Quasi-Primary Operators in Embedding Space}\label{SSecOES}

It is now a trivial matter to lift quasi-primary operators from position space to embedding space.  Using quasi-primary operators with spinor indices only, the action of the conformal algebra on a quasi-primary operator $\mathcal{O}(\eta)$ in embedding space is simply
\eqn{[L_{AB},\mathcal{O}(\eta)]=-i\left(\eta_A\frac{\partial}{\partial\eta^B}-\eta_B\frac{\partial}{\partial\eta^A}\right)\mathcal{O}(\eta)-(\Sigma_{AB}\mathcal{O})(\eta),}[EqAlgOES]
where the generators $\Sigma_{AB}$ are appropriate sums of the generators in the fundamental spinor representation, as in position space.  The action of the conformal algebra on quasi-primary operators in embedding space \eqref{EqAlgOES} is manifestly covariant, contrary to the action of the conformal algebra on quasi-primary operators in position space \eqref{EqAlgOPS}.  Although both quasi-primary operators in position space and in embedding space are denoted by $\mathcal{O}$, the superscript $(x)$ used for quasi-primary operators in position space $\mathcal{O}^{(x)}$ is removed for quasi-primary operators in embedding space $\mathcal{O}$.

The connection between quasi-primary operators in embedding space and quasi-primary operators in position space is obtained by imposing the following supplementary conditions on the quasi-primary operators in embedding space (homogeneity and transversality respectively) \cite{Mack:1969rr,Ferrara:1971vh,Ferrara:1971zy,Ferrara:1972cq,Ferrara:1973eg,Ferrara:1973yt}
\eqn{\eta^A\frac{\partial}{\partial\eta^A}\mathcal{O}(\eta)=-\tau_\mathcal{O}\mathcal{O}(\eta),\quad\quad\eta_A\Gamma^A\mathcal{O}(\eta)=\eta_A\tilde{\Gamma}^A\mathcal{O}(\eta)=0,}[EqSupp]
where $\tau_\mathcal{O}=\Delta_\mathcal{O}-S_\mathcal{O}$ is the twist of the quasi-primary operator.  Here $\Delta_\mathcal{O}$ and $S_\mathcal{O}$ are the conformal dimension and spin of the quasi-primary operator in position space, respectively.  The homogeneity condition is consistent since it commutes with the conformal algebra in embedding space and the hypercone $\eta^2=0$ is preserved when $\eta^A\to\lambda\eta^A$ with $\lambda>0$.  Therefore, quasi-primary operators in embedding space may be required to satisfy the homogeneity condition.  Moreover, note that the homogeneity condition forces $(-\eta^{d+1}+\eta^{d+2})^{\tau_\mathcal{O}}\mathcal{O}(\eta)$ to be a function of the ratios of the $\eta^A$'s, \textit{i.e.} functions of $x^\mu$'s only.  The transversality condition, which acts on each spinor index independently, reduces the number of degrees of freedom to match the appropriate one in position space, \textit{e.g.}
\eqn{\mathcal{O}(\eta)=\left(\begin{array}{c}\mathcal{O}_+(\eta)\\\mathcal{O}_-(\eta)\end{array}\right)=\left(\begin{array}{c}\mathcal{O}_+(\eta)\\-\alpha^{-1}x^\mu\tilde{\gamma}_\mu\mathcal{O}_+(\eta)\end{array}\right),}
and equivalently for all the other spinor indices of the quasi-primary operator in embedding space.

Therefore, on the light-cone $\eta^2=0$, the quasi-primary operator in position space
\eqn{\mathcal{O}(x)\equiv(-\eta^{d+1}+\eta^{d+2})^{\tau_\mathcal{O}}\mathcal{O}(\eta),}[EqQP]
restricted to the first half of the possible values of the spinor indices, is a function of the coordinates $x^\mu$ only and has the appropriate number of degrees of freedom to describe a quasi-primary operator in position space, \textit{i.e.} $\mathcal{O}^{(x)}(x)=\mathcal{O}_+(x)=(-\eta^{d+1}+\eta^{d+2})^{\tau_\mathcal{O}}\mathcal{O}_+(\eta)$.  Most importantly though, the action of the conformal algebra corresponds to \eqref{EqAlgOPS} since
\eqna{
[L_{AB},\mathcal{O}(x)]&=[L_{AB},(-\eta^{d+1}+\eta^{d+2})^{\tau_\mathcal{O}}\mathcal{O}(\eta)]\\
&=(-\eta^{d+1}+\eta^{d+2})^{\tau_\mathcal{O}}[L_{AB},\mathcal{O}(\eta)]\\
&=-i(-\eta^{d+1}+\eta^{d+2})^{\tau_\mathcal{O}}\left(\eta_A\frac{\partial}{\partial\eta^B}-\eta_B\frac{\partial}{\partial\eta^A}\right)\mathcal{O}(\eta)-(\Sigma_{AB}\mathcal{O})(x)\\
&=-i(-\eta^{d+1}+\eta^{d+2})^{\tau_\mathcal{O}}\left(\eta_A\frac{\partial}{\partial\eta^B}-\eta_B\frac{\partial}{\partial\eta^A}\right)(-\eta^{d+1}+\eta^{d+2})^{-\tau_\mathcal{O}}\mathcal{O}(x)-(\Sigma_{AB}\mathcal{O})(x)\\
&=-i\left(\eta_A\frac{\partial x^\mu}{\partial\eta^B}-\eta_B\frac{\partial x^\mu}{\partial\eta^A}\right)\frac{\partial}{\partial x^\mu}\mathcal{O}(x)-(\Sigma_{AB}\mathcal{O})(x)\\
&\phantom{=}\qquad+i\tau_\mathcal{O}\frac{1}{-\eta^{d+1}+\eta^{d+2}}[\eta_A(-g_B^{\phantom{B}d+1}+g_B^{\phantom{B}d+2})-\eta_B(-g_A^{\phantom{A}d+1}+g_A^{\phantom{A}d+2})]\mathcal{O}(x),
}
and $(\1\mathcal{O})(x)=2S_\mathcal{O}\mathcal{O}(x)$ due to the implicit sum on the spinor indices.  Thus \eqref{EqQP} restricted to the first half of the possible values of the spinor indices defines a quasi-primary operator in position space as long as the restriction is implemented \textit{last}.  Denoting spinor indices in embedding space by lower case roman letters ($a,b,\ldots$) and spinor indices in position space by lower case greek letters ($\alpha,\beta,\ldots$), and using the following double index notation $a_i=\spin{\alpha_i^+}{\tilde{\alpha}_i^-}$ to map spinor indices in embedding space to spinor indices in position space, this restriction can be rewritten as
\eqn{\mathcal{O}_{a_1\cdots a_{n_\mathcal{O}}|+}(\eta)=\mathcal{O}_{\spin{\alpha_1^+}{-}\cdots\spin{\alpha_{n_\mathcal{O}}^+}{-}}(\eta),}
where $-$ indicates the corresponding indices are no longer considered.  In this notation, the transversality condition leads to, \textit{e.g.}
\eqn{\mathcal{O}_{a_1\cdots\spin{-}{\tilde{\alpha}_i^-}\cdots a_{n_\mathcal{O}}}(\eta)=-\alpha^{-1}x^\mu(\tilde{\gamma}_\mu)_{\tilde{\alpha}_i^-}^{\phantom{\tilde{\alpha}_i^-}\alpha_i^+}\mathcal{O}_{a_1\cdots\spin{\alpha_i^+}{-}\cdots a_{n_\mathcal{O}}}(\eta).}

To complete the lift of quasi-primary operators from position space to embedding space it is necessary to determine how the irreducible representations of the Lorentz group are encoded in embedding space.  For scalar and spinor representations the lift is trivial.  For $i$-index antisymmetric vector representations the lift is accomplished by noting that $\mathcal{O}_{\alpha\beta}^{(x)}(x)=(\mathcal{P}_{\boldsymbol{e}_i}^{(-,-)})_{\alpha\beta}^{\phantom{\alpha\beta}\beta'\alpha'}\mathcal{O}_{\alpha'\beta'}^{(x)}(x)$ (see Appendix \ref{SecIrrep}) and thus
\begingroup\makeatletter\def\f@size{10}\check@mathfonts\def\maketag@@@#1{\hbox{\m@th\large\normalfont#1}}%
\eqna{
\mathcal{O}_{ab}(x)&=\left(\begin{array}{cc}\mathcal{O}_{\spin{\alpha^+}{-}\spin{\beta^+}{-}}(x)&\mathcal{O}_{\spin{\alpha^+}{-}\spin{-}{\tilde{\beta}^-}}(x)\\\mathcal{O}_{\spin{-}{\tilde{\alpha}^-}\spin{\beta^+}{-}}(x)&\mathcal{O}_{\spin{-}{\tilde{\alpha}^-}\spin{-}{\tilde{\beta}^-}}(x)\end{array}\right)=\left(\begin{array}{cc}\mathcal{O}_{\alpha\beta}^{(x)}(x)&-\alpha^{-1}x^\mu(\tilde{\gamma}_\mu)_{\tilde{\beta}}^{\phantom{\tilde{\beta}}\beta}\mathcal{O}_{\alpha\beta}^{(x)}(x)\\-\alpha^{-1}x^\mu(\tilde{\gamma}_\mu)_{\tilde{\alpha}}^{\phantom{\tilde{\alpha}}\alpha}\mathcal{O}_{\alpha\beta}^{(x)}(x)&\alpha^2x^\mu x^\nu(\tilde{\gamma}_\mu)_{\tilde{\alpha}}^{\phantom{\tilde{\alpha}}\alpha}(\tilde{\gamma}_\nu)_{\tilde{\beta}}^{\phantom{\tilde{\beta}}\beta}\mathcal{O}_{\alpha\beta}^{(x)}(x)\end{array}\right)\\
&=\left(\begin{array}{cc}(\mathcal{P}_{\boldsymbol{e}_i}^{(-,-)})_{\alpha\beta}^{\phantom{\alpha\beta}\beta'\alpha'}\mathcal{O}_{\alpha'\beta'}^{(x)}(x)&-\alpha^{-1}x^\mu(\tilde{\gamma}_\mu)_{\tilde{\beta}}^{\phantom{\tilde{\beta}}\beta}(\mathcal{P}_{\boldsymbol{e}_i}^{(-,-)})_{\alpha\beta}^{\phantom{\alpha\beta}\beta'\alpha'}\mathcal{O}_{\alpha'\beta'}^{(x)}(x)\\-\alpha^{-1}x^\mu(\tilde{\gamma}_\mu)_{\tilde{\alpha}}^{\phantom{\tilde{\alpha}}\alpha}(\mathcal{P}_{\boldsymbol{e}_i}^{(-,-)})_{\alpha\beta}^{\phantom{\alpha\beta}\beta'\alpha'}\mathcal{O}_{\alpha'\beta'}^{(x)}(x)&\alpha^2x^\mu x^\nu(\tilde{\gamma}_\mu)_{\tilde{\alpha}}^{\phantom{\tilde{\alpha}}\alpha}(\tilde{\gamma}_\nu)_{\tilde{\beta}}^{\phantom{\tilde{\beta}}\beta}(\mathcal{P}_{\boldsymbol{e}_i}^{(-,-)})_{\alpha\beta}^{\phantom{\alpha\beta}\beta'\alpha'}\mathcal{O}_{\alpha'\beta'}^{(x)}(x)\end{array}\right)\\
&=\left(\begin{array}{cc}(\mathcal{T}^{\boldsymbol{e}_i})_{\alpha\beta}^{\mu_i\cdots\mu_1}&-\alpha^{-1}x^\mu(\tilde{\gamma}_\mu)_{\tilde{\beta}}^{\phantom{\tilde{\beta}}\beta}(\mathcal{T}^{\boldsymbol{e}_i})_{\alpha\beta}^{\mu_i\cdots\mu_1}\\-\alpha^{-1}x^\mu(\tilde{\gamma}_\mu)_{\tilde{\alpha}}^{\phantom{\tilde{\alpha}}\alpha}(\mathcal{T}^{\boldsymbol{e}_i})_{\alpha\beta}^{\mu_i\cdots\mu_1}&\alpha^2x^\mu x^\nu(\tilde{\gamma}_\mu)_{\tilde{\alpha}}^{\phantom{\tilde{\alpha}}\alpha}(\tilde{\gamma}_\nu)_{\tilde{\beta}}^{\phantom{\tilde{\beta}}\beta}(\mathcal{T}^{\boldsymbol{e}_i})_{\alpha\beta}^{\mu_i\cdots\mu_1}\end{array}\right)\mathcal{O}_{\mu_1\cdots\mu_i}^{(x)}(x)\\
&=(\mathcal{P}_{\boldsymbol{e}_{i+1}}^{(-,-)})_{ab}^{\phantom{ab}b'a'}\mathcal{O}_{a'b'}(x),
}
\endgroup
where for simplicity the indices are chosen for even dimensions with $r$ and $i$ even.  Thus $i$-index antisymmetric vector representations are lifted to $(i+1)$-index antisymmetric vector representations in embedding space.  In other words, since the previous proof also holds for (anti-)self-dual vector representations, quasi-primary operators in defining representations of the Lorentz group uplift to quasi-primary operators in defining representations of the conformal group.

Since more general irreducible representations in position space are simple symmetrizations of the defining representations, this seems to show that a quasi-primary operator $\mathcal{O}^{(x)}$ in position space in a general irreducible $SO(1,d-1)$ representation given by $\boldsymbol{N}^\mathcal{O}=\{N_1^\mathcal{O},\ldots,N_r^\mathcal{O}\}$ is lifted to a quasi-primary operator $\mathcal{O}$ in embedding space in an irreducible representation $\boldsymbol{N}_E^\mathcal{O}=\{0,N_1^\mathcal{O},\ldots,N_r^\mathcal{O}\}$ of $SO(2,d)$.  If it were not for the tracelessness condition of irreducible representations in position space, the previous statement would be correct.  Indeed, since the tracelessness condition of irreducible representations in embedding space is different than the tracelessness condition of irreducible representations in position space due to the trace of the $SO(2,d)$ metric being larger than the trace of the $SO(1,d-1)$ metric, the previous statement cannot be right and must be modified accordingly.  The necessary modification will be made precise in the OPE with the introduction of a new metric, better suited to implement the position space tracelessness condition directly in embedding space.

%%%%%%%%%%%%%%%%%%%%%%%%%%%%%%%%%%%%%%%%%%%%%%%%%%
%%%%%%%%%%%%%%%%%%%%%%%%%%%%%%%%%%%%%%%%%%%%%%%%%%

\section{Differential Operators in Embedding Space}\label{SecDO}

The appropriate metric necessary to uplift quasi-primary operators in general irreducible $SO(1,d-1)$ representations appears naturally from the study of the different differential operators in embedding space.  Since the OPE \eqref{EqSOPE} represents the product of two quasi-primary operators in terms of a series of quasi-primary operators, differential operators are needed to generate conformal descendants.  Thus the OPE must contain well-defined differential operators $\DOPE{a}{i}{j}{k}(\eta_1,\eta_2)$ which act on the quasi-primary operators.

%%%%%%%%%%%%%%%%%%%%%%%%%%%%%%%%%%%%%%%%%%%%%%%%%%

\subsection{A Differential Operator}

In \cite{Fortin:2016dlj}, it was shown that the differential operator $\DOPE{a}{i}{j}{k}(\eta_1,\eta_2)$ appearing in the OPE \eqref{EqSOPE} is built solely with the help of a specific differential operator and its square.  For convenience with correlation function computations from the OPE, another differential operator will be introduced in this section.  For completeness, this section first reviews some of the results of \cite{Fortin:2016dlj} (see also \cite{Bailey:1994,Eastwood:2002su} for related results).

The only consistent differential operators with one derivative which are well defined on the light-cone are
\eqn{\Theta=\eta^A\frac{\partial}{\partial\eta^A},\qquad\mathcal{L}_{AB}=i\left(\eta_A\frac{\partial}{\partial\eta^B}-\eta_B\frac{\partial}{\partial\eta^A}\right),}[EqDiff1]
where $\Theta$ is the homogeneity operator defined in \eqref{EqSupp}, while $\mathcal{L}_{AB}$ is the conformal generator in embedding space and thus satisfies the conformal algebra \eqref{EqAlgES}.  With two derivatives, there is only one consistent differential operator which is not built from \eqref{EqDiff1}.  It is the well-known Thomas-Todorov operator \cite{Dobrev:1975ru,Bailey:1994,Eastwood:2002su} given by
\eqn{\mathcal{K}_A=\left(\eta^B\frac{\partial}{\partial\eta^B}+\frac{d}{2}\right)\frac{\partial}{\partial\eta^A}-\frac{1}{2}\eta_A\frac{\partial}{\partial\eta_B}\frac{\partial}{\partial\eta^B}.}[EqDiff2]
The non-trivial commutation relations satisfied by the homogeneity operator, the conformal generators, and the Thomas-Todorov operator \eqref{EqDiff1} and \eqref{EqDiff2} form a closed system and are
\eqn{
\begin{gathered}
{}[\mathcal{L}_{AB},\mathcal{L}_{CD}]=-i(g_{AC}\mathcal{L}_{BD}-g_{BC}\mathcal{L}_{AD}+g_{AD}\mathcal{L}_{CB}-g_{BD}\mathcal{L}_{CA}),\\
[\Theta,\mathcal{K}_A]=-\mathcal{K}_A,\qquad[\mathcal{L}_{AB},\mathcal{K}_C]=i(g_{BC}\mathcal{K}_A-g_{AC}\mathcal{K}_B).
\end{gathered}
}

From the embedding space coordinates $\eta^A=(\eta^\mu,\eta^{d+1},\eta^{d+2})$ and the new embedding space coordinates $\tilde{\eta}^A=(x^\mu,k,\eta^2)$ where
\eqn{\eta^\mu=kx^\mu,\quad\quad\eta^{d+1}=\frac{\eta^2-k^2(1+x^2)}{2k},\quad\quad\eta^{d+2}=\frac{\eta^2+k^2(1-x^2)}{2k},}
or equivalently
\eqn{x^\mu=\frac{\eta^\mu}{-\eta^{d+1}+\eta^{d+2}},\quad\quad k=-\eta^{d+1}+\eta^{d+2},\quad\quad\eta^2=\eta^\mu\eta_\mu-(\eta^{d+1})^2+(\eta^{d+2})^2,}
it is easy to verify that
\eqna{
\Theta&=k\frac{\partial}{\partial k}+2\eta^2\frac{\partial}{\partial\eta^2},\\
\mathcal{L}_{AB}&=\frac{i}{k}\{\eta_A[g_B^{\phantom{B}\mu}-(-g_B^{\phantom{B}d+1}+g_B^{\phantom{B}d+2})x^\mu]-\eta_B[g_A^{\phantom{A}\mu}-(-g_A^{\phantom{A}d+1}+g_A^{\phantom{A}d+2})x^\mu]\}\frac{\partial}{\partial x^\mu}\\
&\phantom{=}\hspace{20pt}+i[\eta_A(-g_B^{\phantom{B}d+1}+g_B^{\phantom{B}d+2})-\eta_B(-g_A^{\phantom{A}d+1}+g_A^{\phantom{A}d+2})]\frac{\partial}{\partial k},\\
\mathcal{K}_A&=-\frac{\eta_A}{2k^2}\frac{\partial}{\partial x_\mu}\frac{\partial}{\partial x^\mu}+\frac{1}{k}[g_A^{\phantom{A}\mu}-(-g_A^{\phantom{A}d+1}+g_A^{\phantom{A}d+2})x^\mu]\left(k\frac{\partial}{\partial k}+2\eta^2\frac{\partial}{\partial\eta^2}+\frac{d}{2}\right)\frac{\partial}{\partial x^\mu}\\
&\phantom{=}\hspace{20pt}+(-g_A^{\phantom{A}d+1}+g_A^{\phantom{A}d+2})\left(k\frac{\partial}{\partial k}+2\eta^2\frac{\partial}{\partial\eta^2}+\frac{d+2}{2}\right)\frac{\partial}{\partial k}+2\eta_A\eta^2\frac{\partial}{\partial\eta^2}\frac{\partial}{\partial\eta^2},
}
Hence $\Theta$, $\mathcal{L}_{AB}$ and $\mathcal{K}_A$ do not have derivatives with respect to $\eta^2$ on the light-cone and are thus well defined on the light-cone.\footnote{A differential operator is well defined on the light-cone if its effect on a smooth function $f(\eta)=\eta^2g(\eta)$ which vanishes on the light-cone is consistent on the light-cone $\eta^2=0$.  For example, $\Theta f(\eta)=2\eta^2g(\eta)+\eta^2\Theta g(\eta)=0$ while $\frac{\partial}{\partial\eta^A}f(\eta)=2\eta_Ag(\eta)+\eta^2\frac{\partial}{\partial\eta^A}g(\eta)\neq0$ on the light-cone $\eta^2=0$.  Thus $\Theta$ (and $\mathcal{L}_{AB}$ and $\mathcal{K}_A$) is a well-defined differential operator in embedding space while $\frac{\partial}{\partial\eta^A}$ is not.}

From the homogeneity condition \eqref{EqSupp}, the differential operator $\Theta$ does not generate conformal descendants and as such it is not a genuine differential operator.  It is however quite useful to simplify the differential operator $\DOPE{a}{i}{j}{k}(\eta_1,\eta_2)$ that will ultimately appear in the OPE.

In order to narrow the possible differential operators to a single candidate, it is convenient now to discuss some properties of the embedding space.  First, since Casimirs are not genuine differential operators, it is straightforward to conclude from \eqref{EqAlgDO} and the results above that all possible non-trivial differential operators can be written as linear combinations of $(\mathcal{L}^n)_{AB}$ for all $n>0$ and $\mathcal{K}_A$.\footnote{Self-contractions of the Thomas-Todorov operator can be discarded since $4\mathcal{K}^A\mathcal{K}_A=\eta^2\partial^2\partial^2$ which vanishes on the light-cone.}  By defining a differential operator $\mathscr{D}_{n,AB}=\eta_A\eta_B\partial^2S_n+\eta_A\partial_BT_n+\eta_B\partial_AU_n+g_{AB}V_n$ where the derivatives act on the right and all $S_n$, $T_n$, $U_n$ and $V_n$ can depend on the homogeneity operator $\Theta$, one has by recurrence
\eqna{
\mathscr{D}_{n+1,AB}&\equiv(\mathcal{L}\mathscr{D}_n)_{AB}=\mathcal{L}_{AC}\mathscr{D}_{n\phantom{C}B}^{\phantom{n}C}\\
&=\eta_A\eta_B\partial^2[i(d_E-2+\Theta)S_n+iU_n]+\eta_A\partial_B[i(d_E-2+\Theta)T_n+iU_n+iV_n]\\
&\phantom{=}\qquad+\eta_B\partial_A[i(1-\Theta)U_n-iV_n]+g_{AB}(-i\Theta U_n),
}
which translates to
\eqna{
S_{n+1}&=i(d_E-2+\Theta)S_n+iU_n,\\
T_{n+1}&=i(d_E-2+\Theta)T_n+iU_n+iV_n,\\
U_{n+1}&=i(1-\Theta)U_n-iV_n,\\
V_{n+1}&=-i\Theta U_n.
}
For $(\mathcal{L}^n)_{AB}=(\mathcal{L}^2)_{AB}A_n+\mathcal{L}_{AB}B_n+g_{AB}C_n$, where $A_n$, $B_n$ and $C_n$ are functions of the homogeneity operator $\Theta$, the initial values are $S_1=0,T_1=i,U_1=-i,V_1=0$ and the recurrence relations are solved for
\eqna{
S_n&=\frac{-i^n[(1+\Theta)(d_E-2+\Theta)^n+(d_E-3+\Theta)(-\Theta)^n-(d_E-2+2\Theta)]}{(1+\Theta)(d_E-3+\Theta)(d_E-2+2\Theta)},\\
T_n&=\frac{i^n[(1+\Theta)(d_E-4+2\Theta)(d_E-2+\Theta)^n-2(d_E-3+\Theta)(-\Theta)^n+(1-\Theta)(d_E-2+2\Theta)]}{(1+\Theta)(d_E-3+\Theta)(d_E-2+2\Theta)},\\
U_n&=\frac{-i^n[1-(-\Theta)^n]}{1+\Theta},\\
V_n&=\frac{i^n\Theta[1-(-\Theta)^{n-1}]}{1+\Theta}.
}
Therefore the differential operator $(\mathcal{L}^n)_{AB}$ is given by
\eqna{
(\mathcal{L}^n)_{AB}&=\eta_A\eta_B\partial^2S_n+\eta_A\partial_BT_n+\eta_B\partial_AU_n+g_{AB}V_n\\
&=\left[\eta_A\eta_B\partial^2+\eta_A\partial_B\frac{T_n+U_n-(1-\Theta)S_n}{S_n}+\eta_B\partial_A(1-\Theta)+g_{AB}(-\Theta)\right]S_n\\
&\phantom{=}\qquad-(\eta_A\partial_B-\eta_B\partial_A)[U_n-(1-\Theta)S_n]+g_{AB}(V_n+\Theta S_n)\\
&=[\eta_A\eta_B\partial^2-\eta_A\partial_B(d_E-3+\Theta)+\eta_B\partial_A(1-\Theta)+g_{AB}(-\Theta)]S_n\\
&\phantom{=}\qquad+i(\eta_A\partial_B-\eta_B\partial_A)[iU_n-i(1-\Theta)S_n]+g_{AB}(V_n+\Theta S_n)\\
&=(\mathcal{L}^2)_{AB}A_n+\mathcal{L}_{AB}B_n+g_{AB}C_n,
}
where $A_n=S_n$, $B_n=iU_n-i(1-\Theta)S_n$ and $C_n=V_n+\Theta S_n$.  Because the homogeneity operator commutes with the conformal generator, one concludes that, apart from $\mathcal{K}_A$, all non-trivial differential operators appearing in the OPE are functions of uncontracted $\mathcal{L}_{AB}$ and $(\mathcal{L}^2)_{AB}$ only, higher powers of $\mathcal{L}_{AB}$ are redundant.  Moreover, since
\eqna{
\mathcal{L}_{AB}\mathcal{K}^B&=-i\Theta\mathcal{K}_A+\frac{i}{2}\eta^2\partial_A\partial^2,\\
(\mathcal{L}^2)_{AB}\mathcal{K}^B&=-\Theta^2\mathcal{K}_A+\eta^2\left(\mathcal{K}_A-\frac{d-2}{2}\partial_A\right)\partial^2,
}
all remaining products which are well defined on the light cone $\eta^2=0$ are linear combinations of uncontracted $\mathcal{L}_{AB}$, $(\mathcal{L}^2)_{AB}$ and $\mathcal{K}_A$ with different factors that are functions of $\Theta$.  An analog result has been obtained in \cite{Eastwood:2002su} by studying the symmetries of the Laplacian operator. 

In \eqref{EqSOPE}, the OPE differential operator $\DOPE{a}{i}{j}{k}(\eta_1,\eta_2)$ involves two embedding space coordinates on the hypercone, $\eta_1$ and $\eta_2$, with derivatives at $\eta_2$.  With the extra embedding space coordinate $\eta_1$, it is straightforward to verify that
\eqn{\mathcal{K}_{2A}=-\frac{1}{2\ee{1}{2}{}}[(\eta_1\cdot\mathcal{L}_2^2)_A+i(\eta_1\cdot\mathcal{L}_2)_A(\Theta_2-1)+\eta_{1A}\Theta_2].}
Hence, with the introduction of another embedding space coordinate, there is thus only one independent differential operator which is well defined on the light-cone.  Therefore, all non-trivial differential operators appearing in the OPE are functions of uncontracted $\mathcal{L}_{AB}$ and $(\mathcal{L}^2)_{AB}$ acting at $\eta_2$.

Following \cite{Ferrara:1971vh,Ferrara:1971zy,Ferrara:1972cq,Ferrara:1973eg,Ferrara:1973yt}, in \cite{Fortin:2016dlj} the candidate differential operator (at $\eta_i$ and $\eta_j$ with derivatives at $\eta_j$ for future convenience) appearing in the OPE was chosen as
\eqna{
\D_{ij}^A&\equiv\frac{1}{\ee{i}{j}{\frac{1}{2}}}[-i(\eta_i\cdot\mathcal{L}_j)^A-\e{i}{A}{}\Theta_j]=\ee{i}{j}{\frac{1}{2}}\A_{ij}^{AB}\partial_{jB},\\
\D_{ij}^2&\equiv\D_{ij}^A\D_{ijA}=\ee{i}{j}{}\partial_j^2-\eta_i\cdot\partial_j(d_E-4+2\Theta_j)=\ee{i}{j}{}\partial_j^2-(d_E-2+2\Theta_j)\eta_i\cdot\partial_j,
}[EqDO]
where\footnote{It is of interest to note that the differential operator $\D_{ij}^2$ first introduced in \cite{Ferrara:1971vh} is proportional to a simple contraction of the Thomas-Todorov operator used in \cite{Dobrev:1975ru}.}
\eqn{\A_{ij}^{AB}=\frac{1}{\ee{i}{j}{}}[\ee{i}{j}{}g^{AB}-\e{i}{A}{}\e{j}{B}{}-\e{i}{B}{}\e{j}{A}{}].}[EqMetric]
Indeed, from \eqref{EqDO} it is straightforward to show that $\mathcal{L}_j^{AB}$ can be expressed as a combination of $\D_{ij}^A$ and $\Theta_j$, \textit{i.e.}\footnote{Note here that the explicit $i$ in the pre-factors is the imaginary number $i^2=-1$, it is not related to $\eta_i$.  Moreover, although $\eta_i$ appears on the right-hand side of the identity, it does not appear on the left-hand side.}
\eqn{\mathcal{L}_j^{AB}=\frac{i}{\ee{i}{j}{\frac{1}{2}}}(\eta_j^A\D_{ij}^B-\eta_j^B\D_{ij}^A)-\frac{i}{\ee{i}{j}{}}(\eta_i^A\eta_j^B-\eta_i^B\eta_j^A)\Theta_j.}
Hence $\mathcal{L}_{jAB}$ and $(\mathcal{L}_j^2)_{AB}$ are formally replaced by $\D_{ij}^A$ and its square $\D_{ij}^2$ respectively.

It is important to note here that although $\Theta$ is the homogeneity operator defined in \eqref{EqSupp}, which acts trivially and does not play a role in generating conformal descendants, it is nevertheless used in the construction of $\D_{ij}^A$.  Including $\Theta$ leads to the metric $\A_{ij}^{AB}$ which is naturally doubly transverse.  Indeed, $\A_{ij}^{AB}$ can be seen as a doubly-transverse metric since it satisfies the following properties,
\eqn{\A_{ij}^{AB}=\A_{ij}^{BA}=\A_{ji}^{AB}=\A_{ji}^{BA},\qquad\eta_{iA}\A_{ij}^{AB}=\eta_{jA}\A_{ij}^{AB}=0,\qquad\A_{ij}^{AC}\A_{ijC}^{\phantom{ijC}B}=\A_{ij}^{AB}.}
In fact, with its trace being $\A_{ijA}^{\phantom{ijA}A}=d$ as in position space, it will be shown later that the metric \eqref{EqMetric} is the appropriate metric needed to uplift quasi-primary operators in general irreducible $SO(1,d-1)$ representations to embedding space.

The differential operators \eqref{EqDO} satisfy the following identities
\eqn{
\begin{gathered}
\e{i}{A}{}\D_{ijA}=\e{j}{A}{}\D_{ijA}=0,\qquad\A_{ij}^{AB}\D_{ijB}=\D_{ij}^A,\\
[\D_{ij}^A,\D_{ij}^B]=\frac{1}{\ee{i}{j}{\frac{1}{2}}}(\e{i}{A}{}\D_{ij}^B-\e{i}{B}{}\D_{ij}^A),\qquad[\Theta_i,\D_{ij}^A]=\frac{1}{2}\D_{ij}^A,\qquad[\Theta_j,\D_{ij}^A]=-\frac{1}{2}\D_{ij}^A,\\
\D_{ij}^B\e{j}{A}{}-\e{j}{A}{}\D_{ij}^B=\ee{i}{j}{\frac{1}{2}}\A_{ij}^{AB},\\
[\D_{ij}^A,\D_{ij}^2]=\frac{2}{\ee{i}{j}{\frac{1}{2}}}\e{i}{A}{}\D_{ij}^2,\qquad[\Theta_i,\D_{ij}^2]=\D_{ij}^2,\qquad[\Theta_j,\D_{ij}^2]=-\D_{ij}^2,\\
\D_{ij}^2\e{j}{A}{}-\e{j}{A}{}\D_{ij}^2=2\ee{i}{j}{\frac{1}{2}}\D_{ij}^A-d\e{i}{A}{}.
\end{gathered}
}[EqDA]
Thus, the differential operators \eqref{EqDO} have well-defined degrees of homogeneity with respect to $\eta_i$ and $\eta_j$ and they commute with $\ee{i}{j}{}$.

Some of the commutation relations in \eqref{EqDA} can be generalized by recurrence to arbitrary powers of the scalar differential operators.  Indeed, for $h\in\mathbb{R}$ one has by recurrence
\eqn{
\begin{gathered}{}
[\D_{ij}^A,\D_{ij}^{2h}]=\frac{2h}{\ee{i}{j}{\frac{1}{2}}}\e{i}{A}{}\D_{ij}^{2h},\qquad[\Theta_i,\D_{ij}^{2h}]=h\D_{ij}^{2h},\qquad[\Theta_j,\D_{ij}^{2h}]=-h\D_{ij}^{2h},\\
\D_{ij}^{2h}\e{j}{A}{}-\e{j}{A}{}\D_{ij}^{2h}=2h\ee{i}{j}{\frac{1}{2}}\D_{ij}^A\D_{ij}^{2(h-1)}-h(d+2h-2)\e{i}{A}{}\D_{ij}^{2(h-1)}.
\end{gathered}
}[EqDAsq2h]
The generalized commutation relations \eqref{EqDAsq2h} suggest a final differential operator with even more interesting properties than $\D_{ij}^A$.

%%%%%%%%%%%%%%%%%%%%%%%%%%%%%%%%%%%%%%%%%%%%%%%%%%

\subsection{A Better Differential Operator}\label{SSecDO}

The most useful differential operator for the OPE is not $\D_{ij}^A$ but rather
\eqn{\D_{ij|h}^A=\frac{\e{j}{A}{}}{\ee{i}{j}{\frac{1}{2}}}\D_{ij}^2+2h\D_{ij}^A-h(d+2h-2)\frac{\e{i}{A}{}}{\ee{i}{j}{\frac{1}{2}}},}[EqDh]
as motivated by the last commutation relation in \eqref{EqDAsq2h}.  It satisfies several identities, for example
\eqn{
\begin{gathered}
\e{iA}{}{}\D_{ij|h}^A=\ee{i}{j}{\frac{1}{2}}\D_{ij}^2,\qquad\e{jA}{}{}\D_{ij|h}^A=-h(d+2h-2)\ee{i}{j}{\frac{1}{2}},\\
\D_{ij|h}^A\e{jA}{}{}=(h-1)(d-2h)\ee{i}{j}{\frac{1}{2}},\qquad\D_{ij}^{2h}\D_{ij|h'}^A=\D_{ij|h+h'}^A\D_{ij}^{2h}.
\end{gathered}
}
It is straightforward to generate extra identities from the definitions above.  However, the most important identity corresponds to the last commutation relation in \eqref{EqDAsq2h} which can be rewritten in terms of the new differential operator \eqref{EqDh} as
\eqn{\D_{ij}^{2h}\e{j}{A}{}=\ee{i}{j}{\frac{1}{2}}\D_{ij|h}^A\D_{ij}^{2(h-1)},\qquad\D_{ij|h}^A\D_{ij}^{2(h-1)}=\frac{1}{\ee{i}{j}{\frac{1}{2}}}\D_{ij}^{2h}\e{j}{A}{}.}[EqDheta]
By applying \eqref{EqDheta} recursively for two $\eta_j$'s, it is clear that
\eqn{\D_{ij|h+1}^A\D_{ij|h}^B=\D_{ij|h+1}^B\D_{ij|h}^A,\qquad\D_{ij|h+1}^A\D_{ij|hA}=0,}
as can also be seen directly from the definition \eqref{EqDh}.  Then clearly one has
\eqn{\D_{ij}^{2h}\e{j}{{A_1}}{}\cdots\e{j}{{A_n}}{}=\ee{i}{j}{\frac{n}{2}}\D_{ij|h}^{A_1}\cdots\D_{ij|h-(n-1)}^{A_n}\D_{ij}^{2(h-n)},}
and the differential operator
\eqn{\D_{ij}^{(d,h,n)A_1\cdots A_n}\equiv\D_{ij|h+n}^{A_n}\cdots\D_{ij|h+1}^{A_1}\D_{ij}^{2h}=\frac{1}{\ee{i}{j}{\frac{n}{2}}}\D_{ij}^{2(h+n)}\e{j}{A}{1}\cdots\e{j}{A}{n},}[EqDOPE]
is fully symmetric and traceless with respect to the metric $g_{AB}$.

It is important to point out that both $\D_{ij}^A$ and $\D_{ij}^2$ can be expressed in terms of $\D_{ij|h}^A$.  Indeed, the scalar differential operator can be expressed in terms of $\D_{ij|h}^A$ by contracting with the metric $\A_{ij}^{AB}$, as in
\eqn{\D_{ij}^2=\frac{1}{2(h+1)(d+2h)}\A_{ijAB}\D_{ij|h+1}^A\D_{ij|h}^B,}
while the vector differential operator $\D_{ij}^A$ is given by
\eqn{\D_{ij}^A=\frac{1}{2h}\D_{ij|h}^A-\frac{1}{4h(h+1)(d+2h)}\frac{\e{j}{A}{}}{\ee{i}{j}{\frac{1}{2}}}\A_{ijBC}\D_{ij|h+1}^B\D_{ij|h}^C+(d/2+h-1)\frac{\e{i}{A}{}}{\ee{i}{j}{\frac{1}{2}}}.}
Since $\mathcal{L}_j^{AB}$ can be written as a function of $\D_{ij}^A$, it can also be written as a function of \eqref{EqDh},
\eqn{\mathcal{L}_j^{AB}=\frac{i}{2h}\frac{1}{\ee{i}{j}{\frac{1}{2}}}(\eta_j^A\D_{ij|h}^B-\eta_j^B\D_{ij|h}^A)-i(d/2+h)\frac{1}{\ee{i}{j}{}}(\eta_i^A\eta_j^B-\eta_i^B\eta_j^A)\Theta_j,}
suggesting to write the OPE in terms of \eqref{EqDh}.

In fact, with its exceptional properties, it is clear that it will be very convenient to define the OPE with the differential operator \eqref{EqDOPE}.  This statement can be made explicit by looking at the differential operator $\D_{ij}^{(d,h,n)}$ acting on free $\eta_j$'s, as in
\eqn{\D_{ij}^{(d,h,n)A_1\cdots A_n}\e{j}{A}{n+1}\cdots\e{j}{A}{n+k}=\ee{i}{j}{\frac{k}{2}}\D_{ij}^{(d,h-k,n+k)A_1\cdots A_{n+k}}=\frac{1}{\ee{i}{j}{\frac{n}{2}}}\D_{ij}^{(d,h+n,0)}\e{j}{A}{1}\cdots\e{j}{A}{n+k}.}
Hence free $\eta_j$'s appearing in computations of correlation functions can be properly taken into account with the differential operator \eqref{EqDOPE}.  Before returning to the uplift of quasi-primary operators in general irreducible representations of the Lorentz group, it is appropriate to study the action of the differential operator $\D_{ij}^{(d,h,n)}$ in $M$-point correlation functions.

%%%%%%%%%%%%%%%%%%%%%%%%%%%%%%%%%%%%%%%%%%%%%%%%%%

\subsection{General Function for \texorpdfstring{$(M>3)$}{(M>3)}-Point Correlation Functions}

In general, the computation of $M$-point correlation functions from the OPE leads to the study of the tensorial functions
\eqn{I_{ij}^{(d,h,n;\boldsymbol{p})A_1\cdots A_n}=\D_{ij}^{(d,h,n)A_1\cdots A_n}\prod_{a\neq i,j}\frac{1}{\ee{j}{a}{p_a}}.}[EqI]
For $M$-point correlation functions at embedding space coordinates $\eta_1$ to $\eta_M$, the product over $a$ runs from $1$ to $M$.  From the fact that powers of $\ee{i}{j}{}$ commute with $\D_{ij}^{(d,h,n)}$ and that $\ee{j}{j}{}=0$, the product can be restricted to all $a\neq i,j$ without loss of generality.

To proceed, it is useful to homogeneize the differential operator \eqref{EqDOPE} and the quantity \eqref{EqI}.  There are several ways to accomplish this for $M>3$.\footnote{The case $M=3$ will be discussed later.}  Without introducing any powers of $\ee{j}{a}{}$ for all $a\neq i,j$ (which would not commute with the derivatives) the simplest expressions are made out of two extra embedding space coordinates $\eta_k$ and $\eta_\ell$, such that
\eqn{\bar{\D}_{ij;k\ell|h}^A=\frac{\ee{i}{j}{\frac{1}{2}}\ee{k}{\ell}{\frac{1}{2}}}{\ee{i}{k}{\frac{1}{2}}\ee{i}{\ell}{\frac{1}{2}}}\D_{ij|h}^A,\qquad\qquad\bar{\D}_{ij;k\ell}^{(d,h,n)}=\frac{\ee{i}{j}{h+\frac{n}{2}}\ee{k}{\ell}{h+\frac{n}{2}}}{\ee{i}{k}{h+\frac{n}{2}}\ee{i}{\ell}{h+\frac{n}{2}}}\D_{ij}^{(d,h,n)},}[EqDbOPE]
with $k<\ell$ and $k,\ell\neq i,j$, and
\eqn{\bar{I}_{ij;k\ell}^{(d,h,n;\boldsymbol{p})}=\frac{\ee{i}{j}{\bar{p}+h+\frac{n}{2}}\ee{k}{\ell}{\bar{p}+h+\frac{n}{2}}}{\ee{i}{k}{\bar{p}+h+\frac{n}{2}}\ee{i}{\ell}{\bar{p}+h+\frac{n}{2}}}\left[\prod_{a\neq i,j}\ee{i}{a}{p_a}\right]I_{ij}^{(d,h,n;\boldsymbol{p})},}[EqIb]
with $\bar{p}=\sum_{a\neq i,j}p_a$.  The entire pre-factor in \eqref{EqIb} commutes through the differential operator, hence suggesting to write the result in terms of conformal cross-ratios as well as homogeneized embedding space coordinates.

Using the two extra embedding space coordinates $\eta_k$ and $\eta_\ell$, the homogeneized embedding space coordinates are defined as
\eqn{
\begin{gathered}
\bar{\eta}_i^A=\frac{\ee{k}{\ell}{\frac{1}{2}}}{\ee{i}{k}{\frac{1}{2}}\ee{i}{\ell}{\frac{1}{2}}}\eta_i^A,\qquad\qquad\bar{\eta}_j^A=\frac{\ee{i}{k}{\frac{1}{2}}\ee{i}{\ell}{\frac{1}{2}}}{\ee{i}{j}{}\ee{k}{\ell}{\frac{1}{2}}}\eta_j^A,\\\bar{\eta}_a^A=\frac{\ee{i}{k}{\frac{1}{2}}\ee{i}{\ell}{\frac{1}{2}}}{\ee{k}{\ell}{\frac{1}{2}}\ee{i}{a}{}}\eta_a^A\qquad\forall\,a\neq i,j,
\end{gathered}
}[Eqetab]
while the conformal cross-ratios are defined as
\eqna{
x_a&=\frac{\ee{i}{j}{}\ee{k}{\ell}{}\ee{i}{a}{}}{\ee{i}{k}{}\ee{i}{\ell}{}\ee{j}{a}{}}\qquad\forall\,a\neq i,j,\\
z_{ab}&=\frac{\ee{i}{k}{}\ee{i}{\ell}{}\ee{a}{b}{}}{\ee{k}{\ell}{}\ee{i}{a}{}\ee{i}{b}{}}\qquad\forall\,a,b\neq i,j.
}[EqCR]
From \eqref{EqCR} one notes that $z_{ab}=z_{ba}$, $z_{aa}=0$ and $z_{k\ell}=1$.  Therefore there are $M-2$ cross-ratios $x_a$ and $(M-2)(M-3)/2-1$ cross-ratios $z_{ab}$ for a total of $M(M-3)/2$, as expected for $M$-point correlation functions.

With the homogeneized free embedding space coordinates \eqref{Eqetab} and the conformal cross-ratios \eqref{EqCR}, \eqref{EqIb} becomes
\eqn{\bar{I}_{ij;k\ell}^{(d,h,n;\boldsymbol{p})}=\bar{\D}_{ij;k\ell}^{(d,h,n)}\prod_{a\neq i,j}x_a^{p_a}.}[EqIbkl]
where the differential operator $\bar{\D}_{ij;k\ell}^{(d,h,n)}$ can be expressed as derivatives with respect to the homogeneized embedding space coordinates $\bar{\eta}_j^A$ as well as the conformal cross-ratios $x_a$ with $\bar{\partial}_j^Ax_a=\partial_{x_a}\bar{\eta}_j^A=0$.  The conformal cross-ratios $z_{ab}$ do not appear as derivatives in the differential operator since they do not involve the embedding space coordinate $\eta_j$.

First, the metric and scalar products of homogeneized embedding space coordinates can be re-expressed in terms of the homogeneized embedding space coordinates and the conformal cross-ratios as
\eqn{
\begin{gathered}
\A_{ij}^{AB}=g^{AB}-\bar{\eta}_i^A\bar{\eta}_j^B-\bar{\eta}_i^B\bar{\eta}_j^A,\qquad(\bar{\eta}_i\cdot\bar{\eta}_j)=1,\\
\A_{ia}^{AB}=g^{AB}-\bar{\eta}_i^A\bar{\eta}_a^B-\bar{\eta}_i^B\bar{\eta}_a^A,\qquad(\bar{\eta}_i\cdot\bar{\eta}_a)=1,\\
\A_{ja}^{AB}=g^{AB}-x_a\bar{\eta}_j^A\bar{\eta}_a^B-x_a\bar{\eta}_j^B\bar{\eta}_a^A,\qquad(\bar{\eta}_j\cdot\bar{\eta}_a)=1/x_a,\\
\A_{ab}^{AB}=g^{AB}-\frac{1}{z_{ab}}\bar{\eta}_a^A\bar{\eta}_b^B-\frac{1}{z_{ab}}\bar{\eta}_a^B\bar{\eta}_b^A,\qquad(\bar{\eta}_a\cdot\bar{\eta}_b)=z_{ab}.
\end{gathered}
}
Then, the basic $\partial_{jA}$ derivative written in terms of the new variables becomes
\eqna{
\partial_{jA}&=(\partial_{jA}\bar{\eta}_j^B)\bar{\partial}_{jB}+\sum_{a\neq i,j}(\partial_{jA}x_a)\partial_{x_a}\\
&=\frac{\ee{i}{k}{\frac{1}{2}}\ee{i}{\ell}{\frac{1}{2}}}{\ee{i}{j}{}\ee{k}{\ell}{\frac{1}{2}}}\left[\bar{\partial}_{jA}-\bar{\eta}_{iA}\bar{\eta}_j\cdot\bar{\partial}_j+\sum_{a\neq i,j}(\bar{\eta}_{iA}-x_a\bar{\eta}_{aA})x_a\partial_{x_a}\right],
}
which implies
\eqn{\bar{\D}_{ij;k\ell}^A=\frac{\ee{i}{j}{\frac{1}{2}}\ee{k}{\ell}{\frac{1}{2}}}{\ee{i}{k}{\frac{1}{2}}\ee{i}{\ell}{\frac{1}{2}}}\D_{ij}^A=\bar{\partial}_j^A-\bar{\eta}_i^A\bar{\eta}_j\cdot\bar{\partial}_j-\bar{\eta}_j^A\bar{\eta}_i\cdot\bar{\partial}_j-\sum_{a\neq i,j}(x_a\bar{\eta}_a^A-\bar{\eta}_i^A-x_a\bar{\eta}_j^A)x_a\partial_{x_a},
}[EqDAb]
as well as
\eqna{
\bar{\D}_{ij;k\ell}^2&=\bar{\D}_{ij;k\ell}^A\bar{\D}_{ij;k\ell A}=\bar{\partial}_j^2-(d+2\bar{\eta}_j\cdot\bar{\partial}_j)\bar{\eta}_i\cdot\bar{\partial}_j-\sum_{a\neq i,j}2(x_a\bar{\eta}_a\cdot\bar{\partial}_j-\bar{\eta}_i\cdot\bar{\partial}_j-x_a\bar{\eta}_j\cdot\bar{\partial}_j)x_a\partial_{x_a}\\
&\phantom{=}\qquad+\sum_{a,b\neq i,j}(z_{ab}x_a^2x_b^2-2x_a^2x_b)\partial_{x_a}\partial_{x_b}+(d-4)\sum_{a\neq i,j}x_a^2\partial_{x_a}.
}[EqDAsqb]
Hence, the differential operator $\bar{\D}_{ij;k\ell|h}^A$ becomes
\eqn{\bar{\D}_{ij;k\ell|h}^A=\bar{\eta}_j^A\bar{\D}_{ij;k\ell}^2+2h\bar{\D}_{ij;k\ell}^A-2h(d/2+h-1)\bar\eta_i^A,}[EqDhAb]
with the definitions \eqref{EqDAb} and \eqref{EqDAsqb}.  Therefore, the differential operator \eqref{EqDbOPE} is finally expressed in terms of the homogeneized embedding space coordinates and the conformal cross-ratios introduced in \eqref{Eqetab} and \eqref{EqCR}.

By using the form $\bar{\D}_{ij;k\ell}^{(d,h,n)}=\bar{\D}_{ij;k\ell|h+n}\cdots\bar{\D}_{ij;k\ell|h+1}\bar{\D}_{ij;k\ell}^{2h}$, the fractional derivative made out of powers of $\bar{\D}_{ij;k\ell}^2$ acts only on the conformal cross-ratios $x_a$ in \eqref{EqIbkl}.  From \eqref{EqDAsqb} in the limit $x_a\to0$ for all $a\neq i,j$ with $x_a/x_b$ fixed for all $a,b\neq i,j$, one has
\eqn{\bar{\D}_{ij;k\ell}^2\to(-2)\left[\sum_{a,b\neq i,j}x_a^2x_b\partial_{x_a}\partial_{x_b}-(d/2-2)\sum_{a\neq i,j}x_a^2\partial_{x_a}\right],}
which acts on powers of $x_m$ with $m\neq i,j$ as
\eqn{\bar{\D}_{ij;k\ell}^2x_m^{p_m}\to(-2)\left[\sum_{a,b\neq i,j}x_a^2x_b\partial_{x_a}\partial_{x_b}-(d/2-2)\sum_{a\neq i,j}x_a^2\partial_{x_a}\right]x_m^{p_m}=(-2)p_m(p_m+1-d/2)x_m^{p_m+1}.}
Thus, in that limit, one has
\eqn{\bar{\D}_{ij;k\ell}^{2h}x_m^{p_m}\to(-2)^h(p_m)_h(p_m+1-d/2)_hx_m^{p_m+h},}
by recurrence.

Hence, from \eqref{EqIbkl}, the scalar quantity behaves in that limit as
\eqna{
\bar{I}_{ij;k\ell}^{(d,h,0;\boldsymbol{p})}&=\bar{\D}_{ij;k\ell}^{(d,h,0)}\prod_{a\neq i,j}x_a^{p_a}=\bar{\D}_{ij;k\ell}^{2h}\prod_{a\neq i,j}x_a^{p_a}\\
&=\bar{\D}_{ij;k\ell}^{2h}x_m^{\bar{p}}\prod_{a\neq i,j,m}\left(\frac{x_a}{x_m}\right)^{p_a}\to(-2)^h(\bar{p})_h(\bar{p}+1-d/2)_hx_m^{\bar{p}+h},
}
for any $m\neq i,j$, which suggests to write
\eqn{\bar{I}_{ij;k\ell}^{(d,h,0;\boldsymbol{p})}=(-2)^h(\bar{p})_h(\bar{p}+1-d/2)_hx_m^{\bar{p}+h}K_{ij;k\ell;m}^{(d,h;\boldsymbol{p})}(x_m;\boldsymbol{y};\textbf{z}),}
with $y_a=1-x_m/x_a$ for all $a\neq i,j,m$ and $K_{ij;k\ell;m}^{(d,h;\boldsymbol{p})}(0;\boldsymbol{0};\textbf{z})=1$, again for any $m\neq i,j$.  Therefore, there exists a series expansion for $K_{ij;k\ell;m}^{(d,h;\boldsymbol{p})}(x_m;\boldsymbol{y};\textbf{z})$ in $x_m$, the vector of $y_a$ denoted by $\boldsymbol{y}$, and the matrix of $z_{ab}$ denoted by $\textbf{z}$.

To obtain this series expansion, it is convenient to change variables from $\boldsymbol{x}$ to $x_m$ and $\boldsymbol{y}$.  This variable change gives
\eqn{\partial_{x_m}=\partial_{x_m}-\sum_{a\neq i,j,m}\frac{1-y_a}{x_m}\partial_{y_a},\qquad\partial_{x_a}=\frac{(1-y_a)^2}{x_m}\partial_{y_a},}
for all $a\neq i,j,m$.  Hence the derivatives \eqref{EqDAb} and \eqref{EqDAsqb} become
\eqna{
\bar{\D}_{ij;k\ell;m}^A&=\bar{\partial}_j^A-\bar{\eta}_i^A\bar{\eta}_j\cdot\bar{\partial}_j-\bar{\eta}_j^A\bar{\eta}_i\cdot\bar{\partial}_j-x_m(x_m\bar{\eta}_m^A-\bar{\eta}_i^A-x_m\bar{\eta}_j^A)\partial_{x_m}\\
&\phantom{=}\qquad+\sum_{a\neq i,j,m}x_m[(1-y_a)\bar{\eta}_m^A-\bar{\eta}_a^A+y_a\bar{\eta}_j^A]\partial_{y_a},
}
and
\eqna{
\bar{\D}_{ij;k\ell;m}^2&=\bar{\partial}_j^2+(-2)\left\{(d/2+\bar{\eta}_j\cdot\bar{\partial}_j)\bar{\eta}_i\cdot\bar{\partial}_j+x_m(x_m\bar{\eta}_m\cdot\bar{\partial}_j-\bar{\eta}_i\cdot\bar{\partial}_j-x_m\bar{\eta}_j\cdot\bar{\partial}_j)\partial_{x_m}\right.\\
&\phantom{=}\qquad\left.-\sum_{a\neq i,j,m}x_m[(1-y_a)\bar{\eta}_m\cdot\bar{\partial}_j-\bar{\eta}_a\cdot\bar{\partial}_j+y_a\bar{\eta}_j\cdot\bar{\partial}_j]\partial_{y_a}\right.\\
&\phantom{=}\qquad\left.+x_m^3\partial_{x_m}^2-\sum_{a\neq i,j,m}x_m^2(x_mz_{am}-y_a)\partial_{x_m}\partial_{y_a}\right.\\
&\phantom{=}\qquad\left.+\sum_{a,b\neq i,j,m}x_m^2[(1-y_b)z_{am}-z_{ab}/2]\partial_{y_a}\partial_{y_b}\right.\\
&\phantom{=}\qquad\left.-(d/2-2)x_m^2\partial_{x_m}-\sum_{a\neq i,j,m}x_m[x_mz_{am}+(d/2-1)y_a]\partial_{y_a}\right\},
}
respectively, with the corresponding definitions for $\bar{\D}_{ij;k\ell;m|h}^A$ and $\bar{\D}_{ij;k\ell;m}^{(d,h,n)}$ [see \eqref{EqDbOPE} and \eqref{EqDhAb}].

This new expression acts on products as
\eqna{
\bar{\D}_{ij;k\ell;m}^2x_m^{n}\prod_{a\neq i,j,m}y_a^{n_a}&=(-2)\left\{(n+\bar{n})(n+1-d/2)-\sum_{a\neq i,j,m}(n+\bar{n})n_a\frac{x_mz_{am}}{y_a}\right.\\
&\phantom{=}\qquad\left.+\sum_{a,b\neq i,j,m}(n_a-\delta_{ab})n_b\frac{x_mz_{am}}{y_ay_b}-\sum_{a,b\neq i,j,m}\frac{n_an_b}{2}\frac{x_mz_{ab}}{y_ay_b}\right\}x_m^{n+1}\prod_{a\neq i,j,m}y_a^{n_a},
}
where $\bar{n}=\sum_{a\neq i,j,m}n_a$, which can be rewritten as
\eqna{
\frac{\bar{\D}_{ij;k\ell;m}^2x_m^{n}\prod_{a\neq i,j,m}y_a^{n_a}}{(-2)x_m^{n+1}\prod_{a\neq i,j,m}y_a^{n_a}}&=\sum_{\{n_a',n_{am}',n_{ab}'\}\geq0}(-1)_{\bar{n}_m'+\bar{\bar{n}}'}(-\bar{n}_m')_{\bar{n}'}(n+\bar{n})_{1-\bar{n}'-\bar{\bar{n}}'}\\
&\phantom{=}\qquad\times(n+\bar{n}_m'+\bar{\bar{n}}'+1-d/2)_{1-\bar{n}_m'-\bar{\bar{n}}'}\left[\prod_{a\neq i,j,m}(-n_a)_{n_a'+n_{am}'+\bar{n}_a'}\right]\\
&\phantom{=}\qquad\times\prod_{a\neq i,j,m}\left(-\frac{1}{y_a}\right)^{n_a'}\prod_{a\neq i,j,m}\left(-\frac{x_mz_{am}}{y_a}\right)^{n_{am}'}\prod_{\substack{a,b\neq i,j,m\\b>a}}\left(\frac{x_mz_{ab}}{y_ay_b}\right)^{n_{ab}'},
}
with
\eqn{
\begin{gathered}
\bar{n}'=\sum_{a\neq i,j,m}n_a',\qquad\qquad\bar{n}_m'=\sum_{a\neq i,j,m}n_{am}',\\
\bar{n}_a'=\sum_{b\neq i,j,m,a}n_{ab}',\qquad\qquad\bar{\bar{n}}'=\sum_{\substack{a,b\neq i,j,m\\b>a}}n_{ab}',
\end{gathered}
}
where it is understood that $n_{ab}'$ is defined such that $a<b$ (for example, in $\bar{n}_a'$ terms with the wrong ordering must be rewritten with the right ordering).  By construction or by recurrence, one then has
\eqna{
\frac{\bar{\D}_{ij;k\ell;m}^{2h}x_m^{n}\prod_{a\neq i,j,m}y_a^{n_a}}{(-2)^hx_m^{n+h}\prod_{a\neq i,j,m}y_a^{n_a}}&=\sum_{\{n_a',n_{am}',n_{ab}'\}\geq0}(-h)_{\bar{n}_m'+\bar{\bar{n}}'}(-\bar{n}_m')_{\bar{n}'}(n+\bar{n})_{h-\bar{n}'-\bar{\bar{n}}'}\\
&\phantom{=}\qquad\times(n+\bar{n}_m'+\bar{\bar{n}}'+1-d/2)_{h-\bar{n}_m'-\bar{\bar{n}}'}\left[\prod_{a\neq i,j,m}(-n_a)_{n_a'+n_{am}'+\bar{n}_a'}\right]\\
&\phantom{=}\qquad\times\prod_{a\neq i,j,m}\frac{1}{n_a'!}\left(-\frac{1}{y_a}\right)^{n_a'}\prod_{a\neq i,j,m}\frac{1}{n_{am}'!}\left(-\frac{x_mz_{am}}{y_a}\right)^{n_{am}'}\\
&\phantom{=}\qquad\times\prod_{\substack{a,b\neq i,j,m\\b>a}}\frac{1}{n_{ab}'!}\left(\frac{x_mz_{ab}}{y_ay_b}\right)^{n_{ab}'}.
}[EqDxy]

Therefore, the scalar $\bar{I}_{ij;k\ell}^{(d,h,0,\boldsymbol{p})}$ \eqref{EqIb} becomes
\eqna{
\bar{I}_{ij;k\ell}^{(d,h,0;\boldsymbol{p})}&=\bar{\D}_{ij;k\ell;m}^{2h}x_m^{\bar{p}}\prod_{a\neq i,j,m}(1-y_a)^{-p_a}=\sum_{\{n_a\}\geq0}\left[\prod_{a\neq i,j,m}\frac{(p_a)_{n_a}}{n_a!}\right]\bar{\D}_{ij;k\ell;m}^{2h}x_m^{\bar{p}}\prod_{a\neq i,j,m}y_a^{n_a}\\
&=\sum_{\{n_a\}\geq0}\left[\prod_{a\neq i,j,m}\frac{(p_a)_{n_a}}{n_a!}\right](-2)^hx_m^{\bar{p}+h}\prod_{a\neq i,j,m}y_a^{n_a}\\
&\phantom{=}\qquad\times\sum_{\{n_a',n_{am}',n_{ab}'\}\geq0}(-h)_{\bar{n}_m'+\bar{\bar{n}}'}(-\bar{n}_m')_{\bar{n}'}(\bar{p}+\bar{n})_{h-\bar{n}'-\bar{\bar{n}}'}\\
&\phantom{=}\qquad\times(\bar{p}+\bar{n}_m'+\bar{\bar{n}}'+1-d/2)_{h-\bar{n}_m'-\bar{\bar{n}}'}\left[\prod_{a\neq i,j,m}(-n_a)_{n_a'+n_{am}'+\bar{n}_a'}\right]\\
&\phantom{=}\qquad\times\prod_{a\neq i,j,m}\frac{1}{n_a'!}\left(-\frac{1}{y_a}\right)^{n_a'}\prod_{a\neq i,j,m}\frac{1}{n_{am}'!}\left(-\frac{x_mz_{am}}{y_a}\right)^{n_{am}'}\prod_{\substack{a,b\neq i,j,m\\b>a}}\frac{1}{n_{ab}'!}\left(\frac{x_mz_{ab}}{y_ay_b}\right)^{n_{ab}'}.
}
Redefining $n_a\to n_a+n_a'$ and removing the primes on the other indices lead to
\eqna{
\bar{I}_{ij;k\ell}^{(d,h,0;\boldsymbol{p})}&=(-2)^hx^{\bar{p}+h}\sum_{\{n_a,n_{am},n_{ab},n_a'\}\geq0}\left[\prod_{a\neq i,j,m}\frac{(p_a)_{n_a+n_a'}}{(n_a+n_a')!}\right]\\
&\phantom{=}\qquad\times(-h)_{\bar{n}_m+\bar{\bar{n}}}(-\bar{n}_m)_{\bar{n}'}(\bar{p}+\bar{n}+\bar{n}')_{h-\bar{n}'-\bar{\bar{n}}}\\
&\phantom{=}\qquad\times(\bar{p}+\bar{n}_m+\bar{\bar{n}}+1-d/2)_{h-\bar{n}_m-\bar{\bar{n}}}\left[\prod_{a\neq i,j,m}(-n_a-n_a')_{n_a'+n_{am}+\bar{n}_a}\right]\\
&\phantom{=}\qquad\times\prod_{a\neq i,j,m}\frac{(-1)^{n_a'}}{n_a'!}y_a^{n_a}\prod_{a\neq i,j,m}\frac{1}{n_{am}!}\left(-\frac{x_mz_{am}}{y_a}\right)^{n_{am}}\prod_{\substack{a,b\neq i,j,m\\b>a}}\frac{1}{n_{ab}!}\left(\frac{x_mz_{ab}}{y_ay_b}\right)^{n_{ab}},
}
Summing over the different $n_a'$ using the Vandermonde's identity recursively gives as a final answer
\eqna{
\bar{I}_{ij;k\ell}^{(d,h,0;\boldsymbol{p})}&=(-2)^hx_m^{\bar{p}+h}\sum_{\{n_a,n_{am},n_{ab}\}\geq0}(-h)_{\bar{n}_m+\bar{\bar{n}}}(\bar{p}+\bar{n}_m+\bar{\bar{n}}+1-d/2)_{h-\bar{n}_m-\bar{\bar{n}}}\\
&\phantom{=}\qquad\times(\bar{p}+\bar{n}+\bar{n}_m)_{h-\bar{n}_m-\bar{\bar{n}}}(p_m)_{\bar{n}_m}\\
&\phantom{=}\qquad\times\prod_{a\neq i,j,m}\frac{(p_a)_{n_a}}{n_{am}!(n_a-n_{am}-\bar{n}_a)!}y_a^{n_a}\left(\frac{x_mz_{am}}{y_a}\right)^{n_{am}}\prod_{\substack{a,b\neq i,j,m\\b>a}}\frac{1}{n_{ab}!}\left(\frac{x_mz_{ab}}{y_ay_b}\right)^{n_{ab}}\\
&=(-2)^h(\bar{p})_h(\bar{p}+1-d/2)_hx_m^{\bar{p}+h}K_{ij;k\ell;m}^{(d,h;\boldsymbol{p})}(x_m;\boldsymbol{y};\textbf{z}),
}[EqIb0Soln]
with
\begingroup\makeatletter\def\f@size{10}\check@mathfonts\def\maketag@@@#1{\hbox{\m@th\large\normalfont#1}}%
\eqna{
K_{ij;k\ell;m}^{(d,h;\boldsymbol{p})}(x_m;\boldsymbol{y};\textbf{z})&=\sum_{\{n_a,n_{am},n_{ab}\}\geq0}\frac{(-h)_{\bar{n}_m+\bar{\bar{n}}}(p_m)_{\bar{n}_m}(\bar{p}+h)_{\bar{n}-\bar{\bar{n}}}}{(\bar{p})_{\bar{n}+\bar{n}_m}(\bar{p}+1-d/2)_{\bar{n}_m+\bar{\bar{n}}}}\\
&\phantom{=}\qquad\times\prod_{a\neq i,j,m}\frac{(p_a)_{n_a}}{n_{am}!(n_a-n_{am}-\bar{n}_a)!}y_a^{n_a}\left(\frac{x_mz_{am}}{y_a}\right)^{n_{am}}\prod_{\substack{a,b\neq i,j,m\\b>a}}\frac{1}{n_{ab}!}\left(\frac{x_mz_{ab}}{y_ay_b}\right)^{n_{ab}}\\
&=\sum_{\{n_a,n_{am},n_{ab}\}\geq0}k_{ij;k\ell;m}^{(d,h;\boldsymbol{p})}(\boldsymbol{n};\boldsymbol{n}_m;\textbf{n})\prod_{a\neq i,j,m}y_a^{n_a}\left(\frac{x_mz_{am}}{y_a}\right)^{n_{am}}\prod_{\substack{a,b\neq i,j,m\\b>a}}\frac{1}{n_{ab}!}\left(\frac{x_mz_{ab}}{y_ay_b}\right)^{n_{ab}}.
}[EqK0]
\endgroup

Returning to the tensor $\bar{I}_{ij;k\ell}^{(d,h,n,\boldsymbol{p})}$ \eqref{EqIb}, which is fully symmetric and traceless, one has
\eqn{\bar{I}_{ij;k\ell}^{(d,h,n;\boldsymbol{p})}=(-2)^h(\bar{p})_h(\bar{p}+1-d/2)_hx_m^{\bar{p}+h}\sum_{\substack{\{q_r\}\geq0\\\bar{q}=n}}S_{(\boldsymbol{q})}x_m^{\bar{q}-q_0-q_i}K_{ij;k\ell;m}^{(d,h;\boldsymbol{p};\boldsymbol{q})}(x_m;\boldsymbol{y};\textbf{z}),}[EqIbSoln]
where the fully-symmetric tensor $S_{(\boldsymbol{q})}$ is given by
\eqn{S_{(\boldsymbol{q})}^{A_1\cdots A_{\bar{q}}}=g^{(A_1A_2}\cdots g^{A_{2q_0-1}A_{2q_0}}\bar{\eta}_1^{A_{2q_0+1}}\cdots\bar{\eta}_1^{A_{2q_0+q_1}}\cdots\bar{\eta}_M^{A_{\bar{q}-q_M+1}}\cdots\bar{\eta}_M^{A_{\bar{q}})},}[EqS]
with $\bar{q}=2q_0+\sum_{r\geq1}q_r$.  By recurrence [using for example $k_{ij;k\ell;m}^{(d,h;\boldsymbol{p})}(\boldsymbol{n};\boldsymbol{n}_m;\textbf{n})$], one finds
\eqna{
K_{ij;k\ell;m}^{(d,h;\boldsymbol{p};\boldsymbol{q})}&=\frac{(-1)^{\bar{q}-q_0-q_i-q_j}(-2)^{\bar{q}-q_0}\bar{q}!}{\prod_{r\geq0}q_r!}\frac{(-h-\bar{q})_{\bar{q}-q_0-q_j}(p_m)_{q_m}(\bar{p}+h)_{\bar{q}-q_0-q_i}}{(\bar{p})_{\bar{q}-2q_0-q_i-q_j}(\bar{p}+1-d/2)_{-q_0-q_i-q_j}}\prod_{a\neq i,j,m}(p_a)_{q_a}\\
&\phantom{=}\qquad\times K_{ij;k\ell;m}^{(d+2\bar{q}-2q_0,h+q_0+q_j;\boldsymbol{p}+\boldsymbol{q})},
}[EqK]
where it is understood that $\boldsymbol{q}$ in $\boldsymbol{p}+\boldsymbol{q}$ does not include the zeroth, the $i$-th and $j$-th components $q_0$, $q_i$ and $q_j$ respectively.  Hence, the relevant functions appearing in the tensor $\bar{I}_{ij;k\ell}^{(d,h,n,\boldsymbol{p})}$ are given by the standard $K$-function $K_{ij;k\ell;m}^{(d,h;\boldsymbol{p})}(x_m;\boldsymbol{y};\textbf{z})$ for the scalar $\bar{I}_{ij;k\ell}^{(d,h,0,\boldsymbol{p})}$ with properly shifted parameters.

In summary, for $M>3$ the action of the fractional scalar derivative on $x_m$ and $\boldsymbol{y}$ is given by \eqref{EqDxy}; the scalar $\bar{I}_{ij;k\ell}^{(d,h,0;\boldsymbol{p})}$ is given by \eqref{EqIb0Soln} with the function \eqref{EqK0}; and the tensor $\bar{I}_{ij;k\ell}^{(d,h,n;\boldsymbol{p})}$ is given by \eqref{EqIbSoln} with the function \eqref{EqK} and the fully-symmetric tensor \eqref{EqS}.  The non-homogeneized quantity $I_{ij}^{(d,h,n;\boldsymbol{p})}$ is therefore easily obtained from \eqref{EqI} and the above results.

The non-homogeneized quantity $I_{ij}^{(d,h,n;\boldsymbol{p})}$ is the building block of $M$-point correlation functions.  It is directly related to conformal blocks for $M=4$ but it does not exactly correspond to conformal blocks for $M>4$.

%%%%%%%%%%%%%%%%%%%%%%%%%%%%%%%%%%%%%%%%%%%%%%%%%%

\subsection{Properties of the \texorpdfstring{$K$}{K}-function}

For $M=4$, the function $K_{ij;k\ell;m}^{(d,h;\boldsymbol{p})}(x_m;\boldsymbol{y};\textbf{z})$ is exactly the Exton $G$-function for four-point correlation functions \cite{Exton_1995}.  For $M>4$, it is the proper generalization of the Exton $G$-function to $M$-point correlation functions.  As a consistency check, setting $p_M=0$ in \eqref{EqIb} with $i,j,k,\ell,m\neq M$ should lead to the result \eqref{EqIb0Soln} but with parameters relevant to the $M-1$ case.  Quantities like $\bar{p}$ trivially collapse to their $M-1$ counterparts.  As for the $K$-function, \eqref{EqK0} demonstrates that setting $p_M=0$ forces $n_M=0$ which in turn leads to $n_{mM}=n_{aM}=0$ for all $a\neq i,j,m$.  As a consequence, the sums over the extra conformal cross-ratios disappear and the $K$-function reduces to the appropriate $K$-function for the $M-1$ case.

Moreover, the function $K_{ij;k\ell;m}^{(d,h;\boldsymbol{p})}(x_m;\boldsymbol{y};\textbf{z})$ satisfies several interesting properties.  For example, since $I_{ij}^{(d,h,n;\boldsymbol{p})}$ is independent of the choice of $k$, $\ell$ and $m$, one has from \eqref{EqIb}
\eqna{
I_{ij}^{(d,h,n;\boldsymbol{p})}&=\frac{\ee{i}{k}{\bar{p}+h+\frac{n}{2}}\ee{i}{\ell}{\bar{p}+h+\frac{n}{2}}}{\ee{i}{j}{\bar{p}+h+\frac{n}{2}}\ee{k}{\ell}{\bar{p}+h+\frac{n}{2}}}\left[\prod_{a\neq i,j}\ee{i}{a}{-p_a}\right]\bar{I}_{ij;k\ell}^{(d,h,n;\boldsymbol{p})}\\
&=(-2)^h(\bar{p})_h(\bar{p}+1-d/2)_hx_m^{\bar{p}+h}\frac{\ee{i}{k}{\bar{p}+h+\frac{n}{2}}\ee{i}{\ell}{\bar{p}+h+\frac{n}{2}}}{\ee{i}{j}{\bar{p}+h+\frac{n}{2}}\ee{k}{\ell}{\bar{p}+h+\frac{n}{2}}}\left[\prod_{a\neq i,j}\ee{i}{a}{-p_a}\right]\\
&\phantom{=}\qquad\times\sum_{\substack{\{q_r\}\geq0\\\bar{q}=n}}S_{(\boldsymbol{q})}x_m^{\bar{q}-q_0-q_i}K_{ij;k\ell;m}^{(d,h;\boldsymbol{p};\boldsymbol{q})}(x_m;\boldsymbol{y};\textbf{z}),
}
and hence
\eqn{K_{ij;k'\ell';m'}^{(d,h;\boldsymbol{p})}(x_{m'}';\boldsymbol{y}';\textbf{z}')=(1-y_{m'})^{\bar{p}+h}K_{ij;k\ell;m}^{(d,h;\boldsymbol{p})}(x_m;\boldsymbol{y};\textbf{z}),}[EqKxyz]
where $x_{m'}'$, $\boldsymbol{y}'$ and $\textbf{z}'$ are the conformal cross-ratios \eqref{EqCR} built from the particular choices $k'$, $\ell'$ and $m'$.  The latter can be re-expressed in terms of the final conformal cross-ratios as (here $x_a'=x_az_{k'\ell'}$ and $z_{ab}'=z_{ab}/z_{k'\ell'}$ for all $a,b\neq i,j$)
\eqn{x_{m'}'=\frac{x_mz_{k'\ell'}}{1-y_{m'}},\qquad y_a'=\frac{y_a-y_{m'}}{1-y_{m'}},\qquad z_{ab}'=\frac{z_{ab}}{z_{k'\ell'}},}
with $y_m=0$.  In terms of the actual $K$-function \eqref{EqK0}, the identity \eqref{EqKxyz} corresponds to
\begingroup\makeatletter\def\f@size{9}\check@mathfonts\def\maketag@@@#1{\hbox{\m@th\large\normalfont#1}}%
\eqna{
&\sum_{\{n_a,n_{am},n_{ab}\}\geq0}\frac{(-h)_{\bar{n}_m+\bar{\bar{n}}}(p_m)_{\bar{n}_m}(\bar{p}+h)_{\bar{n}-\bar{\bar{n}}}}{(\bar{p})_{\bar{n}+\bar{n}_m}(\bar{p}+1-d/2)_{\bar{n}_m+\bar{\bar{n}}}}\\
&\qquad\times\prod_{a\neq i,j,m}\frac{(p_a)_{n_a}}{n_{am}!(n_a-n_{am}-\bar{n}_a)!}y_a^{n_a}\left(\frac{x_mz_{am}}{y_a}\right)^{n_{am}}\prod_{\substack{a,b\neq i,j,m\\b>a}}\frac{1}{n_{ab}!}\left(\frac{x_mz_{ab}}{y_ay_b}\right)^{n_{ab}}\\
&\qquad=(1-y_{m'})^{-\bar{p}-h}\sum_{\{n_a,n_{am'},n_{ab}\}\geq0}\frac{(-h)_{\bar{n}_{m'}+\bar{\bar{n}}}(p_{m'})_{\bar{n}_{m'}}(\bar{p}+h)_{\bar{n}-\bar{\bar{n}}}}{(\bar{p})_{\bar{n}+\bar{n}_{m'}}(\bar{p}+1-d/2)_{\bar{n}_{m'}+\bar{\bar{n}}}}\\
&\phantom{=}\qquad\times\prod_{a\neq i,j,m'}\frac{(p_a)_{n_a}}{n_{am'}!(n_a-n_{am'}-\bar{n}_a)!}\left(\frac{y_a-y_{m'}}{1-y_{m'}}\right)^{n_a}\left(\frac{x_mz_{am'}}{y_a-y_{m'}}\right)^{n_{am'}}\prod_{\substack{a,b\neq i,j,m'\\b>a}}\frac{1}{n_{ab}!}\left[\frac{x_m(1-y_{m'})z_{ab}}{(y_a-y_{m'})(y_b-y_{m'})}\right]^{n_{ab}},
}
\endgroup
which is obviously trivial for $m'=m$ but highly non-trivial for $m'\neq m$.

It is interesting to note that the dependence on $k$, $\ell$, $k'$ and $\ell'$ disappears completely in the previous expression, implying an identity for all $m'\neq m$ irrespective of the choice for $k$, $\ell$, $k'$ and $\ell'$, \textit{i.e.} \eqref{EqKxyz} corresponds to $M-3$ different identities for $M$-point correlation functions.  In fact, the independence of the $K$-function \eqref{EqK0} on the choice of $k$ and $\ell$ is obvious from its form and the definitions of the conformal cross-ratios \eqref{EqCR} since
\eqn{y_a=1-\frac{\ee{i}{m}{}\ee{j}{a}{}}{\ee{j}{m}{}\ee{i}{a}{}}\qquad\forall\,a\neq i,j,m,\qquad x_mz_{ab}=\frac{\ee{i}{m}{}\ee{i}{j}{}\ee{a}{b}{}}{\ee{j}{m}{}\ee{i}{a}{}\ee{i}{b}{}}\qquad\forall\,a,b\neq i,j.}
This observation suggests that there exists a result equivalent to \eqref{EqIbSoln} for $M=3$, result that will be obtained shortly.

The function $K_{ij;k\ell;m}^{(d,h;\boldsymbol{p})}(x_m;\boldsymbol{y};\textbf{z})$ satisfies also a number of contiguous relations.  Indeed by using the form $\bar{\D}_{ij;k\ell}^{(d,h,n)A_1\cdots A_n}=\bar{\D}_{ij;k\ell}^{2(h+n)}\bar{\eta}_j^{A_1}\cdots\bar{\eta}_j^{A_n}$, $\bar{I}_{ij;k\ell}^{(d,h,n;\boldsymbol{p})}$ contractions with $\bar{\eta}_i$ and $\bar{\eta}_a$ with $a\neq i,j$ are straightforward.  By using the form $\bar{\D}_{ij;k\ell}^{(d,h,n)A_1\cdots A_n}=\bar{\D}_{ij;k\ell|h+n}^{A_n}\cdots\bar{\D}_{ij;k\ell|h+1}^{A_1}\bar{\D}_{ij;k\ell}^{2h}$ instead, $\bar{I}_{ij;k\ell}^{(d,h,n;\boldsymbol{p})}$ contractions with $\bar{\eta}_j$ are straightforward.  Putting everything together leads to
\eqna{
g\cdot\bar{I}_{ij;k\ell}^{(d,h,n;\boldsymbol{p})}&=0,\\
\bar{\eta}_i\cdot\bar{I}_{ij;k\ell}^{(d,h,n;\boldsymbol{p})}&=\bar{I}_{ij;k\ell}^{(d,h+1,n-1;\boldsymbol{p})},\\
\bar{\eta}_j\cdot\bar{I}_{ij;k\ell}^{(d,h,n;\boldsymbol{p})}&=(-2)(-h-n)(-h-n+1-d/2)\bar{I}_{ij;k\ell}^{(d,h,n-1;\boldsymbol{p})},\\
\bar{\eta}_a\cdot\bar{I}_{ij;k\ell}^{(d,h,n;\boldsymbol{p})}&=\bar{I}_{ij;k\ell}^{(d,h+1,n-1;\boldsymbol{p}-\boldsymbol{e}_a)},
}[EqContRel]
where the contracted index does not have to be made explicit due to the fact that $\bar{I}_{ij;k\ell}^{(d,h,n;\boldsymbol{p})}$ is fully symmetric.  The first identity, involving a full contraction with the metric $g$, vanishes due to the tracelessness property of $\bar{I}_{ij;k\ell}^{(d,h,n;\boldsymbol{p})}$.  Using the explicit definitions \eqref{EqIbSoln} with \eqref{EqK} for the left-hand side of \eqref{EqContRel} leads to the aforementioned contiguous relations for the $K$-function.  Indeed, since
\eqna{
S_{(\boldsymbol{q})}^{A_1\cdots A_{\bar{q}}}&=\frac{2q_0}{\bar{q}}g^{A_{\bar{q}}(A_1}S_{(\boldsymbol{q}-\boldsymbol{e}_0)}^{A_2\cdots A_{\bar{q}-1})}+\sum_{r\neq0}\frac{q_r}{\bar{q}}\bar{\eta}_r^{A_{\bar{q}}}S_{(\boldsymbol{q}-\boldsymbol{e}_r)}^{A_1\cdots A_{\bar{q}-1}}\\
&=\frac{2q_0}{\bar{q}(\bar{q}-1)}g^{A_{\bar{q}}A_{\bar{q}-1}}S_{(\boldsymbol{q}-\boldsymbol{e}_0)}^{A_1\cdots A_{\bar{q}-2}}+\frac{4q_0(q_0-1)}{\bar{q}(\bar{q}-1)}g^{A_{\bar{q}}(A_1}g^{|A_{\bar{q}-1}|A_2}S_{(\boldsymbol{q}-2\boldsymbol{e}_0)}^{A_3\cdots A_{\bar{q}-2})}\\
&\phantom{=}\qquad+\sum_{r\neq0}\frac{4q_0q_r}{\bar{q}(\bar{q}-1)}\bar{\eta}_r^{(A_{\bar{q}}}g^{A_{\bar{q}-1})(A_1}S_{(\boldsymbol{q}-\boldsymbol{e}_0-\boldsymbol{e}_r)}^{A_2\cdots A_{\bar{q}-2})}+\sum_{r,s\neq0}\frac{q_r(q_s-\delta_{rs})}{\bar{q}(\bar{q}-1)}\bar{\eta}_r^{A_{\bar{q}}}\bar{\eta}_s^{A_{\bar{q}-1}}S_{(\boldsymbol{q}-\boldsymbol{e}_r-\boldsymbol{e}_s)}^{A_1\cdots A_{\bar{q}-2}},
}
the contiguous relations are given explicitly by (with $y_m=0$)
\eqna{
K_{ij;k\ell;m}^{(d,h;\boldsymbol{p})}(x_m;\boldsymbol{y};\textbf{z})&=-\frac{(\bar{p}-d/2)}{(d/2)}K_{ij;k\ell;m}^{(d+2,h;\boldsymbol{p})}(x_m;\boldsymbol{y};\textbf{z})-\sum_{a\neq i,j}\frac{hp_a}{(d/2)\bar{p}}K_{ij;k\ell;m}^{(d+2,h-1;\boldsymbol{p}+\boldsymbol{e}_a)}(x_m;\boldsymbol{y};\textbf{z})\\
&\phantom{=}\qquad+\sum_{a\neq i,j}\frac{p_a(\bar{p}+h)}{(d/2)\bar{p}}(1-y_a)K_{ij;k\ell;m}^{(d+2,h;\boldsymbol{p}+\boldsymbol{e}_a)}(x_m;\boldsymbol{y};\textbf{z})\\
&\phantom{=}\qquad+\sum_{\substack{a,b\neq i,j\\b>a}}\frac{hp_ap_b(\bar{p}+h)}{(d/2)\bar{p}(\bar{p}+1)(\bar{p}+1-d/2)}xz_{ab}K_{ij;k\ell;m}^{(d+2,h-1;\boldsymbol{p}+\boldsymbol{e}_a+\boldsymbol{e}_b)}(x_m;\boldsymbol{y};\textbf{z}),
}
as well as
\eqna{
K_{ij;k\ell;m}^{(d,h;\boldsymbol{p})}(x_m;\boldsymbol{y};\textbf{z})&=\frac{(\bar{p}-d/2)}{(\bar{p}+h-d/2)}K_{ij;k\ell;m}^{(d+2,h;\boldsymbol{p})}(x_m;\boldsymbol{y};\textbf{z})\\
&\phantom{=}\qquad+\sum_{a\neq i,j}\frac{hp_a}{\bar{p}(\bar{p}+h-d/2)}K_{ij;k\ell;m}^{(d+2,h-1;\boldsymbol{p}+\boldsymbol{e}_a)}(x_m;\boldsymbol{y};\textbf{z}),
}
and
\eqna{
K_{ij;k\ell;m}^{(d,h;\boldsymbol{p})}(x_m;\boldsymbol{y};\textbf{z})&=-\frac{(\bar{p}-d/2)}{(h+d/2)}K_{ij;k\ell;m}^{(d+2,h;\boldsymbol{p})}(x_m;\boldsymbol{y};\textbf{z})\\
&\qquad+\sum_{a\neq i,j}\frac{p_a(\bar{p}+h)}{\bar{p}(h+d/2)}(1-y_a)K_{ij;k\ell;m}^{(d+2,h;\boldsymbol{p}+\boldsymbol{e}_a)}(x_m;\boldsymbol{y};\textbf{z}),
}
and finally, for all $a\neq i,j$,
\eqna{
K_{ij;k\ell;m}^{(d,h;\boldsymbol{p})}(x_m;\boldsymbol{y};\textbf{z})&=-\frac{h}{\bar{p}}K_{ij;k\ell;m}^{(d+2,h-1;\boldsymbol{p}+\boldsymbol{e}_a)}(x_m;\boldsymbol{y};\textbf{z})+\frac{(\bar{p}+h)}{\bar{p}}(1-y_a)K_{ij;k\ell;m}^{(d+2,h;\boldsymbol{p}+\boldsymbol{e}_a)}(x_m;\boldsymbol{y};\textbf{z})\\
&\qquad+\sum_{b\neq i,j,a}\frac{hp_b(\bar{p}+h)}{\bar{p}(\bar{p}+1)(\bar{p}+1-d/2)}x_mz_{ab}K_{ij;k\ell;m}^{(d+2,h-1;\boldsymbol{p}+\boldsymbol{e}_a+\boldsymbol{e}_b)}(x_m;\boldsymbol{y};\textbf{z}).
}
These relations can be verified by recurrence with the help of $k_{ij;k\ell;m}^{(d,h,\boldsymbol{p})}(\boldsymbol{n};\boldsymbol{n}_m;\textbf{n})$.  They can be used as an indirect proof of \eqref{EqIbSoln}.

In terms of $I_{ij}^{(d,h,n;\boldsymbol{p})}$ and non-homogeneized embedding space coordinates, the relations \eqref{EqContRel} can be rewritten as
\eqna{
g\cdot I_{ij}^{(d,h,n;\boldsymbol{p})}&=0,\\
\eta_i\cdot I_{ij}^{(d,h,n;\boldsymbol{p})}&=\ee{i}{j}{\frac{1}{2}}I_{ij}^{(d,h+1,n-1;\boldsymbol{p})},\\
\eta_j\cdot I_{ij}^{(d,h,n;\boldsymbol{p})}&=(-2)(-h-n)(-h-n+1-d/2)\ee{i}{j}{\frac{1}{2}}I_{ij}^{(d,h,n-1;\boldsymbol{p})},\\
\eta_a\cdot I_{ij}^{(d,h,n;\boldsymbol{p})}&=\ee{i}{j}{-\frac{1}{2}}I_{ij}^{(d,h+1,n-1;\boldsymbol{p}-\boldsymbol{e}_a)}.
}

It is obvious that \eqref{EqContRel} can be generalized to any number of contractions, leading to
\eqna{
\left(g^{q_0}\prod_{r\geq1}\bar{\eta}_r^{q_r}\right)\cdot\bar{I}_{ij;k\ell}^{(d,h,n;\boldsymbol{p})}&=\delta_{0q_0}(-2)^{q_j}(-h-n)_{q_j}(-h-n+1-d/2)_{q_j}\\
&\phantom{=}\qquad\times\bar{I}_{ij;k\ell}^{(d,h+\sum_{r\geq1}q_r-q_j,n-\sum_{r\geq1}q_r;\boldsymbol{p}-\boldsymbol{q})},
}
and an analog result for the non-homogeneized quantity $I_{ij}^{(d,h,n;\boldsymbol{p})}$.  As in the previous case, it is understood that $\boldsymbol{q}$ in $\boldsymbol{p}-\boldsymbol{q}$ does not include the zeroth, the $i$-th and the $j$-th components.

%%%%%%%%%%%%%%%%%%%%%%%%%%%%%%%%%%%%%%%%%%%%%%%%%%

\subsection{General Function for Three-Point Correlation Functions}

For three-point correlation functions it is not possible to choose two extra embedding space coordinates to properly homogeneize \eqref{EqI}.  However, the scalar differential operator in this specific case is trivial, and as such it is possible to homogeneize the result appropriately as 
\eqna{
\bar{I}_{ij}^{(d,h,n;p_a)}&=\frac{\ee{j}{a}{p_a+h+\frac{n}{2}}}{\ee{i}{a}{h+\frac{n}{2}}}\D_{ij}^{(d,h,n)}\frac{1}{\ee{j}{a}{p_a}}\\
&=(-2)^h(p_a)_h(p_a+1-d/2)_h\frac{\ee{j}{a}{p_a+h+\frac{n}{2}}}{\ee{i}{a}{\frac{n}{2}}}\D_{ij|h+n}\cdots\D_{ij|h+1}\frac{1}{\ee{j}{a}{p_a+h}}.
}
By recurrence, the action of the vector differential operators can also be found and it gives
\eqna{
\bar{I}_{ij}^{(d,h,n;p_a)}&=(-2)^h(p_a)_h(p_a+1-d/2)_h\sum_{\substack{\{q_r\}\geq0\\\bar{q}=n}}S_{(\boldsymbol{q})}\frac{(-1)^{\bar{q}-q_0-q_i-q_j}(-2)^{\bar{q}-q_0}\bar{q}!}{\prod_{r\geq0}q_r!}\\
&\phantom{=}\qquad\times\frac{(-h-\bar{q})_{\bar{q}-q_0-q_j}(p_a+h)_{\bar{q}-q_0-q_i}}{(p_a+1-d/2)_{-q_0-q_i-q_j}}\\
&=(-2)^h(p_a)_h(p_a+1-d/2)_h\sum_{\substack{\{q_r\}\geq0\\\bar{q}=n}}S_{(\boldsymbol{q})}K_{ij}^{(d,h;p_a;\boldsymbol{q})},
}[EqIb3]
where the fully-symmetric tensor $S_{(\boldsymbol{q})}$ is defined as in \eqref{EqS} (but restricted to three coordinates),
\eqn{S_{(\boldsymbol{q})}^{A_1\cdots A_{\bar{q}}}=g^{(A_1A_2}\cdots g^{A_{2q_0-1}A_{2q_0}}\bar{\eta}_1^{A_{2q_0+1}}\cdots\bar{\eta}_1^{A_{2q_0+q_1}}\bar{\eta}_2^{A_{2q_0+q_1+1}}\cdots\bar{\eta}_2^{A_{2q_0+q_1+q_2}}\bar{\eta}_3^{A_{2q_0+q_1+q_2+1}}\cdots\bar{\eta}_3^{A_{\bar{q}})},}
the $K$-function is
\eqn{K_{ij}^{(d,h;p_a;\boldsymbol{q})}=\frac{(-1)^{\bar{q}-q_0-q_i-q_j}(-2)^{\bar{q}-q_0}\bar{q}!}{\prod_{r\geq0}q_r!}\frac{(-h-\bar{q})_{\bar{q}-q_0-q_j}(p_a+h)_{\bar{q}-q_0-q_i}}{(p_a+1-d/2)_{-q_0-q_i-q_j}},}[EqK3]
and the homogeneous embedding space coordinates are
\eqn{\bar{\eta}_i^A=\frac{\ee{j}{a}{\frac{1}{2}}}{\ee{i}{j}{\frac{1}{2}}\ee{i}{a}{\frac{1}{2}}}\e{i}{A}{},\qquad\bar{\eta}_j^A=\frac{\ee{i}{a}{\frac{1}{2}}}{\ee{i}{j}{\frac{1}{2}}\ee{j}{a}{\frac{1}{2}}}\e{j}{A}{},\qquad\bar{\eta}_a^A=\frac{\ee{i}{j}{\frac{1}{2}}}{\ee{i}{a}{\frac{1}{2}}\ee{j}{a}{\frac{1}{2}}}\e{a}{A}{}.}

The results obtained here for three-point correlation functions show that the general idea used for $(M>3)$-point correlation functions works also in this special case.  The similarities between \eqref{EqIb3} and \eqref{EqIbSoln} as well as \eqref{EqK3} and \eqref{EqK} (with $K_{ij}^{(d,h;p_a)}=1$ for three-point correlation functions) cannot be overlooked.

Moreover, \eqref{EqIb3} satisfies identities identical to the contiguous relations \eqref{EqContRel} by construction.

%%%%%%%%%%%%%%%%%%%%%%%%%%%%%%%%%%%%%%%%%%%%%%%%%%
%%%%%%%%%%%%%%%%%%%%%%%%%%%%%%%%%%%%%%%%%%%%%%%%%%

\section{Operator Product Expansion in Embedding Space}\label{SecOPE}

Equipped with a metric in embedding space that is doubly transverse and has the same trace than the usual Lorentz metric in position space, it is now possible to uplift quasi-primary operators in general $SO(1,d-1)$ irreducible representations to embedding space.  Once the machinery necessary to perform the lift to embedding space is presented, it will be straightforward to write the OPE.

%%%%%%%%%%%%%%%%%%%%%%%%%%%%%%%%%%%%%%%%%%%%%%%%%%

\subsection{Transversality and General Quasi-Primary Operators in Embedding Space}

In section \ref{SSecOES} it was shown that quasi-primary operators in defining representations of the Lorentz group uplift to quasi-primary operators in defining representations of the conformal group such that $\boldsymbol{N}_E^{\mathcal{O}}=\{0,\boldsymbol{N}^{\mathcal{O}}\}$.  Hence scalars uplift to scalars, spinors uplift to spinors and $i$-index antisymmetric tensors uplift to $(i+1)$-index antisymmetric tensors (including the self-dual and anti-self-dual tensors).  Moreover, for defining representations, the lift works at the level of quasi-primary operators.

For general irreducible representations of the Lorentz group however, it is necessary to symmetrize and remove traces as explained in Appendix \ref{SecIrrep}.  The former does not lead to any problem, but the latter is trickier since the tracelessness condition in embedding space must correspond to the tracelessness condition in position space.  That can only occur if the trace of the to-be-used embedding space metric is equal to the trace of the position space metric.  Fortunately, such a candidate exists and is given by $\A_{ij}^{AB}$ \eqref{EqMetric}.  This metric is built with the help of two embedding space coordinates and it is doubly transverse with respect to these two embedding space coordinates.  Hence, if the metric $\A_{ij}^{AB}$ is to be used to remove traces in the lift of general irreducible representations of the Lorentz group to embedding space, then an extra embedding space coordinate is needed to achieve such a lift.

At first glance, this observation might seem problematic since the lift cannot be done at the level of quasi-primary operators.  However in a CFT the most fundamental building block is the OPE and the OPE always necessitates using two embedding space coordinates [see \eqref{EqSOPE}].  Hence it is conceivable to use the $\A_{ij}^{AB}$ metric to uplift general $SO(1,d-1)$ irreducible representations to general $SO(2,d)$ irreducible representations.  How this can be done can be inferred from the link between quasi-primary operators and half-projectors \eqref{EqT} as well as the transversality condition \eqref{EqSupp}.

For the defining representations and from the analysis in Appendix \ref{SecIrrep}, it is clear that in embedding space $\mathcal{O}^{\boldsymbol{N}}(\eta)\sim\mathcal{T}^{\boldsymbol{N}_E}$ as far as the behavior under conformal transformations is concerned.  Since $\eta\cdot\Gamma\eta\cdot\Gamma=\eta\cdot\eta=0$, demanding that the transversality condition \eqref{EqSupp} be satisfied forces the introduction of the doubly-transverse metric such that
\eqn{
\begin{gathered}
\mathcal{O}_a^{\boldsymbol{e}_r}(\eta_i)\sim(\mathcal{T}^{\boldsymbol{e}_{r_E}}\eta_i\cdot\Gamma)_a^{a'},\qquad\mathcal{O}_{ab}^{2\boldsymbol{e}_r}(\eta_i)\sim(\mathcal{T}^{2\boldsymbol{e}_{r_E}}\eta_i\A_{ij}\cdots\A_{ij})_{ab}^{A_1\cdots A_r},\\
\mathcal{O}_{ab}^{\boldsymbol{e}_n}(\eta_i)\sim(\mathcal{T}^{\boldsymbol{e}_{n+1}}\eta_i\A_{ij}\cdots\A_{ij})_{ab}^{A_1\cdots A_n}\qquad\forall\,n\in\{1,\ldots,r-1\},
\end{gathered}
}
with
\eqn{(\mathcal{T}^{\boldsymbol{e}_{n+1}}\eta_i\A_{ij}\cdots\A_{ij})_{ab}^{A_1\cdots A_n}\equiv(\mathcal{T}^{\boldsymbol{e}_{n+1}})_{ab}^{A_0'\cdots A_n'}\A_{ijA_n'}^{\phantom{ijA_n'}A_n}\cdots\A_{ijA_1'}^{\phantom{ijA_1'}A_1}\eta_{iA_0'}.}
Straightforward analogous equations exist in even dimensions.  The transversality property of the metric $\A_{ij}$ with respect the embedding space coordinate $\eta_i$ was necessary to satisfy the transversality condition of quasi-primary operators in embedding space \eqref{EqSupp}.  Hence, with the lift to embedding space introduced here, transversality demands a transverse metric and thus a second embedding space coordinate.\footnote{There are no transverse metrics built from the usual embedding space metric and only one embedding space coordinate.}  It is important to note that the number of dummy indices on the half-projectors for the defining representations in embedding space corresponds to the number of dummy indices on the half-projectors for the defining representations in position space even though $\boldsymbol{N}_E^{\mathcal{O}}=\{0,\boldsymbol{N}^{\mathcal{O}}\}$.  This fact is paramount and will be used profusely below.

Indeed, this observation and \eqref{EqT} suggest the following generalization for quasi-primary operators in general irreducible representations of the Lorentz group, \textit{i.e.} $\mathcal{O}^{\boldsymbol{N}}(\eta_i)\sim(\mathcal{T}_{ij}^{\boldsymbol{N}}\Gamma)$ where
\begingroup\makeatletter\def\f@size{10}\check@mathfonts\def\maketag@@@#1{\hbox{\m@th\large\normalfont#1}}%
\eqna{
(\mathcal{T}_{ij}^{\boldsymbol{N}}\Gamma)&\equiv\left(\left(\frac{\sqrt{2}}{\ee{i}{j}{\frac{1}{2}}}\mathcal{T}^{\boldsymbol{e}_2}\eta_i\A_{ij}\right)^{N_1}\cdots\left(\frac{\sqrt{r}}{\ee{i}{j}{\frac{1}{2}}}\mathcal{T}^{\boldsymbol{e}_{r_E-1}}\eta_i\A_{ij}\cdots\A_{ij}\right)^{N_{r-1}}\right.\\
&\phantom{=}\qquad\times\left.\left(\frac{\sqrt{r+1}}{\ee{i}{j}{\frac{1}{2}}}\mathcal{T}^{2\boldsymbol{e}_{r_E}}\eta_i\A_{ij}\cdots\A_{ij}\right)^{\lfloor N_r/2\rfloor}\left(\frac{1}{\sqrt{2}\ee{i}{j}{}}\mathcal{T}^{\boldsymbol{e}_{r_E}}\eta_i\cdot\Gamma\eta_j\cdot\Gamma\right)^{N_r-2\lfloor N_r/2\rfloor}\right)\cdot\hat{\mathcal{P}}_{ij}^{\boldsymbol{N}}.
}[EqHPP]
\endgroup
The hatted projection operator, which acts on the dummy indices whose number corresponds to the number of indices for the general irreducible representation of the Lorentz group $\boldsymbol{N}$, is naturally the hatted projection operator for the general irreducible representation $\boldsymbol{N}$ in position space (not $\boldsymbol{N}_E$ in embedding space).  It is however built from the metric $\A_{ij}^{AB}$, $\Gamma_{ij}^A\equiv\Gamma^B\A_{ijB}^{\phantom{ijB}A}$ and the proper epsilon tensor, see below, to properly implement the tracelessness condition expected in position space.  Once again, there exist obvious analog formula for the even-dimensional case [see \eqref{EqTeven} for the proper substitutions].

As for its transformation properties under the conformal group, the embedding space spinor indices of $(\mathcal{T}_{ij}^{\boldsymbol{N}}\Gamma)$ transform as expected for a quasi-primary operator in embedding space.  Moreover, \eqref{EqHPP} clearly verifies the transversality condition \eqref{EqSupp} even with the extra $\Gamma$-matrix for the spinor part of the general irreducible representation.  Indeed, the dummy (spinor) index for the spinor part of the general irreducible representation can be contracted without affecting the behavior under conformal transformations as long as the contraction leads to another dummy (spinor) index (to match to the number of dummy indices in position space).  The extra $\Gamma$-matrix is included to allow proper spinor contraction with the hatted projection operator in the appropriate irreducible representation.  Finally, the pre-factors as well as the power of $\ee{i}{j}{}$ are included for future convenience.

%%%%%%%%%%%%%%%%%%%%%%%%%%%%%%%%%%%%%%%%%%%%%%%%%%

\subsection{Transversality and the Differential Operator}

As shown in the previous section, the transversality condition \eqref{EqSupp} suggests that the differential operator appearing in the OPE \eqref{EqSOPE} is proportional to the half-projectors \eqref{EqHPP} for the two quasi-primary operators in embedding space on the left-hand side of the OPE, \textit{i.e.}
\eqn{\DOPE{a}{i}{j}{k}(\eta_1,\eta_2)\propto(\mathcal{T}_{12}^{\boldsymbol{N}_i}\Gamma)(\mathcal{T}_{21}^{\boldsymbol{N}_j}\Gamma).}
This relationship implies that the right-hand side of the OPE \eqref{EqSOPE} transforms as the left-hand side of the OPE \eqref{EqSOPE} under conformal transformations---the number of embedding space spinor indices on both sides match.  Moreover, the transversality conditions for the two quasi-primary operators in embedding space on the left-hand side of the OPE \eqref{EqSOPE} are satisfied automatically by construction.  One is thus left with two sets of dummy indices, one in the $\boldsymbol{N}_i$ irreducible representations of the Lorentz group and one in the $\boldsymbol{N}_j$ irreducible representations of the Lorentz group (but with $1$ and $2$ exchanged, see below), that must be contracted properly.

From the discussion in section \ref{SSecDO}, it is clear that a convenient differential operator to include in the OPE \eqref{EqSOPE} to generate conformal descendants is $\D_{12}^{(d,h,n)}$ \eqref{EqDOPE}, hence
\eqn{\DOPE{a}{i}{j}{k}(\eta_1,\eta_2)\propto\D_{12}^{(d,h_{ijk}-n_a/2,n_a)}.}
for appropriate parameters $h_{ijk}$ and $n_a$ to be determined shortly.  Combining the two latest results leads to three sets of dummy indices: the two sets of dummy indices for the two irreducible representations of the Lorentz group of the quasi-primary operators on the left-hand side of the OPE and the set of ``dummy'' indices in the symmetric-traceless irreducible representation of the conformal group (since the differential operator $\D_{12}^{(d,h_{ijk}-n_a/2,n_a)}$ is traceless with respect to the usual metric $g$ in embedding space).  This last observation might seem troublesome as the leftover sets of dummy indices do not belong to the same group.  However, it is trivial to extract a symmetric-traceless differential operator with respect to the metric $\A_{12}$, \textit{i.e.} with respect to the Lorentz group, by simply projecting unto the appropriate irreducible representation, as in $\D_{12}^{(d,h_{ijk}-n_a/2,n_a)}\to\hat{\mathcal{P}}_{12}^{n_a\boldsymbol{e}_1}\cdot\D_{12}^{(d,h_{ijk}-n_a/2,n_a)}$.

To preserve the transversality condition with respect to the quasi-primary operator at $\eta_2$, it is necessary to put the differential operator to the right of the second half-projector.  As a rule, the differential operator is chosen to be to the right of both half-projectors, hence
\eqn{\DOPE{a}{i}{j}{k}(\eta_1,\eta_2)\propto(\mathcal{T}_{12}^{\boldsymbol{N}_i}\Gamma)(\mathcal{T}_{21}^{\boldsymbol{N}_j}\Gamma)\hat{\mathcal{P}}_{12}^{n_a\boldsymbol{e}_1}\cdot\D_{12}^{(d,h_{ijk}-n_a/2,n_a)},}
where all three sets of dummy indices, which correspond to the Dynkin indices for the irreducible representations $\boldsymbol{N}_i$, $\boldsymbol{N}_j$ and $n_a\boldsymbol{e}_1$ in position space respectively, must be contracted properly.

At this point, it is necessary to focus on the quasi-primary operator on the right-hand side of the OPE \eqref{EqSOPE}.  This quasi-primary operator has embedding space spinor indices which must be contracted properly since there already is the proper number of embedding space spinor indices to match the left-hand side coming from the two half-projectors.  Following Appendix \ref{SecIrrep}, to contract the superfluous embedding space spinor indices, it is appropriate to introduce
\eqna{
(\mathcal{T}_{ij\boldsymbol{N}}\Gamma)&\equiv\hat{\mathcal{P}}_{ji}^{\boldsymbol{N}}\cdot\left(\left(\frac{1}{\sqrt{2}\ee{i}{j}{}}\eta_j\cdot\Gamma\eta_i\cdot\Gamma\mathcal{T}_{\boldsymbol{e}_{r_E}}\right)^{N_r-2\lfloor N_r/2\rfloor}\left(\frac{\sqrt{r+1}}{\ee{i}{j}{\frac{1}{2}}}\A_{ij}\cdots\A_{ij}\eta_i\mathcal{T}_{2\boldsymbol{e}_{r_E}}\right)^{\lfloor N_r/2\rfloor}\right.\\
&\phantom{=}\qquad\times\left.\left(\frac{\sqrt{r}}{\ee{i}{j}{\frac{1}{2}}}\A_{ij}\cdots\A_{ij}\eta_i\mathcal{T}_{\boldsymbol{e}_{r_E-1}}\right)^{N_{r-1}}\cdots\left(\frac{\sqrt{2}}{\ee{i}{j}{\frac{1}{2}}}\A_{ij}\eta_i\mathcal{T}_{\boldsymbol{e}_2}\right)^{N_1}\right),
}[EqHPOPE]
with
\eqn{\A_{ij}\cdots\A_{ij}\eta_i(\mathcal{T}_{\boldsymbol{e}_{n+1}})_{A_n\cdots A_1}^{ba}\equiv\eta_i^{A_0'}\A_{ijA_1}^{\phantom{ijA_1}A_1'}\cdots\A_{ijA_n}^{\phantom{ijA_n}A_n'}(\mathcal{T}^{\boldsymbol{e}_{n+1}})_{A_n'\cdots A_0'}^{ba},}
and an analogous equation for the even-dimensional case.  There are no real differences between \eqref{EqHPP} and \eqref{EqHPOPE}.  The position of the embedding space spinor indices and the dummy indices are simply exchanged.  Note, however, that for fermionic quasi-primary operators $\Op{k}{2}$, the extra $\eta_j\cdot\Gamma\eta_i\cdot\Gamma$ is not needed due to transversality.  It is nevertheless kept in the definition \eqref{EqHPOPE} to avoid introducing another quantity.

From \eqref{EqHPOPE} and the discussion above, it is possible to assert that a practical form for the differential operator in the OPE \eqref{EqSOPE} is
\eqn{\DOPE{a}{i}{j}{k}(\eta_1,\eta_2)\propto(\mathcal{T}_{12}^{\boldsymbol{N}_i}\Gamma)(\mathcal{T}_{21}^{\boldsymbol{N}_j}\Gamma)\hat{\mathcal{P}}_{12}^{n_a\boldsymbol{e}_1}\cdot\D_{12}^{(d,h_{ijk}-n_a/2,n_a)}(\mathcal{T}_{12\boldsymbol{N}_k}\Gamma)*,}
where the half-projector \eqref{EqHPOPE} transforms the embedding space spinor indices of the quasi-primary operator on the right-hand side into a set of dummy indices.  Note that the product denoted by a dot corresponds to a contraction of the dummy variables while a product denoted by a star corresponds to a contraction of the embedding space spinor indices.

It must be stressed that the choice and order of the embedding space coordinates in \eqref{EqHPOPE} are very important.  Indeed, it is impossible to contract directly the quasi-primary operator on the right-hand side of the OPE with $\eta_2\cdot\Gamma$ since the latter is located at $\eta_2$.  By the transversality condition \eqref{EqSupp}, such a contraction would lead to a vanishing contribution to the OPE \eqref{EqSOPE}.

Finally, the differential operator appearing in the OPE \eqref{EqSOPE} has four different sets of dummy indices that must be contracted properly.  With the help of the metric $\A_{12}$ \eqref{EqMetric}, the four sets of dummy indices can be seen as originating from irreducible representations of the Lorentz group (not the conformal group).  To contract them properly, an extra object combining four Lorentz group irreducible representations into a Lorentz group singlet must be introduced.  In the following, such objects are termed the tensor structures $\tOPE{a}{i}{j}{k}{1}{2}$.  These structures will be studied in more detail below.

%%%%%%%%%%%%%%%%%%%%%%%%%%%%%%%%%%%%%%%%%%%%%%%%%%

\subsection{Tensor Structures}

The different tensor structures appear to contract the four sets of dummy variables remaining in the differential operator $\DOPE{a}{i}{j}{k}(\eta_1,\eta_2)$, which can be seen as originating from four different irreducible representations of the Lorentz group with the help of the metric $\A_{12}$, into a Lorentz singlet.  Consequently, to generate a one-to-one map to the expected result for the OPE in position space, the tensor structures $\tOPE{a}{i}{j}{k}{1}{2}$ should be built from the metric $\A_{12}$.

Moreover, since the hatted projection operators in the half-projectors \eqref{EqHPP} appearing in the differential operator $\DOPE{a}{i}{j}{k}(\eta_1,\eta_2)$ can always be replaced by their squares, the tensor structures should satisfy
\eqn{\tOPE{a}{i}{j}{k}{1}{2}=(\hat{\mathcal{P}}_{12}^{\boldsymbol{N}_i})(\hat{\mathcal{P}}_{21}^{\boldsymbol{N}_j})\cdot\tOPE{a}{i}{j}{k}{1}{2}.}
This identity implies that the hatted projection operators appearing in the half-projectors \eqref{EqHPP} used in the differential operator $\DOPE{a}{i}{j}{k}(\eta_1,\eta_2)$ can be moved to the tensor structures $\tOPE{a}{i}{j}{k}{1}{2}$, removing the hatted projection operators from the half-projectors \eqref{EqHPP}.  This observation suggests that group-theoretical properties are taken into account by the tensor structures.

Decomposing the symmetric-traceless differential operator $\D_{12}^{(d,h_{ijk}-n_a/2,n_a)}$ with respect to the metric $g$ into symmetric-traceless irreducible representations with respect to the metric $\A_{12}$ through $\D_{12}^{(d,h_{ijk}-n_a/2,n_a)}\to\hat{\mathcal{P}}_{12}^{n_a\boldsymbol{e}_1}\cdot\D_{12}^{(d,h_{ijk}-n_a/2,n_a)}$ as before and using the previous property for the tensor structures, it is natural to demand also that
\eqn{\tOPE{a}{i}{j}{k}{1}{2}=(\hat{\mathcal{P}}_{12}^{\boldsymbol{N}_i})(\hat{\mathcal{P}}_{21}^{\boldsymbol{N}_j})\cdot\tOPE{a}{i}{j}{k}{1}{2}\cdot(\hat{\mathcal{P}}_{12}^{n_a\boldsymbol{e}_1}).}
From this observation, the explicit hatted projection operator $(\hat{\mathcal{P}}_{12}^{n_a\boldsymbol{e}_1})$ can also be moved from the differential operator $(\hat{\mathcal{P}}_{12}^{n_a\boldsymbol{e}_1})\cdot\D_{12}^{(d,h_{ijk}-n_a/2,n_a)}$ to the tensor structures.

Finally, using the argument just discussed to project unto the proper irreducible representation for the quasi-primary operators on the left-hand side of the OPE \eqref{EqSOPE}, it is reasonable to demand that the tensor structures satisfy
\eqn{\tOPE{a}{i}{j}{k}{1}{2}=(\hat{\mathcal{P}}_{12}^{\boldsymbol{N}_i})(\hat{\mathcal{P}}_{21}^{\boldsymbol{N}_j})\cdot\tOPE{a}{i}{j}{k}{1}{2}\cdot(\hat{\mathcal{P}}_{12}^{n_a\boldsymbol{e}_1})(\hat{\mathcal{P}}_{21}^{\boldsymbol{N}_k}),}[EqTensorStruct]
to project unto the proper irreducible representation for the quasi-primary operator on the right-hand side of the OPE.  Note, however, that one cannot remove the hatted projection operator from the half-projector \eqref{EqHPOPE} since the latter does not commute with the differential operator $\D_{12}^{(d,h_{ijk}-n_a/2,n_a)}$.

To summarize, the tensor structures $\tOPE{a}{i}{j}{k}{1}{2}$ are built from the metric $\A_{12}$ and project unto the proper irreducible representations of the Lorentz group of the quasi-primary operators as well as the symmetric-traceless differential operator as in \eqref{EqTensorStruct}.  Moreover, they merge the four different sets of dummy variables of the irreducible representations of the Lorentz group appearing in the differential operator $\DOPE{a}{i}{j}{k}(\eta_1,\eta_2)$ into a Lorentz singlet, as would be expected of tensor structures for the OPE in position space, although the order of the embedding space coordinates appearing in the different hatted projection operators must be treated with care (see below).  Hence
\eqn{\DOPE{a}{i}{j}{k}(\eta_1,\eta_2)\propto(\mathcal{T}_{12}^{\boldsymbol{N}_i}\Gamma)(\mathcal{T}_{21}^{\boldsymbol{N}_j}\Gamma)\cdot\tOPE{a}{i}{j}{k}{1}{2}\cdot\D_{12}^{(d,h_{ijk}-n_a/a,n_a)}(\mathcal{T}_{12\boldsymbol{N}_k}\Gamma)*,}
where all the dummy variables as well as the extra embedding space spinor indices corresponding to the quasi-primary operator on the right-hand side of the OPE \eqref{EqSOPE} are properly contracted.  By construction, it is straightforward to verify that $\DOPE{a}{i}{j}{k}(\eta_1,\eta_2)$ has the proper behavior under conformal transformations and is transverse with respect to the remaining embedding space spinor indices as expected from the supplemental condition \eqref{EqSupp} of the two quasi-primary operators on the left-hand side of the OPE \eqref{EqSOPE}.

From \eqref{EqTensorStruct} it is now clear that for three quasi-primary operators in specific irreducible representations of the Lorentz group $\boldsymbol{N}_i$, $\boldsymbol{N}_j$ and $\boldsymbol{N}_k$, the number of tensor structures $\tOPE{a}{i}{j}{k}{1}{2}$, \textit{i.e.} the number $N_{ijk}$ denoting the range of the sum over $a$ in \eqref{EqSOPE}, is equal to the number of symmetric-traceless irreducible representations of the Lorentz group appearing in the tensor product decomposition of $\boldsymbol{N}_i\otimes\boldsymbol{N}_j\otimes\boldsymbol{N}_k$ (being careful with the ordering of the embedding space coordinates).  Furthermore, the parameter $n_a$ appearing in the differential operator $\DOPE{a}{i}{j}{k}(\eta_1,\eta_2)$ for each specific tensor structure corresponds to the Dynkin index $n_a\boldsymbol{e}_1$ of the appropriate symmetric-traceless irreducible representation under consideration.  From the OPE \eqref{EqSOPE}, this fact implies that there are $N_{ijk}$ independent OPE coefficients $\cOPE{a}{i}{j}{k}$.

At this point, it is worth pointing out that the extra $\Gamma$-matrices in the half-projectors \eqref{EqHPP} and \eqref{EqHPOPE} were introduced for consistency.  Indeed, without the extra $\Gamma$-matrices, the OPE would vanish when both quasi-primary operators $\mathcal{O}_i(\eta_1)$ and $\mathcal{O}_k(\eta_2)$ are in fermionic irreducible representations of the Lorentz group.

Since all hatted projection operators can be moved to the tensor structures $\tOPE{a}{i}{j}{k}{1}{2}$, the tensor structures can be seen as the proper objects projecting unto the appropriate irreducible representations of the Lorentz group.\footnote{The hatted projection operator in \eqref{EqHPOPE} acts trivially on the quasi-primary operator on the right-hand side of the OPE and as such does not carry non-trivial group-theoretical information.}  Therefore, all non-trivial group-theoretical information present in the OPE can be solely encoded in the tensor structures.

It is noteworthy to stress here that all irreducible representations involved in the discussion about tensor structures are Lorentz group representations, not conformal group representations.  As such, there exists a one-to-one map between tensor structures in embedding space and tensor structures in position space---tensor structures in embedding space are given by the equivalent tensor structures in position space where all metrics are converted to the metric $\A_{12}$ with appropriate substitutions for $\Gamma$-matrices and the epsilon tensor.  This map is the tensor structure analog of the map for hatted projection operators from position space to embedding space.

To be more precise, the tensor structures $\tOPE{a}{i}{j}{k}{1}{2}$ are made out of the following group-theoretical metric, epsilon tensor and spinor quantities,
\eqn{
\begin{gathered}
\A_{12}^{AB}\equiv g^{AB}-\frac{\eta_1^A\eta_2^B}{\ee{1}{2}{}}-\frac{\eta_1^B\eta_2^A}{\ee{1}{2}{}},\\
\epsilon_{12}^{A_1\cdots A_d}\equiv\frac{1}{\ee{1}{2}{}}\eta_{1A_0'}\epsilon^{A_0'A_1'\cdots A_d'A_{d+1}'}\eta_{2A_{d+1}'}\A_{12A_d'}^{\phantom{12A_d'}A_d}\cdots\A_{12A_1'}^{\phantom{12A_1'}A_1},\\
\Gamma_{12}^{A_1\cdots A_n}\equiv\Gamma^{A_1'\cdots A_n'}\A_{12A_n'}^{\phantom{12A_n'}A_n}\cdots\A_{12A_1'}^{\phantom{12A_1'}A_1}\qquad\forall\,n\in\{0,\ldots,r\},
\end{gathered}
}[EqTSPStoES]
which have the appropriate properties (trace, number of vector indices, etc.) as expected for irreducible representations in position space (see Appendix \ref{SecPtoE}).

The modification from the position space metric to the metric $\A_{12}$ is straightforward from arguments related to the proper removal of traces.  The change from the position space epsilon tensor to $\epsilon_{12}$ is warranted since the number of free vector indices must match to generate a proper map between position space tensor structures and embedding space tensor structures.\footnote{Another argument why the epsilon tensor must have two indices contracted with the two embedding space coordinates comes from the form of the $\Gamma$-matrices in embedding space.  Indeed, their form demands that the dummy indices are contracted such that they sum through some of the position space values, effectively forcing contractions with the two embedding space coordinates.}$^,$\footnote{The order of the embedding space coordinates in $\epsilon_{12}$ is also important since $\epsilon_{21}=-\epsilon_{12}$, in contrast with $\A_{12}$ and $\Gamma_{12}$.  This observation is related to the asymmetry in \eqref{EqTensorStruct} [the OPE \eqref{EqSOPE} is clearly asymmetric] and is necessary for introducing the proper contragredient-reflected representations in the tensor structures which lead to singlets.  The contragredient-reflected representation $\boldsymbol{N}^{CR}$ of the irreducible representation $\boldsymbol{N}$ is the contragredient representation of $\boldsymbol{N}$ in odd dimensions or the contragredient representation of $\boldsymbol{N}$ with the last two Dynkin indices (responsible for distinguishing the two inequivalent spinor representations) interchanged in even dimensions.  This definition matches the contractions implied by the tensor structures taking into account the order of the embedding space coordinates as in \eqref{EqTensorStruct}.  In Lorentzian signature, the contragredient-reflected representation is the conjugate representation.}  This last argument can also be used to determine the range of allowed values for $n$ for the antisymmetric spinor quantities $\Gamma_{12}$.\footnote{A careful study, not reproduced here, using a complete basis of matrices as in Appendix \ref{SecIrrep}, demonstrates that the case $n=r_E$ is superfluous.}

%%%%%%%%%%%%%%%%%%%%%%%%%%%%%%%%%%%%%%%%%%%%%%%%%%

\subsection{Homogeneity and the Differential Operator}

Having constructed a differential operator that has the appropriate transformation properties under the conformal group, satisfies the supplemental condition of transversality, and does not have free dummy indices left uncontracted, it is now time to use the remaining supplemental condition, \textit{i.e.} homogeneity \eqref{EqSupp}, to constrain completely the differential operator appearing in the OPE \eqref{EqSOPE}.

The dot product $\ee{1}{2}{}$ is the only scalar quantity remaining\footnote{This is obvious from the fact that the OPE \eqref{EqSOPE} is a function of only two embedding space coordinates.} that could change the homogeneity properties of the differential operator $\DOPE{a}{i}{j}{k}(\eta_1,\eta_2)$ without spoiling any other properties.  Extending the previous discussion, one has
\eqn{\DOPE{a}{i}{j}{k}(\eta_1,\eta_2)\propto\frac{1}{\ee{1}{2}{p_{ijk}}}(\mathcal{T}_{12}^{\boldsymbol{N}_i}\Gamma)(\mathcal{T}_{21}^{\boldsymbol{N}_j}\Gamma)\cdot\tOPE{a}{i}{j}{k}{1}{2}\cdot\D_{12}^{(d,h_{ijk}-n_a/2,n_a)}(\mathcal{T}_{12\boldsymbol{N}_k}\Gamma)*,}
where the power $p_{ijk}$ (as well as $h_{ijk}$) depends on the different quasi-primary operators appearing in the OPE.  To compute the power $p_{ijk}$ and the parameter $h_{ijk}$, it is necessary to determine the degrees of homogeneity of the different quantities appearing in the differential operator $\DOPE{a}{i}{j}{k}(\eta_1,\eta_2)$.

From the definitions \eqref{EqHPP} and \eqref{EqHPOPE}, one has
\eqn{\Theta_i(\mathcal{T}_{ij}^{\boldsymbol{N}^\mathcal{O}}\Gamma)=\frac{1}{2}(S_\mathcal{O}-\xi_\mathcal{O})(\mathcal{T}_{ij}^{\boldsymbol{N}^\mathcal{O}}\Gamma),\qquad\Theta_j(\mathcal{T}_{ij}^{\boldsymbol{N}^\mathcal{O}}\Gamma)=-\frac{1}{2}\left(S_\mathcal{O}-\xi_\mathcal{O}\right)(\mathcal{T}_{ij}^{\boldsymbol{N}^\mathcal{O}}\Gamma),}
where $S_\mathcal{O}$ is the ``spin'' of the irreducible representation $\boldsymbol{N}^\mathcal{O}$ and $\xi_\mathcal{O}=S_\mathcal{O}-\lfloor S_\mathcal{O}\rfloor$.  In other words, $\xi_\mathcal{O}=0$ for bosonic representations while $\xi_\mathcal{O}=1/2$ for fermionic representations.

Since the tensor structures are built with the help of the group-theoretical quantities \eqref{EqTSPStoES}, it is clear that the tensor structures have vanishing degrees of homogeneity with respect to both embedding space coordinates.

For these reasons and the fact that
\eqn{
\begin{gathered}
\Theta_i\D_{ij}^{(d,h-n/2,n)}=h\D_{ij}^{(d,h-n/2,n)},\qquad\Theta_j\D_{ij}^{(d,h-n/2,n)}=-h\D_{ij}^{(d,h-n/2,n)},\\
\Theta_i\ee{i}{j}{}=\Theta_j\ee{i}{j}{}=\ee{i}{j}{},
\end{gathered}
}
the supplemental conditions of homogeneity \eqref{EqSupp} imply that $p_{ijk}=\frac{1}{2}(\tau_i+\tau_j-\tau_k)$ and $h_{ijk}=-\frac{1}{2}(\chi_i-\chi_j+\chi_k)$ with the twist $\tau_{\mathcal{O}}=\Delta_{\mathcal{O}}-S_{\mathcal{O}}$ and $\chi_{\mathcal{O}}=\Delta_{\mathcal{O}}-\xi_{\mathcal{O}}$.

Hence, the differential operator appearing in the OPE \eqref{EqSOPE} can be chosen to be
\eqn{
\begin{gathered}
\DOPE{a}{i}{j}{k}(\eta_1,\eta_2)=\frac{1}{\ee{1}{2}{p_{ijk}}}(\mathcal{T}_{12}^{\boldsymbol{N}_i}\Gamma)(\mathcal{T}_{21}^{\boldsymbol{N}_j}\Gamma)\cdot\tOPE{a}{i}{j}{k}{1}{2}\cdot\D_{12}^{(d,h_{ijk}-n_a/2,n_a)}(\mathcal{T}_{12\boldsymbol{N}_k}\Gamma)*,\\
p_{ijk}=\frac{1}{2}(\tau_i+\tau_j-\tau_k),\qquad h_{ijk}=-\frac{1}{2}(\chi_i-\chi_j+\chi_k),\\
\tau_{\mathcal{O}}=\Delta_{\mathcal{O}}-S_{\mathcal{O}},\qquad\chi_{\mathcal{O}}=\Delta_{\mathcal{O}}-\xi_{\mathcal{O}},\qquad\xi_{\mathcal{O}}=S_{\mathcal{O}}-\lfloor S_{\mathcal{O}}\rfloor.
\end{gathered}
}[EqDOPEFinal]
Here the proportionality factor is taken to be $1$.  Moreover, the extra powers of $\ee{i}{j}{}$ introduced in the half-projectors \eqref{EqHPP} and \eqref{EqHPOPE} were selected to simplify the differential operator \eqref{EqDOPEFinal}.  For example, the extra powers lead to the appearance of the spin in the homogeneity conditions which is needed to get from the conformal dimension to the twist relevant for homogeneity.

%%%%%%%%%%%%%%%%%%%%%%%%%%%%%%%%%%%%%%%%%%%%%%%%%%

\subsection{Operator Product Expansion in Embedding Space}

Combining the result \eqref{EqDOPEFinal} for the differential operator, the definitions \eqref{EqHPP} and \eqref{EqHPOPE} for the half-projectors, as well as the relevant Lorentz group tensor structures \eqref{EqTensorStruct} built from the group-theoretical quantities \eqref{EqTSPStoES}, the OPE \eqref{EqSOPE} is given explicitly by
\eqn{
\begin{gathered}
\Op{i}{1}\Op{j}{2}=(\mathcal{T}_{12}^{\boldsymbol{N}_i}\Gamma)(\mathcal{T}_{21}^{\boldsymbol{N}_j}\Gamma)\cdot\sum_k\sum_{a=1}^{N_{ijk}}\frac{\cOPE{a}{i}{j}{k}\tOPE{a}{i}{j}{k}{1}{2}}{\ee{1}{2}{p_{ijk}}}\cdot\D_{12}^{(d,h_{ijk}-n_a/2,n_a)}(\mathcal{T}_{12\boldsymbol{N}_k}\Gamma)*\Op{k}{2},\\
p_{ijk}=\frac{1}{2}(\tau_i+\tau_j-\tau_k),\qquad h_{ijk}=-\frac{1}{2}(\chi_i-\chi_j+\chi_k),\\
\tau_{\mathcal{O}}=\Delta_{\mathcal{O}}-S_{\mathcal{O}},\qquad\chi_{\mathcal{O}}=\Delta_{\mathcal{O}}-\xi_{\mathcal{O}},\qquad\xi_{\mathcal{O}}=S_{\mathcal{O}}-\lfloor S_{\mathcal{O}}\rfloor,
\end{gathered}
}[EqOPE]
which is the most practical OPE found by the authors.  Again, the generalization to even dimensions is straightforward.

In \eqref{EqOPE}, the non-negative integer $n_a$ is the Dynkin index of the specific symmetric-traceless irreducible representation $n_a\boldsymbol{e}_1$ appearing in the tensor product decomposition of $\boldsymbol{N}_i\otimes\boldsymbol{N}_j\otimes\boldsymbol{N}_k$ associated with the tensor structure $\tOPE{a}{i}{j}{k}{1}{2}$ indexed by $a\in\{1,\ldots,N_{ijk}\}$.  This is the symmetric-traceless irreducible representation necessary to contract the Lorentz group irreducible representations into a singlet, being careful with the order of the embedding space coordinates, as in \eqref{EqTensorStruct}.  Finally, the sum over $k$ runs over all independent quasi-primary operators (including the conjugate operators if they are independent).  Meanwhile, $N_{ijk}$ is the number of tensor structures with OPE coefficients $\cOPE{a}{i}{j}{k}$ which corresponds to the number of symmetric-traceless irreducible representations in the same tensor product decomposition, \textit{i.e.} $\boldsymbol{N}_i\otimes\boldsymbol{N}_j\otimes\boldsymbol{N}_k$, again being careful with the order of the embedding space coordinates, as in \eqref{EqTensorStruct}.

By construction, the OPE in embedding space \eqref{EqOPE} is naturally covariant under the full conformal group $SO(d_E)\equiv SO(2,d)$ and it satisfies appropriate consistency conditions, from \textit{e.g.} the supplemental conditions \eqref{EqSupp} of homogeneity and transversality.  Mathematically, the OPE \eqref{EqOPE} is conformally invariant since it satisfies the following identity
\eqna{
[L_{AB},\Op{i}{1}\Op{j}{2}]&=(\mathcal{T}_{12}^{\boldsymbol{N}_i}\Gamma)(\mathcal{T}_{21}^{\boldsymbol{N}_j}\Gamma)\cdot\sum_k\sum_{a=1}^{N_{ijk}}\frac{\cOPE{a}{i}{j}{k}\tOPE{a}{i}{j}{k}{1}{2}}{\ee{1}{2}{p_{ijk}}}\cdot\D_{12}^{(d,h_{ijk}-n_a/2,n_a)}(\mathcal{T}_{12\boldsymbol{N}_k}\Gamma)\\
&\phantom{=}\qquad*[L_{AB},\Op{k}{2}].
}[EqLABOPE]
The identity \eqref{EqLABOPE} is easily proved by noticing that
\eqn{
\begin{gathered}
{}[L_{AB},\Op{k}{2}]=-\mathcal{L}_{2AB}\Op{k}{2}-(\Sigma_{AB}\mathcal{O}_k)(\eta_2)=-\mathcal{L}_{1AB}\Op{k}{2}-\mathcal{L}_{2AB}\Op{k}{2}-(\Sigma_{AB}\mathcal{O}_k)(\eta_2),\\
[\mathcal{L}_{1AB}+\mathcal{L}_{2AB},\D_{12}^{(d,h,n)C_1\cdots C_n}]=-(S_{AB}\D_{12}^{(d,h,n)})^{C_1\cdots C_n},\qquad\mathcal{L}_{AB}\eta^C-\eta^C\mathcal{L}_{AB}=-(S_{AB}\eta)^C,
\end{gathered}
}
where $\mathcal{L}_{1AB}\Op{k}{2}=0$ is added in the first equation to allow the use of the subsequent commutation relations.  This seemingly trivial property is useful in the study of $M$-point correlation functions since, already at the level of the OPE, it implies that for a Casimir operator $C_{2n}$ or $C_\epsilon$ acting on the left-hand side of the OPE each quasi-primary operator $\Op{k}{2}$ on the right-hand side of the OPE is multiplied by its eigenvalue $c_{2n}^k$ or $c_\epsilon^k$ for the corresponding Casimir operator.

In \eqref{EqOPE} the dummy indices on the half-projectors and the differential operators are contracted through the tensor structures $\tOPE{a}{i}{j}{k}{1}{2}$ (contractions denoted by the $\cdot$ symbol) while the embedding space spinor indices are free for the half-projectors of $\Op{i}{1}$ and $\Op{j}{2}$ but fully contracted with $\Op{k}{2}$ for the last half-projector (contractions denoted by the $*$ symbol).  Concretely, the OPE \eqref{EqOPE} in all its glory is written as
\eqna{
\Op{ia_1\cdots a_{n_i}}{1}\Op{jb_1\cdots b_{n_j}}{2}&=(\mathcal{T}_{12}^{\boldsymbol{N}_i}\Gamma)_{a_1\cdots a_{n_i}}^{\{Aa\}}(\mathcal{T}_{21}^{\boldsymbol{N}_j}\Gamma)_{b_1\cdots b_{n_j}}^{\{Bb\}}\\
&\phantom{=}\qquad\cdot\sum_k\sum_{a=1}^{N_{ijk}}\frac{\cOPE{a}{i}{j}{k}(\tOPE{a}{i}{j}{k}{1}{2})_{\{aA\}\{bB\}}^{\phantom{\{aA\}\{bB\}}{\{Cc\}\{D\}}}}{\ee{1}{2}{p_{ijk}}}\\
&\phantom{=}\qquad\cdot\D_{12\{D\}}^{(d,h_{ijk}-n_a/2,n_a)}(\mathcal{T}_{12\boldsymbol{N}_k}\Gamma)_{\{cC\}}^{c_{n_k}\cdots c_1}*\Op{kc_1\cdots c_{n_k}}{2},
}[EqOPEindices]
with an equivalent equation for the even-dimensional case.  In \eqref{EqOPEindices} the four different sets of dummy indices contracted into a Lorentz group singlet through the tensor structures are $\{Aa\}$, $\{Bb\}$, $\{cC\}$ and $\{D\}$ for the quasi-primary operators $\Op{i}{1}$, $\Op{j}{2}$, $\Op{k}{2}$ and the differential operator $\D_{12}^{(d,h_{ijk}-n_a/2,n_a)}$, respectively.

Several remarks are in order.  First of all, the scalar part of differential operator $\D_{12}^{(d,h_{ijk}-n_a/2,n_a)}$ is applied $2h_{ijk}-n_a$ times on the quasi-primary operator $\Op{k}{2}$ where $h_{ijk}\in\mathbb{R}$.  This is necessary due to the non-compactness of the conformal group.  This could potentially lead to issues deriving conformal blocks yet from the results obtained in Section \ref{SecDO} it is clear that it is not a problem.

Second, although the OPE is clearly asymmetric, it was mentioned that the OPE \eqref{EqOPE} should not depend on the choice of the embedding space coordinate for the quasi-primary operator on the right-hand side.  This fact is not obvious from the OPE \eqref{EqOPE} and it leads to relations between the different tensor structures.

Third, the OPE \eqref{EqOPE} is completely known up to the OPE coefficients $\cOPE{a}{i}{j}{k}$ and the tensor structures $\tOPE{a}{i}{j}{k}{1}{2}$.  In fact, the latter can be seen to encode all the information about the Lorentz group irreducible representations of the quasi-primary operators.  They are built directly from the equivalent tensor structures in position space with the proper modifications as in \eqref{EqTSPStoES} but following \eqref{EqTensorStruct}.  Hence, for specific choices of the quasi-primary operators in irreducible representations $\boldsymbol{N}_i$, $\boldsymbol{N}_j$ and $\boldsymbol{N}_k$, respectively, group-theoretical arguments allow the explicit construction of all the tensor structures $\tOPE{a}{i}{j}{k}{1}{2}$.  As a consequence, only the OPE coefficients $\cOPE{a}{i}{j}{k}$ are unknown in the OPE \eqref{EqOPE} and those might be constrained from conformal bootstrap arguments as described later.

The last remark suggests interesting properties for the tensor structures.  For example, for a fixed set of irreducible representations $\boldsymbol{N}_i$, $\boldsymbol{N}_j$ and $\boldsymbol{N}_k$, the tensor structures form a basis.  Indeed, they take those three general irreducible representations and add an extra symmetric-traceless irreducible representation to form singlets (again being careful about the order of the embedding space coordinates).  Being a basis, they can be orthonormalized such as
\eqn{(\tOPE{a}{i}{j}{k}{1}{2},\tOPE{b}{i}{j}{k}{1}{2})=\delta_{ab},}[EqTSOrtho]
where it is understood that the indices in the inner product $(\cdot,\cdot)$ are contracted appropriately (the second tensor structure should have all its irreducible representations contragredient-reflected to insure proper contractions).  In fact, there is a grading for the associated vector space which is indexed by the set of possible integers $\{n_a|a=1,\ldots,N_{ijk}\}$ related to the different allowed symmetric-traceless irreducible representations $n_a\boldsymbol{e}_1$.  From the existence of the grading, the orthonormality condition \eqref{EqTSOrtho} is trivially satisfied if $n_a\neq n_b$.

As a vector space, one has for ${}_at'={}_a(Ot)$ (dropping all fixed indices on the tensor structures for simplicity) that the product \eqref{EqTSOrtho} satisfies
\eqn{({}_at',{}_bt')=\left({}_a(Ot),{}_{b}(Ot)\right)=(OO^T)_{ab}=\delta_{ab},}
which is still equal to $\delta_{ab}$ for any orthogonal matrix $O$ which satisfies the grading.  Obviously, this statement is nothing else than a simple rotation of the orthonormal basis of tensor structures into another orthonormal basis of tensor structures.  This observation might be useful to simplify computations of conformal blocks by choosing a convenient basis, assuming such a basis exists.  It is also worth mentioning that under such a change of basis, the OPE coefficients transform as ${}_ac'={}_a(cO^T)$.

As a final comment regarding the tensor structures, it is important to realize that the tensor structures which are well defined for the OPE will not necessarily be so for the three-point correlation functions due to the appearance of an extra embedding space coordinate.  Clearly, the tensor structures in the OPE \eqref{EqOPE} depend on only two embedding space coordinates and can be obtained from the position space tensor structures by the appropriate modifications \eqref{EqTSPStoES}, in particular by changing the metric to $\A_{12}$.  It could perhaps be of interest to obtain three-point correlation functions with well-behaved tensor structures directly from the OPE, but it is clear that if it is not already straightforward, then it implies combining different OPE tensor structures together.  In other words, a simple change of the differential operator $\D_{12}^{(d,h_{ijk}-n_a/2,n_a)}$ appearing in the OPE \eqref{EqOPE} is not sufficient.  Indeed, the only plausible change to the differential operator one can make is to introduce fewer free derivatives (\textit{i.e.} decreasing $n_a$) but compensate with group-theoretical invariants like the metric $g$ or $\A_{12}$.  However, such differential operators vanish once contracted with the OPE tensor structures due to their tracelessness property.  On the other hand, combining OPE tensor structures to get three-point correlation functions with well-behaved tensor structures might not be possible while at the same time keeping the orthonormality property \eqref{EqTSOrtho} of the OPE tensor structures.  Therefore it is not clear if such an endeavor has any merit.

It is worth pointing out that the OPE \eqref{EqOPE} is a choice.  Any change in the differential operator, the ordering of the embedding space coordinates and the differential operators, or the form of the half-projectors are allowed as long as the covariance under the conformal group and the two supplemental conditions of homogeneity and transversality are satisfied.  From the analysis in section \eqref{EqDO} and the following sections, it is quite clear though that the choice \eqref{EqOPE} is very powerful, allowing complete analytic control of the computation of $M$-point correlation functions.

%%%%%%%%%%%%%%%%%%%%%%%%%%%%%%%%%%%%%%%%%%%%%%%%%%

\subsection{Properties of the Half-Projectors}

With the computation of $M$-point correlation functions in mind, it is convenient now to analyze some of the properties of the half-projectors \eqref{EqHPP} and \eqref{EqHPOPE}.

Since half-projectors in general irreducible representations $\boldsymbol{N}$ satisfy $\mathcal{T}_{\boldsymbol{N}}*\mathcal{T}^{\boldsymbol{N}}=\hat{\mathcal{P}}^{\boldsymbol{N}}$ where the star product corresponds to contractions of the spinor indices,\footnote{From Appendix \ref{SecIrrep}, the hatted projection operators for the defining irreducible representations in odd dimensions are
\eqn{
\begin{gathered}
(\hat{\mathcal{P}}^{\boldsymbol{e}_r})_\alpha^{\phantom{\alpha}\beta}=\delta_\alpha^{\phantom{\alpha}\beta},\qquad(\hat{\mathcal{P}}^{\boldsymbol{e}_{i\neq r}})_{\mu_i\cdots\mu_1}^{\phantom{\mu_i\cdots\mu_1}\nu_1\cdots\nu_i}=\delta_{[\mu_1}^{\phantom{[\mu_1}\nu_1}\cdots\delta_{\mu_i]}^{\phantom{\mu_i]}\nu_i},\qquad(\hat{\mathcal{P}}^{2\boldsymbol{e}_r})_{\mu_r\cdots\mu_1}^{\phantom{\mu_r\cdots\mu_1}\nu_1\cdots\nu_r}=\delta_{[\mu_1}^{\phantom{[\mu_1}\nu_1}\cdots\delta_{\mu_r]}^{\phantom{\mu_r]}\nu_r},
\end{gathered}
}
while in even dimensions they are
\eqn{
\begin{gathered}
(\hat{\mathcal{P}}^{\boldsymbol{e}_{r-1}})_\alpha^{\phantom{\alpha}\beta}=\delta_\alpha^{\phantom{\alpha}\beta},\qquad(\hat{\mathcal{P}}^{\boldsymbol{e}_r})_{\tilde{\alpha}}^{\phantom{\tilde{\alpha}}\tilde{\beta}}=\delta_{\tilde{\alpha}}^{\phantom{\tilde{\alpha}}\tilde{\beta}},\qquad(\hat{\mathcal{P}}^{\boldsymbol{e}_{i\neq r-1,r}})_{\mu_i\cdots\mu_1}^{\phantom{\mu_i\cdots\mu_1}\nu_1\cdots\nu_i}=\delta_{[\mu_1}^{\phantom{[\mu_1}\nu_1}\cdots\delta_{\mu_i]}^{\phantom{\mu_i]}\nu_i},\\
(\hat{\mathcal{P}}^{\boldsymbol{e}_{r-1}+\boldsymbol{e}_r})_{\mu_{r-1}\cdots\mu_1}^{\phantom{\mu_{r-1}\cdots\mu_1}\nu_1\cdots\nu_{r-1}}=\delta_{[\mu_1}^{\phantom{[\mu_1}\nu_1}\cdots\delta_{\mu_{r-1}]}^{\phantom{\mu_{r-1}]}\nu_{r-1}},\\
(\hat{\mathcal{P}}^{2\boldsymbol{e}_{r-1}})_{\mu_r\cdots\mu_1}^{\phantom{\mu_r\cdots\mu_1}\nu_1\cdots\nu_r}=\frac{1}{2}\delta_{[\mu_1}^{\phantom{[\mu_1}\nu_1}\cdots\delta_{\mu_r]}^{\phantom{\mu_r]}\nu_r}+(-1)^r\frac{\mathscr{K}}{2r!}\epsilon_{\mu_1\cdots\mu_r}^{\phantom{\mu_1\cdots\mu_r}\nu_r\cdots\nu_1},\\
(\hat{\mathcal{P}}^{2\boldsymbol{e}_r})_{\mu_r\cdots\mu_1}^{\phantom{\mu_r\cdots\mu_1}\nu_1\cdots\nu_r}=\frac{1}{2}\delta_{[\mu_1}^{\phantom{[\mu_1}\nu_1}\cdots\delta_{\mu_r]}^{\phantom{\mu_r]}\nu_r}-(-1)^r\frac{\mathscr{K}}{2r!}\epsilon_{\mu_1\cdots\mu_r}^{\phantom{\mu_1\cdots\mu_r}\nu_r\cdots\nu_1}.
\end{gathered}
}
} one has the analogous relation in embedding space given by
\eqn{(\mathcal{T}_{ij\boldsymbol{N}}\Gamma)_{\{aA\}}*(\mathcal{T}_{jk}^{\boldsymbol{N}}\Gamma)^{\{Bb\}}=\frac{\ee{i}{j}{\frac{1}{2}(S-\xi)}}{\ee{j}{k}{\frac{1}{2}(S-\xi)}}\left(\frac{\eta_j\cdot\Gamma\,\hat{\mathcal{P}}_{ji}^{\boldsymbol{N}}\cdot\hat{\mathcal{P}}_{jk}^{\boldsymbol{N}}\,\eta_k\cdot\Gamma}{\ee{j}{k}{}}\right)_{\{aA\}}^{\phantom{\{aA\}}\{Bb\}}.}[EqTSTS]
Note that explicit $\eta\cdot\Gamma$-matrices in \eqref{EqTSTS} can be commuted through the hatted projection operators $\hat{\mathcal{P}}$ when the embedding space coordinates of the explicit $\eta\cdot\Gamma$-matrices and one of the embedding space coordinates of the hatted projection operators correspond.  From \eqref{EqTSTS} it is now clear how the pre-factors in the half-projectors \eqref{EqHPP} and \eqref{EqHPOPE} were chosen.

Furthermore, since
\eqn{(\hat{\mathcal{P}}_{ij}^{\boldsymbol{N}})_{\{aA\}}^{\phantom{\{aA\}}\{B'b'\}}[(C_\Gamma^{-1})_{b'b}]^{2\xi}(g_{B'B})^{n_v}=[(C_\Gamma^{-1})_{ab'}]^{2\xi}(g_{AB'})^{n_v}(\hat{\mathcal{P}}_{ji}^{\boldsymbol{N}^{CR}})_{\{bB\}}^{\phantom{\{bB\}}\{B'b'\}},}
where $\boldsymbol{N}^{CR}$ is the contragredient-reflected representation of $\boldsymbol{N}$, the following product is given by
\eqna{
(\mathcal{T}_{ij\boldsymbol{N}}\Gamma)_{\{aA\}}*(\mathcal{T}_{jk}^{\boldsymbol{N}}\Gamma)\cdot(\mathcal{T}_{kj}^{\boldsymbol{N}^{CR}}\Gamma)&=\frac{\ee{i}{j}{\frac{1}{2}(S-\xi)}}{\ee{j}{k}{\frac{1}{2}(S-\xi)}}\left(\frac{\eta_j\cdot\Gamma\,\hat{\mathcal{P}}_{ji}^{\boldsymbol{N}}\cdot\hat{\mathcal{P}}_{jk}^{\boldsymbol{N}}\,\eta_k\cdot\Gamma}{\ee{j}{k}{}}\right)_{\{aA\}}^{\phantom{\{aA\}}\{Bb\}}\\
&\phantom{=}\qquad\times[(C_\Gamma^{-1})_{bb'}]^{2\xi}(g_{BB'})^{n_v}(\mathcal{T}_{kj}^{\boldsymbol{N}^{CR}}\Gamma)^{\{B'b'\}}\\
&=\frac{\ee{i}{j}{\frac{1}{2}(S-\xi)}}{\ee{j}{k}{\frac{1}{2}(S-\xi)}}[2(C_\Gamma^{-1})_{ab}]^{2\xi}(g_{AB})^{n_v}(\mathcal{T}_{kj}^{\boldsymbol{N}^{CR}}\Gamma\cdot\hat{\mathcal{P}}_{ij}^{\boldsymbol{N}^{CR}})^{\{Bb\}}.
}[EqTSTSTS]

Therefore, in an index-free notation the two products of half-projectors \eqref{EqTSTS} and \eqref{EqTSTSTS} are given by
\eqn{
\begin{gathered}
(\mathcal{T}_{ij\boldsymbol{N}}\Gamma)*(\mathcal{T}_{jk}^{\boldsymbol{N}}\Gamma)=\frac{\ee{i}{j}{\frac{1}{2}(S-\xi)}}{\ee{j}{k}{\frac{1}{2}(S-\xi)}}\left(\frac{\eta_j\cdot\Gamma\,\hat{\mathcal{P}}_{ji}^{\boldsymbol{N}}\cdot\hat{\mathcal{P}}_{jk}^{\boldsymbol{N}}\,\eta_k\cdot\Gamma}{\ee{j}{k}{}}\right),\\
(\mathcal{T}_{ij\boldsymbol{N}}\Gamma)*(\mathcal{T}_{jk}^{\boldsymbol{N}}\Gamma)\cdot(\mathcal{T}_{kj}^{\boldsymbol{N}^{CR}}\Gamma)=\frac{\ee{i}{j}{\frac{1}{2}(S-\xi)}}{\ee{j}{k}{\frac{1}{2}(S-\xi)}}(\mathcal{T}_{kj}^{\boldsymbol{N}^{CR}}\Gamma)\cdot(\hat{\mathcal{P}}_{ij}^{\boldsymbol{N}^{CR}})[2(C_\Gamma^{-T})]^{2\xi}(g)^{n_v},
\end{gathered}
}[EqTSProd]
where all products are written using the matrix convention of Appendix \ref{SecPtoE}.  The products \eqref{EqTSProd} appear naturally when the OPE \eqref{EqOPE} is used in the computation of $M$-point correlation functions.  Moreover, the identities \eqref{EqTSProd} are quite useful in carrying out explicit computations.

In fact, the half-projectors \eqref{EqHPP} satisfy another important identity, given by
\eqn{(\mathcal{T}_{ij}^{\boldsymbol{N}}\Gamma)=\frac{\ee{i}{k}{\frac{1}{2}(S-\xi)}}{\ee{i}{j}{\frac{1}{2}(S-\xi)}}(\mathcal{T}_{ik}^{\boldsymbol{N}}\Gamma)\cdot\left(\frac{\eta_i\cdot\Gamma\,\hat{\mathcal{P}}_{ij}^{\boldsymbol{N}}\,\eta_j\cdot\Gamma}{2^{2\xi}\ee{i}{j}{}}\right),}[EqTkjTkiOPE]
which can be proven from the different properties of the embedding space quantities shown in Appendix \ref{SecPtoE}, leading to
\eqna{
(\mathcal{T}_{ij\boldsymbol{N}}\Gamma)*(\mathcal{T}_{jk}^{\boldsymbol{N}}\Gamma)\cdot(\mathcal{T}_{kj}^{\boldsymbol{N}^{CR}}\Gamma)&=\frac{\ee{i}{j}{\frac{1}{2}(S-\xi)}\ee{i}{k}{\frac{1}{2}(S-\xi)}}{\ee{j}{k}{S-\xi}}\\
&\phantom{=}\qquad\times(\mathcal{T}_{ki}^{\boldsymbol{N}^{CR}}\Gamma)\cdot\left(\frac{\eta_k\cdot\Gamma\,\hat{\mathcal{P}}_{kj}^{\boldsymbol{N}^{CR}}\cdot\hat{\mathcal{P}}_{ij}^{\boldsymbol{N}^{CR}}\,\eta_j\cdot\Gamma}{\ee{j}{k}{}}\right)[(C_\Gamma^{-T})]^{2\xi}(g)^{n_v}.
}[EqTSProdOPE]
With the help of \eqref{EqTkjTkiOPE}, the second embedding space coordinate of any half-projector, which is of no importance for transversality, can be replaced by another embedding space coordinate, allowing the replacement of the half-projector in the second equation of \eqref{EqTSProd} to a different half-projector in \eqref{EqTSProdOPE}.  This property is useful in disentangling the Lorentz contributions from the OPE differential operators in the computations of $M$-point correlation functions.

%%%%%%%%%%%%%%%%%%%%%%%%%%%%%%%%%%%%%%%%%%%%%%%%%%
%%%%%%%%%%%%%%%%%%%%%%%%%%%%%%%%%%%%%%%%%%%%%%%%%%

\section{\texorpdfstring{$M$}{M}-Point Correlation Functions}\label{SecnptCF}

Now that a convenient OPE has been defined, from which analytical results can be extracted, it is finally time to investigate $M$-point correlation functions computed from the OPE.  This section explores the implications of the OPE for $M$-point correlation functions as well as the associativity constraints (\textit{i.e.} the conformal bootstrap) obtained from the $M$-point correlation functions.

%%%%%%%%%%%%%%%%%%%%%%%%%%%%%%%%%%%%%%%%%%%%%%%%%%

\subsection{Conformal Algebra and Correlation Functions}\label{SSecConfAlgCF}

Before attacking the computation of $M$-point correlation functions with the help of the OPE, it is of interest to investigate the action of Casimir operators \eqref{EqCasimirs} on $M$-point correlation functions.  From the knowledge that the OPE is conformally covariant [see \eqref{EqLABOPE}], it is already known that the action of Casimir operators on $M$-point correlation functions is well defined.

Indeed, it is straightforward to verify that correlation functions are conformally invariant (up to possible contact terms) due to the invariance of the vacuum, \textit{i.e.}
\eqna{
\sum_{j=1}^ML_{jAB}\Vev{\Op{i_1}{1}\cdots\Op{i_M}{M}}&\equiv-\sum_{j=1}^M(\mathcal{L}_{jAB}+\Sigma_{jAB})\Vev{\Op{i_1}{1}\cdots\Op{i_M}{M}}\\
&=\sum_{j=1}^M\Vev{\Op{i_1}{1}\cdots[L_{AB},\Op{i_j}{j}]\cdots\Op{i_M}{M}}\\
&=\Vev{[L_{AB},\Op{i_1}{1}\cdots\Op{i_M}{M}]}=0.
}
Therefore, for any correlation function one has
\eqna{
\prod_{k=1}^m\sum_{j\in H}L_{jA_kB_k}\Vev{\Op{i_1}{1}\cdots\Op{i_M}{M}}&=(-1)^m\prod_{k=1}^m\sum_{j\in H^c}L_{jA_{m+1-k}B_{m+1-k}}\Vev{\Op{i_1}{1}\cdots\Op{i_M}{M}}\\
&=\prod_{k=1}^m\sum_{j\in H^c}L_{jB_{m+1-k}A_{m+1-k}}\Vev{\Op{i_1}{1}\cdots\Op{i_M}{M}},
}
where $H\subseteq G=\{1,\ldots,N\}$ and $H^c$ is the complement of the subset $H$ in $G$.  This identity implies
\eqn{\left|\left(\sum_{j\in H}L_j\right)^m\right|\Vev{\Op{i_1}{1}\cdots\Op{i_M}{M}}=\left|\left(\sum_{j\in H^c}L_j\right)^m\right|\Vev{\Op{i_1}{1}\cdots\Op{i_M}{M}},}[EqC2n]
for any integer $m$ and
\eqna{
&\epsilon^{A_1\cdots A_{d_E}}\sum_{j\in H}L_{jA_1A_2}\cdots\sum_{j\in H}L_{jA_{d_E-1}A_{d_E}}\Vev{\Op{i_1}{1}\cdots\Op{i_M}{M}}\\
&\qquad=(-1)^{d_E/2}\epsilon^{A_1\cdots A_{d_E}}\sum_{j\in H^c}L_{jA_1A_2}\cdots\sum_{j\in H^c}L_{jA_{d_E-1}A_{d_E}}\Vev{\Op{i_1}{1}\cdots\Op{i_M}{M}},
}[EqCepsilon]
for $d_E$ even.

The identities \eqref{EqC2n} and \eqref{EqCepsilon} simply state that the action of a Casimir operator on a subset of the quasi-primary operators appearing in a correlation function is equal (up to a possible minus sign) to the action of the same Casimir operator on the complement of the quasi-primary operators.  The possible minus sign in $C_\epsilon$ \eqref{EqCepsilon} appears as a consequence of the contragredient-reflected property of spinors in even dimensions.  Indeed, for $d_E=4n+2$ and $n\in\mathbb{N}$ the two irreducible spinor (Weyl) representations are contragredient-reflected of each other while for $d_E=4n$ and $n\in\mathbb{N}$ the two irreducible spinor (Weyl) representations are self-contragredient-reflected (and vice versa in position space).

The discussion above also implies that the conformal blocks relevant to $M$-point correlation functions are eigenfunctions of the Casimir operators $C_{2n}$ and $C_\epsilon$ with eigenvalues $c_{2n}^\mathcal{O}$ and $c_\epsilon^\mathcal{O}$ for the exchanged quasi-primary operator $\mathcal{O}$, where $C\mathcal{O}=c^\mathcal{O}\mathcal{O}$.  This observation allows to compute and/or check explicit conformal blocks for $M$-point correlation functions.

%%%%%%%%%%%%%%%%%%%%%%%%%%%%%%%%%%%%%%%%%%%%%%%%%%

\subsection{\texorpdfstring{$M$}{M}-Point Correlation Functions from the OPE}

$M$-point correlation functions can be obtained from the OPE in many different ways.  From the OPE \eqref{EqOPE}, $M$-point correlation functions can be computed by recurrence as follows,
\eqna{
\Vev{\Op{i_1}{1}\cdots\Op{i_M}{M}}&=(-1)^{2\xi_{i_1}}\Vev{\Op{i_2}{2}\cdots\Op{i_M}{M}\Op{i_1}{1}},\\
&=(\mathcal{T}_{M1}^{\boldsymbol{N}_{i_M}}\Gamma)(\mathcal{T}_{1M}^{\boldsymbol{N}_{i_1}}\Gamma)\cdot\sum_k\sum_a(-1)^{2(\xi_{i_1}+\xi_k)}\frac{\cOPE{a}{i_M}{i_1}{k}\tOPE{a}{i_M}{i_1}{k}{M}{1}}{\ee{1}{M}{p_{i_Mi_1k}}}\\
&\phantom{=}\qquad\cdot\D_{M1}^{(d,h_{i_Mi_1k}-n_a/2,n_a)}(\mathcal{T}_{M1\boldsymbol{N}_k}\Gamma)*\Vev{\Op{k}{1}\Op{i_2}{2}\cdots\Op{i_{M-1}}{M-1}}.
}[EqCFOPERec]
Clearly, iterating \eqref{EqCFOPERec} $M-1$ times leads to $M$-point correlation functions written in terms of differential operators at the embedding space coordinate $\eta_1$ acting on one-point correlation functions at the same embedding space coordinate $\eta_1$,
\eqna{
\Vev{\Op{i_1}{1}\cdots\Op{i_M}{M}}&=(-1)^{2\xi_{i_1}}\left[\prod_{j=1}^{M-1}(\mathcal{T}_{M-j+1,1}^{\boldsymbol{N}_{i_{M-j+1}}}\Gamma)(\mathcal{T}_{1,M-j+1}^{\boldsymbol{N}_{k_{M-j+1}}}\Gamma)\right.\\
&\phantom{=}\qquad\left.\cdot\sum_{k_{M-j}}\sum_{a_{M-j}}\frac{\cOPE{a_{M-j}}{i_{M-j+1}}{k_{M-j+1}}{k_{M-j}}\tOPE{a_{M-j}}{i_{M-j+1}}{k_{M-j+1}}{k_{M-j}}{M-j+1,}{1}}{\ee{1}{M-j+1}{p_{M-j+1}}}\right.\\
&\phantom{=}\qquad\left.\cdot\D_{M-j+1,1}^{(d,h_{M-j}-n_{a_{M-j}}/2,n_{a_{M-j}})}(\mathcal{T}_{M-j+1,1\boldsymbol{N}_{k_{M-j}}}\Gamma)*\right]\Vev{\Op{k_1}{1}},
}[EqCFOPE]
where
\eqn{p_{M-j+1}=\frac{1}{2}(\tau_{i_{M-j+1}}+\tau_{k_{M-j+1}}-\tau_{k_{M-j}}),\qquad h_{M-j}=-\frac{1}{2}(\chi_{i_{M-j+1}}-\chi_{k_{M-j+1}}+\chi_{k_{M-j}}),}
and $k_M=i_1$.  In \eqref{EqCFOPE} the OPE is being used recursively with all differential operators acting on the embedding space coordinate $\eta_1$.  Moreover, it is understood that the terms in the product are ordered with the $j=1$ term acting last on the one-point correlation functions.

Since the identity operator $\1$ is a special quasi-primary operator with conformal dimension $\Delta_\1=0$ and spin $S_\1=0$, it is the only quasi-primary operator with non-vanishing vacuum expectation value $\vev{\1}=1$.  Hence, the only non-trivial one-point correlation function appearing in the iterated version \eqref{EqCFOPE} is $\vev{\1}$.

Therefore, using results like \eqref{EqK0}, \eqref{EqIbSoln}, \eqref{EqTkjTkiOPE} and \eqref{EqTSProdOPE}, it is straightforward to compute analytically all $M$-point correlation functions from \eqref{EqCFOPE} by simply extracting all the free embedding space coordinates $\eta_1$ and applying the differential operators.  In other words, it is now clear from \eqref{EqCFOPE} and the results of Sections \ref{SecDO}, \ref{SecOPE} and \ref{SecnptCF} that any $M$-point correlation function can be constructed recursively in terms of unknown OPE coefficients.  Furthermore, for $(M<5)$-point correlation functions, the results are directly obtained in terms of linear combinations of the functions $\bar{I}_{ij}^{(d,h,n;p_a)}$ \eqref{EqIb3} and $\bar{I}_{ij;k\ell}^{(d,h,n;\boldsymbol{p})}$ \eqref{EqIbSoln}, the latter being built from the $K$-function \eqref{EqK0}.  Indeed, it is obvious that two-point and three-point correlation functions are simple sums of powers of the products of embedding space coordinates at different points, leading to simple four-point correlation functions written in terms of $\bar{I}_{ij;k\ell}^{(d,h,n;\boldsymbol{p})}$.  This is however not the case for $(M>4)$-point correlation functions since the four-point correlation functions are now functions of the conformal cross-ratios \eqref{EqCR} through the $K$-function \eqref{EqK0}.

%%%%%%%%%%%%%%%%%%%%%%%%%%%%%%%%%%%%%%%%%%%%%%%%%%

\subsection{Two and Three-Point Correlation Functions from the OPE}

To give a general idea on how to use \eqref{EqCFOPE}, this section computes generically two and three-point correlation functions.  In the specific case of two-point correlation functions, \eqref{EqCFOPE} leads to
\eqna{
\Vev{\Op{i_1}{1}\Op{i_2}{2}}&=(-1)^{2\xi_{i_1}}(\mathcal{T}_{21}^{\boldsymbol{N}_{i_2}}\Gamma)(\mathcal{T}_{12}^{\boldsymbol{N}_{k_2}}\Gamma)\cdot\frac{\cOPE{}{i_2}{k_2}{\1}\tOPE{}{i_2}{k_2}{\1}{2}{1}}{\ee{1}{2}{p_2}}\\
&=(-1)^{2\xi_{i_1}}(\mathcal{T}_{21}^{\boldsymbol{N}}\Gamma)(\mathcal{T}_{12}^{\boldsymbol{N}^{CR}}\Gamma)\cdot\frac{\lambda_{\boldsymbol{N}}\cOPE{}{i_2}{i_1}{\1}\hat{\mathcal{P}}_{21}^{\boldsymbol{N}}}{\ee{1}{2}{\tau}},
}[EqCF2pt]
since the only quasi-primary operator with a non-vanishing one-point correlation function is the identity operator for which the power of the differential operator $\D_{21}^{(d,h_1-n_{a_1}/2,n_{a_1})}$ must vanish, forcing $n_{a_1}=h_1=0$ and $p_2=\tau_{i_1}=\tau_{i_2}\equiv\tau$.  Moreover, since the hatted projection operators contract only if the two irreducible representations are contragredient-reflected to each other, then the two quasi-primary operators in the correlation function must be in contragredient-reflected representations, \textit{i.e.} $\boldsymbol{N}_{i_2}=\boldsymbol{N}_{i_1}^{CR}\equiv\boldsymbol{N}$.  Furthermore, simple group-theoretical arguments dictate that the only possible tensor structure must be proportional to the hatted projection operator, $\tOPE{}{i_2}{i_1}{\1}{2}{1}=\lambda_{\boldsymbol{N}}\hat{\mathcal{P}}_{21}^{\boldsymbol{N}}$ where $\lambda_{\boldsymbol{N}}$ is a normalization constant, leading to the general two-point correlation function \eqref{EqCF2pt}.

From this discussion it is straightforward to then conclude that \eqref{EqCFOPE} can be simplified to
\eqna{
\Vev{\Op{i_1}{1}\cdots\Op{i_M}{M}}&=(-1)^{2\xi_{i_1}}\left[\prod_{j=1}^{M-2}(\mathcal{T}_{M-j+1,1}^{\boldsymbol{N}_{i_{M-j+1}}}\Gamma)(\mathcal{T}_{1,M-j+1}^{\boldsymbol{N}_{k_{M-j+1}}}\Gamma)\right.\\
&\phantom{=}\qquad\left.\cdot\sum_{k_{M-j}}\sum_{a_{M-j}}\frac{\cOPE{a_{M-j}}{i_{M-j+1}}{k_{M-j+1}}{k_{M-j}}\tOPE{a_{M-j}}{i_{M-j+1}}{k_{M-j+1}}{k_{M-j}}{M-j+1,}{1}}{\ee{1}{M-j+1}{p_{M-j+1}}}\right.\\
&\phantom{=}\qquad\left.\cdot\D_{M-j+1,1}^{(d,h_{M-j}-n_{a_{M-j}}/2,n_{a_{M-j}})}(\mathcal{T}_{M-j+1,1\boldsymbol{N}_{k_{M-j}}}\Gamma)*\right]\\
&\phantom{=}\qquad\times(\mathcal{T}_{21}^{\boldsymbol{N}_{i_2}}\Gamma)(\mathcal{T}_{12}^{\boldsymbol{N}_{k_2}}\Gamma)\cdot\frac{\cOPE{}{i_2}{k_2}{\1}\tOPE{}{i_2}{k_2}{\1}{2}{1}}{\ee{1}{2}{p_2}},
}[EqCFOPE2]
where $k_1=k_\1$ and $a_1=a_\1$ which imply $\tau_{k_1}=h_{k_1}=n_{a_1}=0$ and thus $\tau_{k_2}=\tau_{i_2}$ and $\boldsymbol{N}_{k_2}=\boldsymbol{N}_{i_2}^{CR}$ with $p_2=\tau_{i_2}$.

Proceeding to three-point correlation functions, \eqref{EqCFOPE2} becomes
\eqna{
\Vev{\Op{i_1}{1}\Op{i_2}{2}\Op{i_3}{3}}&=(-1)^{2\xi_{i_1}}(\mathcal{T}_{31}^{\boldsymbol{N}_{i_3}}\Gamma)(\mathcal{T}_{13}^{\boldsymbol{N}_{k_3}}\Gamma)\cdot\sum_{k_2}\sum_{a_2}\frac{\cOPE{a_2}{i_3}{k_3}{k_2}\tOPE{a_2}{i_3}{k_3}{k_2}{3}{1}}{\ee{1}{3}{p_3}}\\
&\phantom{=}\qquad\cdot\D_{31}^{(d,h_2-n_{a_2}/2,n_{a_2})}(\mathcal{T}_{31\boldsymbol{N}_{k_2}}\Gamma)*(\mathcal{T}_{21}^{\boldsymbol{N}_{i_2}}\Gamma)(\mathcal{T}_{12}^{\boldsymbol{N}_{k_2}}\Gamma)\cdot\frac{\cOPE{}{i_2}{k_2}{\1}\tOPE{}{i_2}{k_2}{\1}{2}{1}}{\ee{1}{2}{p_2}}.
}
Using the identity \eqref{EqTSProdOPE} leads to
\eqna{
\Vev{\Op{i_1}{1}\Op{i_2}{2}\Op{i_3}{3}}&=(-1)^{2\xi_{i_1}}(\mathcal{T}_{31}^{\boldsymbol{N}_{i_3}}\Gamma)(\mathcal{T}_{13}^{\boldsymbol{N}_{k_3}}\Gamma)\cdot\sum_{k_2}\sum_{a_2}\frac{\cOPE{a_2}{i_3}{k_3}{k_2}\tOPE{a_2}{i_3}{k_3}{k_2}{3}{1}}{\ee{1}{3}{p_3}}\\
&\phantom{=}\qquad\cdot\D_{31}^{(d,h_2-n_{a_2}/2,n_{a_2})}\frac{\lambda_{\boldsymbol{N}_{i_2}}\cOPE{}{i_2}{k_2}{\1}}{\ee{1}{2}{p_2}}\frac{\ee{1}{3}{\frac{1}{2}(S_{k_2}-\xi_{k_2})}\ee{2}{3}{\frac{1}{2}(S_{k_2}-\xi_{k_2})}}{\ee{1}{2}{S_{k_2}-\xi_{k_2}}}\\
&\phantom{=}\qquad\cdot(\mathcal{T}_{23}^{\boldsymbol{N}_{k_2}^{CR}}\Gamma)\cdot\left(\frac{\eta_2\cdot\Gamma\,\hat{\mathcal{P}}_{21}^{\boldsymbol{N}_{i_2}}\cdot\hat{\mathcal{P}}_{31}^{\boldsymbol{N}_{i_2}}\,\eta_1\cdot\Gamma}{\ee{1}{2}{}}\right)[(C_\Gamma^{-T})]^{2\xi_{i_2}}(g)^{n_v^{i_2}}\\
&=(-1)^{2\xi_{i_1}}(\mathcal{T}_{31}^{\boldsymbol{N}_{i_3}}\Gamma)(\mathcal{T}_{13}^{\boldsymbol{N}_{i_1}}\Gamma)(\mathcal{T}_{23}^{\boldsymbol{N}_{i_2}}\Gamma)\cdot\sum_{a_2}\frac{\lambda_{\boldsymbol{N}_{i_2}}\cCF{a_2}{i_3}{i_1}{i_2}\,\tCF{a_2}{i_3}{i_1}{i_2}{3}{1}}{\ee{1}{3}{p_3-\frac{1}{2}(S_{i_2}-\xi_{i_2})}}\\
&\phantom{=}\qquad\cdot\D_{31}^{(d,h_2-n_{a_2}/2,n_{a_2})}\frac{\ee{2}{3}{\frac{1}{2}(S_{i_2}-\xi_{i_2})}}{\ee{1}{2}{p_2+S_{i_2}-\xi_{i_2}}}\left(\frac{\eta_2\cdot\Gamma\,\hat{\mathcal{P}}_{21}^{\boldsymbol{N}_{i_2}}\cdot\hat{\mathcal{P}}_{31}^{\boldsymbol{N}_{i_2}}\,\eta_1\cdot\Gamma}{\ee{1}{2}{}}\right),
}[EqCF3pt]
where all quantities in \eqref{EqCF3pt} with explicit $k_2$ can be replaced by $i_2^{CR}$, including in the OPE coefficients and tensor structures, by redefining
\eqn{\sum_{k_2}\cOPE{}{i_2}{k_2}{\1}\,\cOPE{a_2}{i_3}{i_1}{k_2}=\cCF{a_2}{i_3}{i_1}{i_2},\qquad\tOPE{a_2}{i_3}{i_1}{i_2^{CR}}{3}{1}[(C_\Gamma^{-1})]^{2\xi_{i_2}}(g)^{n_v^{i_2}}=\tCF{a_2}{i_3}{i_1}{i_2}{3}{1},}
as new quantities.

At this point, it is sufficient to extract the free embedding space coordinates $\eta_1$ from the factor which is acted upon by the differential operator [the last line of \eqref{EqCF3pt}] and use $\bar{I}_{ij}^{(d,h,n;p_a)}$ \eqref{EqIb3} to get the explicit three-point correlation functions from \eqref{EqCF3pt}.

Obviously, since $\bar{I}_{ij}^{(d,h,n;p_a)}$ \eqref{EqIb3} is a simple sum of powers of products of embedding space coordinates, the four-point correlation functions are directly computable in terms of $\bar{I}_{ij;k\ell}^{(d,h,n;\boldsymbol{p})}$ \eqref{EqIbSoln}.  As mentioned before, the presence of the $K$-function starting at $M=4$ implies that $(M>4)$-point correlation functions are not given straightforwardly in terms of $\bar{I}_{ij;k\ell}^{(d,h,n;\boldsymbol{p})}$ \eqref{EqIbSoln}.  Nevertheless, they can be built from $\bar{I}_{ij;k\ell}^{(d,h,n;\boldsymbol{p})}$ \eqref{EqIbSoln} with appropriate summations.

%%%%%%%%%%%%%%%%%%%%%%%%%%%%%%%%%%%%%%%%%%%%%%%%%%

\subsection{\texorpdfstring{$(M>3)$}{M>3}-Point Seed Conformal Blocks from the OPE}

To point out the difference between $M=4$ and $M>4$, it is of interest to study the scalar functions for $(M>3)$-point correlation functions generalizing the role of the $K$-function in four-point correlation functions.  To proceed, it is thus convenient to focus on the fully scalar case where all exchanged quasi-primary operators appearing in the different OPEs in \eqref{EqCFOPE2} are Lorentz scalars, in which case the seed conformal block contributions $G_M^{(d,\boldsymbol{h};\boldsymbol{p})}$ to $(M>3)$-point correlation functions simplify to
\eqna{
G_M^{(d,\boldsymbol{h};\boldsymbol{p})}(x_2^M;\boldsymbol{y}^M;\textbf{z}^M)&=\frac{(x_2^M)^{-\bar{p}_{M-1}-\bar{h}_{M-1}}\bar{\D}_{M1;23;2}^{2h_{M-1}}}{(-2)^{h_{M-1}}(\bar{p}_{M-1}+\bar{h}_{M-2})_{h_{M-1}}(\bar{p}_{M-1}+\bar{h}_{M-2}+1-d/2)_{h_{M-1}}}\\
&\phantom{=}\qquad\times\frac{(x_2^M)^{\bar{p}_{M-1}+\bar{h}_{M-2}}}{(1-y_{M-1}^M)^{p_{M-1}}}G_{M-1}^{(d,\boldsymbol{h};\boldsymbol{p})}(x_2^{M-1};\boldsymbol{y}^{M-1};\textbf{z}^{M-1}),
}[EqCBRec]
where $\bar{p}_k=\sum_{i=2}^kp_i$ and $\bar{h}_k=\sum_{i=2}^kh_i$.  Here \eqref{EqCBRec} is obtained from \eqref{EqCFOPERec} by taking all external quasi-primary operators as being scalars, by taking only one contribution from a scalar quasi-primary operator in the sum over $k$, and by discarding all remaining constants to match the normalization in \eqref{EqCBRec}.  Moreover, \eqref{EqCBRec} is homogeneized appropriately using the embedding space coordinates $\eta_1$, $\eta_2$ and $\eta_3$.  In other words, the conformal cross-ratios \eqref{EqCR} are
\eqna{
x_a^i&=\frac{\ee{1}{i}{}\ee{2}{3}{}\ee{a}{i}{}}{\ee{2}{i}{}\ee{3}{i}{}\ee{1}{a}{}}\qquad\forall\,a\neq 1,i,\\
z_{ab}^i&=\frac{\ee{2}{i}{}\ee{3}{i}{}\ee{a}{b}{}}{\ee{2}{3}{}\ee{a}{i}{}\ee{b}{i}{}}\qquad\forall\,a,b\neq 1,i,
}
with $z_{23}^i=1$ for all $i$ (note that $i>3$).  The link between the conformal cross-ratios at $M-1$ and the conformal cross-ratios at $M$ is given by
\eqn{x_2^{M-1}=\frac{1-y_{M-1}^M}{z_{3,M-1}^M},\qquad y_a^{M-1}=1-(1-y_a^M)\frac{z_{2,M-1}^M}{z_{a,M-1}^M},\qquad z_{ab}^{M-1}=\frac{z_{2,M-1}^Mz_{3,M-1}^Mz_{ab}^M}{z_{a,M-1}^Mz_{b,M-1}^M},}
for all $2<a<M-1$ and $1<a<b<M-1$ respectively.

Iterating \eqref{EqCBRec} as in \eqref{EqCFOPE} leads to
\eqna{
G_M^{(d,\boldsymbol{h};\boldsymbol{p})}(x_2^M;\boldsymbol{y}^M;\textbf{z}^M)&=\left[\prod_{j=1}^{M-4}\frac{(x_2^{M-j+2})^{\bar{p}_{M-j+1}+\bar{h}_{M-j}}}{(1-y_{M-j+1}^{M-j+2})^{p_{M-j+1}}}\right.\\
&\phantom{=}\qquad\left.\times\frac{(x_2^{M-j+1})^{-\bar{p}_{M-j}-\bar{h}_{M-j}}\bar{\D}_{M-j+1,1;23;2}^{2h_{M-j}}}{(-2)^{h_{M-j}}(\bar{p}_{M-j}+\bar{h}_{M-j-1})_{h_{M-j}}(\bar{p}_{M-j}+\bar{h}_{M-j-1}+1-d/2)_{h_{M-j}}}\right]\\
&\phantom{=}\qquad\times\frac{(x_2^5)^{\bar{p}_4+\bar{h}_3}}{(1-y_4^5)^{p_4}}\frac{(x_2^4)^{-\bar{p}_3-\bar{h}_3}\bar{\D}_{41;23;2}^{2h_3}}{(-2)^{h_3}(\bar{p}_3+\bar{h}_2)_{h_3}(\bar{p}_3+\bar{h}_2+1-d/2)_{h_3}}\frac{(x_2^4)^{\bar{p}_3+\bar{h}_2}}{(1-y_3^4)^{p_3}},
}[EqCB]
where the four-point seed conformal block has been factored out and $x_2^{M+1}=1$, $y_M^{M+1}=0$.  A look at \eqref{EqCB} shows that the four-point seed conformal block is simply the $K$-function \eqref{EqK0}, \textit{i.e.} $G_4^{(d,h_2,h_3;p_2,p_3)}(x_2^4;y_3^4)=K_{41;23;2}^{(d,h_3;p_2+h_2,p_3)}(x_2^4;y_3^4)$.  From \eqref{EqCBRec} and the boundary condition at $M=4$ given by the $K$-function, the $(M>4)$-point seed conformal blocks are clearly not $K$-functions, nevertheless they can be computed analytically by recurrence from the four-point seed conformal block as in \eqref{EqCBRec}.  In fact, from the knowledge of the action of the scalar differential operator \eqref{EqIb0Soln}, it is only necessary to expand the $(M-1)$-point seed conformal block in the appropriate variables, act with the scalar differential operator as in \eqref{EqIb0Soln}, and finally re-sum the extra sums to get the $M$-point seed conformal block (only the last step is not explicitly done here, see \cite{Rosenhaus:2018zqn} for recent results on five-point correlation functions).\footnote{The tensorial $M$-point conformal block is straightforwardly obtained from the $M$-point seed conformal block and \eqref{EqIbSoln}.}

It is interesting to note that the recurrence relation \eqref{EqCBRec} can be extended down such that the boundary condition corresponds to three-point correlation functions.  Indeed, from \eqref{EqK3} it was argued that the $K$-function for $M=3$ is simply $1$.  Therefore, the recurrence relation \eqref{EqCBRec} can be used with the boundary condition $G_3^{(d,h_2;p_2)}=1$.

The $M$-point seed conformal blocks are thus the fundamental building blocks of the conformal blocks appearing in $M$-point correlation functions of quasi-primary operators in general irreducible representations of the Lorentz group.

%%%%%%%%%%%%%%%%%%%%%%%%%%%%%%%%%%%%%%%%%%%%%%%%%%

\subsection{Conformal Bootstrap and Four-Point Correlation Functions}

The formalism developed in this paper allows to compute analytically any $M$-point correlation function in terms of unknown OPE coefficients (up to re-summations).  This section discusses in general terms the associativity properties of $M$-point correlation functions that can constrain the unknown OPE coefficients, giving rise to the conformal bootstrap (see also \cite{Rychkov:2016iqz}).

From the two and three-point correlation functions \eqref{EqCF2pt} and \eqref{EqCF3pt} respectively, it is unambiguous that the OPE coefficients satisfy some symmetry properties on their indices.  These symmetry properties correspond simply to the freedom in the choice of the two quasi-primary operators used in the OPE.  For $(M>3)$-point correlation functions this freedom puts constraints on the unknown OPE coefficients, since different exchanged quasi-primary operators occur with different OPE contractions, contrary to three-point correlation functions.

Although there are constraints on the OPE coefficients from all $(M>3)$-point correlation functions, it is possible to argue that only the constraints originating from four-point correlation functions are independent.

Schematically, a specific $M$-point correlation function can be visualized as a set of $M$ one-vertices.  Depicting the OPE as a three-vertex, the freedom in the choice of the quasi-primary operators used in the OPE leads to different representations of the same $M$-point correlation function, and thus constraints on the OPE coefficients.

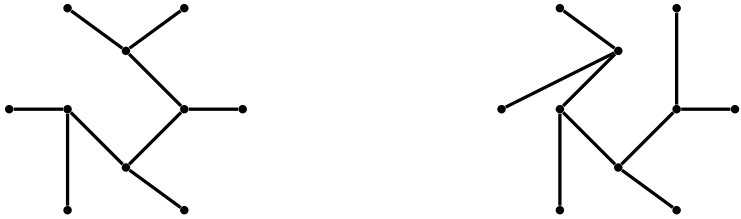
\begin{figure}[t]
\centering
\resizebox{10cm}{!}{%
\begin{tikzpicture}[thick]
\begin{scope}[xshift=-60]
\node[fill,circle,scale=0.2] (node1) at (1,0) {};
\node[fill,circle,scale=0.2] (node2) at (0.5,0.866) {};
\node[fill,circle,scale=0.2] (node3) at (-0.5,0.866) {};
\node[fill,circle,scale=0.2] (node4) at (-1,0) {};
\node[fill,circle,scale=0.2] (node5) at (-0.5,-0.866) {};
\node[fill,circle,scale=0.2] (node6) at (0.5,-0.866) {};
\node[fill,circle,scale=0.2] (node1p) at (0.5,0) {};
\node[fill,circle,scale=0.2] (node2p) at (0,0.5) {};
\node[fill,circle,scale=0.2] (node3p) at (-0.5,0) {};
\node[fill,circle,scale=0.2] (node4p) at (0,-0.5) {};
\draw[-] (node1)--(node1p);
\draw[-] (node2)--(node2p);
\draw[-] (node3)--(node2p);
\draw[-] (node4)--(node3p);
\draw[-] (node5)--(node3p);
\draw[-] (node6)--(node4p);
\draw[-] (node1p)--(node2p);
\draw[-] (node1p)--(node4p);
\draw[-] (node3p)--(node4p);
\end{scope}
\begin{scope}[xshift=60]
\node[fill,circle,scale=0.2] (node1) at (1,0) {};
\node[fill,circle,scale=0.2] (node2) at (0.5,0.866) {};
\node[fill,circle,scale=0.2] (node3) at (-0.5,0.866) {};
\node[fill,circle,scale=0.2] (node4) at (-1,0) {};
\node[fill,circle,scale=0.2] (node5) at (-0.5,-0.866) {};
\node[fill,circle,scale=0.2] (node6) at (0.5,-0.866) {};
\node[fill,circle,scale=0.2] (node1p) at (0.5,0) {};
\node[fill,circle,scale=0.2] (node2p) at (0,0.5) {};
\node[fill,circle,scale=0.2] (node3p) at (-0.5,0) {};
\node[fill,circle,scale=0.2] (node4p) at (0,-0.5) {};
\draw[-] (node1)--(node1p);
\draw[-] (node2)--(node1p);
\draw[-] (node3)--(node2p);
\draw[-] (node4)--(node2p);
\draw[-] (node5)--(node3p);
\draw[-] (node6)--(node4p);
\draw[-] (node1p)--(node4p);
\draw[-] (node2p)--(node3p);
\draw[-] (node3p)--(node4p);
\end{scope}
\end{tikzpicture}
}
\caption{Two different representations of the same six-point correlation function leading to seemingly new constraints on the OPE coefficients.}
\label{Fig6pt}
\end{figure}

\begin{figure}[t]
\centering
\resizebox{10cm}{!}{%
\begin{tikzpicture}[thick]
\begin{scope}[xshift=-60]
\node[fill,circle,scale=0.2] (node1) at (1,0) {};
\node[fill,circle,scale=0.2] (node2) at (0.5,0.866) {};
\node[fill,circle,scale=0.2] (node3) at (-0.5,0.866) {};
\node[fill,circle,scale=0.2] (node4) at (-1,0) {};
\node[fill,circle,scale=0.2] (node5) at (-0.5,-0.866) {};
\node[fill,circle,scale=0.2] (node6) at (0.5,-0.866) {};
\node[fill,circle,scale=0.2] (node1p) at (0.5,0) {};
\node[fill,circle,scale=0.2] (node2p) at (0,0.5) {};
\node[fill,circle,scale=0.2] (node3p) at (-0.5,0) {};
\node[fill,circle,scale=0.2] (node4p) at (0,-0.5) {};
\draw[-] (node1)--(node1p);
\draw[-] (node2)--(node2p);
\draw[-] (node3)--(node2p);
\draw[-] (node4)--(node3p);
\draw[-] (node5)--(node3p);
\draw[-] (node6)--(node4p);
\draw[-] (node1p)--(node2p);
\draw[-] (node1p)--(node4p);
\draw[-] (node3p)--(node4p);
\ellipsebyfoci{draw,dashed,red}{node1p}{node2p}{1.4}
\end{scope}
\node at (0,0) {$=$};
\begin{scope}[xshift=60]
\node[fill,circle,scale=0.2] (node1) at (1,0) {};
\node[fill,circle,scale=0.2] (node2) at (0.5,0.866) {};
\node[fill,circle,scale=0.2] (node3) at (-0.5,0.866) {};
\node[fill,circle,scale=0.2] (node4) at (-1,0) {};
\node[fill,circle,scale=0.2] (node5) at (-0.5,-0.866) {};
\node[fill,circle,scale=0.2] (node6) at (0.5,-0.866) {};
\node[fill,circle,scale=0.2] (node1p) at (0.5,0) {};
\node[fill,circle,scale=0.2] (node2p) at (0,0.5) {};
\node[fill,circle,scale=0.2] (node3p) at (-0.5,0) {};
\node[fill,circle,scale=0.2] (node4p) at (0,-0.5) {};
\draw[-] (node1)--(node1p);
\draw[-] (node2)--(node1p);
\draw[-] (node3)--(node2p);
\draw[-] (node4)--(node3p);
\draw[-] (node5)--(node3p);
\draw[-] (node6)--(node4p);
\draw[-] (node1p)--(node2p);
\draw[-] (node2p)--(node4p);
\draw[-] (node3p)--(node4p);
\end{scope}
\end{tikzpicture}
}
\\\vspace{1cm}
\resizebox{10cm}{!}{%
\begin{tikzpicture}[thick]
\begin{scope}[xshift=-60]
\node[fill,circle,scale=0.2] (node1) at (1,0) {};
\node[fill,circle,scale=0.2] (node2) at (0.5,0.866) {};
\node[fill,circle,scale=0.2] (node3) at (-0.5,0.866) {};
\node[fill,circle,scale=0.2] (node4) at (-1,0) {};
\node[fill,circle,scale=0.2] (node5) at (-0.5,-0.866) {};
\node[fill,circle,scale=0.2] (node6) at (0.5,-0.866) {};
\node[fill,circle,scale=0.2] (node1p) at (0.5,0) {};
\node[fill,circle,scale=0.2] (node2p) at (0,0.5) {};
\node[fill,circle,scale=0.2] (node3p) at (-0.5,0) {};
\node[fill,circle,scale=0.2] (node4p) at (0,-0.5) {};
\draw[-] (node1)--(node1p);
\draw[-] (node2)--(node1p);
\draw[-] (node3)--(node2p);
\draw[-] (node4)--(node3p);
\draw[-] (node5)--(node3p);
\draw[-] (node6)--(node4p);
\draw[-] (node1p)--(node2p);
\draw[-] (node2p)--(node4p);
\draw[-] (node3p)--(node4p);
\ellipsebyfoci{draw,dashed,red}{node2p}{node4p}{1.2}
\end{scope}
\node at (0,0) {$=$};
\begin{scope}[xshift=60]
\node[fill,circle,scale=0.2] (node1) at (1,0) {};
\node[fill,circle,scale=0.2] (node2) at (0.5,0.866) {};
\node[fill,circle,scale=0.2] (node3) at (-0.5,0.866) {};
\node[fill,circle,scale=0.2] (node4) at (-1,0) {};
\node[fill,circle,scale=0.2] (node5) at (-0.5,-0.866) {};
\node[fill,circle,scale=0.2] (node6) at (0.5,-0.866) {};
\node[fill,circle,scale=0.2] (node1p) at (0.5,0) {};
\node[fill,circle,scale=0.2] (node2p) at (0,0.5) {};
\node[fill,circle,scale=0.2] (node3p) at (-0.5,0) {};
\node[fill,circle,scale=0.2] (node4p) at (0,-0.5) {};
\draw[-] (node1)--(node1p);
\draw[-] (node2)--(node1p);
\draw[-] (node3)--(node2p);
\draw[-] (node4)--(node3p);
\draw[-] (node5)--(node3p);
\draw[-] (node6)--(node4p);
\draw[-] (node1p)--(node4p);
\draw[-] (node2p)--(node4p);
\draw[-] (node2p)--(node3p);
\end{scope}
\end{tikzpicture}
}
\\\vspace{1cm}
\resizebox{10cm}{!}{%
\begin{tikzpicture}[thick]
\begin{scope}[xshift=-60]
\node[fill,circle,scale=0.2] (node1) at (1,0) {};
\node[fill,circle,scale=0.2] (node2) at (0.5,0.866) {};
\node[fill,circle,scale=0.2] (node3) at (-0.5,0.866) {};
\node[fill,circle,scale=0.2] (node4) at (-1,0) {};
\node[fill,circle,scale=0.2] (node5) at (-0.5,-0.866) {};
\node[fill,circle,scale=0.2] (node6) at (0.5,-0.866) {};
\node[fill,circle,scale=0.2] (node1p) at (0.5,0) {};
\node[fill,circle,scale=0.2] (node2p) at (0,0.5) {};
\node[fill,circle,scale=0.2] (node3p) at (-0.5,0) {};
\node[fill,circle,scale=0.2] (node4p) at (0,-0.5) {};
\draw[-] (node1)--(node1p);
\draw[-] (node2)--(node1p);
\draw[-] (node3)--(node2p);
\draw[-] (node4)--(node3p);
\draw[-] (node5)--(node3p);
\draw[-] (node6)--(node4p);
\draw[-] (node1p)--(node4p);
\draw[-] (node2p)--(node4p);
\draw[-] (node2p)--(node3p);
\ellipsebyfoci{draw,dashed,red}{node2p}{node3p}{1.4}
\end{scope}
\node at (0,0) {$=$};
\begin{scope}[xshift=60]
\node[fill,circle,scale=0.2] (node1) at (1,0) {};
\node[fill,circle,scale=0.2] (node2) at (0.5,0.866) {};
\node[fill,circle,scale=0.2] (node3) at (-0.5,0.866) {};
\node[fill,circle,scale=0.2] (node4) at (-1,0) {};
\node[fill,circle,scale=0.2] (node5) at (-0.5,-0.866) {};
\node[fill,circle,scale=0.2] (node6) at (0.5,-0.866) {};
\node[fill,circle,scale=0.2] (node1p) at (0.5,0) {};
\node[fill,circle,scale=0.2] (node2p) at (0,0.5) {};
\node[fill,circle,scale=0.2] (node3p) at (-0.5,0) {};
\node[fill,circle,scale=0.2] (node4p) at (0,-0.5) {};
\draw[-] (node1)--(node1p);
\draw[-] (node2)--(node1p);
\draw[-] (node3)--(node2p);
\draw[-] (node4)--(node2p);
\draw[-] (node5)--(node3p);
\draw[-] (node6)--(node4p);
\draw[-] (node1p)--(node4p);
\draw[-] (node2p)--(node3p);
\draw[-] (node3p)--(node4p);
\end{scope}
\end{tikzpicture}
}
\caption{Associativity properties of four-point correlation functions linking the two different representations of the six-point correlation function of Figure \ref{Fig6pt}.}
\label{Fig6ptCB}
\end{figure}
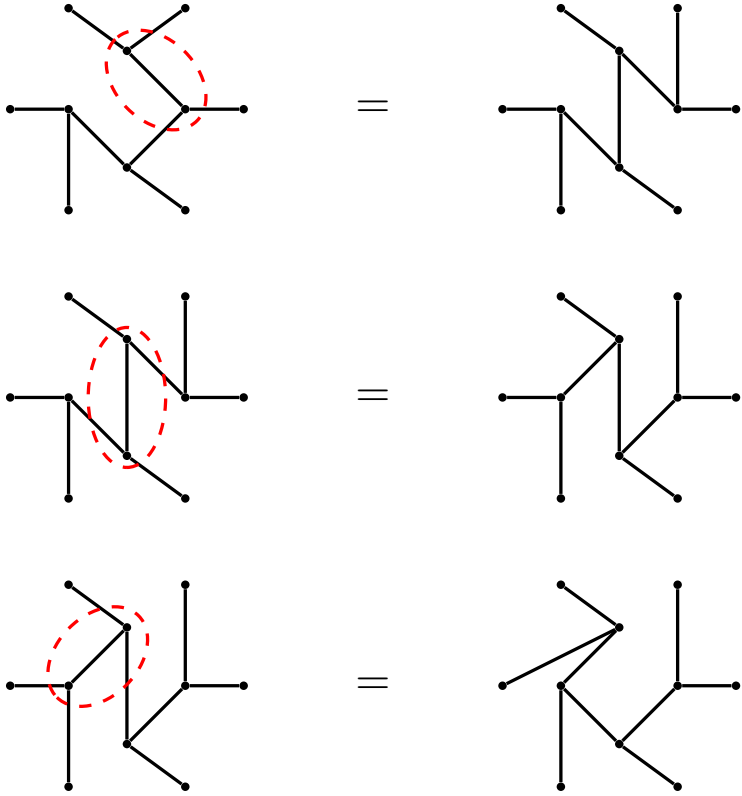

For example, Figure \ref{Fig6pt} shows two different representations of a specific six-point correlation function.  Their equality would seem to imply extra constraints on the OPE coefficients on top of the constraints coming from the associativity of the four-point correlation functions.
This is however not the case as shown in Figure \ref{Fig6ptCB}.  Indeed, in Figure \ref{Fig6ptCB} the associativity properties of four-point correlation functions are used recursively to relate the two different representations of the six-point correlation function of Figure \ref{Fig6pt}.  It is not hard to argue that such behavior generalizes to any $M$-point correlation function, implying that only the associativity constraints on four-point correlation functions lead to independent constraints on the OPE coefficients.

Therefore, to implement the full conformal bootstrap it is only necessary to use associativity properties, or crossing symmetry, on all four-point correlation functions.  The latter are computable analytically in terms of the $K$-function with the help of the formalism introduced here.  More explicit formulas for $M$-point correlation functions will be discussed elsewhere.

%%%%%%%%%%%%%%%%%%%%%%%%%%%%%%%%%%%%%%%%%%%%%%%%%%
%%%%%%%%%%%%%%%%%%%%%%%%%%%%%%%%%%%%%%%%%%%%%%%%%%

\section{Summary}\label{SecSum}

The new embedding space formalism developed here, mainly the new uplift of quasi-primary operators to embedding space, leads to an explicit formula for the OPE in embedding space \eqref{EqOPE}.  This specific form of embedding space OPE is particularly useful since the action of the OPE differential operator is explicitly known \eqref{EqIbSoln}.\footnote{All the ingredients to construct the specific OPE differential operator used here were already known in \cite{Fortin:2016dlj}.  However, the action of an OPE differential operator built arbitrarily from the basic differential operators $\D_{12}^A$ and $\D_{12}^2$ is not necessarily easily tractable, contrary to the OPE differential operator introduced here.  Its power relies on the ease of use in computations of correlation functions, as demonstrated in \eqref{EqIbSoln}.}  The embedding space OPE \eqref{EqOPE} is also expressed in terms of tensor structures that encode all the non-trivial information about the Lorentz group.  These tensor structures are thus completely determined by group-theoretical arguments.

The embedding space OPE is then used recursively in the computation of $M$-point correlation functions.  The computation of two-point and three-point correlation functions shows that they are completely determined from the OPE, \textit{i.e} from conformal invariance, up to the OPE coefficients.  The tensor structures, which properly contract the dummy indices to singlets, are thus also fundamental quantities in the computation of $M$-point correlation functions.  As a consequence, the embedding space OPE \eqref{EqOPE} is completely fixed by conformal invariance and group-theoretical arguments up to the OPE coefficients.\footnote{It is important to mention that any technique used in the computation of correlation functions must deal with the equivalent of the tensor structures introduced here.  Hence, the tensor structures are fundamental to any approach attempting to compute correlation functions.}  As a result, all $M$-point correlation functions can be computed in terms of the OPE coefficients (up to re-summations for $M>4$) from the OPE in embedding space introduced in this work.

Thus there is a set of explicit rules allowing the computations of $M$-point correlation functions from $(M-1)$-point correlation functions directly in embedding space where the action of the conformal group is homogeneous.  Assuming that the $(M-1)$-point correlation functions are known, the steps are
\begin{itemize}
\item Choose two specific quasi-primary operators in the $M$-point correlation function of interest;
\item Determine the tensor structures by finding all exchanged quasi-primary operators from group-theoretical arguments [see \eqref{EqTensorStruct}];
\item Use the embedding space OPE \eqref{EqOPE} for the chosen two quasi-primary operators;
\item Re-express the known $(M-1)$-point correlation function in terms of free homogeneized embedding space coordinates \eqref{Eqetab} and conformal cross-ratios \eqref{EqCR};
\item Apply the OPE differential operator using \eqref{EqIbSoln};
\item For $M>4$, re-sum the extra sums.
\end{itemize}
With these results, OPE coefficients can then be constrained from crossing symmetry directly in embedding space.  Moreover, since the full set of constraints on the OPE coefficients comes from crossing symmetry applied on all four-point correlation functions and all four-point correlation functions are explicitly known (no re-summations are required), the full conformal bootstrap can now be implemented (either the analytical bootstrap as in \textit{e.g.} \cite{Dobrev:1977qv,Cornalba:2007fs,Cornalba:2009ax,Pappadopulo:2012jk,Costa:2012cb,Hogervorst:2013sma,Hartman:2015lfa,Li:2015itl,Hartman:2016dxc,Simmons-Duffin:2016gjk,Hofman:2016awc,Hartman:2016lgu,Afkhami-Jeddi:2016ntf,Hogervorst:2017sfd,Caron-Huot:2017vep,Kulaxizi:2017ixa,Li:2017lmh,Cuomo:2017wme,Dymarsky:2017yzx,Dey:2017oim,Simmons-Duffin:2017nub,Elkhidir:2017iov,Kravchuk:2018htv,Karateev:2018oml,Liendo:2019jpu,Albayrak:2019gnz} or the numerical bootstrap as in \textit{e.g.} \cite{Rattazzi:2008pe,Rychkov:2009ij,Caracciolo:2009bx,Poland:2010wg,Rattazzi:2010gj,Rattazzi:2010yc,Vichi:2011ux,Poland:2011ey,Rychkov:2011et,ElShowk:2012ht,Liendo:2012hy,ElShowk:2012hu,Gliozzi:2013ysa,Kos:2013tga,Alday:2013opa,Gaiotto:2013nva,Berkooz:2014yda,El-Showk:2014dwa,Nakayama:2014lva,Nakayama:2014yia,Chester:2014fya,Kos:2014bka,Caracciolo:2014cxa,Nakayama:2014sba,Paulos:2014vya,Bae:2014hia,Beem:2014zpa,Chester:2014gqa,Simmons-Duffin:2015qma,Alday:2015eya,Bobev:2015jxa,Kos:2015mba,Alday:2015ota,Chester:2015qca,Beem:2015aoa,Iliesiu:2015qra,Poland:2015mta,Lemos:2015awa,Alday:2015ewa,Kim:2015oca,Lin:2015wcg,Chester:2015lej,Chester:2016wrc,Rychkov:2016iqz,Behan:2016dtz,Dey:2016zbg,Alday:2016mxe,Nakayama:2016knq,El-Showk:2016mxr,Li:2016wdp,Pang:2016xno,Lin:2016gcl,Alday:2016njk,Alday:2016jfr,Lemos:2016xke,Beem:2016wfs,Li:2017ddj,Collier:2017shs,Cornagliotto:2017dup,Gadde:2017sjg,Hogervorst:2017kbj,Rychkov:2017tpc,Nakayama:2017vdd,Dymarsky:2017xzb,Chang:2017xmr,Dymarsky:2017yzx,Stergiou:2018gjj,Poland:2018epd,Kousvos:2018rhl,Karateev:2019pvw,Stergiou:2019dcv}).

%%%%%%%%%%%%%%%%%%%%%%%%%%%%%%%%%%%%%%%%%%%%%%%%%%
%%%%%%%%%%%%%%%%%%%%%%%%%%%%%%%%%%%%%%%%%%%%%%%%%%

\section{Discussion and Conclusion}\label{SecConc}

This paper delivered on the program outlined in \cite{Fortin:2016lmf,Fortin:2016dlj,Comeau:2019xco} on the operator product expansion in conformal field theories.  Using the embedding space formalism and a new, more convenient, quasi-primary operator uplift valid for all irreducible representations of the Lorentz group, the most useful operator product expansion in embedding space was found.  Consistency conditions allowed to determine its explicit form in terms of tensor structures (which contain all non-trivial information about the Lorentz group irreducible representations), the operator product expansion differential operator (which takes care of conformal invariance), and the operator product expansion coefficients (which differentiate between conformal field theories).

Then, the proper action of the operator product expansion differential operator on embedding space coordinates was determined, leading to tensorial functions which are the correct generalizations of the Exton function to any $M$-point correlation function.\footnote{Note that the $H$-function obtained in \cite{Fortin:2016dlj} and studied in \cite{Comeau:2019xco} is not of any use here.  Indeed, in the current work the embedding space OPE is applied recursively instead of simultaneously to two pairs of quasi-primary operators as in \cite{Fortin:2016dlj,Comeau:2019xco}.  Since the tensorial functions are explicitly known, the $H$-function studied there can be bypassed.  With its interesting invariance under $D_6$, it might nevertheless be of interest to develop the formalism to simultaneous applications of the embedding space operator product expansion.}  Using the embedding space operator product expansion recursively and the action of the operator product expansion differential operator, $M$-point correlation functions were then computed schematically from the trivial one-point correlation functions.  All two-point, three-point and (schematically from the recurrence relation) four-point correlation functions were obtained in terms of the tensorial functions mentioned above.  For $(M>4)$-point correlation functions, extra re-summations are needed to determine the seed conformal blocks.

At this point, all well-behaved conformal quantities that were previously known in principle are now known explicitly (up to re-summations for $M>4$).  They are given in terms of the tensorial functions (originating from the operator product expansion differential operator) and the group-theoretical tensor structures.  Only appropriate contractions between the tensorial functions and the tensor structures remains to be done to evaluate all correlation functions explicitly.  This is to be compared with existing techniques aiming at computing conformal blocks where the remaining steps (compared to the method introduced here) needed to evaluate correlation functions are of other types (\textit{e.g.} in the shadow formalism approach \cite{SimmonsDuffin:2012uy} it is necessary to remove shadow contributions, in the weight shifting approach \cite{Karateev:2017jgd} it is necessary to compute $6j$ symbols and solve recurrence relations, while in the Casimir approach \cite{Isachenkov:2016gim,Schomerus:2016epl,Schomerus:2017eny,Isachenkov:2017qgn} it is necessary to compute the eigenfunctions of matrix differential operators).  The main advantages of the method proposed here are that all irreducible representations are treated uniformly through the tensor structures and that the operator product expansion, being the most fundamental quantity of a conformal field theory, allows the computation of all correlation functions.

Crossing symmetry can then be implemented directly in the new embedding space, which is the natural setting for conformal field theories, by singling out all contributions with the same Lorentz group structure.  There is absolutely no reason to project back unto usual position space to implement the conformal bootstrap approach.

Since the full power of crossing symmetry can be harnessed from the knowledge of all four-point correlation functions and that all four-point correlation functions can be computed from the set of systematic rules developed here (without requiring any re-summation), it is clear that the conformal bootstrap approach can now be fully implemented.\footnote{It is interesting to note that the four-point seed conformal block is simply the Exton function.  Hence the analytic conformal bootstrap might be easily expandable to all quasi-primary operators in general irreducible representations of the Lorentz group with already existing inversion formulas \cite{Dobrev:1977qv,Cornalba:2007fs,Cornalba:2009ax,Pappadopulo:2012jk,Costa:2012cb,Hogervorst:2013sma,Hartman:2015lfa,Li:2015itl,Hartman:2016dxc,Simmons-Duffin:2016gjk,Hofman:2016awc,Hartman:2016lgu,Afkhami-Jeddi:2016ntf,Hogervorst:2017sfd,Caron-Huot:2017vep,Kulaxizi:2017ixa,Li:2017lmh,Cuomo:2017wme,Dymarsky:2017yzx,Dey:2017oim,Simmons-Duffin:2017nub,Elkhidir:2017iov,Kravchuk:2018htv,Karateev:2018oml,Liendo:2019jpu,Albayrak:2019gnz}.}  The necessary quantities for its implementation being the tensor structures (which are fundamental to any technique aiming to compute conformal blocks) and the four-point tensorial functions (which are now known exactly).  Although higher-point correlation functions are not necessary for the conformal bootstrap, they are nevertheless of interest, for example in multi-particle scatterings in anti-de Sitter space through the correspondence between conformal field theory and anti-de Sitter space.

In a series of follow-up papers, the machinery developed here will be used to compute $M$-point correlation functions of quasi-primary operators in general irreducible representations of the Lorentz group.  Definite and easily-manipulable expressions in terms of the tensorial functions, the tensor structures, and the operator product expansion coefficients will be given for two-point, three-point and four-point correlation functions.  Manifest expressions for the seed conformal blocks will also be computed for higher-point correlation functions.  Consistency of the formalism will be checked as well by projecting back to position space for some simple cases where analytical results exist.

Moreover, one interesting future avenue of research would be to extend the embedding space operator product expansion to the supersymmetric embedding space \cite{Goldberger:2011yp,Goldberger:2012xb}, realizing a supersymmetric embedding space operator product expansion and deriving the related supersymmetric tensorial functions encoding all the information about superconformal invariance.  Such a generalization would allow the study of generic superconformal field theories from their correlation functions.

%%%%%%%%%%%%%%%%%%%%%%%%%%%%%%%%%%%%%%%%%%%%%%%%%%
%%%%%%%%%%%%%%%%%%%%%%%%%%%%%%%%%%%%%%%%%%%%%%%%%%

\ack{
The authors would like to thank the CERN Theory Group, where this work was conceived, for its hospitality.  The authors would like to thank Walter Goldberger, Wenjie Ma and Valentina Prilepina for useful discussions.  The work of JFF is supported by NSERC and FRQNT.
}

%%%%%%%%%%%%%%%%%%%%%%%%%%%%%%%%%%%%%%%%%%%%%%%%%%
%%%%%%%%%%%%%%%%%%%%%%%%%%%%%%%%%%%%%%%%%%%%%%%%%%

\setcounter{section}{0}
\renewcommand{\thesection}{\Alph{section}}

\section{Irreducible Representations of \texorpdfstring{$SO(p,q)$}{SO(p,q)}}\label{SecIrrep}

This appendix discusses how to construct irreducible representations of $SO(p,q)$ from the fundamental spinor representations.  Several important properties of the half-projectors, which appear copiously in the OPE, are also shown.\footnote{In this appendix all operators $\mathcal{O}$ are assumed to transform in the same space than the $SO(p,q)$ group under consideration.  As such they should not be confused with quasi-primary operators in embedding space.  Moreover, vector indices are chosen to start at $1$ instead of $0$ to simplify the notation.}

%%%%%%%%%%%%%%%%%%%%%%%%%%%%%%%%%%%%%%%%%%%%%%%%%%

\subsection{Real Pseudo-Orthogonal Group \texorpdfstring{$SO(p,q)$}{SO(p,q)}}\label{SSecSO}

The real pseudo-orthogonal group $SO(p,q)$ is the connected proper group of real linear transformations of $d$-dimensional spacetime preserving a non-degenerate but indefinite metric $g_{\mu\nu}=\text{diag}(+1,\ldots,+1,-1,\ldots,-1)$ where $\mu,\nu\in\{1,2,\ldots,p+q\}$ with ``$p$'' plus signs, ``$q$'' minus signs and $d=p+q$.  The corresponding Lie algebra $so(p,q)$ is simple [except for $p+q=4$] and non-compact, thus all its non-trivial unitary representations are infinite-dimensional.  For physics purposes, the representations of interest are finite-dimensional and non-unitary.

%%%%%%%%%%%%%%%%%%%%%%%%%%%%%%%%%%%%%%%%%%%%%%%%%%

\subsubsection{Basics of \texorpdfstring{$SO(p,q)$}{SO(p,q)}}\label{SSSecSO}

The group $SO(p,q)$ consists of all real $(p+q)$-dimensional matrices $\Lambda\equiv\Lambda_\mu^{\phantom{\mu}\nu}$, where $\Lambda=\exp(\tfrac{i}{2}\omega^{\mu\nu}s_{\mu\nu})$ with real antisymmetric $\omega^{\mu\nu}$, preserving the non-degenerate but indefinite metric $g_{\mu\nu}$, \textit{i.e.} $\Lambda_\mu^{\phantom{\mu}\mu'}\Lambda_\mu^{\phantom{\nu}\nu'}g_{\mu'\nu'}=g_{\mu\nu}$, with $\det\Lambda=+1$.  Its Lie algebra $so(p,q)$, spanned by $M_{\mu\nu}=-M_{\nu\mu}$, satisfies
\eqn{[M_{\mu\nu},M_{\lambda\rho}]=-(s_{\mu\nu})_\lambda^{\phantom{\lambda}\lambda'}M_{\lambda'\rho}-(s_{\mu\nu})_\rho^{\phantom{\rho}\rho'}M_{\lambda\rho'}.}
Here $(s_{\mu\nu})^{\lambda\rho}=i(\delta_\mu^{\phantom{\mu}\lambda}\delta_\nu^{\phantom{\nu}\rho}-\delta_\mu^{\phantom{\mu}\rho}\delta_\nu^{\phantom{\nu}\lambda})$
are the Lie algebra generators in the fundamental vector representation which also satisfy the Lie algebra (by construction), \textit{i.e.}
\eqn{[s_{\mu\nu},s_{\lambda\rho}]=-(s_{\mu\nu})_{\lambda}^{\phantom{\lambda}\lambda'}s_{\lambda'\rho}-(s_{\mu\nu})_{\rho}^{\phantom{\rho}\rho'}s_{\lambda\rho'}.}
The fully antisymmetric tensor $\epsilon^{\mu_1\cdots\mu_{p+q}}$ where $\epsilon^{1\cdots(p+q)}=1$ is another invariant tensor of $SO(p,q)$ since
\eqn{\epsilon^{\mu_1\cdots\mu_{p+q}}\to\epsilon^{\nu_1\cdots\nu_{p+q}}\Lambda_{\nu_{p+q}}^{\phantom{\nu_{p+q}}\mu_{p+q}}\cdots\Lambda_{\nu_1}^{\phantom{\nu_1}\mu_1}=(\det\Lambda)\epsilon^{\mu_1\cdots\mu_{p+q}}.}
Defining $\epsilon_{\mu_1\cdots\mu_{p+q}}=g_{\mu_1\nu_1}\cdots g_{\mu_{p+q}\nu_{p+q}}\epsilon^{\nu_1\cdots\nu_{p+q}}$, the fully antisymmetric tensor satisfies
\eqn{\epsilon_{\mu_1\cdots\mu_n\lambda_{d-n}\cdots\lambda_1}\epsilon^{\lambda_1\cdots\lambda_{d-n}\nu_n\cdots\nu_1}=(-1)^{q+d(d-1)/2}n!(d-n)!\delta_{[\mu_1}^{\phantom{[\mu_1}\nu_1}\cdots\delta_{\mu_n]}^{\phantom{\mu_n]}\nu_n}.}
Here the generalized identity is
\eqn{\delta_{[\mu_1}^{\phantom{[\mu_1}\nu_1}\cdots\delta_{\mu_n]}^{\phantom{\mu_n]}\nu_n}=\frac{1}{n!}\sum_{\sigma\in S_n}(-1)^\sigma\delta_{\mu_{\sigma(1)}}^{\phantom{\mu_{\sigma(1)}}\nu_1}\cdots\delta_{\mu_{\sigma(n)}}^{\phantom{\mu_{\sigma(n)}}\nu_n},}
where $S_n$ is the symmetric group with element $\sigma$ and sign $(-1)^\sigma$.

%%%%%%%%%%%%%%%%%%%%%%%%%%%%%%%%%%%%%%%%%%%%%%%%%%

\subsection{Irreducible Spinor Representations of \texorpdfstring{$SO(p,q)$}{SO(p,q)}}\label{SSecSpinor}

A (matrix) representation of $SO(p,q)$ on a vector space $V$ is a homomorphism $S:SO(p,q)\to GL(V)$ where $GL(V)$ is the general linear group on $V$.  An irreducible representation is a representation with exactly two subrepresentations, the zero-dimensional subspace and $V$ itself.  As explained below, the real pseudo-orthogonal group has fundamental spinor representations from which all other irreducible representations can be constructed.  Since there is one (two) irreducible spinor representation(s) in odd (even) dimensions, each case is studied separately.

%%%%%%%%%%%%%%%%%%%%%%%%%%%%%%%%%%%%%%%%%%%%%%%%%%

\subsubsection{Odd Dimensions: \texorpdfstring{$p+q=2r+1$}{p+q=2r+1}}\label{SSSecOddSpinor}

The irreducible spinor representation $S\equiv S_\alpha^{\phantom{\alpha}\beta}$ is given by the exponentiation of the generators $\sigma_{\mu\nu}\equiv(\sigma_{\mu\nu})_\alpha^{\phantom{\alpha}\beta}$ in the fundamental spinor representation, $S=\exp(\tfrac{i}{2}\omega^{\mu\nu}\sigma_{\mu\nu})$, with real antisymmetric $\omega^{\mu\nu}$.  The generators in the fundamental spinor representation are constructed from the square $2^r$-dimensional $\gamma$-matrices $\gamma_\mu\equiv(\gamma_\mu)_\alpha^{\phantom{\alpha}\beta}$ which are elements of the Clifford algebra,
\eqn{\gamma_\mu\gamma_\nu+\gamma_\nu\gamma_\mu=2g_{\mu\nu}\mathds{1},}
as
\eqn{\sigma_{\mu\nu}=\frac{i}{4}(\gamma_\mu\gamma_\nu-\gamma_\nu\gamma_\mu).}
By definition, they satisfy the Lie algebra
\eqn{[\sigma_{\mu\nu},\sigma_{\lambda\rho}]=-(s_{\mu\nu})_{\lambda}^{\phantom{\lambda}\lambda'}\sigma_{\lambda'\rho}-(s_{\mu\nu})_{\rho}^{\phantom{\rho}\rho'}\sigma_{\lambda\rho'}.}
Note that $\Lambda_\mu^{\phantom{\mu}\mu'}\gamma_{\mu'}$ satisfy the Clifford algebra and thus should be related by a similarity transformation to $\gamma_\mu$.  This similarity transformation is nothing else than the fundamental spinor representation $S$, \textit{i.e.} $\Lambda_\mu^{\phantom{\mu}\mu'}\gamma_{\mu'}=S\gamma_\mu S^{-1}$.  Note also that although there are two inequivalent representations of the Clifford algebra in odd dimensions, only one is discussed here since both lead to the same fundamental spinor representation of the Lie algebra.  Therefore there is only one irreducible spinor representation of $SO(p,q)$ in odd dimensions.

Since $S$ is a representation of $SO(p,q)$, the adjoint $S^{-\dagger}$ with generators $\sigma_{\mu\nu}^\dagger$, contragredient\footnote{The tensor product decomposition of a representation and its contragredient contains a singlet.} $S^{-T}$ with generators $-\sigma_{\mu\nu}^T$ and conjugate $S^*$ with generators $-\sigma_{\mu\nu}^*$ are also representations.  However unicity of $S$ implies that it is self-adjoint, self-contragredient and self-conjugate.  Thus there exist similarity transformations which relate the adjoint, contragredient and conjugate representations to $S$.  Defining the matrices $A\equiv A_\alpha^{\phantom{\alpha}\beta}$, $C\equiv C^{\alpha\beta}$ and $B=C^{-\dagger}A\propto A^{-T}C$ (with $B\equiv B^{\alpha\beta}$) such that
\eqna{
\gamma_\mu^\dagger&=(-1)^qA\gamma_\mu A^{-1},\\
\gamma_\mu^T&=(-1)^rC\gamma_\mu C^{-1},\\
\gamma_\mu^*&=(-1)^{r+q}B\gamma_\mu B^{-1},
}[EqSymgammaOdd]
one thus has
\eqna{
\sigma_{\mu\nu}^\dagger=A\sigma_{\mu\nu}A^{-1}\qquad&\Rightarrow\qquad S^{-\dagger}=ASA^{-1},\\
-\sigma_{\mu\nu}^T=C\sigma_{\mu\nu}C^{-1}\qquad&\Rightarrow\qquad S^{-T}=CSC^{-1},\\
-\sigma_{\mu\nu}^*=B\sigma_{\mu\nu}B^{-1}\qquad&\Rightarrow\qquad S^*=BSB^{-1},
}
showing the fundamental spinor representation is indeed self-adjoint with $S^\dagger\equiv(S^\dagger)_\beta^{\phantom{\beta}\alpha}$, self-contragredient with $S^T\equiv(S^T)_{\phantom{\beta}\alpha}^\beta$ and self-conjugate with $S^*\equiv(S^*)_{\phantom{\alpha}\beta}^\alpha$.  The matrix $A$ can be given in explicit form, $A=\gamma_{p+1}\cdots\gamma_{p+q}$ (hence the index position), and the matrix $C$ is unitary ($C^\dagger C=\mathds{1}$) with $C^T=(-1)^{r(r+1)/2}C$.  From its definition, the Clifford algebra and \eqref{EqSymgammaOdd}, $A$ satisfies special properties given by
\eqna{
A^{-1}&=(-1)^{q(q+1)/2}A,\\
A^\dagger&=(-1)^{q(q+1)/2}A,\\
A^T&=(-1)^{rq+q(q-1)/2}CAC^{-1},\\
A^*&=(-1)^{q(r+1)}CAC^{-1}.
}
Thus both $A$ and $C$ are unitary, and hence $B=CA$, which satisfies similar properties like
\eqn{B^T=(-1)^{r(r+1)/2+rq+q(q-1)/2}B,}
is also unitary.  These properties are invariant under unitary similarity transformations $\gamma_\mu\to U\gamma_\mu U^\dagger$ where $U^\dagger U=\mathds{1}$ as long as the matrices are changed into $A\to UAU^\dagger$, $C\to U^*CU^\dagger$ and $B\to U^*BU^\dagger$ as dictated by \eqref{EqSymgammaOdd}.

Any element $\psi\equiv\psi_\alpha$ in the linear space $V$ transforms as $\psi\to S\psi$.  The related representations act appropriately, \textit{i.e.} $\psi^\dagger\equiv(\psi^\dagger)^\alpha$ transforms as $\psi^\dagger\to\psi^\dagger S^\dagger=\psi^\dagger AS^{-1}A^{-1}$, $\psi^T\equiv(\psi^T)_\alpha$ transforms as $\psi^T\to\psi^TS^T=\psi^TCS^{-1}C^{-1}$ and $\psi^*\equiv(\psi^*)^\alpha$ transforms as $\psi^*\to S^*\psi^*=BSB^{-1}\psi^*$.

Since $\psi^C=B^{-1}\psi^*$ transforms also as $\psi^C\to S\psi^C$, $\psi^C$ could be proportional to $\psi$ and hence halves the (real) dimension of the fundamental spinor representation, leading to the irreducible Majorana representation.  For this reality condition to exist, the consistency condition $(\psi^C)^C=\psi$ must be satisfied, which enforces $B^{-1}B^T=\mathds{1}$.  Therefore the irreducible Majorana representation exists only for $r(r+1)/2+rq+q(q-1)/2=0\text{ mod }2$.

More complicated irreducible representations can be studied from the product of fundamental spinor representations with the help of the matrices $\gamma^{[n]}$, given by
\eqn{\gamma^{[n]}\equiv\gamma^{\mu_1\cdots\mu_n}=\gamma^{[\mu_1}\cdots\gamma^{\mu_n]}=\frac{1}{n!}\sum_{\sigma\in S_n}(-1)^\sigma\gamma^{\mu_{\sigma(1)}}\cdots\gamma^{\mu_{\sigma(n)}},}
where $S_n$ is the symmetric group with element $\sigma$ and sign $(-1)^\sigma$.  Since there are $\genfrac{(}{)}{0pt}{}{2r+1}{n}$ fully antisymmetric (on their vector indices) matrices $\gamma^{[n]}$, they form a complete basis of square $2^r$-dimensional matrices for $n\in\{0,1,\ldots,r\}$.  Indeed there are $2^{2r}$ such matrices which corresponds to the number of components of square $2^r$-dimensional matrices.  Note that the remaining matrices with $n\in\{r+1,\ldots,2r+1\}$ are related to these through the $SO(p,q)$-invariant epsilon tensor $\epsilon^{\mu_1\cdots\mu_{2r+1}}$,
\eqn{\gamma^{\mu_1\cdots\mu_n}=\frac{\mathscr{K}}{(d-n)!}\epsilon_{\phantom{\mu_1\cdots\mu_n}\nu_{d-n}\cdots\nu_1}^{\mu_1\cdots\mu_n}\gamma^{\nu_1\cdots\nu_{d-n}},}
where $\mathscr{K}$ is the proportionality constant in $\gamma^{\mu_1\cdots\mu_d}=\epsilon^{\mu_1\cdots\mu_d}\gamma^{1\cdots d}=\epsilon^{\mu_1\cdots\mu_d}\gamma^1\cdots\gamma^d=\mathscr{K}\epsilon^{\mu_1\cdots\mu_d}\1$ which satifies $\mathscr{K}^2=(-1)^{r+q}$.

The matrices $\gamma^{[n]}$ for $n\in\{0,1,\ldots,r\}$ satisfy the following important identities,
\eqna{
\text{tr}(\gamma_{\mu_n\cdots\mu_1}\gamma^{\nu_1\cdots\nu_m})&=2^rn!\delta_{[\mu_1}^{\phantom{[\mu_1}\nu_1}\cdots\delta_{\mu_n]}^{\phantom{\mu_i]}\nu_n}\delta_{nm},\\
\gamma_{\mu_1\cdots\mu_n\nu}&=\gamma_{\mu_1\cdots\mu_n}\gamma_\nu+\sum_{i=1}^n(-1)^{n+1-i}g_{\nu\mu_i}\gamma_{\mu_1\cdots\widehat{\mu_i}\cdots\mu_n},\\
\gamma_{\nu\mu_1\cdots\mu_n}&=\gamma_\nu\gamma_{\mu_1\cdots\mu_n}+\sum_{i=1}^n(-1)^ig_{\nu\mu_i}\gamma_{\mu_1\cdots\widehat{\mu_i}\cdots\mu_n},
}
which lead to Fierz identities given by
\eqna{
(\gamma^{\mu_1\cdots\mu_i}C^{-1})_{\alpha_1\beta_1}(\gamma^{\nu_1\cdots\nu_j}C^{-1})_{\alpha_2\beta_2}&=\mathcal{K}_{\phantom{\mu_1\cdots\mu_i\nu_1\cdots\nu_j}\lambda_k\cdots\lambda_1\rho_\ell\cdots\rho_1}^{\mu_1\cdots\mu_i\nu_1\cdots\nu_j}(\gamma^{\lambda_1\cdots\lambda_k}C^{-1})_{\alpha_1\beta_2}(\gamma^{\rho_1\cdots\rho_\ell}C^{-1})_{\alpha_2\beta_1},\\
\mathcal{K}_{\phantom{\mu_1\cdots\mu_i\nu_1\cdots\nu_j}\lambda_1\cdots\lambda_k\rho_1\cdots\rho_\ell}^{\mu_1\cdots\mu_i\nu_1\cdots\nu_j}&=\frac{1}{4^rk!\ell!}\text{tr}(\gamma_{\lambda_1\cdots\lambda_k}\gamma^{\mu_1\cdots\mu_i}\gamma_{\rho_1\cdots\rho_\ell}\gamma^{\nu_1\cdots\nu_j}),
}
where the matrix $C^{-1}$ has been used to lower indices.  Indeed, since $C$ has two upper indices, it is possible to raise and lower indices using $C$ and $C^{-1}$, respectively.  This allows the study of the symmetry properties of $C\gamma^{[n]}$, which are
\eqn{\left(C\gamma^{[n]}\right)^T=(-1)^{n(n-1)/2+nr+r(r+1)/2}C\gamma^{[n]}.}
This equation has a periodicity under $n\to n+4$.  Thus for $i\in\{0,1,\ldots,\lfloor r/4\rfloor\}$, all $C\gamma^{[r-4i]}$ and $C\gamma^{[r-3-4i]}$ are symmetric while all $C\gamma^{[r-1-4i]}$ and $C\gamma^{[r-2-4i]}$ are antisymmetric.  This is consistent with the numbers of symmetric and antisymmetric matrices which are $2^{r-1}(2^r+1)$ and $2^{r-1}(2^r-1)$ respectively.  Furthermore, these matrices (rewritten with lower spinor indices) satisfy
\eqn{B^{-1}(\gamma^{[n]}C^{-1})^*B^{-T}=(-1)^{n(r+q)+q(r+1)}\gamma^{[n]}C^{-1},}
which is useful in investigating conjugate representations.

A general irreducible representation of $SO(p,q)$ is given by a set of non-negative integers $\boldsymbol{N}=\{N_1,\ldots,N_r\}=\sum_{i=1}^rN_i\boldsymbol{e}_i$ where $r$ is the rank of the Lie algebra and $\boldsymbol{e}_i\equiv(\boldsymbol{e}_i)_j=\delta_{ij}$ is the $i$-th unit vector.  Denoting operators in irreducible representations of $SO(p,q)$ by $\mathcal{O}_{\alpha_1\cdots\alpha_n}^{\boldsymbol{N}}$ with
\eqn{n=2S=2\sum_{i=1}^{r-1}N_i+N_r,}
where $S$ is the ``spin'' of the operator, the behavior of the operators under $SO(p,q)$ tranformations can be encoded as
\eqn{\mathcal{O}_{\alpha_1\cdots\alpha_n}^{\boldsymbol{N}}\sim\mathcal{T}_{\alpha_1\cdots\alpha_n}^{\boldsymbol{N}},}
where $\sim$ indicates that both quantities transform in the same way under $SO(p,q)$.  The fundamental quantities $\mathcal{T}_{\alpha_1\cdots\alpha_n}^{\boldsymbol{N}}$, henceforth called half-projectors, are building blocks for irreducible representations and can be constructed recursively (with respect to $n$) as will be shown below.  They correspond to ``square roots'' of the relevant projection operators to the appropriate irreducible representations,
\eqn{(\mathcal{P}_{\boldsymbol{N}})_{\alpha_1\cdots\alpha_n}^{\phantom{\alpha_1\cdots\alpha_n}\alpha_n'\cdots\alpha_1'}=\mathcal{T}_{\alpha_1\cdots\alpha_n}^{\boldsymbol{N}}\mathcal{T}_{\boldsymbol{N}}^{\alpha_n'\cdots\alpha_1'},}
such that $\mathcal{P}_{\boldsymbol{N'}}\mathcal{T}^{\boldsymbol{N}}=\delta_{\boldsymbol{N'}\boldsymbol{N}}\mathcal{T}^{\boldsymbol{N}}$ and $\mathcal{P}_{\boldsymbol{N}}\mathcal{P}_{\boldsymbol{N'}}=\delta_{\boldsymbol{N'}\boldsymbol{N}}\mathcal{P}_{\boldsymbol{N}}$.  The previous contractions vanish automatically when $S'\neq S$ and for a given fixed $S$, the projection operators satisfy $\sum_{\boldsymbol{N}|S\,\text{fixed}}\mathcal{P}_{\boldsymbol{N}}=\mathds{1}$.

The trivial representation $\boldsymbol{N}=\boldsymbol{0}=\{0,\ldots,0\}$, for which $n=0$, is simply a scalar, hence $\mathcal{O}^{\boldsymbol{0}}\sim\mathcal{T}^{\boldsymbol{0}}=1$.

The irreducible spinor representation is the first non-trivial irreducible representation, for which $n=1$, and is denoted by $\boldsymbol{N}=\boldsymbol{e}_r$.  Operators in irreducible spinor representations satisfy $\mathcal{O}_\alpha^{\boldsymbol{e}_r}\sim(\mathcal{T}^{\boldsymbol{e}_r})_\alpha^\beta=\delta_\alpha^{\phantom{\alpha}\beta}$ where the index $\beta$ is a dummy index which must be contracted properly.  The corresponding projection operator is simply $(\mathcal{P}_{\boldsymbol{e}_r})_\alpha^{\phantom{\alpha}\alpha'}=(\mathcal{T}^{\boldsymbol{e}_r})_\alpha^\beta(\mathcal{T}_{\boldsymbol{e}_r})_\beta^{\alpha'}=\delta_\alpha^{\phantom{\alpha}\alpha'}$.  Here the projection operator is the same as the half-projector since both sets of indices on the latter belong to the same space.  The irreducible spinor representation is a defining representation from which more complicated irreducible representations can be constructed.

Indeed, for $n=2$, it is possible to build the irreducible representations by properly symmetrizing the product of two irreducible spinor representations.  This is easily achieved by introducing the fundamental matrices
\eqn{(\mathcal{T}^{\boldsymbol{e}_i})_{\alpha\beta}^{\mu_1\cdots\mu_i}=\frac{1}{\sqrt{2^ri!}}(\gamma^{\mu_1\cdots\mu_i}C^{-1})_{\alpha\beta},\qquad(\mathcal{T}^{2\boldsymbol{e}_r})_{\alpha\beta}^{\mu_1\cdots\mu_r}=\frac{1}{\sqrt{2^rr!}}(\gamma^{\mu_1\cdots\mu_r}C^{-1})_{\alpha\beta}.}
The symmetry properties of the matrices $\gamma^{[n]}C^{-1}$, which form a complete basis of square matrices, combined with the identities they satisfy, demonstrate that they are the proper quantities leading to the projection operators
\begingroup\makeatletter\def\f@size{10}\check@mathfonts\def\maketag@@@#1{\hbox{\m@th\large\normalfont#1}}%
\eqn{(\mathcal{P}_{\boldsymbol{e}_i})_{\alpha\beta}^{\phantom{\alpha\beta}\beta'\alpha'}=\frac{1}{2^ri!}(\gamma^{\mu_1\cdots\mu_i}C^{-1})_{\alpha\beta}(C\gamma_{\mu_i\cdots\mu_1})^{\beta'\alpha'},\qquad(\mathcal{P}_{2\boldsymbol{e}_r})_{\alpha\beta}^{\phantom{\alpha\beta}\beta'\alpha'}=\frac{1}{2^rr!}(\gamma^{\mu_1\cdots\mu_r}C^{-1})_{\alpha\beta}(C\gamma_{\mu_r\cdots\mu_1})^{\beta'\alpha'}.}
\endgroup
Hence operators in irreducible representations with $n=2$ verify $\mathcal{O}_{\alpha\beta}^{\boldsymbol{e}_i}\sim(\mathcal{T}^{\boldsymbol{e}_i})_{\alpha\beta}^{\mu_1\cdots\mu_i}$ for $i\neq r$ and $\mathcal{O}_{\alpha\beta}^{2\boldsymbol{e}_r}\sim(\mathcal{T}^{2\boldsymbol{e}_r})_{\alpha\beta}^{\mu_1\cdots\mu_r}$.  Again, the $\mu$ indices are dummy indices which must be contracted properly.

The irreducible representations constructed up to now are called defining representations.  They are given by $\boldsymbol{N}=\boldsymbol{e}_i$ for all $i\in\{1,\ldots,r\}$ and $\boldsymbol{N}=2\boldsymbol{e}_r$.  The irreducible spinor representation is $\boldsymbol{N}=\boldsymbol{e}_r$ and is the only defining representation with $n=1$.  At $n=2$, the defining representations are the irreducible $i$-index antisymmetric vector representations $\boldsymbol{N}=\boldsymbol{e}_i$ for $i\in\{1,\ldots,r-1\}$ and the irreducible $r$-index antisymmetric vector representation given by $\boldsymbol{N}=2\boldsymbol{e}_r$.

It is now more convenient to obtain the irreducible representations at $n=2$ from another point of view since it will be easier to generalize it to irreducible representations for which $n>2$.  Indeed, since $\boldsymbol{e}_r\otimes\boldsymbol{e}_r=\boldsymbol{0}\oplus_{i=1}^{r-1}\boldsymbol{e}_i\oplus2\boldsymbol{e}_r$, it is clearly possible to construct an irreducible representation at $n=2$ by multiplying two irreducible representations at $n=1$ and properly symmetrizing the dummy indices.  Thus one would expect that
\eqn{\mathcal{T}_{\alpha\beta}^{\boldsymbol{N}}=\hat{\mathcal{T}}^{\boldsymbol{N}}\mathcal{T}_\alpha^{\boldsymbol{e}_r}\mathcal{T}_\beta^{\boldsymbol{e}_r},}
where $\hat{\mathcal{T}}^{\boldsymbol{N}}$ are appropriate symmetrizing operators acting on the dummy indices appearing on the right hand side.  In the case at hand, these symmetrizing operators are trivially the same as the fundamental building blocks defined above, $\hat{\mathcal{T}}^{\boldsymbol{N}}=\mathcal{T}^{\boldsymbol{N}}$, since
\eqn{(\mathcal{T}^{\boldsymbol{N}})_{\alpha\beta}^{\mu_1\cdots\mu_i}=(\hat{\mathcal{T}}^{\boldsymbol{N}})_{\gamma\delta}^{\mu_1\cdots\mu_i}(\mathcal{T}^{\boldsymbol{e}_r})_\alpha^\gamma(\mathcal{T}^{\boldsymbol{e}_r})_\beta^\delta,}
for $\boldsymbol{N}=\boldsymbol{e}_{i\neq r}$ and $\boldsymbol{N}=2\boldsymbol{e}_r$.  Clearly, they mix the two irreducible spinor representations into irreducible representations at $n=2$, changing the dummy indices from $\gamma$ and $\delta$ to $\mu$ indices.  Although it is not relevant here, they also lead to projection operators
\eqn{(\hat{\mathcal{P}}^{\boldsymbol{N}})_{\delta'\gamma'}^{\phantom{\delta'\gamma'}\gamma\delta}=(\hat{\mathcal{T}}_{\boldsymbol{N}})_{\mu_1\cdots\mu_i}^{\gamma\delta}(\hat{\mathcal{T}}^{\boldsymbol{N}})_{\delta'\gamma'}^{\mu_i\cdots\mu_1}.}

This technique is particularly useful for larger irreducible representations.  For $n=3$, which can be obtained from the products $\boldsymbol{e}_{i\neq r}\otimes\boldsymbol{e}_r$ and $2\boldsymbol{e}_r\otimes\boldsymbol{e}_r$, the fundamental building blocks for the irreducible representations can be obtained as before.  By recursion, the first product\footnote{For explicit tensor product decompositions, it is assumed that $r$ is larger than the explicit indices of the other irreducible representations.  Hence for $\boldsymbol{e}_1\otimes\boldsymbol{e}_r$ one assumes $r>1$.} to study is $\boldsymbol{e}_1\otimes\boldsymbol{e}_r=(\boldsymbol{e}_1+\boldsymbol{e}_r)\oplus\boldsymbol{e}_r$ since it only leads to one new irreducible representation, namely $\boldsymbol{e}_1+\boldsymbol{e}_r$.  To obtain it, the irreducible representation $\boldsymbol{e}_r$ must first be subtracted properly.  To subtract it, it is necessary to construct it from the product of $\boldsymbol{e}_1$ and $\boldsymbol{e}_r$, which can be done as follows,
\eqn{(\mathcal{T}^{\boldsymbol{e}_r\subset\boldsymbol{e}_1\otimes\boldsymbol{e}_r})_{\alpha\beta\gamma}^{\lambda}=(\hat{\mathcal{T}}^{\boldsymbol{e}_r\subset\boldsymbol{e}_1\otimes\boldsymbol{e}_r})_{\mu\delta}^\lambda(\mathcal{T}^{\boldsymbol{e}_1})_{\alpha\beta}^\mu(\mathcal{T}^{\boldsymbol{e}_r})_\gamma^\delta=\frac{1}{\sqrt{2^rd}}(\gamma^\mu C^{-1})_{\alpha\beta}(\gamma_\mu)_\gamma^{\phantom{\gamma}\lambda},}
where
\eqn{(\hat{\mathcal{T}}^{\boldsymbol{e}_r\subset\boldsymbol{e}_1\otimes\boldsymbol{e}_r})_{\mu\delta}^\lambda=\sqrt{\frac{2^r}{d}}(\mathcal{T}_{\boldsymbol{e}_1})_\mu^{\beta\lambda}(\mathcal{T}_{\boldsymbol{e}_r})_\delta^\alpha(C^{-1})_{\alpha\beta}=\frac{1}{\sqrt{d}}(\gamma_\mu)_\delta^{\phantom{\delta}\lambda},}
with $\lambda$ the dummy index in the irreducible spinor representation, demonstrating the fundamental building block is in the right irreducible representation.  It is important to note that the hatted half-projector is the only independent quantity which leads to the proper dummy index associated to an irreducible representation $\boldsymbol{e}_r$ and which can be built from the half-projectors for $\boldsymbol{e}_1$ and $\boldsymbol{e}_r$, where spinor indices are contracted with the help of $C^{-1}$.  Since the irreducible representation $\boldsymbol{e}_r$ is embedded in a non-minimal tensor product, the (hatted) half-projectors and projectors are different than the minimal ones.  This difference is indicated by explicitly showing which tensor product they originate from.

It is now possible to proceed in two different but equivalent ways.  Both ways involve computing the corresponding projection operator in the relevant tensor product, either with the half-projector or the hatted half-projector,
\eqna{
(\mathcal{P}_{\boldsymbol{e}_r\subset\boldsymbol{e}_1\otimes\boldsymbol{e}_r})_{\alpha\beta\gamma}^{\phantom{\alpha\beta\gamma}\gamma'\beta'\alpha'}&=(\mathcal{T}^{\boldsymbol{e}_r\subset\boldsymbol{e}_1\otimes\boldsymbol{e}_r})_{\alpha\beta\gamma}^\lambda(\mathcal{T}_{\boldsymbol{e}_r\subset\boldsymbol{e}_1\otimes\boldsymbol{e}_r})_\lambda^{\gamma'\beta'\alpha'}\\
&=\frac{1}{2^rd}(\gamma^\mu C^{-1})_{\alpha\beta}(\gamma_\mu)_\gamma^{\phantom{\gamma}\lambda}(C\gamma^\nu)^{\beta'\alpha'}(\gamma_\nu)_\lambda^{\phantom{\lambda}\gamma'}=\frac{1}{2^rd}(\gamma^\mu C^{-1})_{\alpha\beta}(\gamma_\mu\gamma_\nu)_\gamma^{\phantom{\gamma}\gamma'}(C\gamma^\nu)^{\beta'\alpha'},\\
(\hat{\mathcal{P}}^{\boldsymbol{e}_r\subset\boldsymbol{e}_1\otimes\boldsymbol{e}_r})_{\delta'\mu'}^{\phantom{\delta'\mu'}\mu\delta}&=(\hat{\mathcal{T}}_{\boldsymbol{e}_r\subset\boldsymbol{e}_1\otimes\boldsymbol{e}_r})_\lambda^{\mu\delta}(\hat{\mathcal{T}}^{\boldsymbol{e}_r\subset\boldsymbol{e}_1\otimes\boldsymbol{e}_r})_{\delta'\mu'}^\lambda=\frac{1}{d}(\gamma^\mu)_\lambda^{\phantom{\lambda}\delta}(\gamma_{\mu'})_{\delta'}^{\phantom{\delta'}\lambda}=\frac{1}{d}(\gamma_{\mu'}\gamma^\mu)_{\delta'}^{\phantom{\delta'}\delta},
}
and using the associated completeness relation.  Indeed, from the tensor product, $\mathcal{P}_{\boldsymbol{e}_1+\boldsymbol{e}_r}+\mathcal{P}_{\boldsymbol{e}_r\subset\boldsymbol{e}_1\otimes\boldsymbol{e}_r}=\mathds{1}$ and $\hat{\mathcal{P}}^{\boldsymbol{e}_1+\boldsymbol{e}_r}+\hat{\mathcal{P}}^{\boldsymbol{e}_r\subset\boldsymbol{e}_1\otimes\boldsymbol{e}_r}=\mathds{1}$, hence the fundamental building block for the mixed irreducible representation is simply
\eqna{
(\mathcal{T}^{\boldsymbol{e}_1+\boldsymbol{e}_r})_{\alpha\beta\gamma}^{\mu\delta}&=(\mathcal{T}^{\boldsymbol{e}_1})_{\alpha\beta}^\mu(\mathcal{T}^{\boldsymbol{e}_r})_\gamma^\delta-(\mathcal{P}_{\boldsymbol{e}_r\subset\boldsymbol{e}_1\otimes\boldsymbol{e}_r})_{\alpha\beta\gamma}^{\phantom{\alpha\beta\gamma}\gamma'\beta'\alpha'}(\mathcal{T}^{\boldsymbol{e}_1})_{\alpha'\beta'}^\mu(\mathcal{T}^{\boldsymbol{e}_r})_{\gamma'}^\delta\\
&=(\mathcal{T}^{\boldsymbol{e}_1})_{\alpha\beta}^\mu(\mathcal{T}^{\boldsymbol{e}_r})_\gamma^\delta-(\hat{\mathcal{P}}^{\boldsymbol{e}_r\subset\boldsymbol{e}_1\otimes\boldsymbol{e}_r})_{\phantom{\mu\delta}\delta'\mu'}^{\mu\delta}(\mathcal{T}^{\boldsymbol{e}_1})_{\alpha\beta}^{\mu'}(\mathcal{T}^{\boldsymbol{e}_r})_\gamma^{\delta'}\\
&=\frac{1}{\sqrt{2^r}}(\gamma^\mu C^{-1})_{\alpha\beta}\delta_\gamma^{\phantom{\gamma}\delta}-\frac{1}{\sqrt{2^r}d}(\gamma^{\mu'}C^{-1})_{\alpha\beta}(\gamma_{\mu'}\gamma^\mu)_\gamma^{\phantom{\gamma}\delta}.
}
Therefore the group properties of the corresponding operators are encoded as $\mathcal{O}_{\alpha\beta\gamma}^{\boldsymbol{e}_1+\boldsymbol{e}_r}\sim(\mathcal{T}^{\boldsymbol{e}_1+\boldsymbol{e}_r})_{\alpha\beta\gamma}^{\mu\delta}$ where $\mu$ and $\delta$ are the dummy indices of the irreducible representation.

It is now convenient to compute the hatted half-projector for the $\boldsymbol{e}_1+\boldsymbol{e}_r$ irreducible representation.  From the definition, $(\mathcal{T}^{\boldsymbol{e}_1+\boldsymbol{e}_r})_{\alpha\beta\gamma}^{\mu\delta}=(\hat{\mathcal{T}}^{\boldsymbol{e}_1+\boldsymbol{e}_r})_{\delta'\mu'}^{\mu\delta}(\mathcal{T}^{\boldsymbol{e}_1})_{\alpha\beta}^{\mu'}(\mathcal{T}^{\boldsymbol{e}_r})_\gamma^{\delta'}$, and thus
\eqn{(\hat{\mathcal{T}}^{\boldsymbol{e}_1+\boldsymbol{e}_r})_{\delta'\mu'}^{\mu\delta}=\delta_{\mu'}^{\phantom{\mu'}\mu}\delta_{\delta'}^{\phantom{\delta'}\delta}-\frac{1}{d}(\gamma_{\mu'}\gamma^\mu)_{\delta'}^{\phantom{\delta'}\delta}.}
The associated hatted projection operator is given by its definition,
\eqn{(\hat{\mathcal{P}}^{\boldsymbol{e}_1+\boldsymbol{e}_r})_{\delta'\mu'}^{\phantom{\delta'\mu'}\mu\delta}=(\hat{\mathcal{T}}_{\boldsymbol{e}_1+\boldsymbol{e}_r})_{\delta''\mu''}^{\mu\delta}(\hat{\mathcal{T}}^{\boldsymbol{e}_1+\boldsymbol{e}_r})_{\delta'\mu'}^{\mu''\delta''}=\delta_{\mu'}^{\phantom{\mu'}\mu}\delta_{\delta'}^{\phantom{\delta'}\delta}-\frac{1}{d}(\gamma_{\mu'}\gamma^\mu)_{\delta'}^{\phantom{\delta'}\delta},}
which is, as expected, identical to $\1-\hat{\mathcal{P}}^{\boldsymbol{e}_r\subset\boldsymbol{e}_1\otimes\boldsymbol{e}_r}$, and moreover the same than the hatted half-projector $(\hat{\mathcal{P}}^{\boldsymbol{e}_1+\boldsymbol{e}_r})_{\delta'\mu'}^{\phantom{\delta'\mu'}\mu\delta}=(\hat{\mathcal{T}}_{\boldsymbol{e}_1+\boldsymbol{e}_r})_{\delta'\mu'}^{\mu\delta}$.  Indeed, since both types of indices on the hatted half-projector live in the same space, the ``square root'' of the hatted projection operator must be equal to itself.

It is interesting to note that proper hatted projection operators $\hat{\mathcal{P}}^{\boldsymbol{N}}$, \textit{i.e.} hatted projection operators for irreducible representations that are the largest in the appropriate products of defining irreducible representations, are trivially equal to $(\mathcal{T}^{\boldsymbol{N}})_{\alpha_1\cdots\alpha_n}(\mathcal{T}_{\boldsymbol{N}})^{\alpha_n\cdots\alpha_1}$, as for example
\eqna{
(\hat{\mathcal{P}}^{\boldsymbol{e}_1+\boldsymbol{e}_r})_{\delta'\mu'}^{\phantom{\delta'\mu'}\mu\delta}&=\left[\frac{1}{\sqrt{2^r}}(\gamma^\mu C^{-1})_{\alpha\beta}\delta_\gamma^{\phantom{\gamma}\delta}-\frac{1}{\sqrt{2^r}d}(\gamma^{\nu}C^{-1})_{\alpha\beta}(\gamma_{\nu}\gamma^\mu)_\gamma^{\phantom{\gamma}\delta}\right]\\
&\phantom{=}\qquad\times\left[\frac{1}{\sqrt{2^r}}(C\gamma_{\mu'})^{\beta\alpha}\delta_{\delta'}^{\phantom{\delta'}\gamma}-\frac{1}{\sqrt{2^r}d}(C\gamma_{\nu'})^{\beta\alpha}(\gamma_{\mu'}\gamma^{\nu'})_{\delta'}^{\phantom{\delta'}\gamma}\right]=(\mathcal{T}^{\boldsymbol{e}_1+\boldsymbol{e}_r})_{\alpha\beta\gamma}^{\mu\delta}(\mathcal{T}_{\boldsymbol{e}_1+\boldsymbol{e}_r})_{\delta'\mu'}^{\gamma\beta\alpha}.
}
The hatted projection operator for the $\boldsymbol{e}_1+\boldsymbol{e}_r$ irreducible representation will be useful to construct the next half-projector.

All remaining irreducible representations at $n=3$ can be constructed similarly.  For example, the next non-trivial irreducible representation at $n=3$, which is $\boldsymbol{e}_2+\boldsymbol{e}_r$, appears in the product $\boldsymbol{e}_2\otimes\boldsymbol{e}_r=(\boldsymbol{e}_2+\boldsymbol{e}_r)\oplus(\boldsymbol{e}_1+\boldsymbol{e}_r)\oplus\boldsymbol{e}_r$.  In order to build this irreducible representation, it is necessary to construct the two other irreducible representations appearing in the tensor product and subtract them as shown above.  The simplest way to accomplish this for the $\boldsymbol{e}_r$ irreducible representation is to follow the procedure highlighted before, \textit{i.e.}
\eqn{(\mathcal{T}^{\boldsymbol{e}_r\subset\boldsymbol{e}_2\otimes\boldsymbol{e}_r})_{\alpha\beta\gamma}^{\lambda}=(\hat{\mathcal{T}}^{\boldsymbol{e}_r\subset\boldsymbol{e}_2\otimes\boldsymbol{e}_r})_{\mu\nu\delta}^\lambda(\mathcal{T}^{\boldsymbol{e}_2})_{\alpha\beta}^{\mu\nu}(\mathcal{T}^{\boldsymbol{e}_r})_\gamma^\delta=\frac{1}{\sqrt{2^{r+1}d(d-1)}}(\gamma^{\mu\nu}C^{-1})_{\alpha\beta}(\gamma_{\mu\nu})_\gamma^{\phantom{\gamma}\lambda},}
where
\eqn{(\hat{\mathcal{T}}^{\boldsymbol{e}_r\subset\boldsymbol{e}_2\otimes\boldsymbol{e}_r})_{\mu\nu\delta}^\lambda=\sqrt{\frac{2^{r+1}}{d(d-1)}}(\mathcal{T}_{\boldsymbol{e}_2})_{\mu\nu}^{\beta\lambda}(\mathcal{T}_{\boldsymbol{e}_r})_\delta^\alpha(C^{-1})_{\alpha\beta}=\frac{1}{\sqrt{d(d-1)}}(\gamma_{\mu\nu})_\delta^{\phantom{\delta}\lambda},}
is the only independent quantity with the proper behavior which can be constructed from the irreducible representations present in the tensor product with spinor indices contracted with the help of $C^{-1}$.  The associated hatted projection operator is then
\eqn{(\hat{\mathcal{P}}^{\boldsymbol{e}_r\subset\boldsymbol{e}_2\otimes\boldsymbol{e}_r})_{\delta'\nu'\mu'}^{\phantom{\delta'\nu'\mu'}\mu\nu\delta}=(\hat{\mathcal{T}}_{\boldsymbol{e}_r\subset\boldsymbol{e}_2\otimes\boldsymbol{e}_r})_\lambda^{\mu\nu\delta}(\hat{\mathcal{T}}^{\boldsymbol{e}_r\subset\boldsymbol{e}_2\otimes\boldsymbol{e}_r})_{\delta'\nu'\mu'}^\lambda=\frac{1}{d(d-1)}(\gamma_{\nu'\mu'}\gamma^{\mu\nu})_{\delta'}^{\phantom{\delta'}\delta}.}
For the $\boldsymbol{e}_1+\boldsymbol{e}_r$ irreducible representation, the quickest path to the hatted half-projector is again to built it from the properly normalized quantities in the original tensor product contracted with $\gamma^\rho C^{-1}$ (this time to single out $\boldsymbol{e}_1$) and projected on $\boldsymbol{e}_1+\boldsymbol{e}_r$, \textit{i.e.}
\eqna{
(\hat{\mathcal{T}}^{\boldsymbol{e}_1+\boldsymbol{e}_r\subset\boldsymbol{e}_2\otimes\boldsymbol{e}_r})_{\mu\nu\delta}^{\rho\lambda}&=\sqrt{\frac{-2^{r-2}d}{(d-2)(d-4)}}(\hat{\mathcal{P}}^{\boldsymbol{e}_1+\boldsymbol{e}_r})_{\lambda'\rho'}^{\phantom{\lambda'\rho'}\rho\lambda}(\mathcal{T}_{\boldsymbol{e}_2})_{\mu\nu}^{\beta\lambda'}(\mathcal{T}_{\boldsymbol{e}_r})_\delta^\alpha(\gamma^{\rho'}C^{-1})_{\alpha\beta}\\
&=\sqrt{\frac{-d}{8(d-2)(d-4)}}\left[(\gamma^\rho\gamma_{\mu\nu})_\delta^{\phantom{\delta}\lambda}-\frac{d-4}{d}(\gamma_{\mu\nu}\gamma^\rho)_\delta^{\phantom{\delta}\lambda}\right].
}
The last step, \textit{i.e.} contracting with $\boldsymbol{e}_1+\boldsymbol{e}_r$ which is necessary to obtain a different hatted projection operator, was implicitly done before since the $\boldsymbol{e}_r$ projector is the identity.  The hatted projection operator is thus
\eqna{
(\hat{\mathcal{P}}^{\boldsymbol{e}_1+\boldsymbol{e}_r\subset\boldsymbol{e}_2\otimes\boldsymbol{e}_r})_{\delta'\nu'\mu'}^{\phantom{\delta'\nu'\mu'}\mu\nu\delta}&=(\hat{\mathcal{T}}_{\boldsymbol{e}_1+\boldsymbol{e}_r\subset\boldsymbol{e}_2\otimes\boldsymbol{e}_r})_{\rho\lambda}^{\mu\nu\delta}(\hat{\mathcal{T}}^{\boldsymbol{e}_1+\boldsymbol{e}_r\subset\boldsymbol{e}_2\otimes\boldsymbol{e}_r})_{\delta'\nu'\mu'}^{\lambda\rho}\\
&=\frac{-2}{d-2}\left[\delta_{[\nu'}^{\phantom{[\nu'}[\mu}(\gamma_{\mu']}\gamma^{\nu]})_{\delta'}^{\phantom{\delta'}\delta}+\frac{1}{d}(\gamma_{\nu'\mu'}\gamma^{\mu\nu})_{\delta'}^{\phantom{\delta'}\delta}\right],
}
and it is orthogonal to the hatted projection operator $\hat{\mathcal{P}}^{\boldsymbol{e}_r\subset\boldsymbol{e}_2\otimes\boldsymbol{e}_r}$ by construction.  From the completeness relation, the fundamental building block for the $\boldsymbol{e}_2+\boldsymbol{e}_r$ irreducible representation is thus
\eqn{(\mathcal{T}^{\boldsymbol{e}_2+\boldsymbol{e}_r})_{\alpha\beta\gamma}^{\mu\nu\delta}=[\delta_{\mu'}^{\phantom{\mu'}\mu}\delta_{\nu'}^{\phantom{\nu'}\nu}\delta_{\delta'}^{\phantom{\delta'}\delta}-(\hat{\mathcal{P}}^{\boldsymbol{e}_r\subset\boldsymbol{e}_2\otimes\boldsymbol{e}_r})_{\delta'\nu'\mu'}^{\phantom{\delta'\nu'\mu'}\mu\nu\delta}-(\hat{\mathcal{P}}^{\boldsymbol{e}_1+\boldsymbol{e}_r\subset\boldsymbol{e}_2\otimes\boldsymbol{e}_r})_{\delta'\nu'\mu'}^{\phantom{\delta'\nu'\mu'}\mu\nu\delta}](\mathcal{T}^{\boldsymbol{e}_2})_{\alpha\beta}^{\mu'\nu'}(\mathcal{T}^{\boldsymbol{e}_r})_\gamma^{\delta'}.}
As a consistency check, it is possible to verify that the $n=3$ projection operators $\mathcal{P}_{\boldsymbol{e}_r\subset\boldsymbol{e}_1\otimes\boldsymbol{e}_r}$ and $\mathcal{P}_{\boldsymbol{e}_r\subset\boldsymbol{e}_2\otimes\boldsymbol{e}_r}$ are orthogonal.  The same can be checked for $\mathcal{P}_{\boldsymbol{e}_1+\boldsymbol{e}_r}$ and $\mathcal{P}_{\boldsymbol{e}_1+\boldsymbol{e}_r\subset\boldsymbol{e}_2\otimes\boldsymbol{e}_r}$.

For a more familiar example at $n=4$, the tensor product $\boldsymbol{e}_1\otimes\boldsymbol{e}_1=2\boldsymbol{e}_1\oplus\boldsymbol{e}_2\oplus\boldsymbol{0}$ leads to the traceless $2$-index symmetric fundamental building block $\mathcal{T}^{2\boldsymbol{e}_1}$ as follows.  First, since it is already known that $\boldsymbol{0}$ is a scalar and $\boldsymbol{e}_2$ is a $2$-index antisymmetric tensor, one has
\eqn{(\hat{\mathcal{T}}^{\boldsymbol{0}\subset\boldsymbol{e}_1\otimes\boldsymbol{e}_1})_{\mu\nu}=\frac{1}{\sqrt{d}}g_{\mu\nu},\qquad(\hat{\mathcal{T}}^{\boldsymbol{e}_2\subset\boldsymbol{e}_1\otimes\boldsymbol{e}_1})_{\mu'\nu'}^{\nu\mu}=\delta_{\mu'}^{\phantom{\mu'}[\mu}\delta_{\nu'}^{\phantom{\nu'}\nu]},}
which lead to
\eqna{
(\hat{\mathcal{P}}^{\boldsymbol{0}\subset\boldsymbol{e}_1\otimes\boldsymbol{e}_1})_{\nu'\mu'}^{\phantom{\nu'\mu'}\mu\nu}&=(\hat{\mathcal{T}}_{\boldsymbol{0}\subset\boldsymbol{e}_1\otimes\boldsymbol{e}_1})^{\mu\nu}(\hat{\mathcal{T}}^{\boldsymbol{0}\subset\boldsymbol{e}_1\otimes\boldsymbol{e}_1})_{\nu'\mu'}=\frac{1}{d}g^{\mu\nu}g_{\nu'\mu'},\\
(\hat{\mathcal{P}}^{\boldsymbol{e}_2\subset\boldsymbol{e}_1\otimes\boldsymbol{e}_1})_{\nu'\mu'}^{\phantom{\nu'\mu'}\mu\nu}=&(\hat{\mathcal{T}}_{\boldsymbol{e}_2\subset\boldsymbol{e}_1\otimes\boldsymbol{e}_1})_{\nu''\mu''}^{\mu\nu}(\hat{\mathcal{T}}^{\boldsymbol{e}_2\subset\boldsymbol{e}_1\otimes\boldsymbol{e}_1})_{\nu'\mu'}^{\mu''\nu''}=\delta_{\mu'}^{\phantom{\mu'}[\mu}\delta_{\nu'}^{\phantom{\nu'}\nu]}.
}
Then, with the help of $\hat{\mathcal{P}}^{2\boldsymbol{e}_1}+\hat{\mathcal{P}}^{\boldsymbol{e}_2\subset\boldsymbol{e}_1\otimes\boldsymbol{e}_1}+\hat{\mathcal{P}}^{\boldsymbol{e}_0\subset\boldsymbol{e}_1\otimes\boldsymbol{e}_1}=\mathds{1}$, the fundamental building block is simply
\eqna{
(\mathcal{T}^{2\boldsymbol{e}_1})_{\alpha_1\beta_1\alpha_2\beta_2}^{\mu\nu}&=\left[\delta_{\mu'}^{\phantom{\mu'}\mu}\delta_{\nu'}^{\phantom{\nu'}\nu}-(\hat{\mathcal{P}}^{\boldsymbol{e}_2\subset\boldsymbol{e}_1\otimes\boldsymbol{e}_1})_{\nu'\mu'}^{\phantom{\nu'\mu'}\mu\nu}-(\hat{\mathcal{P}}^{\boldsymbol{0}\subset\boldsymbol{e}_1\otimes\boldsymbol{e}_1})_{\nu'\mu'}^{\phantom{\nu'\mu'}\mu\nu}\right](\mathcal{T}^{\boldsymbol{e}_1})_{\alpha_1\beta_1}^{\mu'}(\mathcal{T}^{\boldsymbol{e}_1})_{\alpha_1\beta_1}^{\nu'}\\
&=(\mathcal{T}^{\boldsymbol{e}_1})_{\alpha_1\beta_1}^{(\mu}(\mathcal{T}^{\boldsymbol{e}_1})_{\alpha_2\beta_2}^{\nu)}-\frac{1}{d}g^{\mu\nu}g_{\nu'\mu'}(\mathcal{T}^{\boldsymbol{e}_1})_{\alpha_1\beta_1}^{\mu'}(\mathcal{T}^{\boldsymbol{e}_1})_{\alpha_2\beta_2}^{\nu'},
}
as expected.  Here the parenthesis denote symmetrization of the corresponding indices.  The corresponding hatted half-projector is simply
\eqn{(\hat{\mathcal{T}}^{2\boldsymbol{e}_1})_{\nu'\mu'}^{\mu\nu}=\delta_{\mu'}^{\phantom{\mu'}(\mu}\delta_{\nu'}^{\phantom{\nu'}\nu)}-\frac{1}{d}g^{\mu\nu}g_{\nu'\mu'},}
with the associated hatted projection operator $(\hat{\mathcal{P}}^{2\boldsymbol{e}_1})_{\nu'\mu'}^{\phantom{\nu'\mu'}\mu\nu}=(\hat{\mathcal{T}}^{2\boldsymbol{e}_1})_{\nu'\mu'}^{\mu\nu}$.  It is now straightforward to see the properties of operators in the traceless $2$-index symmetric irreducible representation $\mathcal{O}_{\alpha_1\beta_1\alpha_2\beta_2}^{2\boldsymbol{e}_1}\sim(\mathcal{T}^{2\boldsymbol{e}_1})_{\alpha_1\beta_1\alpha_2\beta_2}^{\mu\nu}$ where $\mu$ and $\nu$ are the dummy indices of the irreducible representation which must be contracted properly with symmetric and traceless quantities.

From this technique, it is now clear that all irreducible representations can be constructed recursively from the fundamental building blocks of the defining representations.  Therefore the fundamental building block for a general irreducible representation $\boldsymbol{N}=\sum_{i=1}^rN_i\boldsymbol{e}_i$ is
\eqna{
(\mathcal{T}^{\boldsymbol{N}})_{\alpha_1\cdots\alpha_n}^{\mu_1\cdots\mu_{n_v}\delta}&=\left((\mathcal{T}^{\boldsymbol{e}_1})^{N_1}\cdots(\mathcal{T}^{\boldsymbol{e}_{r-1}})^{N_{r-1}}(\mathcal{T}^{2\boldsymbol{e}_r})^{\lfloor N_r/2\rfloor}(\mathcal{T}^{\boldsymbol{e}_r})^{N_r-2\lfloor N_r/2\rfloor}\right)_{\alpha_1\cdots\alpha_n}^{\mu_1\cdots\mu_{n_v}\delta}\\
&\qquad-\text{smaller irreducible representations},
}
where the $N_i$ half-projectors $\mathcal{T}^{\boldsymbol{e}_i}$ are symmetrized, the number of vector indices is $n_v=\sum_{i=1}^{r-1}iN_i+r\lfloor N_r/2\rfloor$, the spinor index $\delta$ appears only if $N_r$ is odd, and the subtracted smaller irreducible representations are completely fixed by recursion.  The relation above can be simplified further with the help of the proper hatted projection operators since
\eqna{
(\mathcal{T}^{\boldsymbol{N}})_{\alpha_1\cdots\alpha_n}^{\mu_1\cdots\mu_{n_v}\delta}&=\left((\mathcal{T}^{\boldsymbol{e}_1})^{N_1}\cdots(\mathcal{T}^{\boldsymbol{e}_{r-1}})^{N_{r-1}}(\mathcal{T}^{2\boldsymbol{e}_r})^{\lfloor N_r/2\rfloor}(\mathcal{T}^{\boldsymbol{e}_r})^{N_r-2\lfloor N_r/2\rfloor}\right)_{\alpha_1\cdots\alpha_n}^{\mu_1'\cdots\mu_{n_v}'\delta'}\\
&\phantom{=}\qquad\times(\hat{\mathcal{P}}^{\boldsymbol{N}})_{\delta'\mu_{n_v}'\cdots\mu_1'}^{\phantom{\delta'\mu_{n_v}'\cdots\mu_1'}\mu_1\cdots\mu_{n_v}\delta},
}[EqTodd]
by definition.  Thus operators in the irreducible $\boldsymbol{N}$ representation can be rewritten as
\eqn{\mathcal{O}_{\alpha_1\cdots\alpha_n}^{\boldsymbol{N}}=(\mathcal{T}^{\boldsymbol{N}})_{\alpha_1\cdots\alpha_n}^{\delta\mu_{n_v}\cdots\mu_1}\mathcal{O}_{\mu_1\cdots\mu_{n_v}\delta}^{\boldsymbol{N}},\qquad\mathcal{O}_{\mu_1\cdots\mu_{n_v}\delta}^{\boldsymbol{N}}=(\mathcal{T}_{\boldsymbol{N}})_{\mu_1\cdots\mu_{n_v}\delta}^{\alpha_n\cdots\alpha_1}\mathcal{O}_{\alpha_1\cdots\alpha_n}^{\boldsymbol{N}},}
where $\mathcal{O}_{\mu_1\cdots\mu_{n_v}\delta}^{\boldsymbol{N}}$ transforms in the irreducible $\boldsymbol{N}$ representation but with respect to the ``dummy'' variables.

Now that the general form of the largest term appearing in the half-projectors \eqref{EqTodd} has been found with the help of the bottom-up approach developed above, it is convenient to discuss a top-down technique to construct the full half-projectors.  Indeed, the top-down approach relies on the standard Young tableau techniques (slightly modified for spinor representations) to obtain the fundamental building blocks for the hatted projection operators with the appropriate symmetry properties over their sets of indices.  This leads to the hatted projection operators without the smaller irreducible representations which are not explicit traces, but there remains smaller irreducible representations which are explicit traces.  These irreducible representations can be subtracted straightforwardly by demanding the tracelessness condition for any pairs of dummy vector indices and any pairs of dummy vector and spinor indices.

Thus, once the smaller irreducible representations which are not explicit traces have been removed, the remaining smaller irreducible representations appearing in \eqref{EqTodd} can be eliminated directly with the help of $(\mathcal{T}^{\boldsymbol{N}})_{\alpha_1\cdots\alpha_n}^{\mu_1\cdots\mu_{n_v}\delta}g_{\mu_i\mu_j}=0$ and $(\mathcal{T}^{\boldsymbol{N}})_{\alpha_1\cdots\alpha_n}^{\mu_1\cdots\mu_{n_v}\delta}(\gamma_{\mu_i})_\delta^{\phantom{\delta}\gamma}=0$ for any $i$ and $j$.  From the definition \eqref{EqTodd}, the tracelessness condition applies also directly to the proper hatted projection operators.  These conditions can be explicitly checked in the examples above since $(\mathcal{T}^{\boldsymbol{e}_2+\boldsymbol{e}_r})_{\alpha\beta\gamma}^{\mu\nu\delta}g_{\mu\nu}=(\mathcal{T}^{\boldsymbol{e}_2+\boldsymbol{e}_r})_{\alpha\beta\gamma}^{\mu\nu\delta}(\gamma_\mu)_\delta^{\phantom{\delta}\gamma}=0$ for the spinor irreducible representation $\boldsymbol{e}_2+\boldsymbol{e}_r$ while $(\mathcal{T}^{2\boldsymbol{e}_1})_{\alpha_1\beta_1\alpha_2\beta_2}^{\mu\nu}g_{\mu\nu}=0$ for the non-spinor irreducible representation $2\boldsymbol{e}_1$.

Finally, conjugate operators can sometimes be discarded when they are linearly dependent with non-conjugate operators.  For example, when the Majorana condition can be imposed, all conjugate operators are proportional to the original operators, they are thus straightforwardly linearly dependent and can be discarded.  When the Majorana condition cannot be imposed, all operators in vector representations (\textit{i.e.} with $N_r$ even) can be reduced to two ``real'' components and as such their conjugate operators are not linearly independent and they can be forgotten altogether.  The same procedure cannot be performed on operators in spinor representations (\textit{i.e.} with $N_r$ odd) which implies their conjugate operators are linearly independent and must be included.

Indeed, the conjugate operators of $\mathcal{O}_{\alpha_1\cdots\alpha_n}^{\boldsymbol{N}}$, denoted by
\eqn{\mathcal{O}^{\boldsymbol{N}C}\equiv(\mathcal{O}^{\boldsymbol{N}C})_{\alpha_1\cdots\alpha_n}=(B^{-1})_{\alpha_1\alpha_1'}\cdots(B^{-1})_{\alpha_n\alpha_n'}(\mathcal{O}^{\boldsymbol{N}*})^{\alpha_1'\cdots\alpha_n'},}
satisfy
\eqna{(\mathcal{O}^{\boldsymbol{N}C})_{\alpha_1\cdots\alpha_n}&=\left(B^{-1}[(\mathcal{T}^{\boldsymbol{N}})^{\delta\mu_{n_v}\cdots\mu_1}]^*B^{-T}\right)_{\alpha_1\cdots\alpha_n}(\mathcal{O}_{\mu_1\cdots\mu_{n_v}\delta}^{\boldsymbol{N}})^*\\
&=\left\{\begin{array}{cc}(-1)^{f(\boldsymbol{N})}(\mathcal{T}^{\boldsymbol{N}^C})^{\delta\mu_{n_v}\cdots\mu_1}_{\alpha_1\cdots\alpha_n}(B^{-1}\mathcal{O}^{\boldsymbol{N}*})_{\mu_1\cdots\mu_{n_v}\delta}&N_r\,\text{odd}\\(-1)^{f(\boldsymbol{N})}(\mathcal{T}^{\boldsymbol{N}^C})^{\mu_{n_v}\cdots\mu_1}_{\alpha_1\cdots\alpha_n}(\mathcal{O}^{\boldsymbol{N}*})_{\mu_1\cdots\mu_{n_v}}&N_r\,\text{even}\end{array}\right.,
}
where
\eqn{f(\boldsymbol{N})=\sum_{i=1}^{r-1}[i(r+q)+q(r+1)]N_i+[r(r+q)+q(r+1)]\lfloor N_r/2\rfloor,}
since $\mathcal{O}_{\alpha_1\cdots\alpha_n}^{\boldsymbol{N}}=(\mathcal{T}^{\boldsymbol{N}})_{\alpha_1\cdots\alpha_n}^{\delta\mu_{n_v}\cdots\mu_1}\mathcal{O}_{\mu_1\cdots\mu_{n_v}\delta}^{\boldsymbol{N}}$ and
\eqna{
\left(B^{-1}[(\mathcal{T}^{\boldsymbol{N}})^{\mu_1\cdots\mu_{n_v}\delta}]^*B^{-T}\right)_{\alpha_1\cdots\alpha_n}&=\left((B^{-1}\mathcal{T}^{\boldsymbol{e}_1*}B^{-T})^{N_1}\cdots(B^{-1}\mathcal{T}^{\boldsymbol{e}_{r-1}*}B^{-T})^{N_{r-1}}\right.\\
&\phantom{=}\qquad\times\left.(B^{-1}\mathcal{T}^{2\boldsymbol{e}_r*}B^{-T})^{\lfloor N_r/2\rfloor}\right.\\
&\phantom{=}\qquad\times\left.(B^{-1}\mathcal{T}^{\boldsymbol{e}_r*})^{N_r-2\lfloor N_r/2\rfloor}\hat{\mathcal{P}}^{\boldsymbol{N}*}\right)_{\alpha_1\cdots\alpha_n}^{\mu_1\cdots\mu_{n_v}\delta}.
}
Therefore, for vector representations where $\boldsymbol{N}^C=\boldsymbol{N}$, it is always possible to trade conjugate operators by considering the operator's real and imaginary parts independently.  For spinor representations, conjugate operators are linearly independent when the Majorana condition cannot be imposed.

%%%%%%%%%%%%%%%%%%%%%%%%%%%%%%%%%%%%%%%%%%%%%%%%%%

\subsubsection{Even Dimensions: \texorpdfstring{$p+q=2r$}{p+q=2r}}\label{SSSecEvenSpinor}

In even dimensions, the two inequivalent irreducible spinor representations $S\equiv S_\alpha^{\phantom{\alpha}\beta}$ and $\tilde{S}\equiv\tilde{S}_{\tilde{\alpha}}^{\phantom{\tilde{\alpha}}\tilde{\beta}}$ are given by the exponentiation of the generators $\sigma_{\mu\nu}\equiv(\sigma_{\mu\nu})_\alpha^{\phantom{\alpha}\beta}$ and $\tilde{\sigma}_{\mu\nu}\equiv(\tilde{\sigma}_{\mu\nu})_{\tilde{\alpha}}^{\phantom{\tilde{\alpha}}\tilde{\beta}}$ in the fundamental spinor representations, $S=\exp(\tfrac{i}{2}\omega^{\mu\nu}\sigma_{\mu\nu})$ and $\tilde{S}=\exp(\tfrac{i}{2}\omega^{\mu\nu}\tilde{\sigma}_{\mu\nu})$ respectively, with real antisymmetric $\omega^{\mu\nu}$.  The generators in the fundamental spinor representations are constructed from the square $2^{r-1}$-dimensional $\gamma$-matrices $\gamma_\mu\equiv(\gamma_\mu)_\alpha^{\phantom{\alpha}\tilde{\alpha}}$ and $\tilde{\gamma}$-matrices $\tilde{\gamma}_\mu\equiv(\tilde{\gamma}_\mu)_{\tilde{\alpha}}^{\phantom{\tilde{\alpha}}\alpha}$ which are elements of the ``Clifford'' algebra,
\eqna{
\gamma_\mu\tilde{\gamma}_\nu+\gamma_\nu\tilde{\gamma}_\mu&=2g_{\mu\nu}\mathds{1},\\
\tilde{\gamma}_\mu\gamma_\nu+\tilde{\gamma}_\nu\gamma_\mu&=2g_{\mu\nu}\mathds{1},
}
as
\eqna{
\sigma_{\mu\nu}&=\frac{i}{4}(\gamma_\mu\tilde{\gamma}_\nu-\gamma_\nu\tilde{\gamma}_\mu),\\
\tilde{\sigma}_{\mu\nu}&=\frac{i}{4}(\tilde{\gamma}_\mu\gamma_\nu-\tilde{\gamma}_\nu\gamma_\mu).
}
By definition, they satisfy the Lie algebra
\eqna{
[\sigma_{\mu\nu},\sigma_{\lambda\rho}]&=-(s_{\mu\nu})_{\lambda}^{\phantom{\lambda}\lambda'}\sigma_{\lambda'\rho}-(s_{\mu\nu})_{\rho}^{\phantom{\rho}\rho'}\sigma_{\lambda\rho'},\\
[\tilde{\sigma}_{\mu\nu},\tilde{\sigma}_{\lambda\rho}]&=-(s_{\mu\nu})_{\lambda}^{\phantom{\lambda}\lambda'}\tilde{\sigma}_{\lambda'\rho}-(s_{\mu\nu})_{\rho}^{\phantom{\rho}\rho'}\tilde{\sigma}_{\lambda\rho'}.
}
Note that $\Lambda_\mu^{\phantom{\mu}\mu'}\gamma_{\mu'}$ and $\Lambda_\mu^{\phantom{\mu}\mu'}\tilde{\gamma}_{\mu'}$ satisfy the ``Clifford'' algebra and thus should be related by a generalized similarity transformation to $\gamma_\mu$ and $\tilde{\gamma}_\mu$.  These generalized similarity transformations are nothing else than the fundamental spinor representations $S$ and $\tilde{S}$, \textit{i.e.} $\Lambda_\mu^{\phantom{\mu}\mu'}\gamma_{\mu'}=S\gamma_\mu\tilde{S}^{-1}$ and $\Lambda_\mu^{\phantom{\mu}\mu'}\tilde{\gamma}_{\mu'}=\tilde{S}\tilde{\gamma}_\mu S^{-1}$.

Since $S$ is a representation of $SO(p,q)$, the adjoint $S^{-\dagger}$ with generators $\sigma_{\mu\nu}^\dagger$, contragredient $S^{-T}$ with generators $-\sigma_{\mu\nu}^T$ and conjugate $S^*$ with generators $-\sigma_{\mu\nu}^*$ are also representations.  The same is true of the tilde representations.  However unicity of $S$ and $\tilde{S}$ implies that these other representations are related to $S$ and $\tilde{S}$.  Thus there exist similarity transformations which relate the adjoint, contragredient and conjugate representations to $S$ or $\tilde{S}$.  Contrary to the odd-dimensional case, there are four different scenarios depending on the dimension and the signature of $SO(p,q)$.  Each case is treated separately below.

Defining the matrices $A$, $C$, $B$ as well as $\tilde{A}$, $\tilde{C}$ and $\tilde{B}$,
\eqn{
\begin{array}{cccccc}
& & \text{$r$ even} & \text{$r$ even} & \text{$r$ odd} & \text{$r$ odd}\\
& & \text{$q$ even} & \text{$q$ odd} & \text{$q$ even} & \text{$q$ odd}\\
A & \equiv & A_\alpha^{\phantom{\alpha}\beta} & A_{\tilde{\alpha}}^{\phantom{\tilde{\alpha}}\beta} & A_\alpha^{\phantom{\alpha}\beta} & A_{\tilde{\alpha}}^{\phantom{\tilde{\alpha}}\beta}\\
C & \equiv & C^{\alpha\beta} & C^{\alpha\beta} & C^{\tilde{\alpha}\beta} & C^{\tilde{\alpha}\beta}\\
B & \equiv & B^{\alpha\beta}=(C^{-\dagger}A)^{\alpha\beta} & B^{\tilde{\alpha}\beta}=(\tilde{C}^{-\dagger}A)^{\tilde{\alpha}\beta} & B^{\tilde{\alpha}\beta}=(C^{-\dagger}A)^{\tilde{\alpha}\beta} & B^{\alpha\beta}=(\tilde{C}^{-\dagger}A)^{\alpha\beta}\\
\tilde{A} & \equiv & \tilde{A}_{\tilde{\alpha}}^{\phantom{\tilde{\alpha}}\tilde{\beta}} & \tilde{A}_\alpha^{\phantom{\alpha}\tilde{\beta}} & \tilde{A}_{\tilde{\alpha}}^{\phantom{\tilde{\alpha}}\tilde{\beta}} & \tilde{A}_\alpha^{\phantom{\alpha}\tilde{\beta}}\\
\tilde{C} & \equiv & \tilde{C}^{\tilde{\alpha}\tilde{\beta}} & \tilde{C}^{\tilde{\alpha}\tilde{\beta}} & \tilde{C}^{\alpha\tilde{\beta}} & \tilde{C}^{\alpha\tilde{\beta}}\\
\tilde{B} & \equiv & \tilde{B}^{\tilde{\alpha}\tilde{\beta}}=(\tilde{C}^{-\dagger}\tilde{A})^{\tilde{\alpha}\tilde{\beta}} & \tilde{B}^{\alpha\tilde{\beta}}=(C^{-\dagger}\tilde{A})^{\alpha\tilde{\beta}} & \tilde{B}^{\alpha\tilde{\beta}}=(\tilde{C}^{-\dagger}\tilde{A})^{\alpha\tilde{\beta}} & \tilde{B}^{\tilde{\alpha}\tilde{\beta}}=(C^{-\dagger}\tilde{A})^{\tilde{\alpha}\tilde{\beta}},
\end{array}
}
such that
\eqn{
\begin{array}{cccccccc}
& & & & \text{$r$ even} & \text{$r$ even} & \text{$r$ odd} & \text{$r$ odd}\\
& & & & \text{$q$ even} & \text{$q$ odd} & \text{$q$ even} & \text{$q$ odd}\\
\gamma_\mu^\dagger & = & (-1)^q & \times & \tilde{A}\tilde{\gamma}_\mu A^{-1} & A\gamma_\mu\tilde{A}^{-1} & \tilde{A}\tilde{\gamma}_\mu A^{-1} & A\gamma_\mu\tilde{A}^{-1}\\
\gamma_\mu^T & = & (-1)^r & \times & \tilde{C}\tilde{\gamma}_\mu C^{-1} & \tilde{C}\tilde{\gamma}_\mu C^{-1} & C\gamma_\mu\tilde{C}^{-1} & C\gamma_\mu\tilde{C}^{-1}\\
\gamma_\mu^* & = & (-1)^{r+q} & \times & B\gamma_\mu\tilde{B}^{-1} & \tilde{B}\tilde{\gamma}_\mu B^{-1} & \tilde{B}\tilde{\gamma}_\mu B^{-1} & B\gamma_\mu\tilde{B}^{-1}\\
\tilde{\gamma}_\mu^\dagger & = & (-1)^q & \times & A\gamma_\mu\tilde{A}^{-1} & \tilde{A}\tilde{\gamma}_\mu A^{-1} & A\gamma_\mu\tilde{A}^{-1} & \tilde{A}\tilde{\gamma}_\mu A^{-1}\\
\tilde{\gamma}_\mu^T & = & (-1)^r & \times & C\gamma_\mu\tilde{C}^{-1} & C\gamma_\mu\tilde{C}^{-1} & \tilde{C}\tilde{\gamma}_\mu C^{-1} & \tilde{C}\tilde{\gamma}_\mu C^{-1}\\
\tilde{\gamma}_\mu^* & = & (-1)^{r+q} & \times & \tilde{B}\tilde{\gamma}_\mu B^{-1} & B\gamma_\mu\tilde{B}^{-1} & B\gamma_\mu\tilde{B}^{-1} & \tilde{B}\tilde{\gamma}_\mu B^{-1},
\end{array}
}[EqSymgammaEven]
one thus has
\eqn{
\begin{array}{cccccc}
& & \text{$r$ even} & \text{$r$ even} & \text{$r$ odd} & \text{$r$ odd}\\
& & \text{$q$ even} & \text{$q$ odd} & \text{$q$ even} & \text{$q$ odd}\\
\sigma_{\mu\nu}^\dagger & = & A\sigma_{\mu\nu}A^{-1} & \tilde{A}\tilde{\sigma}_{\mu\nu}\tilde{A}^{-1} & A\sigma_{\mu\nu}A^{-1} & \tilde{A}\tilde{\sigma}_{\mu\nu}\tilde{A}^{-1}\\
-\sigma_{\mu\nu}^T & = & C\sigma_{\mu\nu}C^{-1} & C\sigma_{\mu\nu}C^{-1} & \tilde{C}\tilde{\sigma}_{\mu\nu}\tilde{C}^{-1} & \tilde{C}\tilde{\sigma}_{\mu\nu}\tilde{C}^{-1}\\
-\sigma_{\mu\nu}^* & = & B\sigma_{\mu\nu}B^{-1} & \tilde{B}\tilde{\sigma}_{\mu\nu}\tilde{B}^{-1} & \tilde{B}\tilde{\sigma}_{\mu\nu}\tilde{B}^{-1} & B\sigma_{\mu\nu}B^{-1}\\
\tilde{\sigma}_{\mu\nu}^\dagger & = & \tilde{A}\tilde{\sigma}_{\mu\nu}\tilde{A}^{-1} & A\sigma_{\mu\nu}A^{-1} & \tilde{A}\tilde{\sigma}_{\mu\nu}\tilde{A}^{-1} & A\sigma_{\mu\nu}A^{-1}\\
-\tilde{\sigma}_{\mu\nu}^T & = & \tilde{C}\tilde{\sigma}_{\mu\nu}\tilde{C}^{-1} & \tilde{C}\tilde{\sigma}_{\mu\nu}\tilde{C}^{-1} & C\sigma_{\mu\nu}C^{-1} & C\sigma_{\mu\nu}C^{-1}\\
-\tilde{\sigma}_{\mu\nu}^* & = & \tilde{B}\tilde{\sigma}_{\mu\nu}\tilde{B}^{-1} & B\sigma_{\mu\nu}B^{-1} & B\sigma_{\mu\nu}B^{-1} & \tilde{B}\tilde{\sigma}_{\mu\nu}\tilde{B}^{-1},
\end{array}
}
or
\eqn{
\begin{array}{cccccc}
& & \text{$r$ even} & \text{$r$ even} & \text{$r$ odd} & \text{$r$ odd}\\
& & \text{$q$ even} & \text{$q$ odd} & \text{$q$ even} & \text{$q$ odd}\\
S^{-\dagger} & = & ASA^{-1} & \tilde{A}\tilde{S}\tilde{A}^{-1} & ASA^{-1} & \tilde{A}\tilde{S}\tilde{A}^{-1}\\
S^{-T} & = & CSC^{-1} & CSC^{-1} & \tilde{C}\tilde{S}\tilde{C}^{-1} & \tilde{C}\tilde{S}\tilde{C}^{-1}\\
S^* & = & BSB^{-1} & \tilde{B}\tilde{S}\tilde{B}^{-1} & \tilde{B}\tilde{S}\tilde{B}^{-1} & BSB^{-1}\\
\tilde{S}^{-\dagger} & = & \tilde{A}\tilde{S}\tilde{A}^{-1} & ASA^{-1} & \tilde{A}\tilde{S}\tilde{A}^{-1} & ASA^{-1}\\
\tilde{S}^{-T} & = & \tilde{C}\tilde{S}\tilde{C}^{-1} & \tilde{C}\tilde{S}\tilde{C}^{-1} & CSC^{-1} & CSC^{-1}\\
\tilde{S}^* & = & \tilde{B}\tilde{S}\tilde{B}^{-1} & BSB^{-1} & BSB^{-1} & \tilde{B}\tilde{S}\tilde{B}^{-1}.
\end{array}
}
The matrices $A$ and $\tilde{A}$ can be given in explicit form, $A=\gamma_{p+1}\cdots\tilde{\gamma}_{p+q}$ and $\tilde{A}=\gamma_{p+1}\cdots\gamma_{p+q}$ where $\gamma$'s and $\tilde{\gamma}$'s alternate and the type of the last one is fixed (hence the index position), and the matrices $C$ and $\tilde{C}$ are unitary ($C^\dagger C=\mathds{1}$, $\tilde{C}^\dagger\tilde{C}=\mathds{1}$) with $C^T=(-1)^{r(r+1)/2}C$, $\tilde{C}^T=(-1)^{r(r+1)/2}\tilde{C}$ for $r$ even and $C^T=(-1)^{r(r+1)/2}\tilde{C}$, $\tilde{C}^T=(-1)^{r(r+1)/2}C$ for $r$ odd.  From their definitions, the ``Clifford'' algebra and \eqref{EqSymgammaEven}, $A$ and $\tilde{A}$ satisfy special properties given by
\eqn{
\begin{array}{cccccccc}
& & & & \text{$r$ even} & \text{$r$ even} & \text{$r$ odd} & \text{$r$ odd}\\
& & & & \text{$q$ even} & \text{$q$ odd} & \text{$q$ even} & \text{$q$ odd}\\
A^{-1} & = & (-1)^{q(q+1)/2} & \times & A & \tilde{A} & A & \tilde{A}\\
A^\dagger & = & (-1)^{q(q+1)/2} & \times & A & \tilde{A} & A & \tilde{A}\\
A^T & = & (-1)^{rq+q(q-1)/2} & \times & CAC^{-1} & C\tilde{A}\tilde{C}^{-1} & \tilde{C}\tilde{A}\tilde{C}^{-1} & \tilde{C}AC^{-1}\\
A^* & = & (-1)^{q(r+1)} & \times & CAC^{-1} & \tilde{C}AC^{-1} & \tilde{C}\tilde{A}\tilde{C}^{-1} & C\tilde{A}\tilde{C}^{-1}\\
\tilde{A}^{-1} & = & (-1)^{q(q+1)/2} & \times & \tilde{A} & A & \tilde{A} & A\\
\tilde{A}^\dagger & = & (-1)^{q(q+1)/2} & \times & \tilde{A} & A & \tilde{A} & A\\
\tilde{A}^T & = & (-1)^{rq+q(q-1)/2} & \times & \tilde{C}\tilde{A}\tilde{C}^{-1} & \tilde{C}AC^{-1} & CAC^{-1} & C\tilde{A}\tilde{C}^{-1}\\
\tilde{A}^* & = & (-1)^{q(r+1)} & \times & \tilde{C}\tilde{A}\tilde{C}^{-1} & C\tilde{A}\tilde{C}^{-1} & CAC^{-1} & \tilde{C}AC^{-1}.
\end{array}
}
Thus $A$, $C$, $\tilde{A}$ and $\tilde{C}$ are unitary, and hence $B$ and $\tilde{B}$, which satisfy similar properties like
\eqn{
\begin{array}{cccccccc}
& & & & \text{$r$ even} & \text{$r$ even} & \text{$r$ odd} & \text{$r$ odd}\\
& & & & \text{$q$ even} & \text{$q$ odd} & \text{$q$ even} & \text{$q$ odd}\\
B^T & = & (-1)^{r(r+1)/2+rq+q(q-1)/2} & \times & B & \tilde{B} & \tilde{B} & B\\
\tilde{B}^T & = & (-1)^{r(r+1)/2+rq+q(q-1)/2} & \times & \tilde{B} & B & B & \tilde{B},
\end{array}
}
are also unitary.  These properties are invariant under generalized unitary similarity transformations $\gamma_\mu\to U\gamma_\mu\tilde{U}^\dagger$, $\tilde{\gamma}_\mu\to\tilde{U}\tilde{\gamma}_\mu U^\dagger$ where $U^\dagger U=\mathds{1}$ and $\tilde{U}^\dagger\tilde{U}=\mathds{1}$ as long as the matrices are changed into
\eqn{
\begin{array}{cccccc}
& & \text{$r$ even} & \text{$r$ even} & \text{$r$ odd} & \text{$r$ odd}\\
& & \text{$q$ even} & \text{$q$ odd} & \text{$q$ even} & \text{$q$ odd}\\
A & \to & UAU^\dagger & \tilde{U}AU^\dagger & UAU^\dagger & \tilde{U}AU^\dagger\\
C & \to & U^*CU^\dagger & U^*CU^\dagger & \tilde{U}^*CU^\dagger & \tilde{U}^*CU^\dagger\\
B & \to & U^*BU^\dagger & \tilde{U}^*BU^\dagger & \tilde{U}^*BU^\dagger & U^*BU^\dagger\\
\tilde{A} & \to & \tilde{U}\tilde{A}\tilde{U}^\dagger & U\tilde{A}\tilde{U}^\dagger & \tilde{U}\tilde{A}\tilde{U}^\dagger & U\tilde{A}\tilde{U}^\dagger\\
\tilde{C} & \to & \tilde{U}^*\tilde{C}\tilde{U}^\dagger & \tilde{U}^*\tilde{C}\tilde{U}^\dagger & U^*\tilde{C}\tilde{U}^\dagger & U^*\tilde{A}\tilde{U}^\dagger\\
\tilde{B} & \to & \tilde{U}^*\tilde{B}\tilde{U}^\dagger & U^*\tilde{B}\tilde{U}^\dagger & U^*\tilde{B}\tilde{U}^\dagger & \tilde{U}^*\tilde{A}\tilde{U}^\dagger,
\end{array}
}
as dictated by \eqref{EqSymgammaEven}.

Any element $\psi\equiv\psi_\alpha$ ($\tilde{\psi}\equiv\tilde{\psi}_{\tilde{\alpha}}$) in the linear space $V$ ($\tilde{V}$) transforms as $\psi\to S\psi$ ($\tilde{\psi}\to\tilde{S}\tilde{\psi}$).  The related representations act appropriately, \textit{i.e.}
\eqn{
\begin{array}{cccccc}
& & \text{$r$ even} & \text{$r$ even} & \text{$r$ odd} & \text{$r$ odd}\\
& & \text{$q$ even} & \text{$q$ odd} & \text{$q$ even} & \text{$q$ odd}\\
\psi^\dagger\equiv(\psi^\dagger)^\alpha\to\psi^\dagger S^\dagger & = & \psi^\dagger AS^{-1}A^{-1} & \psi^\dagger\tilde{A}\tilde{S}^{-1}\tilde{A}^{-1} & \psi^\dagger AS^{-1}A^{-1} & \psi^\dagger\tilde{A}\tilde{S}^{-1}\tilde{A}^{-1}\\
\psi^T\equiv(\psi^T)_\alpha\to\psi^TS^T & = & \psi^TCS^{-1}C^{-1} & \psi^TCS^{-1}C^{-1} & \psi^T\tilde{C}\tilde{S}^{-1}\tilde{C}^{-1} & \psi^T\tilde{C}\tilde{S}^{-1}\tilde{C}^{-1}\\
\psi^*\equiv(\psi^*)^\alpha\to S^*\psi^* & = & BSB^{-1}\psi^* & \tilde{B}\tilde{S}\tilde{B}^{-1}\psi^* & \tilde{B}\tilde{S}\tilde{B}^{-1}\psi^* & BSB^{-1}\psi^*\\
\tilde{\psi}^\dagger\equiv(\tilde{\psi}^\dagger)^{\tilde{\alpha}}\to\tilde{\psi}^\dagger\tilde{S}^\dagger & = & \tilde{\psi}^\dagger\tilde{A}\tilde{S}^{-1}\tilde{A}^{-1} & \tilde{\psi}^\dagger AS^{-1}A^{-1} & \tilde{\psi}^\dagger\tilde{A}\tilde{S}^{-1}\tilde{A}^{-1} & \tilde{\psi}^\dagger AS^{-1}A^{-1}\\
\tilde{\psi}^T\equiv(\tilde{\psi}^T)_{\tilde{\alpha}}\to\tilde{\psi}^T\tilde{S}^T & = & \tilde{\psi}^T\tilde{C}\tilde{S}^{-1}\tilde{C}^{-1} & \tilde{\psi}^T\tilde{C}\tilde{S}^{-1}\tilde{C}^{-1} & \tilde{\psi}^TCS^{-1}C^{-1} & \tilde{\psi}^TCS^{-1}C^{-1}\\
\tilde{\psi}^*\equiv(\tilde{\psi}^*)^{\tilde{\alpha}}\to\tilde{S}^*\tilde{\psi}^* & = & \tilde{B}\tilde{S}\tilde{B}^{-1}\tilde{\psi}^* & BSB^{-1}\tilde{\psi}^* & BSB^{-1}\tilde{\psi}^* & \tilde{B}\tilde{S}\tilde{B}^{-1}\tilde{\psi}^*.
\end{array}
}

By defining $\psi^C$ and $\tilde{\psi}^C$ such as
\eqn{
\begin{array}{cccccc}
& & \text{$r$ even} & \text{$r$ even} & \text{$r$ odd} & \text{$r$ odd}\\
& & \text{$q$ even} & \text{$q$ odd} & \text{$q$ even} & \text{$q$ odd}\\
\psi^C & = & B^{-1}\psi^*\sim\psi & \tilde{B}^{-1}\psi^*\sim\tilde{\psi} & \tilde{B}^{-1}\psi^*\sim\tilde{\psi} & B^{-1}\psi^*\sim\psi\\
\tilde{\psi}^C & = & \tilde{B}^{-1}\tilde{\psi}^*\sim\tilde{\psi} & B^{-1}\tilde{\psi}^*\sim\psi & B^{-1}\tilde{\psi}^*\sim\psi & \tilde{B}^{-1}\tilde{\psi}^*\sim\tilde{\psi},
\end{array}
}
where $\sim$ indicates that the spinors transform identically, $\psi^C$ and $\tilde{\psi}^C$ could be proportional to $\psi$ and $\tilde{\psi}$ respectively when $r+q=0\text{ mod }2$ and hence halves the (real) dimension of the fundamental spinor representations, leading to the irreducible Weyl-Majorana representations.  For this reality condition to exist, $r$ and $q$ must be both even or both odd and the consistency conditions $(\psi^C)^C=\psi$ and $(\tilde{\psi}^C)^C=\tilde{\psi}$ must be satisfied, which enforce $B^{-1}B^T=1$ and $\tilde{B}^{-1}\tilde{B}^T=1$.  Therefore the irreducible Weyl-Majorana representation exists only for $r+q=0\text{ mod }2$ and $r(r+1)/2+rq+q(q-1)/2=0\text{ mod }2$.

More complicated irreducible representations can be studied from the product of fundamental spinor representations with the help of the antisymmetrized $\gamma$-matrices.  In the even-dimensional case, there are four types of matrices depending on the type of indices they carry.  Since there are $\genfrac{(}{)}{0pt}{}{2r}{2n}$ fully antisymmetric (on their vector indices) matrices $\gamma^{[2n]}$ and $\tilde{\gamma}^{[2n]}$, given by
\eqna{
\gamma^{[2n]}&\equiv\gamma^{\mu_1\cdots\mu_{2n}}=\gamma^{[\mu_1}\cdots\tilde{\gamma}^{\mu_{2n}]}=\frac{1}{(2n)!}\sum_{\sigma\in S_{2n}}(-1)^\sigma\gamma^{\mu_{\sigma(1)}}\cdots\tilde{\gamma}^{\mu_{\sigma(2n)}},\\
\tilde{\gamma}^{[2n]}&\equiv\tilde{\gamma}^{\mu_1\cdots\mu_{2n}}=\tilde{\gamma}^{[\mu_1}\cdots\gamma^{\mu_{2n}]}=\frac{1}{(2n)!}\sum_{\sigma\in S_{2n}}(-1)^\sigma\tilde{\gamma}^{\mu_{\sigma(1)}}\cdots\gamma^{\mu_{\sigma(2n)}},
}
where $S_{2n}$ is the symmetric group with element $\sigma$ and sign $(-1)^\sigma$, they form a complete basis of square $2^{r-1}$-dimensional matrices for $n\in\{0,1,\ldots,\lfloor r/2\rfloor\}$.  Indeed there are $2^{2(r-1)}$ such matrices (when the duality condition is taken into account, see below) which corresponds to the number of components of square $2^{r-1}$-dimensional matrices.  There are also $\genfrac{(}{)}{0pt}{}{2r}{2n+1}$ fully antisymmetric (on their vector indices) matrices $\gamma^{[2n+1]}$ and $\tilde{\gamma}^{[2n+1]}$, given by
\eqna{
\gamma^{[2n+1]}&\equiv\gamma^{\mu_1\cdots\mu_{2n+1}}=\gamma^{[\mu_1}\cdots\gamma^{\mu_{2n+1}]}=\frac{1}{(2n+1)!}\sum_{\sigma\in S_{2n+1}}(-1)^\sigma\gamma^{\mu_{\sigma(1)}}\cdots\gamma^{\mu_{\sigma(2n+1)}},\\
\tilde{\gamma}^{[2n+1]}&\equiv\tilde{\gamma}^{\mu_1\cdots\mu_{2n+1}}=\tilde{\gamma}^{[\mu_1}\cdots\tilde{\gamma}^{\mu_{2n+1}]}=\frac{1}{(2n+1)!}\sum_{\sigma\in S_{2n+1}}(-1)^\sigma\tilde{\gamma}^{\mu_{\sigma(1)}}\cdots\tilde{\gamma}^{\mu_{\sigma(2n+1)}},
}
where $S_{2n+1}$ is the symmetric group with element $\sigma$ and sign $(-1)^\sigma$, which form a complete basis of square $2^{r-1}$-dimensional matrices for $n\in\{0,1,\ldots,\lfloor(r-1)/2\rfloor\}$.  Indeed there are $2^{2(r-1)}$ such matrices (when the duality condition is taken into account, see below) which corresponds to the number of components of square $2^{r-1}$-dimensional matrices.

Note that the remaining matrices $\gamma^{[2n]}$ and $\tilde{\gamma}^{[2n]}$ with $n\in\{\lfloor r/2\rfloor+1,\ldots,r\}$ and $\gamma^{[2n+1]}$ and $\tilde{\gamma}^{[2n+1]}$ with $n\in\{\lfloor(r-1)/2\rfloor+1,\ldots,r-1\}$ are related to these through the $SO(p,q)$-invariant epsilon tensor $\epsilon^{\mu_1\cdots\mu_{2r}}$,
\eqna{
\gamma^{\mu_1\cdots\mu_n}&=\frac{\mathscr{K}}{(d-n)!}\epsilon_{\phantom{\mu_1\cdots\mu_n}\nu_{d-n}\cdots\nu_1}^{\mu_1\cdots\mu_n}\gamma^{\nu_1\cdots\nu_{d-n}},\\
\tilde{\gamma}^{\mu_1\cdots\mu_n}&=-\frac{\mathscr{K}}{(d-n)!}\epsilon_{\phantom{\mu_1\cdots\mu_n}\nu_{d-n}\cdots\nu_1}^{\mu_1\cdots\mu_n}\tilde{\gamma}^{\nu_1\cdots\nu_{d-n}},
}
where $\mathscr{K}$ is the proportionality constant in $\gamma^{\mu_1\cdots\mu_d}=\epsilon^{\mu_1\cdots\mu_d}\gamma^{1\cdots d}=\epsilon^{\mu_1\cdots\mu_d}\gamma^1\cdots\tilde{\gamma}^d=\mathscr{K}\epsilon^{\mu_1\cdots\mu_d}\1$ which satifies $\mathscr{K}^2=(-1)^{r+q}$.

Therefore, the epsilon tensor is responsible for an extra factor of one half in the number of matrices $\gamma^{[r]}$ and $\tilde{\gamma}^{[r]}$, \textit{i.e.} $\tfrac{1}{2}\genfrac{(}{)}{0pt}{}{2r}{r}$ instead of $\genfrac{(}{)}{0pt}{}{2r}{r}$, due to their self-duality and anti-self-duality properties, leading to the right counting mentioned above.

The $\gamma$-matrices for $n\in\{0,1,\ldots,r\}$ satisfy the following important identities,
\eqna{
\text{tr}(\gamma_{\mu_n\cdots\mu_1}\gamma^{\nu_1\cdots\nu_m})&=2^{r-1}n!\left(\delta_{[\mu_1}^{\phantom{[\mu_1}\nu_1}\cdots\delta_{\mu_n]}^{\phantom{\mu_i]}\nu_n}\pm(-1)^r\frac{\mathscr{K}}{r!}\epsilon_{\mu_1\cdots\mu_r}^{\phantom{\mu_1\cdots\mu_r}\nu_r\cdots\nu_1}\delta_{nr}\right)\delta_{nm},\\
\gamma_{\mu_1\cdots\mu_n\nu}&=\gamma_{\mu_1\cdots\mu_n}\gamma_\nu+\sum_{i=1}^n(-1)^{n+1-i}g_{\nu\mu_i}\gamma_{\mu_1\cdots\widehat{\mu_i}\cdots\mu_n},\\
\gamma_{\nu\mu_1\cdots\mu_n}&=\gamma_\nu\tilde{\gamma}_{\mu_1\cdots\mu_n}+\sum_{i=1}^n(-1)^ig_{\nu\mu_i}\gamma_{\mu_1\cdots\widehat{\mu_i}\cdots\mu_n},
}
where the appropriate tilde and untilde matrices must be used to ensure matrix multiplication consistency and the $\pm$ distinguishes between self-dual ($\gamma^{\nu_1\cdots\nu_m}$ with $+$) and anti-self-dual ($\tilde{\gamma}^{\nu_1\cdots\nu_m}$ with $-$) irreducible representations.  These properties also lead to Fierz identities of the type
\eqna{
(\gamma^{\mu_1\cdots\mu_i}C^{-1})_{\alpha_1\alpha_2}(\gamma^{\nu_1\cdots\nu_j}C^{-1})_{\beta_1\beta_2}&=\mathcal{K}_{\phantom{\mu_1\cdots\mu_i\nu_1\cdots\nu_j}\lambda_1\cdots\lambda_k\rho_1\cdots\rho_\ell}^{\mu_1\cdots\mu_i\nu_1\cdots\nu_j}(\gamma^{\lambda_1\cdots\lambda_k}C^{-1})_{\alpha_1\beta_2}(\gamma^{\rho_1\cdots\rho_\ell}C^{-1})_{\beta_1\alpha_2},\\
\mathcal{K}_{\phantom{\mu_1\cdots\mu_i\nu_1\cdots\nu_j}\lambda_1\cdots\lambda_k\rho_1\cdots\rho_\ell}^{\mu_1\cdots\mu_i\nu_1\cdots\nu_j}&=\frac{1}{4^{r-1}k!\ell!}\text{tr}(\gamma_{\lambda_k\cdots\lambda_1}\gamma^{\mu_1\cdots\mu_i}\gamma_{\rho_\ell\cdots\rho_1}\gamma^{\nu_1\cdots\nu_j}).
}
where the matrices $C^{-1}$ and $\tilde{C}^{-1}$ have been used to lower indices and the appropriate indices must be used.  Indeed, since $C$ and $\tilde{C}$ have two upper indices, it is possible to raise and lower indices using $C$, $\tilde{C}$ and $C^{-1}$, $\tilde{C}^{-1}$ respectively.  This allows the study of the symmetry properties of $C\gamma^{[n]}$ and $\tilde{C}\tilde{\gamma}^{[n]}$, which are
\eqn{
\begin{array}{cccccccc}
& & & & \text{$r$ even} & \text{$r$ even} & \text{$r$ odd} & \text{$r$ odd}\\
& & & & \text{$n$ even} & \text{$n$ odd} & \text{$n$ even} & \text{$n$ odd}\\
\left(C\gamma^{[n]}\right)^T & = & (-1)^{n(n-1)/2+nr+r(r+1)/2} & \times & C\gamma^{[n]} & \tilde{C}\tilde{\gamma}^{[n]} & \tilde{C}\tilde{\gamma}^{[n]} & C\gamma^{[n]}\\
\left(\tilde{C}\tilde{\gamma}^{[n]}\right)^T & = & (-1)^{n(n-1)/2+nr+r(r+1)/2} & \times & \tilde{C}\tilde{\gamma}^{[n]} & C\gamma^{[n]} & C\gamma^{[n]} & \tilde{C}\tilde{\gamma}^{[n]}.
\end{array}
}
These equations have a periodicity under $n\to n+4$.  Thus for $i\in\{0,1,\ldots,\lfloor r/4\rfloor\}$, the symmetry properties of the different matrices are $\left(C\gamma^{[r-4i]}\right)^T=C\gamma^{[r-4i]}$ and $\left(C\gamma^{[r-2-4i]}\right)^T=-C\gamma^{[r-2-4i]}$ while $\left(C\gamma^{[r-3-4i]}\right)^T=\tilde{C}\tilde{\gamma}^{[r-3-4i]}$ and $\left(C\gamma^{[r-1-4i]}\right)^T=-\tilde{C}\tilde{\gamma}^{[r-1-4i]}$ with the corresponding symmetry properties for the tilde matrices.  This is consistent with the numbers of symmetric and antisymmetric matrices which are $2^{r-2}(2^{r-1}+1)$ and $2^{r-2}(2^{r-1}-1)$ respectively.  As for the odd dimensional case, these matrices (rewritten with lower spinor indices) satisfy
\eqna{
B^{-1}(\gamma^{[n]}C^{-1})^*B^{-T}&=(-1)^{n(r+q)+q(r+1)}\gamma^{[n]}C^{-1},\\
B^{-1}(\tilde{\gamma}^{[n]}C^{-1})^*B^{-T}&=(-1)^{n(r+q)+q(r+1)}\tilde{\gamma}^{[n]}C^{-1},
}
where the necessary tilde and untilde matrices must be used, which is useful in investigating conjugate representations.

As in odd dimension, an irreducible representation of $SO(p,q)$ in even dimension is given by a set of non-negative integers $\boldsymbol{N}=\{N_1,\ldots,N_r\}=\sum_{i=1}^rN_i\boldsymbol{e}_i$ where $r$ is the rank of the Lie algebra and $\boldsymbol{e}_i\equiv(\boldsymbol{e}_i)_j=\delta_{ij}$ is the $i$-th unit vector.  Operators in irreducible representations of $SO(p,q)$, denoted by $\mathcal{O}_{\alpha_1\cdots\alpha_n;\tilde{\alpha}_1\cdots\tilde{\alpha}_{\tilde{n}}}^{\boldsymbol{N}}$ with ``spin'' $S$ such that
\eqn{n+\tilde{n}=2S=2\sum_{i=1}^{r-2}N_i+N_{r-1}+N_r,}
behave as
\eqn{\mathcal{O}_{\alpha_1\cdots\alpha_n;\tilde{\alpha}_1\cdots\tilde{\alpha}_{\tilde{n}}}^{\boldsymbol{N}}\sim\mathcal{T}_{\alpha_1\cdots\alpha_n;\tilde{\alpha}_1\cdots\tilde{\alpha}_{\tilde{n}}}^{\boldsymbol{N}},}
under $SO(p,q)$ transformations.  Here $\sim$ indicates that both the operator $\mathcal{O}_{\alpha_1\cdots\alpha_n;\tilde{\alpha}_1\cdots\tilde{\alpha}_{\tilde{n}}}^{\boldsymbol{N}}$ and the half-projector $\mathcal{T}_{\alpha_1\cdots\alpha_n;\tilde{\alpha}_1\cdots\tilde{\alpha}_{\tilde{n}}}^{\boldsymbol{N}}$ have the same behavior under $SO(p,q)$ transformations.  The half-projectors are fundamental group theory quantities associated to irreducible representations and, as in odd dimension, they can be built recursively.  They also corresponds to ``square roots'' of the associated projection operators which project to the appropriate irreducible representation,
\eqn{(\mathcal{P}_{\boldsymbol{N}})_{\alpha_1\cdots\alpha_n;\tilde{\alpha}_1\cdots\tilde{\alpha}_{\tilde{n}}}^{\phantom{\alpha_1\cdots\alpha_n;\tilde{\alpha}_1\cdots\tilde{\alpha}_{\tilde{n}}}\alpha_n'\cdots\alpha_1';\tilde{\alpha}_{\tilde{n}}'\cdots\tilde{\alpha}_1'}=\mathcal{T}_{\alpha_1\cdots\alpha_n;\tilde{\alpha}_1\cdots\tilde{\alpha}_{\tilde{n}}}^{\boldsymbol{N}}\mathcal{T}_{\boldsymbol{N}}^{\alpha_n'\cdots\alpha_1';\tilde{\alpha}_{\tilde{n}}'\cdots\tilde{\alpha}_1'}.}
such that the usual properties of projection operators are satisfied, \textit{i.e} $\mathcal{P}_{\boldsymbol{N'}}\mathcal{T}^{\boldsymbol{N}}=\delta_{\boldsymbol{N'}\boldsymbol{N}}\mathcal{T}^{\boldsymbol{N}}$ and $\mathcal{P}_{\boldsymbol{N}}\mathcal{P}_{\boldsymbol{N'}}=\delta_{\boldsymbol{N'}\boldsymbol{N}}\mathcal{P}_{\boldsymbol{N}}$.  The previous contractions vanish automatically when $S'\neq S$ and for a given fixed $S$, the projection operators satisfy $\sum_{\boldsymbol{N}|S\,\text{fixed}}\mathcal{P}_{\boldsymbol{N}}=\mathds{1}$ as expected.

The trivial representation $\boldsymbol{N}=\boldsymbol{0}=\{0,\ldots,0\}$ is simply a scalar, hence $\mathcal{O}^{\boldsymbol{0}}\sim\mathcal{T}^{\boldsymbol{0}}=1$.

Contrary to odd dimension, there are two inequivalent irreducible spinor representations which are denoted by $\boldsymbol{N}=\boldsymbol{e}_{r-1}$ and $\boldsymbol{N}=\boldsymbol{e}_r$ respectively.  Operators in irreducible spinor representations transform as $\mathcal{O}_\alpha^{\boldsymbol{e}_{r-1}}\sim(\mathcal{T}^{\boldsymbol{e}_{r-1}})_\alpha^\beta=\delta_\alpha^{\phantom{\alpha}\beta}$ and $\mathcal{O}_{\tilde{\alpha}}^{\boldsymbol{e}_r}\sim(\mathcal{T}^{\boldsymbol{e}_r})_{\tilde{\alpha}}^{\tilde{\beta}}=\delta_{\tilde{\alpha}}^{\phantom{\tilde{\alpha}}\tilde{\beta}}$ respectively.  As in odd dimension, $\beta$ and $\tilde{\beta}$ are dummy indices to be contracted properly.  The projection operators corresponding to the irreducible spinor representations are given by $(\mathcal{P}_{\boldsymbol{e}_{r-1}})_\alpha^{\phantom{\alpha}\alpha'}=(\mathcal{T}^{\boldsymbol{e}_{r-1}})_\alpha^\beta(\mathcal{T}_{\boldsymbol{e}_{r-1}})_\beta^{\alpha'}=\delta_\alpha^{\phantom{\alpha}\alpha'}$ and $(\mathcal{P}_{\boldsymbol{e}_r})_{\tilde{\alpha}}^{\phantom{\tilde{\alpha}}\tilde{\alpha}'}=(\mathcal{T}^{\boldsymbol{e}_r})_{\tilde{\alpha}}^{\tilde{\beta}}(\mathcal{T}_{\boldsymbol{e}_r})_{\tilde{\beta}}^{\tilde{\alpha}'}=\delta_{\tilde{\alpha}}^{\phantom{\tilde{\alpha}}\tilde{\alpha}'}$ and they correspond to the half-projectors.  The two irreducible spinor representations are defining representations from which all the remaining representations can be built.

Contrary to odd dimension though, there are two different ways of constructing some of the remaining defining representations.  For example, the half-projectors for the irreducible $i$-index antisymmetric vector representations $\boldsymbol{N}=\boldsymbol{e}_i$ for $i\in\{1,\ldots,r-2\}$ are given by
\eqn{
\begin{array}{cccccc}
& & & & \text{$r$ even} & \text{$r$ odd}\\
(\mathcal{T}^{\boldsymbol{e}_{r-2i}})_{\alpha\beta}^{\mu_1\cdots\mu_{r-2i}} & = & \frac{1}{\sqrt{2^{r-1}(r-2i)!}} & \times & (\gamma^{\mu_1\cdots\mu_{r-2i}}C^{-1})_{\alpha\beta} & (\gamma^{\mu_1\cdots\mu_{r-2i}}\tilde{C}^{-1})_{\alpha\beta},\\
(\mathcal{T}^{\boldsymbol{e}_{r-2i}})_{\tilde{\alpha}\tilde{\beta}}^{\mu_1\cdots\mu_{r-2i}} & = & \frac{1}{\sqrt{2^{r-1}(r-2i)!}} & \times & (\tilde{\gamma}^{\mu_1\cdots\mu_{r-2i}}\tilde{C}^{-1})_{\tilde{\alpha}\tilde{\beta}} & (\tilde{\gamma}^{\mu_1\cdots\mu_{r-2i}}C^{-1})_{\tilde{\alpha}\tilde{\beta}},
\end{array}
}
and
\eqn{
\begin{array}{cccccc}
& & & & \text{$r$ even} & \text{$r$ odd}\\
(\mathcal{T}^{\boldsymbol{e}_{r-2i-1}})_{\alpha\tilde{\beta}}^{\mu_1\cdots\mu_{r-2i-1}} & = & \frac{1}{\sqrt{2^{r-1}(r-2i-1)!}} & \times & (\gamma^{\mu_1\cdots\mu_{r-2i-1}}\tilde{C}^{-1})_{\alpha\tilde{\beta}} & (\gamma^{\mu_1\cdots\mu_{r-2i-1}}C^{-1})_{\alpha\tilde{\beta}},\\
(\mathcal{T}^{\boldsymbol{e}_{r-2i-1}})_{\tilde{\alpha}\beta}^{\mu_1\cdots\mu_{r-2i-1}} & = & \frac{1}{\sqrt{2^{r-1}(r-2i-1)!}} & \times & (\tilde{\gamma}^{\mu_1\cdots\mu_{r-2i-1}}C^{-1})_{\tilde{\alpha}\beta} & (\tilde{\gamma}^{\mu_1\cdots\mu_{r-2i-1}}\tilde{C}^{-1})_{\tilde{\alpha}\beta},
\end{array}
}
while
\eqn{
\begin{array}{cccccc}
& & & & \text{$r$ even} & \text{$r$ odd}\\
(\mathcal{T}^{\boldsymbol{e}_{r-1}+\boldsymbol{e}_r})_{\alpha\tilde{\beta}}^{\mu_1\cdots\mu_{r-1}} & = & \frac{1}{\sqrt{2^{r-1}(r-1)!}} & \times & (\gamma^{\mu_1\cdots\mu_{r-1}}\tilde{C}^{-1})_{\alpha\tilde{\beta}} & (\gamma^{\mu_1\cdots\mu_{r-1}}C^{-1})_{\alpha\tilde{\beta}},\\
(\mathcal{T}^{\boldsymbol{e}_{r-1}+\boldsymbol{e}_r})_{\tilde{\alpha}\beta}^{\mu_1\cdots\mu_{r-1}} & = & \frac{1}{\sqrt{2^{r-1}(r-1)!}} & \times & (\tilde{\gamma}^{\mu_1\cdots\mu_{r-1}}C^{-1})_{\tilde{\alpha}\beta} & (\tilde{\gamma}^{\mu_1\cdots\mu_{r-1}}\tilde{C}^{-1})_{\tilde{\alpha}\beta},\\
(\mathcal{T}^{2\boldsymbol{e}_{r-1}})_{\alpha\beta}^{\mu_1\cdots\mu_r} & = & \frac{1}{\sqrt{2^rr!}} & \times & (\gamma^{\mu_1\cdots\mu_r}C^{-1})_{\alpha\beta} & (\gamma^{\mu_1\cdots\mu_r}\tilde{C}^{-1})_{\alpha\beta},\\
(\mathcal{T}^{2\boldsymbol{e}_r})_{\tilde{\alpha}\tilde{\beta}}^{\mu_1\cdots\mu_r} & = & \frac{1}{\sqrt{2^rr!}} & \times & (\tilde{\gamma}^{\mu_1\cdots\mu_r}\tilde{C}^{-1})_{\tilde{\alpha}\tilde{\beta}} & (\tilde{\gamma}^{\mu_1\cdots\mu_r}C^{-1})_{\tilde{\alpha}\tilde{\beta}}.
\end{array}
}

The associated projectors are thus
\begingroup\makeatletter\def\f@size{7}\check@mathfonts\def\maketag@@@#1{\hbox{\m@th\large\normalfont#1}}%
\eqn{
\begin{array}{cccccc}
& & & & \text{$r$ even} & \text{$r$ odd}\\
\mathcal{P}_{\boldsymbol{e}_{r-2i}}^{(-,-)} & = & \frac{1}{2^{r-1}(r-2i)!} & \times & (\gamma^{\mu_1\cdots\mu_{r-2i}}C^{-1})_{\alpha\beta}(C\gamma_{\mu_{r-2i}\cdots\mu_1})^{\beta'\alpha'} & (\gamma^{\mu_1\cdots\mu_{r-2i}}\tilde{C}^{-1})_{\alpha\beta}(\tilde{C}\tilde{\gamma}_{\mu_{r-2i}\cdots\mu_1})^{\beta'\alpha'},\\
\mathcal{P}_{\boldsymbol{e}_{r-2i}}^{(\sim,\sim)} & = & \frac{1}{2^{r-1}(r-2i)!} & \times & (\tilde{\gamma}^{\mu_1\cdots\mu_{r-2i}}\tilde{C}^{-1})_{\tilde{\alpha}\tilde{\beta}}(\tilde{C}\tilde{\gamma}_{\mu_{r-2i}\cdots\mu_1})^{\tilde{\beta}'\tilde{\alpha}'} & (\tilde{\gamma}^{\mu_1\cdots\mu_{r-2i}}C^{-1})_{\tilde{\alpha}\tilde{\beta}}(C\gamma_{\mu_{r-2i}\cdots\mu_1})^{\tilde{\beta}'\tilde{\alpha}'},\\
\mathcal{P}_{\boldsymbol{e}_{r-2i-1}}^{(-,\sim)} & = & \frac{1}{2^{r-1}(r-2i-1)!} & \times & (\gamma^{\mu_1\cdots\mu_{r-2i-1}}\tilde{C}^{-1})_{\alpha\tilde{\beta}}(\tilde{C}\tilde{\gamma}_{\mu_{r-2i-1}\cdots\mu_1})^{\tilde{\beta}'\alpha'} & (\gamma^{\mu_1\cdots\mu_{r-2i-1}}C^{-1})_{\alpha\tilde{\beta}}(C\gamma_{\mu_{r-2i-1}\cdots\mu_1})^{\tilde{\beta}'\alpha'},\\
\mathcal{P}_{\boldsymbol{e}_{r-2i-1}}^{(\sim,-)} & = & \frac{1}{2^{r-1}(r-2i-1)!} & \times & (\tilde{\gamma}^{\mu_1\cdots\mu_{r-2i-1}}C^{-1})_{\tilde{\alpha}\beta}(C\gamma_{\mu_{r-2i-1}\cdots\mu_1})^{\beta'\tilde{\alpha}'} & (\tilde{\gamma}^{\mu_1\cdots\mu_{r-2i-1}}\tilde{C}^{-1})_{\tilde{\alpha}\beta}(\tilde{C}\tilde{\gamma}_{\mu_{r-2i-1}\cdots\mu_1})^{\beta'\tilde{\alpha}'},
\end{array}
}
\endgroup
and
\begingroup\makeatletter\def\f@size{10}\check@mathfonts\def\maketag@@@#1{\hbox{\m@th\large\normalfont#1}}%
\eqn{
\begin{array}{cccccc}
& & & & \text{$r$ even} & \text{$r$ odd}\\
\mathcal{P}_{\boldsymbol{e}_{r-1}+\boldsymbol{e}_r}^{(-,\sim)} & = & \frac{1}{2^{r-1}(r-1)!} & \times & (\gamma^{\mu_1\cdots\mu_{r-1}}\tilde{C}^{-1})_{\alpha\tilde{\beta}}(\tilde{C}\tilde{\gamma}_{\mu_{r-1}\cdots\mu_1})^{\tilde{\beta}'\alpha'} & (\gamma^{\mu_1\cdots\mu_{r-1}}C^{-1})_{\alpha\tilde{\beta}}(C\gamma_{\mu_{r-1}\cdots\mu_1})^{\tilde{\beta}'\alpha'},\\
\mathcal{P}_{\boldsymbol{e}_{r-1}+\boldsymbol{e}_r}^{(\sim,-)} & = & \frac{1}{2^{r-1}(r-1)!} & \times & (\tilde{\gamma}^{\mu_1\cdots\mu_{r-1}}C^{-1})_{\tilde{\alpha}\beta}(C\gamma_{\mu_{r-1}\cdots\mu_1})^{\beta'\tilde{\alpha}'} & (\tilde{\gamma}^{\mu_1\cdots\mu_{r-1}}\tilde{C}^{-1})_{\tilde{\alpha}\beta}(\tilde{C}\tilde{\gamma}_{\mu_{r-1}\cdots\mu_1})^{\beta'\tilde{\alpha}'},\\
\mathcal{P}_{2\boldsymbol{e}_{r-1}}^{(-,-)} & = & \frac{1}{2^rr!} & \times & (\gamma^{\mu_1\cdots\mu_r}C^{-1})_{\alpha\beta}(C\gamma_{\mu_r\cdots\mu_1})^{\beta'\alpha'} & (\gamma^{\mu_1\cdots\mu_r}\tilde{C}^{-1})_{\alpha\beta}(\tilde{C}\tilde{\gamma}_{\mu_r\cdots\mu_1})^{\beta'\alpha'},\\
\mathcal{P}_{2\boldsymbol{e}_r}^{(\sim,\sim)} & = & \frac{1}{2^rr!} & \times & (\tilde{\gamma}^{\mu_1\cdots\mu_r}\tilde{C}^{-1})_{\tilde{\alpha}\tilde{\beta}}(\tilde{C}\tilde{\gamma}_{\mu_r\cdots\mu_1})^{\tilde{\beta}'\tilde{\alpha}'} & (\tilde{\gamma}^{\mu_1\cdots\mu_r}C^{-1})_{\tilde{\alpha}\tilde{\beta}}(C\gamma_{\mu_r\cdots\mu_1})^{\tilde{\beta}'\tilde{\alpha}'}.
\end{array}
}
\endgroup
Obviously, by choosing an ordering for the untilde and tilde spinor indices, it is possible to get rid of some of the redundant quantities introduced above and focus only on the independent ones.  At the end, operators in irreducible representations with $S=1$ verify $\mathcal{O}^{\boldsymbol{e}_i}\sim(\mathcal{T}^{\boldsymbol{e}_i})^{\mu_1\cdots\mu_i}$ for $i\neq r-1,r$ as well as $\mathcal{O}_{\alpha\tilde{\beta}}^{\boldsymbol{e}_{r-1}+\boldsymbol{e}_r}\sim(\mathcal{T}^{\boldsymbol{e}_{r-1}+\boldsymbol{e}_r})_{\alpha\tilde{\beta}}^{\mu_1\cdots\mu_r}$, $\mathcal{O}_{\alpha\beta}^{2\boldsymbol{e}_{r-1}}\sim(\mathcal{T}^{2\boldsymbol{e}_{r-1}})_{\alpha\beta}^{\mu_1\cdots\mu_r}$ and $\mathcal{O}_{\tilde{\alpha}\tilde{\beta}}^{2\boldsymbol{e}_r}\sim(\mathcal{T}^{2\boldsymbol{e}_r})_{\tilde{\alpha}\tilde{\beta}}^{\mu_1\cdots\mu_r}$.  Here again the $\mu$ indices are dummy indices which must be contracted properly.

The irreducible representations given by $\boldsymbol{N}\in\{\boldsymbol{0},\boldsymbol{e}_i,\boldsymbol{e}_{r-1}+\boldsymbol{e}_r,2\boldsymbol{e}_{r-1},2\boldsymbol{e}_r\}$ for all $i$ are called the defining representations.  They correspond respectively to the trivial representation $\boldsymbol{0}$, the two spinor representations $\boldsymbol{e}_{r-1}$ and $\boldsymbol{e}_r$, the $i$-index antisymmetric representations $\boldsymbol{e}_{i\neq r-1,r}$ with the $(r-1)$-index antisymmetric $\boldsymbol{e}_{r-1}+\boldsymbol{e}_r$ and the two $r$-index antisymmetric $2\boldsymbol{e}_{r-1}$ (self-dual) and $2\boldsymbol{e}_r$ (anti-self-dual) representations singled out.

From the discussion in odd dimension, one could construct more general irreducible representations in even dimension following either the bottom-up technique or the top-down method introduced before.  In any case the simplest way is to use the hatted projection operators constrained for example with the top-down method by appropriately symmetrizing tensor products of the defining representations and then removing smaller irreducible representations, by relying in part on the tracelessness condition.  Therefore, the half-projector for a general irreducible representation $\boldsymbol{N}=\sum_{i=1}^rN_i\boldsymbol{e}_i$ is
\eqna{
(\mathcal{T}^{\boldsymbol{N}})_{\alpha_1\cdots\alpha_n;\tilde{\alpha}_1\cdots\tilde{\alpha}_{\tilde{n}}}^{\mu_1\cdots\mu_{n_v}\delta}&=\left((\mathcal{T}^{\boldsymbol{e}_1})^{N_1}\cdots(\mathcal{T}^{\boldsymbol{e}_{r-2}})^{N_{r-2}}(\mathcal{T}^{2\boldsymbol{e}_{r-1}})^{\lfloor(N_{r-1}-\text{min}\{N_{r-1},N_r\})/2\rfloor}\right.\\
&\qquad\left.\times(\mathcal{T}^{2\boldsymbol{e}_r})^{\lfloor(N_r-\text{min}\{N_{r-1},N_r\})/2\rfloor}(\mathcal{T}^{\boldsymbol{e}_{r-1}+\boldsymbol{e}_r})^{\text{min}\{N_{r-1},N_r\}}\right.\\
&\qquad\left.\times(\mathcal{T}^{\boldsymbol{e}_{r-1}})^{N_{r-1}-\text{min}\{N_{r-1},N_r\}-2\lfloor(N_{r-1}-\text{min}\{N_{r-1},N_r\})/2\rfloor}\right.\\
&\qquad\left.\times(\mathcal{T}^{\boldsymbol{e}_r})^{N_r-\text{min}\{N_{r-1},N_r\}-2\lfloor(N_r/2-\text{min}\{N_{r-1},N_r\})\rfloor}\right)_{\alpha_1\cdots\alpha_n;\tilde{\alpha}_1\cdots\tilde{\alpha}_{\tilde{n}}}^{\mu_1'\cdots\mu_{n_v}'\delta'}\\
&\qquad\times(\hat{\mathcal{P}}^{\boldsymbol{N}})_{\delta'\mu_{n_v}'\cdots\mu_1'}^{\phantom{\delta'\mu_{n_v}'\cdots\mu_1'}\mu_1\cdots\mu_{n_v}\delta},
}[EqTeven]
where the number of vector indices is
\eqn{n_v=\sum_{i=1}^{r-2}iN_i+(r-1)\text{min}\{N_{r-1},N_r\}+r\lfloor(N_{r-1}-\text{min}\{N_{r-1},N_r\})/2\rfloor+r\lfloor(N_r-\text{min}\{N_{r-1},N_r\})/2\rfloor,}
the spinor index $\delta$ or $\tilde{\delta}$ appears only if $N_{r-1}-\text{min}\{N_{r-1},N_r\}-2\lfloor(N_{r-1}-\text{min}\{N_{r-1},N_r\})/2\rfloor=1$ or $N_r-\text{min}\{N_{r-1},N_r\}-2\lfloor(N_r-\text{min}\{N_{r-1},N_r\})/2\rfloor=1$ respectively, and the proper hatted projection operator subtracting smaller irreducible representations is partly fixed by the tracelessness condition $(\mathcal{T}^{\boldsymbol{N}})_{\alpha_1\cdots\alpha_n;\tilde{\alpha}_1\cdots\tilde{\alpha}_{\tilde{n}}}^{\mu_1\cdots\mu_{n_v}\delta}g_{\mu_i\mu_j}=0$ and $(\mathcal{T}^{\boldsymbol{N}})_{\alpha_1\cdots\alpha_n;\tilde{\alpha}_1\cdots\tilde{\alpha}_{\tilde{n}}}^{\mu_1\cdots\mu_{n_v}\delta}(\gamma_{\mu_i})_\delta^{\phantom{\delta}\tilde{\gamma}}=0$ or $(\mathcal{T}^{\boldsymbol{N}})_{\alpha_1\cdots\alpha_n;\tilde{\alpha}_1\cdots\tilde{\alpha}_{\tilde{n}}}^{\mu_1\cdots\mu_{n_v}\tilde{\delta}}(\tilde{\gamma}_{\mu_i})_{\tilde{\delta}}^{\phantom{\tilde{\delta}}\gamma}=0$ for any $i$ and $j$.  As in the odd dimensional case, using \eqref{EqTeven} the tracelessness condition applies also directly to the proper hatted projection operators.

As in odd dimension, operators in the irreducible $\boldsymbol{N}$ representation can thus be rewritten as
\eqn{\mathcal{O}_{\alpha_1\cdots\alpha_n;\tilde{\alpha}_1\cdots\tilde{\alpha}_{\tilde{n}}}^{\boldsymbol{N}}=(\mathcal{T}^{\boldsymbol{N}})_{\alpha_1\cdots\alpha_n;\tilde{\alpha}_1\cdots\tilde{\alpha}_{\tilde{n}}}^{\delta\mu_{n_v}\cdots\mu_1}\mathcal{O}_{\mu_1\cdots\mu_{n_v}\delta}^{\boldsymbol{N}},\qquad\mathcal{O}_{\mu_1\cdots\mu_{n_v}\delta}^{\boldsymbol{N}}=(\mathcal{T}_{\boldsymbol{N}})_{\mu_1\cdots\mu_{n_v}\delta}^{\alpha_n\cdots\alpha_1;\tilde{\alpha}_{\tilde{n}}\cdots\tilde{\alpha}_1}\mathcal{O}_{\alpha_1\cdots\alpha_n;\tilde{\alpha}_1\cdots\tilde{\alpha}_{\tilde{n}}}^{\boldsymbol{N}},}
with the help of \eqref{EqTeven} where $\mathcal{O}_{\mu_1\cdots\mu_{n_v}\delta}^{\boldsymbol{N}}$ transforms in the irreducible $\boldsymbol{N}$ representation but with respect to the ``dummy'' variables.

Finally, the conjugate operators are denoted by
\eqna{
\mathcal{O}^{\boldsymbol{N}C}&\equiv(\mathcal{O}^{\boldsymbol{N}C})_{\alpha_1\cdots\alpha_n;\tilde{\alpha}_1\cdots\tilde{\alpha}_{\tilde{n}}}=(B^{-1})_{\alpha_1\alpha_1'}\cdots(B^{-1})_{\alpha_n\alpha_n'}(\tilde{B}^{-1})_{\tilde{\alpha}_1\tilde{\alpha}_1'}\cdots(\tilde{B}^{-1})_{\tilde{\alpha}_{\tilde{n}}\tilde{\alpha}_{\tilde{n}}'}(\mathcal{O}^{\boldsymbol{N}*})^{\alpha_1'\cdots\alpha_n';\tilde{\alpha}_1'\cdots\tilde{\alpha}_{\tilde{n}}'},\\
\mathcal{O}^{\boldsymbol{N}C}&\equiv(\mathcal{O}^{\boldsymbol{N}C})_{\alpha_1\cdots\alpha_{\tilde{n}};\tilde{\alpha}_1\cdots\tilde{\alpha}_n}=(B^{-1})_{\alpha_1\tilde{\alpha}_1'}\cdots(B^{-1})_{\alpha_{\tilde{n}}\tilde{\alpha}_{\tilde{n}}'}(\tilde{B}^{-1})_{\tilde{\alpha}_1\alpha_1'}\cdots(\tilde{B}^{-1})_{\tilde{\alpha}_n\alpha_n'}(\mathcal{O}^{\boldsymbol{N}*})^{\alpha_1'\cdots\alpha_n';\tilde{\alpha}_1'\cdots\tilde{\alpha}_{\tilde{n}}'},
}
for $r+q$ even and odd respectively.  With the proper spinor indices and $B$ matrices for the two different cases, they satisfy
\eqna{(\mathcal{O}^{\boldsymbol{N}C})&=\left(B^{-1}[(\mathcal{T}^{\boldsymbol{N}})^{\delta\mu_{n_v}\cdots\mu_1}]^*B^{-T}\right)(\mathcal{O}_{\mu_1\cdots\mu_{n_v}\delta}^{\boldsymbol{N}})^*\\
&=\left\{\begin{array}{cc}(-1)^{f(\boldsymbol{N})}(\mathcal{T}^{\boldsymbol{N}^C})^{\delta\mu_{n_v}\cdots\mu_1}(B^{-1}\mathcal{O}^{\boldsymbol{N}*})_{\mu_1\cdots\mu_{n_v}\delta}&N_{r-1}+N_r\,\text{odd}\\(-1)^{f(\boldsymbol{N})}(\mathcal{T}^{\boldsymbol{N}^C})^{\mu_{n_v}\cdots\mu_1}(\mathcal{O}^{\boldsymbol{N}*})_{\mu_1\cdots\mu_{n_v}}&N_{r-1}+N_r\,\text{even}\end{array}\right.,
}
where
\eqna{
f(\boldsymbol{N})&=\sum_{i=1}^{r-2}[i(r+q)+q(r+1)]N_i+[(r-1)(r+q)+q(r+1)]\text{min}\{N_{r-1},N_r\}\\
&\phantom{=}\qquad+[r(r+q)+q(r+1)]\lfloor|N_{r-1}-N_r|/2\rfloor,
}
Therefore, as in odd dimension, for vector representations it is (almost) always possible to trade conjugate operators by considering the operator's real and imaginary parts independently.  Indeed, since these representations can be reduced to real components with $(\mathcal{O}^{\boldsymbol{N}*})_{\mu_1\cdots\mu_{n_v}}=\mathcal{O}_{\mu_1\cdots\mu_{n_v}}^{\boldsymbol{N}}$, one obtains that $\mathcal{O}^{\boldsymbol{N}C}\propto\mathcal{O}^{\boldsymbol{N}}$ showing that the conjugate is not linearly independent.  Note however that the condition $(\mathcal{O}^{\boldsymbol{N}*})_{\mu_1\cdots\mu_{n_v}}=\mathcal{O}_{\mu_1\cdots\mu_{n_v}}^{\boldsymbol{N}}$ cannot be imposed on general (anti-)self-dual representations if $r+q$ is odd (as can also be seen from the fact that $\boldsymbol{N}^C\neq\boldsymbol{N}$ when $r+q$ is odd), in which case their conjugate representations are not linearly independent.  Thus, apart from general (anti-)self-dual representations, general vector representations can always be reduced to real and imaginary components such that the conjugate operators are not linearly independent.  For general spinor representations on the other hand, conjugate operators are linearly independent when the Weyl-Majorana condition cannot be imposed.

%%%%%%%%%%%%%%%%%%%%%%%%%%%%%%%%%%%%%%%%%%%%%%%%%%

\subsection{Summary}

In summary, the transformation properties of operators $\mathcal{O}^{\boldsymbol{N}}$ in general irreducible representations of $SO(p,q)$ denoted by non-negative Dynkin indices $\boldsymbol{N}=\{N_1,\ldots,N_r\}$ can be encoded in appropriate half-projectors $\mathcal{T}^{\boldsymbol{N}}$.  Here $r$ is the rank of $SO(p,q)$ and it corresponds to $p+q=2r+1$ for odd dimensions or $p+q=2r$ for even dimensions.  Moreover, the ``spin'' $S$ of the operator is given by
\eqna{
\text{$d$ odd:}&\qquad2S=2\sum_{i=1}^{r-1}N_i+N_r,\\
\text{$d$ even:}&\qquad2S=2\sum_{i=1}^{r-2}N_i+N_{r-1}+N_r.
}

More precisely, in correlation functions, $\mathcal{O}^{\boldsymbol{N}}\sim\mathcal{T}^{\boldsymbol{N}}$ where the dummy indices on the half-projectors must be contracted properly (usually with functions of spacetime coordinates).  The half-projectors are built from the half-projectors for the defining irreducible representations and are given by
\eqn{(\mathcal{T}^{\boldsymbol{N}})=\left((\mathcal{T}^{\boldsymbol{e}_1})^{N_1}\cdots\right)\cdot(\hat{\mathcal{P}}^{\boldsymbol{N}}),}[EqT]
where an index-free notation was used.  The hatted projection operators $\hat{\mathcal{P}}^{\boldsymbol{N}}$ intertwine the dummy variables and combine them in the proper irreducible representations.  The hatted projection operators satisfy the tracelessness conditions
\eqn{(\hat{\mathcal{P}}^{\boldsymbol{N}})\cdot g=(\hat{\mathcal{P}}^{\boldsymbol{N}})\cdot(\gamma)=(\hat{\mathcal{P}}^{\boldsymbol{N}})\cdot(\tilde{\gamma})=0,}
where all contractions are on dummy variables.

Finally, conjugate operators for vector representations can (almost) always be discarded by considering real and imaginary parts as independent operators.  It is also the case for operators in spinor representations as long as the Majorana(-Weyl) condition is satisfied.  Otherwise conjugate operators for spinor representations are linearly independent and must be included.

%%%%%%%%%%%%%%%%%%%%%%%%%%%%%%%%%%%%%%%%%%%%%%%%%%
%%%%%%%%%%%%%%%%%%%%%%%%%%%%%%%%%%%%%%%%%%%%%%%%%%

\section{Properties of Embedding Space Quantities}\label{SecPtoE}

This appendix presents several identities satisfied by the embedding space quantities $\A_{ij}$, $\epsilon_{ij}$ and $\Gamma_{ij}$.  These identities originate from their position space counterparts obtained in Appendix \ref{SecIrrep}, showing the rationale behind the substitutions \eqref{EqTSPStoES} and proving the half-projector identities of Section \ref{SecOPE}.  As in Appendix \ref{SecIrrep}, the Lorentz group is chosen to be $SO(p,q)$ with the conformal group being $SO(p+1,q+1)$.

%%%%%%%%%%%%%%%%%%%%%%%%%%%%%%%%%%%%%%%%%%%%%%%%%%

\subsection{Identities for Three Embedding Space Coordinates}

In embedding space, the OPE metric, epsilon tensor and $\Gamma$-matrices depend on two embedding space coordinates.  As such, the most general identities depend on more than two embedding space coordinates.  For the most important identities, only three embedding space coordinates are necessary.

For relations involving two OPE metrics or two epsilon tensors, the identities are
\eqn{
\begin{gathered}
\A_{ijkA}^{\phantom{ijkA}B}\equiv\A_{jiA}^{\phantom{jiA}C}\A_{jkC}^{\phantom{jkC}B}=\A_{jiA}^{\phantom{jiA}B}-\frac{(\A_{ji}\cdot\eta_k)_A\eta_j^B}{\ee{j}{k}{}}=\A_{jkA}^{\phantom{jkA}B}-\frac{\eta_{jA}(\eta_i\cdot\A_{jk})^B}{\ee{i}{j}{}},\\
\epsilon_{jiA_1\cdots A_nC_{d-n}\cdots C_1}\epsilon_{jk}^{C_1\cdots C_{d-n}B_n\cdots B_1}=(-1)^{q+d(d-1)/2}n!(d-n)!\A_{ijk[A_1}^{\phantom{ijk[A_1}B_1}\cdots\A_{ijkA_n]}^{\phantom{ijkA_n]}B_n}.
\end{gathered}
}
Here both $q$ and $d$ are position space quantities, not the embedding space ones.  These identities are straightforwardly obtained by using the properties of the metric $g_{AB}$ and the epsilon tensor $\epsilon^{A_1\cdots A_{d+2}}$.  Analogous identities requiring two OPE $\Gamma$-matrices $\Gamma_{ji}$ and $\Gamma_{jk}$ exist, leading to \eqref{EqTSTS}.  They are
\eqn{\text{tr}[(\Gamma_{jiA_n\cdots A_1}\eta_i\cdot\Gamma)(\eta_j\cdot\Gamma\Gamma_{jk}^{B_1\cdots B_m})]=2^{r+1}n!\ee{i}{j}{}\A_{ijk[A_1}^{\phantom{ijk[A_1}B_1}\cdots\A_{ijkA_n]}^{\phantom{ijkA_n]}B_n}\delta_{nm},}
in odd dimensions and
\eqna{
\text{tr}[(\Gamma_{jiA_n\cdots A_1}\eta_i\cdot\Gamma)(\eta_j\cdot\Gamma\Gamma_{jk}^{B_1\cdots B_m})]&=2^rn!\ee{i}{j}{}\left(\A_{ijk[A_1}^{\phantom{ijk[A_1}B_1}\cdots\A_{ijkA_n]}^{\phantom{ijkA_n]}B_n}\right.\\
&\phantom{=}\qquad\left.\pm(-1)^r\frac{\mathscr{K}}{r!}\epsilon_{jiA_1\cdots A_r}^{\phantom{jiA_1\cdots A_r}B'_r\cdots B'_1}\A_{jkB'_1}^{\phantom{jkB'_1}B_1}\cdots\A_{jkB'_r}^{\phantom{jkB'_r}B_r}\delta_{nr}\right)\delta_{nm},
}
in even dimensions.  Here, the $\mathscr{K}$ is the one in position space (see below) and $\pm$ differentiates between self-dual ($\eta_j\cdot\Gamma\tilde{\Gamma}_{jk}^{B_1\cdots B_m}$ with $+$) and anti-self-dual ($\eta_j\cdot\tilde{\Gamma}\Gamma_{jk}^{B_1\cdots B_m}$ with $-$) irreducible representations.

There are also important identities concerning mixed quantities.  For example, for the OPE metric and the epsilon tensor one has
\begingroup\makeatletter\def\f@size{10}\check@mathfonts\def\maketag@@@#1{\hbox{\m@th\large\normalfont#1}}%
\eqna{
&\epsilon_{ji}^{A'_1\cdots A'_d}\mathcal{A}_{jkA'_d}^{\phantom{jkA'_d}A_d}\cdots\mathcal{A}_{jkA'_1}^{\phantom{jkA'_1}A_1}=\frac{1}{\ee{i}{j}{}}\eta_{jA'_0}\epsilon^{A'_0\cdots A'_{d+1}}\eta_{iA'_{d+1}}\mathcal{A}_{jkA'_d}^{\phantom{jkA'_d}A_d}\cdots\mathcal{A}_{jkA'_1}^{\phantom{jkA'_1}A_1}\\
&\qquad=\frac{1}{\ee{i}{j}{}}\eta_{jA'_0}\epsilon^{A'_0\cdots A'_{d+1}}\eta_{iA_{d+1}}g_{A'_{d+1}}^{\phantom{A'_{d+1}}A_{d+1}}\mathcal{A}_{jkA'_d}^{\phantom{jkA'_d}A_d}\cdots\mathcal{A}_{jkA'_1}^{\phantom{jkA'_1}A_1}\\
&\qquad=\frac{1}{\ee{i}{j}{}}\eta_{jA'_0}\epsilon^{A'_0\cdots A'_{d+1}}\eta_{iA_{d+1}}\left(\mathcal{A}_{jkA'_{d+1}}^{\phantom{jkA'_{d+1}}A_{d+1}}+\frac{\eta_{jA'_{d+1}}\eta_k^{A_{d+1}}}{\ee{j}{k}{}}+\frac{\eta_{kA'_{d+1}}\eta_j^{A_{d+1}}}{\ee{j}{k}{}}\right)\mathcal{A}_{jkA'_d}^{\phantom{jkA'_d}A_d}\cdots\mathcal{A}_{jkA'_1}^{\phantom{jkA'_1}A_1}\\
&\qquad=\frac{1}{\ee{i}{j}{}}\eta_{jA'_0}\epsilon^{A'_0\cdots A'_{d+1}}\eta_{iA_{d+1}}\left(\mathcal{A}_{jkA'_{d+1}}^{\phantom{jkA'_{d+1}}A_{d+1}}+\frac{\eta_{kA'_{d+1}}\eta_j^{A_{d+1}}}{\ee{j}{k}{}}\right)\mathcal{A}_{jkA'_d}^{\phantom{jkA'_d}A_d}\cdots\mathcal{A}_{jkA'_1}^{\phantom{jkA'_1}A_1}\\
&\qquad=\frac{1}{\ee{i}{j}{}\ee{j}{k}{}}\eta_{jA'_0}\epsilon^{A'_0\cdots A'_{d+1}}\eta_{iA_{d+1}}\eta_{kA'_{d+1}}\eta_j^{A_{d+1}}\mathcal{A}_{jkA'_d}^{\phantom{jkA'_d}A_d}\cdots\mathcal{A}_{jkA'_1}^{\phantom{jkA'_1}A_1}\\
&\qquad=\epsilon_{jk}^{A_1\cdots A_d}.
}
\endgroup
Here, the fact that $\mathcal{A}_{jk}$ is a $(d+2)\times(d+2)$ matrix with rank $d$ (it has two vanishing eigenvalues with eigenvectors $\eta_j$ and $\eta_k$ respectively) was used to prove that the antisymmetrization of $d+1$ $\mathcal{A}_{jk}$ vanishes.  Moreover, for the OPE metric and the $\Gamma$-matrix one has
\eqn{\eta_j\cdot\Gamma\Gamma_{ji}^{B_1\cdots B_n}\A_{jkB_n}^{\phantom{jkB_n}A_n}\cdots\A_{jkB_1}^{\phantom{jkB_1}A_1}=\eta_j\cdot\Gamma\Gamma_{jk}^{A_1\cdots A_n}.}
These last identities lead to a proof of \eqref{EqTkjTkiOPE} for all irreducible representations of the Lorentz group, including the (anti-)self-dual ones.

Finally, identities involving both the OPE epsilon tensor and the $\Gamma$-matrix are
\eqn{\eta_j\cdot\Gamma\Gamma_{ji}^{A_1\cdots A_n}=\frac{\mathscr{K}}{(d-n)!}\epsilon_{ji}^{A_1\cdots A_n}{}_{B_{d-n}\cdots B_1}\eta_j\cdot\Gamma\Gamma_{jk}^{B_1\cdots B_{d-n}},}
in odd dimensions and
\eqna{
\eta_j\cdot\Gamma\tilde{\Gamma}_{ji}^{A_1\cdots A_n}&=\frac{\mathscr{K}}{(d-n)!}\epsilon_{ji}^{A_1\cdots A_n}{}_{B_{d-n}\cdots B_1}\eta_j\cdot\Gamma\tilde{\Gamma}_{jk}^{B_1\cdots B_{d-n}},\\
\eta_j\cdot\tilde{\Gamma}\Gamma_{ji}^{A_1\cdots A_n}&=-\frac{\mathscr{K}}{(d-n)!}\epsilon_{ji}^{A_1\cdots A_n}{}_{B_{d-n}\cdots B_1}\eta_j\cdot\tilde{\Gamma}\Gamma_{jk}^{B_1\cdots B_{d-n}},
}
in even dimensions.  Here, from the definition of the $\Gamma$-matrices one has $\Gamma^{1\cdots d+2}=\mathscr{K}\left(\begin{array}{cc}\1&0\\0&\1\end{array}\right)$ with $\mathscr{K}$ the same than in position space.  This observation supports the substitution rules \eqref{EqTSPStoES} $g^{\mu\nu}\to\A_{12}^{AB}$, $\epsilon^{\mu_1\cdots\mu_d}\to\epsilon_{12}^{A_1\cdots A_d}$ and $\gamma^{\mu_1\cdots\mu_n}\to\Gamma_{12}^{A_1\cdots A_n}$ for the OPE tensor structures $\tOPE{a}{i}{j}{k}{1}{2}$ from the standard ones in position space.

All the identities presented here can be used to replace one of the hatted projection operators in \eqref{EqTSTS},
\eqn{\left(\frac{\eta_j\cdot\Gamma\,\hat{\mathcal{P}}_{ji}^{\boldsymbol{N}}\cdot\hat{\mathcal{P}}_{jk}^{\boldsymbol{N}}\,\eta_k\cdot\Gamma}{\ee{j}{k}{}}\right),}
by the corresponding product of metrics.  Indeed, one can either replace $\hat{\mathcal{P}}_{ji}^{\boldsymbol{N}}\to\A_{ji}\cdots\A_{ji}$ or $\hat{\mathcal{P}}_{jk}^{\boldsymbol{N}}\to\A_{jk}\cdots\A_{jk}$ but not both, somewhat simplifying the subsequent analysis.

%%%%%%%%%%%%%%%%%%%%%%%%%%%%%%%%%%%%%%%%%%%%%%%%%%

\subsection{Rules for Products of Metrics}

As mentioned above, the most general identities involving embedding space quantities exhibit several embedding space coordinates.  They can be found with the help of arguments analogous to the ones used for three embedding space coordinates.

Here the only necessary rule for the matrix products of several metrics $\A_{ij}$ at different embedding space coordinates is presented.  It takes two forms given by
\eqn{\A_{ij}\cdot\A_{k\ell}=\A_{ij}-\frac{(\A_{ij}\cdot\eta_k)\eta_\ell}{\ee{k}{\ell}{}}-\frac{(\A_{ij}\cdot\eta_\ell)\eta_k}{\ee{k}{\ell}{}}=\A_{k\ell}-\frac{\eta_i(\A_{k\ell}\cdot\eta_j)}{\ee{i}{j}{}}-\frac{\eta_j(\A_{k\ell}\cdot\eta_i)}{\ee{i}{j}{}}.}[EqRule]
Obviously, it is not necessary to go to a higher number of metrics since the above rule is already closed under subsequent matrix multiplication once they are supplemented with
\eqn{\eta_\ell\cdot(\A_{ij}\cdot\eta_k)=(\eta_\ell\cdot\A_{ij}\cdot\eta_k),}
where $(\A_{ij}\cdot\eta_k)=0$ if $k=i$ or $k=j$, which is quite trivial.  The rule \eqref{EqRule} works for any $M$-point correlation functions and it simplifies when some of the embedding space coordinates are the same.

For example, since there are no more than four different embedding space coordinates appearing in the four-point correlation functions, the rule \eqref{EqRule} can be simplified since several matrix products appearing on the right-hand side vanish explicitly.  Indeed, with up to four different embedding space coordinates denoted by $i$, $j$, $k$ and $\ell$ respectively, the non-vanishing rules are
\eqn{
\begin{gathered}
\A_{ij}\cdot\A_{k\ell}=\A_{ij}-\frac{(\A_{ij}\cdot\eta_k)\eta_\ell}{\ee{k}{\ell}{}}-\frac{(\A_{ij}\cdot\eta_\ell)\eta_k}{\ee{k}{\ell}{}}=\A_{k\ell}-\frac{\eta_i(\A_{k\ell}\cdot\eta_j)}{\ee{i}{j}{}}-\frac{\eta_j(\A_{k\ell}\cdot\eta_i)}{\ee{i}{j}{}},\\
\A_{ij}\cdot\A_{jk}=\A_{ij}-\frac{(\A_{ij}\cdot\eta_k)\eta_j}{\ee{j}{k}{}}=\A_{jk}-\frac{\eta_j(\A_{jk}\cdot\eta_i)}{\ee{i}{j}{}},\qquad\A_{ij}\cdot\A_{ij}=\A_{ij},\\
(\eta_\ell\cdot\A_{ij}\cdot\eta_k)=\ee{k}{\ell}{}\left[1-\frac{\ee{i}{k}{}\ee{j}{\ell}{}}{\ee{i}{j}{}\ee{k}{\ell}{}}-\frac{\ee{i}{\ell}{}\ee{j}{k}{}}{\ee{i}{j}{}\ee{k}{\ell}{}}\right],
\end{gathered}
}
where the last square bracket is a function of the conformal cross-ratios only.

Obvious generalizations exist for $(M>4)$-point correlation functions.

%%%%%%%%%%%%%%%%%%%%%%%%%%%%%%%%%%%%%%%%%%%%%%%%%%
%%%%%%%%%%%%%%%%%%%%%%%%%%%%%%%%%%%%%%%%%%%%%%%%%%

\bibliography{OPE}

%bibliography generated by nb.bst v1.01 (C) 2003-2010 Niklas Beisert
\begin{thebibliography}{100}
\ifx\href\asklfhas\newcommand{\href}[2]{#2}\fi
\ifx\arxivref\asklfhas\newcommand{\arxivref}[2]{\href{http://arxiv.org/abs/#1}{#2}}\fi
\ifx\doiref\asklfhas\newcommand{\doiref}[2]{\href{http://dx.doi.org/#1}{#2}}\fi
\parskip 0pt
\normalsize

\bibitem{Zamolodchikov:1986gt}
A.~B. Zamolodchikov,
\textit{``{Irreversibility of the Flux of the Renormalization Group in a 2D
  Field Theory}''},
JETP~Lett. \textbf{43}, 730 (1986)\ignorespaces\ignorespaces,
[Pisma Zh. Eksp. Teor. Fiz.43,565(1986)]\ignorespaces
\bibitem{Polchinski:1987dy}
J.~Polchinski,
\textit{``{Scale and Conformal Invariance in Quantum Field Theory}''},
\doiref{10.1016/0550-3213(88)90179-4}{Nucl.~Phys. \textbf{B303}, 226
  (1988)\ignorespaces}\ignorespaces
\bibitem{Cardy:1988cwa}
J.~L. Cardy,
\textit{``{Is There a c Theorem in Four-Dimensions?}''},
\doiref{10.1016/0370-2693(88)90054-8}{Phys.~Lett. \textbf{B215}, 749
  (1988)\ignorespaces}\ignorespaces
\bibitem{Jack:1990eb}
I.~Jack \& H.~Osborn,
\textit{``{Analogs for the $c$ Theorem for Four-dimensional Renormalizable
  Field Theories}''},
\doiref{10.1016/0550-3213(90)90584-Z}{Nucl.~Phys. \textbf{B343}, 647
  (1990)\ignorespaces}\ignorespaces
\bibitem{Osborn:1991gm}
H.~Osborn,
\textit{``{Weyl consistency conditions and a local renormalization group
  equation for general renormalizable field theories}''},
\doiref{10.1016/0550-3213(91)80030-P}{Nucl.~Phys. \textbf{B363}, 486
  (1991)\ignorespaces}\ignorespaces
\bibitem{Komargodski:2011vj}
Z.~Komargodski \& A.~Schwimmer,
\textit{``{On Renormalization Group Flows in Four Dimensions}''},
\doiref{10.1007/JHEP12(2011)099}{JHEP \textbf{1112}, 099
  (2011)\ignorespaces}\ignorespaces,
\normalsize{\texttt{\arxivref{1107.3987}{arXiv:1107.3987}}}\ignorespaces
\bibitem{Luty:2012ww}
M.~A. Luty, J.~Polchinski \& R.~Rattazzi,
\textit{``{The $a$-theorem and the Asymptotics of 4D Quantum Field Theory}''},
\doiref{10.1007/JHEP01(2013)152}{JHEP \textbf{1301}, 152
  (2013)\ignorespaces}\ignorespaces,
\normalsize{\texttt{\arxivref{1204.5221}{arXiv:1204.5221}}}\ignorespaces
\bibitem{Fortin:2012hn}
J.-F. Fortin, B.~Grinstein \& A.~Stergiou,
\textit{``{Limit Cycles and Conformal Invariance}''},
\doiref{10.1007/JHEP01(2013)184}{JHEP \textbf{1301}, 184
  (2013)\ignorespaces}\ignorespaces,
\normalsize{\texttt{\arxivref{1208.3674}{arXiv:1208.3674}}}\ignorespaces
\bibitem{Ferrara:1971vh}
S.~Ferrara, A.~F. Grillo \& R.~Gatto,
\textit{``{Manifestly conformal covariant operator-product expansion}''},
\doiref{10.1007/BF02770435}{Lett.~Nuovo~Cim. \textbf{2S2}, 1363
  (1971)\ignorespaces}\ignorespaces,
[Lett. Nuovo Cim.2,1363(1971)]\ignorespaces
\bibitem{Ferrara:1971zy}
S.~Ferrara, R.~Gatto \& A.~F. Grillo,
\textit{``{Conformal invariance on the light cone and canonical dimensions}''},
\doiref{10.1016/0550-3213(71)90333-6}{Nucl.~Phys. \textbf{B34}, 349
  (1971)\ignorespaces}\ignorespaces
\bibitem{Ferrara:1972cq}
S.~Ferrara, A.~F. Grillo \& R.~Gatto,
\textit{``{Manifestly conformal-covariant expansion on the light cone}''},
\doiref{10.1103/PhysRevD.5.3102}{Phys.~Rev. \textbf{D5}, 3102
  (1972)\ignorespaces}\ignorespaces
\bibitem{Ferrara:1973eg}
S.~Ferrara, P.~Gatto \& A.~F. Grilla,
\textit{``{Conformal algebra in space-time and operator product expansion}''},
\doiref{10.1007/BFb0111104}{Springer~Tracts~Mod.~Phys. \textbf{67}, 1
  (1973)\ignorespaces}\ignorespaces
\bibitem{Dobrev:1975ru}
V.~K. Dobrev, V.~B. Petkova, S.~G. Petrova \& I.~T. Todorov,
\textit{``{Dynamical Derivation of Vacuum Operator Product Expansion in
  Euclidean Conformal Quantum Field Theory}''},
\doiref{10.1103/PhysRevD.13.887}{Phys.~Rev. \textbf{D13}, 887
  (1976)\ignorespaces}\ignorespaces
\bibitem{Mack:1976pa}
G.~Mack,
\textit{``{Convergence of Operator Product Expansions on the Vacuum in
  Conformal Invariant Quantum Field Theory}''},
\doiref{10.1007/BF01609130}{Commun.~Math.~Phys. \textbf{53}, 155
  (1977)\ignorespaces}\ignorespaces
\bibitem{Ferrara:1973yt}
S.~Ferrara, A.~F. Grillo \& R.~Gatto,
\textit{``{Tensor representations of conformal algebra and conformally
  covariant operator product expansion}''},
\doiref{10.1016/0003-4916(73)90446-6}{Annals~Phys. \textbf{76}, 161
  (1973)\ignorespaces}\ignorespaces
\bibitem{Polyakov:1974gs}
A.~M. Polyakov,
\textit{``{Nonhamiltonian approach to conformal quantum field theory}''},
Zh.~Eksp.~Teor.~Fiz. \textbf{66}, 23 (1974)\ignorespaces\ignorespaces,
[Sov. Phys. JETP39,9(1974)]\ignorespaces
\bibitem{Dobrev:1977qv}
V.~K. Dobrev, G.~Mack, V.~B. Petkova, S.~G. Petrova \& I.~T. Todorov,
\textit{``{Harmonic Analysis on the n-Dimensional Lorentz Group and Its
  Application to Conformal Quantum Field Theory}''},
\doiref{10.1007/BFb0009678}{Lect.~Notes~Phys. \textbf{63}, 1
  (1977)\ignorespaces}\ignorespaces
\bibitem{Cornalba:2007fs}
L.~Cornalba,
\textit{``{Eikonal methods in AdS/CFT: Regge theory and multi-reggeon
  exchange}''},
\normalsize{\texttt{\arxivref{0710.5480}{arXiv:0710.5480}}}\ignorespaces
\bibitem{Cornalba:2009ax}
L.~Cornalba, M.~S. Costa \& J.~Penedones,
\textit{``{Deep Inelastic Scattering in Conformal QCD}''},
\doiref{10.1007/JHEP03(2010)133}{JHEP \textbf{1003}, 133
  (2010)\ignorespaces}\ignorespaces,
\normalsize{\texttt{\arxivref{0911.0043}{arXiv:0911.0043}}}\ignorespaces
\bibitem{Pappadopulo:2012jk}
D.~Pappadopulo, S.~Rychkov, J.~Espin \& R.~Rattazzi,
\textit{``{OPE Convergence in Conformal Field Theory}''},
\doiref{10.1103/PhysRevD.86.105043}{Phys.~Rev. \textbf{D86}, 105043
  (2012)\ignorespaces}\ignorespaces,
\normalsize{\texttt{\arxivref{1208.6449}{arXiv:1208.6449}}}\ignorespaces
\bibitem{Costa:2012cb}
M.~S. Costa, V.~Goncalves \& J.~Penedones,
\textit{``{Conformal Regge theory}''},
\doiref{10.1007/JHEP12(2012)091}{JHEP \textbf{1212}, 091
  (2012)\ignorespaces}\ignorespaces,
\normalsize{\texttt{\arxivref{1209.4355}{arXiv:1209.4355}}}\ignorespaces
\bibitem{Hogervorst:2013sma}
M.~Hogervorst \& S.~Rychkov,
\textit{``{Radial Coordinates for Conformal Blocks}''},
\doiref{10.1103/PhysRevD.87.106004}{Phys.~Rev. \textbf{D87}, 106004
  (2013)\ignorespaces}\ignorespaces,
\normalsize{\texttt{\arxivref{1303.1111}{arXiv:1303.1111}}}\ignorespaces
\bibitem{Hartman:2015lfa}
T.~Hartman, S.~Jain \& S.~Kundu,
\textit{``{Causality Constraints in Conformal Field Theory}''},
\doiref{10.1007/JHEP05(2016)099}{JHEP \textbf{1605}, 099
  (2016)\ignorespaces}\ignorespaces,
\normalsize{\texttt{\arxivref{1509.00014}{arXiv:1509.00014}}}\ignorespaces
\bibitem{Kim:2015oca}
H.~Kim, P.~Kravchuk \& H.~Ooguri,
\textit{``{Reflections on Conformal Spectra}''},
\doiref{10.1007/JHEP04(2016)184}{JHEP \textbf{1604}, 184
  (2016)\ignorespaces}\ignorespaces,
\normalsize{\texttt{\arxivref{1510.08772}{arXiv:1510.08772}}}\ignorespaces,
[,322(2015)]\ignorespaces
\bibitem{Li:2015itl}
D.~Li, D.~Meltzer \& D.~Poland,
\textit{``{Conformal Collider Physics from the Lightcone Bootstrap}''},
\doiref{10.1007/JHEP02(2016)143}{JHEP \textbf{1602}, 143
  (2016)\ignorespaces}\ignorespaces,
\normalsize{\texttt{\arxivref{1511.08025}{arXiv:1511.08025}}}\ignorespaces
\bibitem{Hartman:2016dxc}
T.~Hartman, S.~Jain \& S.~Kundu,
\textit{``{A New Spin on Causality Constraints}''},
\doiref{10.1007/JHEP10(2016)141}{JHEP \textbf{1610}, 141
  (2016)\ignorespaces}\ignorespaces,
\normalsize{\texttt{\arxivref{1601.07904}{arXiv:1601.07904}}}\ignorespaces
\bibitem{Simmons-Duffin:2016gjk}
D.~Simmons-Duffin,
\textit{``{The Conformal Bootstrap}''},
\normalsize{\texttt{\arxivref{1602.07982}{arXiv:1602.07982}}}\ignorespaces,
in \textit{``{Proceedings, Theoretical Advanced Study Institute in Elementary
  Particle Physics: New Frontiers in Fields and Strings (TASI 2015): Boulder,
  CO, USA, June 1-26, 2015}''},
1-74\ignorespaces
\bibitem{Hofman:2016awc}
D.~M. Hofman, D.~Li, D.~Meltzer, D.~Poland \& F.~Rejon-Barrera,
\textit{``{A Proof of the Conformal Collider Bounds}''},
\doiref{10.1007/JHEP06(2016)111}{JHEP \textbf{1606}, 111
  (2016)\ignorespaces}\ignorespaces,
\normalsize{\texttt{\arxivref{1603.03771}{arXiv:1603.03771}}}\ignorespaces
\bibitem{Hartman:2016lgu}
T.~Hartman, S.~Kundu \& A.~Tajdini,
\textit{``{Averaged Null Energy Condition from Causality}''},
\doiref{10.1007/JHEP07(2017)066}{JHEP \textbf{1707}, 066
  (2017)\ignorespaces}\ignorespaces,
\normalsize{\texttt{\arxivref{1610.05308}{arXiv:1610.05308}}}\ignorespaces
\bibitem{Afkhami-Jeddi:2016ntf}
N.~Afkhami-Jeddi, T.~Hartman, S.~Kundu \& A.~Tajdini,
\textit{``{Einstein gravity 3-point functions from conformal field theory}''},
\doiref{10.1007/JHEP12(2017)049}{JHEP \textbf{1712}, 049
  (2017)\ignorespaces}\ignorespaces,
\normalsize{\texttt{\arxivref{1610.09378}{arXiv:1610.09378}}}\ignorespaces
\bibitem{Gadde:2017sjg}
A.~Gadde,
\textit{``{In search of conformal theories}''},
\normalsize{\texttt{\arxivref{1702.07362}{arXiv:1702.07362}}}\ignorespaces
\bibitem{Hogervorst:2017sfd}
M.~Hogervorst \& B.~C. van~Rees,
\textit{``{Crossing symmetry in alpha space}''},
\doiref{10.1007/JHEP11(2017)193}{JHEP \textbf{1711}, 193
  (2017)\ignorespaces}\ignorespaces,
\normalsize{\texttt{\arxivref{1702.08471}{arXiv:1702.08471}}}\ignorespaces
\bibitem{Caron-Huot:2017vep}
S.~Caron-Huot,
\textit{``{Analyticity in Spin in Conformal Theories}''},
\doiref{10.1007/JHEP09(2017)078}{JHEP \textbf{1709}, 078
  (2017)\ignorespaces}\ignorespaces,
\normalsize{\texttt{\arxivref{1703.00278}{arXiv:1703.00278}}}\ignorespaces
\bibitem{Hogervorst:2017kbj}
M.~Hogervorst,
\textit{``{Crossing Kernels for Boundary and Crosscap CFTs}''},
\normalsize{\texttt{\arxivref{1703.08159}{arXiv:1703.08159}}}\ignorespaces
\bibitem{Kulaxizi:2017ixa}
M.~Kulaxizi, A.~Parnachev \& A.~Zhiboedov,
\textit{``{Bulk Phase Shift, CFT Regge Limit and Einstein Gravity}''},
\doiref{10.1007/JHEP06(2018)121}{JHEP \textbf{1806}, 121
  (2018)\ignorespaces}\ignorespaces,
\normalsize{\texttt{\arxivref{1705.02934}{arXiv:1705.02934}}}\ignorespaces
\bibitem{Li:2017lmh}
D.~Li, D.~Meltzer \& D.~Poland,
\textit{``{Conformal Bootstrap in the Regge Limit}''},
\doiref{10.1007/JHEP12(2017)013}{JHEP \textbf{1712}, 013
  (2017)\ignorespaces}\ignorespaces,
\normalsize{\texttt{\arxivref{1705.03453}{arXiv:1705.03453}}}\ignorespaces
\bibitem{Cuomo:2017wme}
G.~F. Cuomo, D.~Karateev \& P.~Kravchuk,
\textit{``{General Bootstrap Equations in 4D CFTs}''},
\doiref{10.1007/JHEP01(2018)130}{JHEP \textbf{1801}, 130
  (2018)\ignorespaces}\ignorespaces,
\normalsize{\texttt{\arxivref{1705.05401}{arXiv:1705.05401}}}\ignorespaces,
[,57(2017)]\ignorespaces
\bibitem{Dey:2017oim}
P.~Dey \& A.~Kaviraj,
\textit{``{Towards a Bootstrap approach to higher orders of epsilon
  expansion}''},
\doiref{10.1007/JHEP02(2018)153}{JHEP \textbf{1802}, 153
  (2018)\ignorespaces}\ignorespaces,
\normalsize{\texttt{\arxivref{1711.01173}{arXiv:1711.01173}}}\ignorespaces
\bibitem{Simmons-Duffin:2017nub}
D.~Simmons-Duffin, D.~Stanford \& E.~Witten,
\textit{``{A spacetime derivation of the Lorentzian OPE inversion formula}''},
\doiref{10.1007/JHEP07(2018)085}{JHEP \textbf{1807}, 085
  (2018)\ignorespaces}\ignorespaces,
\normalsize{\texttt{\arxivref{1711.03816}{arXiv:1711.03816}}}\ignorespaces
\bibitem{Elkhidir:2017iov}
E.~Elkhidir \& D.~Karateev,
\textit{``{Scalar-Fermion Analytic Bootstrap in 4D}''},
\normalsize{\texttt{\arxivref{1712.01554}{arXiv:1712.01554}}}\ignorespaces
\bibitem{Kravchuk:2018htv}
P.~Kravchuk \& D.~Simmons-Duffin,
\textit{``{Light-ray operators in conformal field theory}''},
\doiref{10.1007/JHEP11(2018)102}{JHEP \textbf{1811}, 102
  (2018)\ignorespaces}\ignorespaces,
\normalsize{\texttt{\arxivref{1805.00098}{arXiv:1805.00098}}}\ignorespaces,
[,236(2018)]\ignorespaces
\bibitem{Karateev:2018oml}
D.~Karateev, P.~Kravchuk \& D.~Simmons-Duffin,
\textit{``{Harmonic Analysis and Mean Field Theory}''},
\normalsize{\texttt{\arxivref{1809.05111}{arXiv:1809.05111}}}\ignorespaces
\bibitem{Liendo:2019jpu}
P.~Liendo, Y.~Linke \& V.~Schomerus,
\textit{``{A Lorentzian inversion formula for defect CFT}''},
\normalsize{\texttt{\arxivref{1903.05222}{arXiv:1903.05222}}}\ignorespaces
\bibitem{Albayrak:2019gnz}
S.~Albayrak, D.~Meltzer \& D.~Poland,
\textit{``{More Analytic Bootstrap: Nonperturbative Effects and Fermions}''},
\normalsize{\texttt{\arxivref{1904.00032}{arXiv:1904.00032}}}\ignorespaces
\bibitem{Rattazzi:2008pe}
R.~Rattazzi, V.~S. Rychkov, E.~Tonni \& A.~Vichi,
\textit{``{Bounding scalar operator dimensions in 4D CFT}''},
\doiref{10.1088/1126-6708/2008/12/031}{JHEP \textbf{0812}, 031
  (2008)\ignorespaces}\ignorespaces,
\normalsize{\texttt{\arxivref{0807.0004}{arXiv:0807.0004}}}\ignorespaces
\bibitem{Rychkov:2009ij}
V.~S. Rychkov \& A.~Vichi,
\textit{``{Universal Constraints on Conformal Operator Dimensions}''},
\doiref{10.1103/PhysRevD.80.045006}{Phys.~Rev. \textbf{D80}, 045006
  (2009)\ignorespaces}\ignorespaces,
\normalsize{\texttt{\arxivref{0905.2211}{arXiv:0905.2211}}}\ignorespaces
\bibitem{Caracciolo:2009bx}
F.~Caracciolo \& V.~S. Rychkov,
\textit{``{Rigorous Limits on the Interaction Strength in Quantum Field
  Theory}''},
\doiref{10.1103/PhysRevD.81.085037}{Phys.~Rev. \textbf{D81}, 085037
  (2010)\ignorespaces}\ignorespaces,
\normalsize{\texttt{\arxivref{0912.2726}{arXiv:0912.2726}}}\ignorespaces
\bibitem{Poland:2010wg}
D.~Poland \& D.~Simmons-Duffin,
\textit{``{Bounds on 4D Conformal and Superconformal Field Theories}''},
\doiref{10.1007/JHEP05(2011)017}{JHEP \textbf{1105}, 017
  (2011)\ignorespaces}\ignorespaces,
\normalsize{\texttt{\arxivref{1009.2087}{arXiv:1009.2087}}}\ignorespaces
\bibitem{Rattazzi:2010gj}
R.~Rattazzi, S.~Rychkov \& A.~Vichi,
\textit{``{Central Charge Bounds in 4D Conformal Field Theory}''},
\doiref{10.1103/PhysRevD.83.046011}{Phys.~Rev. \textbf{D83}, 046011
  (2011)\ignorespaces}\ignorespaces,
\normalsize{\texttt{\arxivref{1009.2725}{arXiv:1009.2725}}}\ignorespaces
\bibitem{Poland:2011ey}
D.~Poland, D.~Simmons-Duffin \& A.~Vichi,
\textit{``{Carving Out the Space of 4D CFTs}''},
\doiref{10.1007/JHEP05(2012)110}{JHEP \textbf{1205}, 110
  (2012)\ignorespaces}\ignorespaces,
\normalsize{\texttt{\arxivref{1109.5176}{arXiv:1109.5176}}}\ignorespaces
\bibitem{Rychkov:2011et}
S.~Rychkov,
\textit{``{Conformal Bootstrap in Three Dimensions?}''},
\normalsize{\texttt{\arxivref{1111.2115}{arXiv:1111.2115}}}\ignorespaces
\bibitem{ElShowk:2012ht}
S.~El-Showk, M.~F. Paulos, D.~Poland, S.~Rychkov, D.~Simmons-Duffin \&
  A.~Vichi,
\textit{``{Solving the 3D Ising Model with the Conformal Bootstrap}''},
\doiref{10.1103/PhysRevD.86.025022}{Phys.~Rev. \textbf{D86}, 025022
  (2012)\ignorespaces}\ignorespaces,
\normalsize{\texttt{\arxivref{1203.6064}{arXiv:1203.6064}}}\ignorespaces
\bibitem{Liendo:2012hy}
P.~Liendo, L.~Rastelli \& B.~C. van~Rees,
\textit{``{The Bootstrap Program for Boundary CFT$_d$}''},
\doiref{10.1007/JHEP07(2013)113}{JHEP \textbf{1307}, 113
  (2013)\ignorespaces}\ignorespaces,
\normalsize{\texttt{\arxivref{1210.4258}{arXiv:1210.4258}}}\ignorespaces
\bibitem{ElShowk:2012hu}
S.~El-Showk \& M.~F. Paulos,
\textit{``{Bootstrapping Conformal Field Theories with the Extremal Functional
  Method}''},
\doiref{10.1103/PhysRevLett.111.241601}{Phys.~Rev.~Lett. \textbf{111}, 241601
  (2013)\ignorespaces}\ignorespaces,
\normalsize{\texttt{\arxivref{1211.2810}{arXiv:1211.2810}}}\ignorespaces
\bibitem{Gliozzi:2013ysa}
F.~Gliozzi,
\textit{``{More constraining conformal bootstrap}''},
\doiref{10.1103/PhysRevLett.111.161602}{Phys.~Rev.~Lett. \textbf{111}, 161602
  (2013)\ignorespaces}\ignorespaces,
\normalsize{\texttt{\arxivref{1307.3111}{arXiv:1307.3111}}}\ignorespaces
\bibitem{Alday:2013opa}
L.~F. Alday \& A.~Bissi,
\textit{``{The superconformal bootstrap for structure constants}''},
\doiref{10.1007/JHEP09(2014)144}{JHEP \textbf{1409}, 144
  (2014)\ignorespaces}\ignorespaces,
\normalsize{\texttt{\arxivref{1310.3757}{arXiv:1310.3757}}}\ignorespaces
\bibitem{Gaiotto:2013nva}
D.~Gaiotto, D.~Mazac \& M.~F. Paulos,
\textit{``{Bootstrapping the 3d Ising twist defect}''},
\doiref{10.1007/JHEP03(2014)100}{JHEP \textbf{1403}, 100
  (2014)\ignorespaces}\ignorespaces,
\normalsize{\texttt{\arxivref{1310.5078}{arXiv:1310.5078}}}\ignorespaces
\bibitem{El-Showk:2014dwa}
S.~El-Showk, M.~F. Paulos, D.~Poland, S.~Rychkov, D.~Simmons-Duffin \&
  A.~Vichi,
\textit{``{Solving the 3d Ising Model with the Conformal Bootstrap II.
  c-Minimization and Precise Critical Exponents}''},
\doiref{10.1007/s10955-014-1042-7}{J.~Stat.~Phys. \textbf{157}, 869
  (2014)\ignorespaces}\ignorespaces,
\normalsize{\texttt{\arxivref{1403.4545}{arXiv:1403.4545}}}\ignorespaces
\bibitem{Chester:2014fya}
S.~M. Chester, J.~Lee, S.~S. Pufu \& R.~Yacoby,
\textit{``{The $ \mathcal{N}=8 $ superconformal bootstrap in three
  dimensions}''},
\doiref{10.1007/JHEP09(2014)143}{JHEP \textbf{1409}, 143
  (2014)\ignorespaces}\ignorespaces,
\normalsize{\texttt{\arxivref{1406.4814}{arXiv:1406.4814}}}\ignorespaces
\bibitem{Kos:2014bka}
F.~Kos, D.~Poland \& D.~Simmons-Duffin,
\textit{``{Bootstrapping Mixed Correlators in the 3D Ising Model}''},
\doiref{10.1007/JHEP11(2014)109}{JHEP \textbf{1411}, 109
  (2014)\ignorespaces}\ignorespaces,
\normalsize{\texttt{\arxivref{1406.4858}{arXiv:1406.4858}}}\ignorespaces
\bibitem{Caracciolo:2014cxa}
F.~Caracciolo, A.~Castedo~Echeverri, B.~von~Harling \& M.~Serone,
\textit{``{Bounds on OPE Coefficients in 4D Conformal Field Theories}''},
\doiref{10.1007/JHEP10(2014)020}{JHEP \textbf{1410}, 020
  (2014)\ignorespaces}\ignorespaces,
\normalsize{\texttt{\arxivref{1406.7845}{arXiv:1406.7845}}}\ignorespaces
\bibitem{Paulos:2014vya}
M.~F. Paulos,
\textit{``{JuliBootS: a hands-on guide to the conformal bootstrap}''},
\normalsize{\texttt{\arxivref{1412.4127}{arXiv:1412.4127}}}\ignorespaces
\bibitem{Beem:2014zpa}
C.~Beem, M.~Lemos, P.~Liendo, L.~Rastelli \& B.~C. van~Rees,
\textit{``{The $ \mathcal{N}=2 $ superconformal bootstrap}''},
\doiref{10.1007/JHEP03(2016)183}{JHEP \textbf{1603}, 183
  (2016)\ignorespaces}\ignorespaces,
\normalsize{\texttt{\arxivref{1412.7541}{arXiv:1412.7541}}}\ignorespaces
\bibitem{Simmons-Duffin:2015qma}
D.~Simmons-Duffin,
\textit{``{A Semidefinite Program Solver for the Conformal Bootstrap}''},
\doiref{10.1007/JHEP06(2015)174}{JHEP \textbf{1506}, 174
  (2015)\ignorespaces}\ignorespaces,
\normalsize{\texttt{\arxivref{1502.02033}{arXiv:1502.02033}}}\ignorespaces
\bibitem{Bobev:2015jxa}
N.~Bobev, S.~El-Showk, D.~Mazac \& M.~F. Paulos,
\textit{``{Bootstrapping SCFTs with Four Supercharges}''},
\doiref{10.1007/JHEP08(2015)142}{JHEP \textbf{1508}, 142
  (2015)\ignorespaces}\ignorespaces,
\normalsize{\texttt{\arxivref{1503.02081}{arXiv:1503.02081}}}\ignorespaces
\bibitem{Beem:2015aoa}
C.~Beem, M.~Lemos, L.~Rastelli \& B.~C. van~Rees,
\textit{``{The (2, 0) superconformal bootstrap}''},
\doiref{10.1103/PhysRevD.93.025016}{Phys.~Rev. \textbf{D93}, 025016
  (2016)\ignorespaces}\ignorespaces,
\normalsize{\texttt{\arxivref{1507.05637}{arXiv:1507.05637}}}\ignorespaces
\bibitem{Iliesiu:2015qra}
L.~Iliesiu, F.~Kos, D.~Poland, S.~S. Pufu, D.~Simmons-Duffin \& R.~Yacoby,
\textit{``{Bootstrapping 3D Fermions}''},
\doiref{10.1007/JHEP03(2016)120}{JHEP \textbf{1603}, 120
  (2016)\ignorespaces}\ignorespaces,
\normalsize{\texttt{\arxivref{1508.00012}{arXiv:1508.00012}}}\ignorespaces
\bibitem{Poland:2015mta}
D.~Poland \& A.~Stergiou,
\textit{``{Exploring the Minimal 4D $\mathcal{N}=1$ SCFT}''},
\doiref{10.1007/JHEP12(2015)121}{JHEP \textbf{1512}, 121
  (2015)\ignorespaces}\ignorespaces,
\normalsize{\texttt{\arxivref{1509.06368}{arXiv:1509.06368}}}\ignorespaces
\bibitem{Lemos:2015awa}
M.~Lemos \& P.~Liendo,
\textit{``{Bootstrapping $ \mathcal{N}=2 $ chiral correlators}''},
\doiref{10.1007/JHEP01(2016)025}{JHEP \textbf{1601}, 025
  (2016)\ignorespaces}\ignorespaces,
\normalsize{\texttt{\arxivref{1510.03866}{arXiv:1510.03866}}}\ignorespaces
\bibitem{Lin:2015wcg}
Y.-H. Lin, S.-H. Shao, D.~Simmons-Duffin, Y.~Wang \& X.~Yin,
\textit{``{$ \mathcal{N} $ = 4 superconformal bootstrap of the K3 CFT}''},
\doiref{10.1007/JHEP05(2017)126}{JHEP \textbf{1705}, 126
  (2017)\ignorespaces}\ignorespaces,
\normalsize{\texttt{\arxivref{1511.04065}{arXiv:1511.04065}}}\ignorespaces
\bibitem{Chester:2016wrc}
S.~M. Chester \& S.~S. Pufu,
\textit{``{Towards bootstrapping QED$_{3}$}''},
\doiref{10.1007/JHEP08(2016)019}{JHEP \textbf{1608}, 019
  (2016)\ignorespaces}\ignorespaces,
\normalsize{\texttt{\arxivref{1601.03476}{arXiv:1601.03476}}}\ignorespaces
\bibitem{Rychkov:2016iqz}
S.~Rychkov,
\textit{``{EPFL Lectures on Conformal Field Theory in D>= 3 Dimensions}''}
\bibitem{Behan:2016dtz}
C.~Behan,
\textit{``{PyCFTBoot: A flexible interface for the conformal bootstrap}''},
\doiref{10.4208/cicp.OA-2016-0107}{Commun.~Comput.~Phys. \textbf{22}, 1
  (2017)\ignorespaces}\ignorespaces,
\normalsize{\texttt{\arxivref{1602.02810}{arXiv:1602.02810}}}\ignorespaces
\bibitem{El-Showk:2016mxr}
S.~El-Showk \& M.~F. Paulos,
\textit{``{Extremal bootstrapping: go with the flow}''},
\doiref{10.1007/JHEP03(2018)148}{JHEP \textbf{1803}, 148
  (2018)\ignorespaces}\ignorespaces,
\normalsize{\texttt{\arxivref{1605.08087}{arXiv:1605.08087}}}\ignorespaces
\bibitem{Lin:2016gcl}
Y.-H. Lin, S.-H. Shao, Y.~Wang \& X.~Yin,
\textit{``{(2, 2) superconformal bootstrap in two dimensions}''},
\doiref{10.1007/JHEP05(2017)112}{JHEP \textbf{1705}, 112
  (2017)\ignorespaces}\ignorespaces,
\normalsize{\texttt{\arxivref{1610.05371}{arXiv:1610.05371}}}\ignorespaces
\bibitem{Lemos:2016xke}
M.~Lemos, P.~Liendo, C.~Meneghelli \& V.~Mitev,
\textit{``{Bootstrapping $\mathcal{N}=3$ superconformal theories}''},
\doiref{10.1007/JHEP04(2017)032}{JHEP \textbf{1704}, 032
  (2017)\ignorespaces}\ignorespaces,
\normalsize{\texttt{\arxivref{1612.01536}{arXiv:1612.01536}}}\ignorespaces
\bibitem{Beem:2016wfs}
C.~Beem, L.~Rastelli \& B.~C. van~Rees,
\textit{``{More ${\mathcal N}=4$ superconformal bootstrap}''},
\doiref{10.1103/PhysRevD.96.046014}{Phys.~Rev. \textbf{D96}, 046014
  (2017)\ignorespaces}\ignorespaces,
\normalsize{\texttt{\arxivref{1612.02363}{arXiv:1612.02363}}}\ignorespaces
\bibitem{Li:2017ddj}
D.~Li, D.~Meltzer \& A.~Stergiou,
\textit{``{Bootstrapping mixed correlators in 4D $ \mathcal{N} $ = 1 SCFTs}''},
\doiref{10.1007/JHEP07(2017)029}{JHEP \textbf{1707}, 029
  (2017)\ignorespaces}\ignorespaces,
\normalsize{\texttt{\arxivref{1702.00404}{arXiv:1702.00404}}}\ignorespaces
\bibitem{Collier:2017shs}
S.~Collier, P.~Kravchuk, Y.-H. Lin \& X.~Yin,
\textit{``{Bootstrapping the Spectral Function: On the Uniqueness of Liouville
  and the Universality of BTZ}''},
\doiref{10.1007/JHEP09(2018)150}{JHEP \textbf{1809}, 150
  (2018)\ignorespaces}\ignorespaces,
\normalsize{\texttt{\arxivref{1702.00423}{arXiv:1702.00423}}}\ignorespaces
\bibitem{Cornagliotto:2017dup}
M.~Cornagliotto, M.~Lemos \& V.~Schomerus,
\textit{``{Long Multiplet Bootstrap}''},
\doiref{10.1007/JHEP10(2017)119}{JHEP \textbf{1710}, 119
  (2017)\ignorespaces}\ignorespaces,
\normalsize{\texttt{\arxivref{1702.05101}{arXiv:1702.05101}}}\ignorespaces
\bibitem{Rychkov:2017tpc}
J.~Qiao \& S.~Rychkov,
\textit{``{Cut-touching linear functionals in the conformal bootstrap}''},
\doiref{10.1007/JHEP06(2017)076}{JHEP \textbf{1706}, 076
  (2017)\ignorespaces}\ignorespaces,
\normalsize{\texttt{\arxivref{1705.01357}{arXiv:1705.01357}}}\ignorespaces
\bibitem{Nakayama:2017vdd}
Y.~Nakayama,
\textit{``{Bootstrap experiments on higher dimensional CFTs}''},
\doiref{10.1142/S0217751X18500367}{Int.~J.~Mod.~Phys. \textbf{A33}, 1850036
  (2018)\ignorespaces}\ignorespaces,
\normalsize{\texttt{\arxivref{1705.02744}{arXiv:1705.02744}}}\ignorespaces
\bibitem{Chang:2017xmr}
C.-M. Chang \& Y.-H. Lin,
\textit{``{Carving Out the End of the World or (Superconformal Bootstrap in Six
  Dimensions)}''},
\doiref{10.1007/JHEP08(2017)128}{JHEP \textbf{1708}, 128
  (2017)\ignorespaces}\ignorespaces,
\normalsize{\texttt{\arxivref{1705.05392}{arXiv:1705.05392}}}\ignorespaces
\bibitem{Dymarsky:2017yzx}
A.~Dymarsky, F.~Kos, P.~Kravchuk, D.~Poland \& D.~Simmons-Duffin,
\textit{``{The 3d Stress-Tensor Bootstrap}''},
\doiref{10.1007/JHEP02(2018)164}{JHEP \textbf{1802}, 164
  (2018)\ignorespaces}\ignorespaces,
\normalsize{\texttt{\arxivref{1708.05718}{arXiv:1708.05718}}}\ignorespaces,
[,343(2017)]\ignorespaces
\bibitem{Poland:2018epd}
D.~Poland, S.~Rychkov \& A.~Vichi,
\textit{``{The Conformal Bootstrap: Theory, Numerical Techniques, and
  Applications}''},
\doiref{10.1103/RevModPhys.91.015002}{Rev.~Mod.~Phys. \textbf{91}, 15002
  (2019)\ignorespaces}\ignorespaces,
\normalsize{\texttt{\arxivref{1805.04405}{arXiv:1805.04405}}}\ignorespaces,
[Rev. Mod. Phys.91,015002(2019)]\ignorespaces
\bibitem{Karateev:2019pvw}
D.~Karateev, P.~Kravchuk, M.~Serone \& A.~Vichi,
\textit{``{Fermion Conformal Bootstrap in 4d}''},
\normalsize{\texttt{\arxivref{1902.05969}{arXiv:1902.05969}}}\ignorespaces
\bibitem{Rattazzi:2010yc}
R.~Rattazzi, S.~Rychkov \& A.~Vichi,
\textit{``{Bounds in 4D Conformal Field Theories with Global Symmetry}''},
\doiref{10.1088/1751-8113/44/3/035402}{J.~Phys. \textbf{A44}, 035402
  (2011)\ignorespaces}\ignorespaces,
\normalsize{\texttt{\arxivref{1009.5985}{arXiv:1009.5985}}}\ignorespaces
\bibitem{Vichi:2011ux}
A.~Vichi,
\textit{``{Improved bounds for CFT's with global symmetries}''},
\doiref{10.1007/JHEP01(2012)162}{JHEP \textbf{1201}, 162
  (2012)\ignorespaces}\ignorespaces,
\normalsize{\texttt{\arxivref{1106.4037}{arXiv:1106.4037}}}\ignorespaces
\bibitem{Kos:2013tga}
F.~Kos, D.~Poland \& D.~Simmons-Duffin,
\textit{``{Bootstrapping the $O(N)$ vector models}''},
\doiref{10.1007/JHEP06(2014)091}{JHEP \textbf{1406}, 091
  (2014)\ignorespaces}\ignorespaces,
\normalsize{\texttt{\arxivref{1307.6856}{arXiv:1307.6856}}}\ignorespaces
\bibitem{Berkooz:2014yda}
M.~Berkooz, R.~Yacoby \& A.~Zait,
\textit{``{Bounds on $\mathcal{N} = 1$ superconformal theories with global
  symmetries}''},
\doiref{10.1007/JHEP01(2015)132, 10.1007/JHEP08(2014)008}{JHEP \textbf{1408},
  008 (2014)\ignorespaces}\ignorespaces,
\normalsize{\texttt{\arxivref{1402.6068}{arXiv:1402.6068}}}\ignorespaces,
[Erratum: JHEP01,132(2015)]\ignorespaces
\bibitem{Nakayama:2014lva}
Y.~Nakayama \& T.~Ohtsuki,
\textit{``{Approaching the conformal window of $O(n)\times O(m)$ symmetric
  Landau-Ginzburg models using the conformal bootstrap}''},
\doiref{10.1103/PhysRevD.89.126009}{Phys.~Rev. \textbf{D89}, 126009
  (2014)\ignorespaces}\ignorespaces,
\normalsize{\texttt{\arxivref{1404.0489}{arXiv:1404.0489}}}\ignorespaces
\bibitem{Nakayama:2014yia}
Y.~Nakayama \& T.~Ohtsuki,
\textit{``{Five dimensional $O(N)$-symmetric CFTs from conformal bootstrap}''},
\doiref{10.1016/j.physletb.2014.05.058}{Phys.~Lett. \textbf{B734}, 193
  (2014)\ignorespaces}\ignorespaces,
\normalsize{\texttt{\arxivref{1404.5201}{arXiv:1404.5201}}}\ignorespaces
\bibitem{Nakayama:2014sba}
Y.~Nakayama \& T.~Ohtsuki,
\textit{``{Bootstrapping phase transitions in QCD and frustrated spin
  systems}''},
\doiref{10.1103/PhysRevD.91.021901}{Phys.~Rev. \textbf{D91}, 021901
  (2015)\ignorespaces}\ignorespaces,
\normalsize{\texttt{\arxivref{1407.6195}{arXiv:1407.6195}}}\ignorespaces
\bibitem{Bae:2014hia}
J.-B. Bae \& S.-J. Rey,
\textit{``{Conformal Bootstrap Approach to O(N) Fixed Points in Five
  Dimensions}''},
\normalsize{\texttt{\arxivref{1412.6549}{arXiv:1412.6549}}}\ignorespaces
\bibitem{Chester:2014gqa}
S.~M. Chester, S.~S. Pufu \& R.~Yacoby,
\textit{``{Bootstrapping $O(N)$ vector models in 4 $< d <$ 6}''},
\doiref{10.1103/PhysRevD.91.086014}{Phys.~Rev. \textbf{D91}, 086014
  (2015)\ignorespaces}\ignorespaces,
\normalsize{\texttt{\arxivref{1412.7746}{arXiv:1412.7746}}}\ignorespaces
\bibitem{Kos:2015mba}
F.~Kos, D.~Poland, D.~Simmons-Duffin \& A.~Vichi,
\textit{``{Bootstrapping the O(N) Archipelago}''},
\doiref{10.1007/JHEP11(2015)106}{JHEP \textbf{1511}, 106
  (2015)\ignorespaces}\ignorespaces,
\normalsize{\texttt{\arxivref{1504.07997}{arXiv:1504.07997}}}\ignorespaces
\bibitem{Chester:2015qca}
S.~M. Chester, S.~Giombi, L.~V. Iliesiu, I.~R. Klebanov, S.~S. Pufu \&
  R.~Yacoby,
\textit{``{Accidental Symmetries and the Conformal Bootstrap}''},
\doiref{10.1007/JHEP01(2016)110}{JHEP \textbf{1601}, 110
  (2016)\ignorespaces}\ignorespaces,
\normalsize{\texttt{\arxivref{1507.04424}{arXiv:1507.04424}}}\ignorespaces
\bibitem{Chester:2015lej}
S.~M. Chester, L.~V. Iliesiu, S.~S. Pufu \& R.~Yacoby,
\textit{``{Bootstrapping $O(N)$ Vector Models with Four Supercharges in $3 \leq
  d \leq4$}''},
\doiref{10.1007/JHEP05(2016)103}{JHEP \textbf{1605}, 103
  (2016)\ignorespaces}\ignorespaces,
\normalsize{\texttt{\arxivref{1511.07552}{arXiv:1511.07552}}}\ignorespaces
\bibitem{Dey:2016zbg}
P.~Dey, A.~Kaviraj \& K.~Sen,
\textit{``{More on analytic bootstrap for O(N) models}''},
\doiref{10.1007/JHEP06(2016)136}{JHEP \textbf{1606}, 136
  (2016)\ignorespaces}\ignorespaces,
\normalsize{\texttt{\arxivref{1602.04928}{arXiv:1602.04928}}}\ignorespaces
\bibitem{Nakayama:2016knq}
Y.~Nakayama,
\textit{``{Bootstrap bound for conformal multi-flavor QCD on lattice}''},
\doiref{10.1007/JHEP07(2016)038}{JHEP \textbf{1607}, 038
  (2016)\ignorespaces}\ignorespaces,
\normalsize{\texttt{\arxivref{1605.04052}{arXiv:1605.04052}}}\ignorespaces
\bibitem{Li:2016wdp}
Z.~Li \& N.~Su,
\textit{``{Bootstrapping Mixed Correlators in the Five Dimensional Critical
  O(N) Models}''},
\doiref{10.1007/JHEP04(2017)098}{JHEP \textbf{1704}, 098
  (2017)\ignorespaces}\ignorespaces,
\normalsize{\texttt{\arxivref{1607.07077}{arXiv:1607.07077}}}\ignorespaces
\bibitem{Pang:2016xno}
Y.~Pang, J.~Rong \& N.~Su,
\textit{``{$\phi^{3}$ theory with F$_{4}$ flavor symmetry in 6 − 2$\epsilon$
  dimensions: 3-loop renormalization and conformal bootstrap}''},
\doiref{10.1007/JHEP12(2016)057}{JHEP \textbf{1612}, 057
  (2016)\ignorespaces}\ignorespaces,
\normalsize{\texttt{\arxivref{1609.03007}{arXiv:1609.03007}}}\ignorespaces
\bibitem{Dymarsky:2017xzb}
A.~Dymarsky, J.~Penedones, E.~Trevisani \& A.~Vichi,
\textit{``{Charting the space of 3D CFTs with a continuous global symmetry}''},
\normalsize{\texttt{\arxivref{1705.04278}{arXiv:1705.04278}}}\ignorespaces
\bibitem{Stergiou:2018gjj}
A.~Stergiou,
\textit{``{Bootstrapping hypercubic and hypertetrahedral theories in three
  dimensions}''},
\doiref{10.1007/JHEP05(2018)035}{JHEP \textbf{1805}, 035
  (2018)\ignorespaces}\ignorespaces,
\normalsize{\texttt{\arxivref{1801.07127}{arXiv:1801.07127}}}\ignorespaces
\bibitem{Kousvos:2018rhl}
S.~R. Kousvos \& A.~Stergiou,
\textit{``{Bootstrapping Mixed Correlators in Three-Dimensional Cubic
  Theories}''},
\normalsize{\texttt{\arxivref{1810.10015}{arXiv:1810.10015}}}\ignorespaces
\bibitem{Stergiou:2019dcv}
A.~Stergiou,
\textit{``{Bootstrapping MN and Tetragonal CFTs in Three Dimensions}''},
\normalsize{\texttt{\arxivref{1904.00017}{arXiv:1904.00017}}}\ignorespaces
\bibitem{Alday:2015eya}
L.~F. Alday, A.~Bissi \& T.~Lukowski,
\textit{``{Large spin systematics in CFT}''},
\doiref{10.1007/JHEP11(2015)101}{JHEP \textbf{1511}, 101
  (2015)\ignorespaces}\ignorespaces,
\normalsize{\texttt{\arxivref{1502.07707}{arXiv:1502.07707}}}\ignorespaces
\bibitem{Alday:2015ota}
L.~F. Alday \& A.~Zhiboedov,
\textit{``{Conformal Bootstrap With Slightly Broken Higher Spin Symmetry}''},
\doiref{10.1007/JHEP06(2016)091}{JHEP \textbf{1606}, 091
  (2016)\ignorespaces}\ignorespaces,
\normalsize{\texttt{\arxivref{1506.04659}{arXiv:1506.04659}}}\ignorespaces
\bibitem{Alday:2015ewa}
L.~F. Alday \& A.~Zhiboedov,
\textit{``{An Algebraic Approach to the Analytic Bootstrap}''},
\doiref{10.1007/JHEP04(2017)157}{JHEP \textbf{1704}, 157
  (2017)\ignorespaces}\ignorespaces,
\normalsize{\texttt{\arxivref{1510.08091}{arXiv:1510.08091}}}\ignorespaces
\bibitem{Alday:2016mxe}
L.~F. Alday \& A.~Bissi,
\textit{``{Crossing symmetry and Higher spin towers}''},
\doiref{10.1007/JHEP12(2017)118}{JHEP \textbf{1712}, 118
  (2017)\ignorespaces}\ignorespaces,
\normalsize{\texttt{\arxivref{1603.05150}{arXiv:1603.05150}}}\ignorespaces
\bibitem{Alday:2016njk}
L.~F. Alday,
\textit{``{Large Spin Perturbation Theory for Conformal Field Theories}''},
\doiref{10.1103/PhysRevLett.119.111601}{Phys.~Rev.~Lett. \textbf{119}, 111601
  (2017)\ignorespaces}\ignorespaces,
\normalsize{\texttt{\arxivref{1611.01500}{arXiv:1611.01500}}}\ignorespaces
\bibitem{Alday:2016jfr}
L.~F. Alday,
\textit{``{Solving CFTs with Weakly Broken Higher Spin Symmetry}''},
\doiref{10.1007/JHEP10(2017)161}{JHEP \textbf{1710}, 161
  (2017)\ignorespaces}\ignorespaces,
\normalsize{\texttt{\arxivref{1612.00696}{arXiv:1612.00696}}}\ignorespaces
\bibitem{Ferrara:1973vz}
S.~Ferrara, A.~F. Grillo, G.~Parisi \& R.~Gatto,
\textit{``{Covariant expansion of the conformal four-point function}''},
\doiref{10.1016/0550-3213(72)90587-1, 10.1016/0550-3213(73)90467-7}{Nucl.~Phys.
  \textbf{B49}, 77 (1972)\ignorespaces}\ignorespaces,
[Erratum: Nucl. Phys.B53,643(1973)]\ignorespaces
\bibitem{Ferrara:1974nf}
S.~Ferrara, A.~F. Grillo, R.~Gatto \& G.~Parisi,
\textit{``{Analyticity properties and asymptotic expansions of conformal
  covariant green's functions}''},
\doiref{10.1007/BF02813413}{Nuovo~Cim. \textbf{A19}, 667
  (1974)\ignorespaces}\ignorespaces
\bibitem{Dolan:2000ut}
F.~A. Dolan \& H.~Osborn,
\textit{``{Conformal four point functions and the operator product
  expansion}''},
\doiref{10.1016/S0550-3213(01)00013-X}{Nucl.~Phys. \textbf{B599}, 459
  (2001)\ignorespaces}\ignorespaces,
\normalsize{\texttt{\arxivref{hep-th/0011040}{hep-th/0011040}}}\ignorespaces
\bibitem{Dolan:2003hv}
F.~A. Dolan \& H.~Osborn,
\textit{``{Conformal partial waves and the operator product expansion}''},
\doiref{10.1016/j.nuclphysb.2003.11.016}{Nucl.~Phys. \textbf{B678}, 491
  (2004)\ignorespaces}\ignorespaces,
\normalsize{\texttt{\arxivref{hep-th/0309180}{hep-th/0309180}}}\ignorespaces
\bibitem{Dolan:2011dv}
F.~A. Dolan \& H.~Osborn,
\textit{``{Conformal Partial Waves: Further Mathematical Results}''},
\normalsize{\texttt{\arxivref{1108.6194}{arXiv:1108.6194}}}\ignorespaces
\bibitem{Giombi:2011rz}
S.~Giombi, S.~Prakash \& X.~Yin,
\textit{``{A Note on CFT Correlators in Three Dimensions}''},
\doiref{10.1007/JHEP07(2013)105}{JHEP \textbf{1307}, 105
  (2013)\ignorespaces}\ignorespaces,
\normalsize{\texttt{\arxivref{1104.4317}{arXiv:1104.4317}}}\ignorespaces
\bibitem{Costa:2011mg}
M.~S. Costa, J.~Penedones, D.~Poland \& S.~Rychkov,
\textit{``{Spinning Conformal Correlators}''},
\doiref{10.1007/JHEP11(2011)071}{JHEP \textbf{1111}, 071
  (2011)\ignorespaces}\ignorespaces,
\normalsize{\texttt{\arxivref{1107.3554}{arXiv:1107.3554}}}\ignorespaces
\bibitem{Costa:2011dw}
M.~S. Costa, J.~Penedones, D.~Poland \& S.~Rychkov,
\textit{``{Spinning Conformal Blocks}''},
\doiref{10.1007/JHEP11(2011)154}{JHEP \textbf{1111}, 154
  (2011)\ignorespaces}\ignorespaces,
\normalsize{\texttt{\arxivref{1109.6321}{arXiv:1109.6321}}}\ignorespaces
\bibitem{SimmonsDuffin:2012uy}
D.~Simmons-Duffin,
\textit{``{Projectors, Shadows, and Conformal Blocks}''},
\doiref{10.1007/JHEP04(2014)146}{JHEP \textbf{1404}, 146
  (2014)\ignorespaces}\ignorespaces,
\normalsize{\texttt{\arxivref{1204.3894}{arXiv:1204.3894}}}\ignorespaces
\bibitem{Costa:2014rya}
M.~S. Costa \& T.~Hansen,
\textit{``{Conformal correlators of mixed-symmetry tensors}''},
\doiref{10.1007/JHEP02(2015)151}{JHEP \textbf{1502}, 151
  (2015)\ignorespaces}\ignorespaces,
\normalsize{\texttt{\arxivref{1411.7351}{arXiv:1411.7351}}}\ignorespaces
\bibitem{Elkhidir:2014woa}
E.~Elkhidir, D.~Karateev \& M.~Serone,
\textit{``{General Three-Point Functions in 4D CFT}''},
\doiref{10.1007/JHEP01(2015)133}{JHEP \textbf{1501}, 133
  (2015)\ignorespaces}\ignorespaces,
\normalsize{\texttt{\arxivref{1412.1796}{arXiv:1412.1796}}}\ignorespaces
\bibitem{Echeverri:2015rwa}
A.~Castedo~Echeverri, E.~Elkhidir, D.~Karateev \& M.~Serone,
\textit{``{Deconstructing Conformal Blocks in 4D CFT}''},
\doiref{10.1007/JHEP08(2015)101}{JHEP \textbf{1508}, 101
  (2015)\ignorespaces}\ignorespaces,
\normalsize{\texttt{\arxivref{1505.03750}{arXiv:1505.03750}}}\ignorespaces
\bibitem{Hijano:2015zsa}
E.~Hijano, P.~Kraus, E.~Perlmutter \& R.~Snively,
\textit{``{Witten Diagrams Revisited: The AdS Geometry of Conformal Blocks}''},
\doiref{10.1007/JHEP01(2016)146}{JHEP \textbf{1601}, 146
  (2016)\ignorespaces}\ignorespaces,
\normalsize{\texttt{\arxivref{1508.00501}{arXiv:1508.00501}}}\ignorespaces
\bibitem{Rejon-Barrera:2015bpa}
F.~Rejon-Barrera \& D.~Robbins,
\textit{``{Scalar-Vector Bootstrap}''},
\doiref{10.1007/JHEP01(2016)139}{JHEP \textbf{1601}, 139
  (2016)\ignorespaces}\ignorespaces,
\normalsize{\texttt{\arxivref{1508.02676}{arXiv:1508.02676}}}\ignorespaces
\bibitem{Penedones:2015aga}
J.~Penedones, E.~Trevisani \& M.~Yamazaki,
\textit{``{Recursion Relations for Conformal Blocks}''},
\doiref{10.1007/JHEP09(2016)070}{JHEP \textbf{1609}, 070
  (2016)\ignorespaces}\ignorespaces,
\normalsize{\texttt{\arxivref{1509.00428}{arXiv:1509.00428}}}\ignorespaces
\bibitem{Iliesiu:2015akf}
L.~Iliesiu, F.~Kos, D.~Poland, S.~S. Pufu, D.~Simmons-Duffin \& R.~Yacoby,
\textit{``{Fermion-Scalar Conformal Blocks}''},
\doiref{10.1007/JHEP04(2016)074}{JHEP \textbf{1604}, 074
  (2016)\ignorespaces}\ignorespaces,
\normalsize{\texttt{\arxivref{1511.01497}{arXiv:1511.01497}}}\ignorespaces
\bibitem{Echeverri:2016dun}
A.~Castedo~Echeverri, E.~Elkhidir, D.~Karateev \& M.~Serone,
\textit{``{Seed Conformal Blocks in 4D CFT}''},
\doiref{10.1007/JHEP02(2016)183}{JHEP \textbf{1602}, 183
  (2016)\ignorespaces}\ignorespaces,
\normalsize{\texttt{\arxivref{1601.05325}{arXiv:1601.05325}}}\ignorespaces
\bibitem{Isachenkov:2016gim}
M.~Isachenkov \& V.~Schomerus,
\textit{``{Superintegrability of $d$-dimensional Conformal Blocks}''},
\doiref{10.1103/PhysRevLett.117.071602}{Phys.~Rev.~Lett. \textbf{117}, 071602
  (2016)\ignorespaces}\ignorespaces,
\normalsize{\texttt{\arxivref{1602.01858}{arXiv:1602.01858}}}\ignorespaces
\bibitem{Costa:2016hju}
M.~S. Costa, T.~Hansen, J.~Penedones \& E.~Trevisani,
\textit{``{Projectors and seed conformal blocks for traceless mixed-symmetry
  tensors}''},
\doiref{10.1007/JHEP07(2016)018}{JHEP \textbf{1607}, 018
  (2016)\ignorespaces}\ignorespaces,
\normalsize{\texttt{\arxivref{1603.05551}{arXiv:1603.05551}}}\ignorespaces
\bibitem{Costa:2016xah}
M.~S. Costa, T.~Hansen, J.~Penedones \& E.~Trevisani,
\textit{``{Radial expansion for spinning conformal blocks}''},
\doiref{10.1007/JHEP07(2016)057}{JHEP \textbf{1607}, 057
  (2016)\ignorespaces}\ignorespaces,
\normalsize{\texttt{\arxivref{1603.05552}{arXiv:1603.05552}}}\ignorespaces
\bibitem{Chen:2016bxc}
H.-Y. Chen \& J.~D. Qualls,
\textit{``{Quantum Integrable Systems from Conformal Blocks}''},
\doiref{10.1103/PhysRevD.95.106011}{Phys.~Rev. \textbf{D95}, 106011
  (2017)\ignorespaces}\ignorespaces,
\normalsize{\texttt{\arxivref{1605.05105}{arXiv:1605.05105}}}\ignorespaces
\bibitem{Nishida:2016vds}
M.~Nishida \& K.~Tamaoka,
\textit{``{Geodesic Witten diagrams with an external spinning field}''},
\doiref{10.1093/ptep/ptx055}{PTEP \textbf{2017}, 053B06
  (2017)\ignorespaces}\ignorespaces,
\normalsize{\texttt{\arxivref{1609.04563}{arXiv:1609.04563}}}\ignorespaces
\bibitem{Cordova:2016emh}
C.~Cordova, T.~T. Dumitrescu \& K.~Intriligator,
\textit{``{Multiplets of Superconformal Symmetry in Diverse Dimensions}''},
\doiref{10.1007/JHEP03(2019)163}{JHEP \textbf{1903}, 163
  (2019)\ignorespaces}\ignorespaces,
\normalsize{\texttt{\arxivref{1612.00809}{arXiv:1612.00809}}}\ignorespaces
\bibitem{Schomerus:2016epl}
V.~Schomerus, E.~Sobko \& M.~Isachenkov,
\textit{``{Harmony of Spinning Conformal Blocks}''},
\doiref{10.1007/JHEP03(2017)085}{JHEP \textbf{1703}, 085
  (2017)\ignorespaces}\ignorespaces,
\normalsize{\texttt{\arxivref{1612.02479}{arXiv:1612.02479}}}\ignorespaces
\bibitem{Kravchuk:2016qvl}
P.~Kravchuk \& D.~Simmons-Duffin,
\textit{``{Counting Conformal Correlators}''},
\doiref{10.1007/JHEP02(2018)096}{JHEP \textbf{1802}, 096
  (2018)\ignorespaces}\ignorespaces,
\normalsize{\texttt{\arxivref{1612.08987}{arXiv:1612.08987}}}\ignorespaces
\bibitem{Gliozzi:2017hni}
F.~Gliozzi, A.~L. Guerrieri, A.~C. Petkou \& C.~Wen,
\textit{``{The analytic structure of conformal blocks and the generalized
  Wilson-Fisher fixed points}''},
\doiref{10.1007/JHEP04(2017)056}{JHEP \textbf{1704}, 056
  (2017)\ignorespaces}\ignorespaces,
\normalsize{\texttt{\arxivref{1702.03938}{arXiv:1702.03938}}}\ignorespaces
\bibitem{Castro:2017hpx}
A.~Castro, E.~Llabrés \& F.~Rejon-Barrera,
\textit{``{Geodesic Diagrams, Gravitational Interactions \& OPE Structures}''},
\doiref{10.1007/JHEP06(2017)099}{JHEP \textbf{1706}, 099
  (2017)\ignorespaces}\ignorespaces,
\normalsize{\texttt{\arxivref{1702.06128}{arXiv:1702.06128}}}\ignorespaces
\bibitem{Dyer:2017zef}
E.~Dyer, D.~Z. Freedman \& J.~Sully,
\textit{``{Spinning Geodesic Witten Diagrams}''},
\doiref{10.1007/JHEP11(2017)060}{JHEP \textbf{1711}, 060
  (2017)\ignorespaces}\ignorespaces,
\normalsize{\texttt{\arxivref{1702.06139}{arXiv:1702.06139}}}\ignorespaces
\bibitem{Sleight:2017fpc}
C.~Sleight \& M.~Taronna,
\textit{``{Spinning Witten Diagrams}''},
\doiref{10.1007/JHEP06(2017)100}{JHEP \textbf{1706}, 100
  (2017)\ignorespaces}\ignorespaces,
\normalsize{\texttt{\arxivref{1702.08619}{arXiv:1702.08619}}}\ignorespaces
\bibitem{Chen:2017yia}
H.-Y. Chen, E.-J. Kuo \& H.~Kyono,
\textit{``{Anatomy of Geodesic Witten Diagrams}''},
\doiref{10.1007/JHEP05(2017)070}{JHEP \textbf{1705}, 070
  (2017)\ignorespaces}\ignorespaces,
\normalsize{\texttt{\arxivref{1702.08818}{arXiv:1702.08818}}}\ignorespaces
\bibitem{Pasterski:2017kqt}
S.~Pasterski \& S.-H. Shao,
\textit{``{Conformal basis for flat space amplitudes}''},
\doiref{10.1103/PhysRevD.96.065022}{Phys.~Rev. \textbf{D96}, 065022
  (2017)\ignorespaces}\ignorespaces,
\normalsize{\texttt{\arxivref{1705.01027}{arXiv:1705.01027}}}\ignorespaces
\bibitem{Cardoso:2017qmj}
V.~Cardoso, T.~Houri \& M.~Kimura,
\textit{``{Mass Ladder Operators from Spacetime Conformal Symmetry}''},
\doiref{10.1103/PhysRevD.96.024044}{Phys.~Rev. \textbf{D96}, 024044
  (2017)\ignorespaces}\ignorespaces,
\normalsize{\texttt{\arxivref{1706.07339}{arXiv:1706.07339}}}\ignorespaces
\bibitem{Karateev:2017jgd}
D.~Karateev, P.~Kravchuk \& D.~Simmons-Duffin,
\textit{``{Weight Shifting Operators and Conformal Blocks}''},
\doiref{10.1007/JHEP02(2018)081}{JHEP \textbf{1802}, 081
  (2018)\ignorespaces}\ignorespaces,
\normalsize{\texttt{\arxivref{1706.07813}{arXiv:1706.07813}}}\ignorespaces,
[,91(2017)]\ignorespaces
\bibitem{Kravchuk:2017dzd}
P.~Kravchuk,
\textit{``{Casimir recursion relations for general conformal blocks}''},
\doiref{10.1007/JHEP02(2018)011}{JHEP \textbf{1802}, 011
  (2018)\ignorespaces}\ignorespaces,
\normalsize{\texttt{\arxivref{1709.05347}{arXiv:1709.05347}}}\ignorespaces,
[,164(2017)]\ignorespaces
\bibitem{Dey:2017fab}
P.~Dey, K.~Ghosh \& A.~Sinha,
\textit{``{Simplifying large spin bootstrap in Mellin space}''},
\doiref{10.1007/JHEP01(2018)152}{JHEP \textbf{1801}, 152
  (2018)\ignorespaces}\ignorespaces,
\normalsize{\texttt{\arxivref{1709.06110}{arXiv:1709.06110}}}\ignorespaces
\bibitem{Hollands:2017chb}
S.~Hollands,
\textit{``{Action principle for OPE}''},
\doiref{10.1016/j.nuclphysb.2017.11.013}{Nucl.~Phys. \textbf{B926}, 614
  (2018)\ignorespaces}\ignorespaces,
\normalsize{\texttt{\arxivref{1710.05601}{arXiv:1710.05601}}}\ignorespaces
\bibitem{Schomerus:2017eny}
V.~Schomerus \& E.~Sobko,
\textit{``{From Spinning Conformal Blocks to Matrix Calogero-Sutherland
  Models}''},
\doiref{10.1007/JHEP04(2018)052}{JHEP \textbf{1804}, 052
  (2018)\ignorespaces}\ignorespaces,
\normalsize{\texttt{\arxivref{1711.02022}{arXiv:1711.02022}}}\ignorespaces
\bibitem{Isachenkov:2017qgn}
M.~Isachenkov \& V.~Schomerus,
\textit{``{Integrability of conformal blocks. Part I. Calogero-Sutherland
  scattering theory}''},
\doiref{10.1007/JHEP07(2018)180}{JHEP \textbf{1807}, 180
  (2018)\ignorespaces}\ignorespaces,
\normalsize{\texttt{\arxivref{1711.06609}{arXiv:1711.06609}}}\ignorespaces
\bibitem{Faller:2017hyt}
J.~Faller, S.~Sarkar \& M.~Verma,
\textit{``{Mellin Amplitudes for Fermionic Conformal Correlators}''},
\doiref{10.1007/JHEP03(2018)106}{JHEP \textbf{1803}, 106
  (2018)\ignorespaces}\ignorespaces,
\normalsize{\texttt{\arxivref{1711.07929}{arXiv:1711.07929}}}\ignorespaces
\bibitem{Rong:2017cow}
J.~Rong \& N.~Su,
\textit{``{Scalar CFTs and Their Large N Limits}''},
\doiref{10.1007/JHEP09(2018)103}{JHEP \textbf{1809}, 103
  (2018)\ignorespaces}\ignorespaces,
\normalsize{\texttt{\arxivref{1712.00985}{arXiv:1712.00985}}}\ignorespaces
\bibitem{Chen:2017xdz}
H.-Y. Chen, E.-J. Kuo \& H.~Kyono,
\textit{``{Towards Spinning Mellin Amplitudes}''},
\doiref{10.1016/j.nuclphysb.2018.04.019}{Nucl.~Phys. \textbf{B931}, 291
  (2018)\ignorespaces}\ignorespaces,
\normalsize{\texttt{\arxivref{1712.07991}{arXiv:1712.07991}}}\ignorespaces
\bibitem{Sleight:2018epi}
C.~Sleight \& M.~Taronna,
\textit{``{Spinning Mellin Bootstrap: Conformal Partial Waves, Crossing Kernels
  and Applications}''},
\doiref{10.1002/prop.201800038}{Fortsch.~Phys. \textbf{66}, 1800038
  (2018)\ignorespaces}\ignorespaces,
\normalsize{\texttt{\arxivref{1804.09334}{arXiv:1804.09334}}}\ignorespaces
\bibitem{Costa:2018mcg}
M.~S. Costa \& T.~Hansen,
\textit{``{AdS Weight Shifting Operators}''},
\doiref{10.1007/JHEP09(2018)040}{JHEP \textbf{1809}, 040
  (2018)\ignorespaces}\ignorespaces,
\normalsize{\texttt{\arxivref{1805.01492}{arXiv:1805.01492}}}\ignorespaces
\bibitem{Kobayashi:2018okw}
N.~Kobayashi \& T.~Nishioka,
\textit{``{Spinning conformal defects}''},
\doiref{10.1007/JHEP09(2018)134}{JHEP \textbf{1809}, 134
  (2018)\ignorespaces}\ignorespaces,
\normalsize{\texttt{\arxivref{1805.05967}{arXiv:1805.05967}}}\ignorespaces
\bibitem{Bhatta:2018gjb}
A.~Bhatta, P.~Raman \& N.~V. Suryanarayana,
\textit{``{Scalar Blocks as Gravitational Wilson Networks}''},
\doiref{10.1007/JHEP12(2018)125}{JHEP \textbf{1812}, 125
  (2018)\ignorespaces}\ignorespaces,
\normalsize{\texttt{\arxivref{1806.05475}{arXiv:1806.05475}}}\ignorespaces
\bibitem{Lauria:2018klo}
E.~Lauria, M.~Meineri \& E.~Trevisani,
\textit{``{Spinning operators and defects in conformal field theory}''},
\normalsize{\texttt{\arxivref{1807.02522}{arXiv:1807.02522}}}\ignorespaces
\bibitem{Liu:2018jhs}
J.~Liu, E.~Perlmutter, V.~Rosenhaus \& D.~Simmons-Duffin,
\textit{``{$d$-dimensional SYK, AdS Loops, and $6j$ Symbols}''},
\doiref{10.1007/JHEP03(2019)052}{JHEP \textbf{1903}, 052
  (2019)\ignorespaces}\ignorespaces,
\normalsize{\texttt{\arxivref{1808.00612}{arXiv:1808.00612}}}\ignorespaces
\bibitem{Gromov:2018hut}
N.~Gromov, V.~Kazakov \& G.~Korchemsky,
\textit{``{Exact Correlation Functions in Conformal Fishnet Theory}''},
\normalsize{\texttt{\arxivref{1808.02688}{arXiv:1808.02688}}}\ignorespaces
\bibitem{Zhou:2018sfz}
X.~Zhou,
\textit{``{Recursion Relations in Witten Diagrams and Conformal Partial
  Waves}''},
\normalsize{\texttt{\arxivref{1812.01006}{arXiv:1812.01006}}}\ignorespaces
\bibitem{Kazakov:2018gcy}
V.~Kazakov, E.~Olivucci \& M.~Preti,
\textit{``{Generalized fishnets and exact four-point correlators in chiral
  CFT$_{4}$}''},
\doiref{10.1007/JHEP06(2019)078}{JHEP \textbf{1906}, 078
  (2019)\ignorespaces}\ignorespaces,
\normalsize{\texttt{\arxivref{1901.00011}{arXiv:1901.00011}}}\ignorespaces
\bibitem{Rosenhaus:2018zqn}
V.~Rosenhaus,
\textit{``{Multipoint Conformal Blocks in the Comb Channel}''},
\doiref{10.1007/JHEP02(2019)142}{JHEP \textbf{1902}, 142
  (2019)\ignorespaces}\ignorespaces,
\normalsize{\texttt{\arxivref{1810.03244}{arXiv:1810.03244}}}\ignorespaces
\bibitem{Dirac:1936fq}
P.~A.~M. Dirac,
\textit{``{Wave equations in conformal space}''},
\doiref{10.2307/1968455}{Annals~Math. \textbf{37}, 429
  (1936)\ignorespaces}\ignorespaces
\bibitem{Mack:1969rr}
G.~Mack \& A.~Salam,
\textit{``{Finite component field representations of the conformal group}''},
\doiref{10.1016/0003-4916(69)90278-4}{Annals~Phys. \textbf{53}, 174
  (1969)\ignorespaces}\ignorespaces
\bibitem{Weinberg:2010fx}
S.~Weinberg,
\textit{``{Six-dimensional Methods for Four-dimensional Conformal Field
  Theories}''},
\doiref{10.1103/PhysRevD.82.045031}{Phys.Rev. \textbf{D82}, 045031
  (2010)\ignorespaces}\ignorespaces,
\normalsize{\texttt{\arxivref{1006.3480}{arXiv:1006.3480}}}\ignorespaces
\bibitem{Weinberg:2012mz}
S.~Weinberg,
\textit{``{Six-dimensional Methods for Four-dimensional Conformal Field
  Theories II: Irreducible Fields}''},
\doiref{10.1103/PhysRevD.86.085013}{Phys.Rev. \textbf{D86}, 085013
  (2012)\ignorespaces}\ignorespaces,
\normalsize{\texttt{\arxivref{1209.4659}{arXiv:1209.4659}}}\ignorespaces
\bibitem{Fortin:2016lmf}
J.-F. Fortin \& W.~Skiba,
\textit{``{Conformal Bootstrap in Embedding Space}''},
\doiref{10.1103/PhysRevD.93.105047}{Phys.~Rev. \textbf{D93}, 105047
  (2016)\ignorespaces}\ignorespaces,
\normalsize{\texttt{\arxivref{1602.05794}{arXiv:1602.05794}}}\ignorespaces
\bibitem{Fortin:2016dlj}
J.-F. Fortin \& W.~Skiba,
\textit{``{Conformal Differential Operator in Embedding Space and its
  Applications}''},
\normalsize{\texttt{\arxivref{1612.08672}{arXiv:1612.08672}}}\ignorespaces
\bibitem{Comeau:2019xco}
V.~Comeau, J.-F. Fortin \& W.~Skiba,
\textit{``{Further Results on a Function Relevant for Conformal Blocks}''},
\normalsize{\texttt{\arxivref{1902.08598}{arXiv:1902.08598}}}\ignorespaces
\bibitem{Bailey:1994}
T.~Bailey, M.~Eastwood \& A.~Gover,
\textit{``Thomas's Structure Bundle for Conformal, Projective and Related
  Structures''},
\doiref{10.1216/rmjm/1181072333}{Rocky~Mountain~J.~Math. \textbf{24}, 1191
  (1994)\ignorespaces}\ignorespaces,
\href{https://doi.org/10.1216/rmjm/1181072333}{\texttt{https://doi.org/10.1216/rmjm/1181072333}}
\bibitem{Eastwood:2002su}
M.~G. Eastwood,
\textit{``{Higher symmetries of the Laplacian}''},
\doiref{10.4007/annals.2005.161.1645}{Annals~Math. \textbf{161}, 1645
  (2005)\ignorespaces}\ignorespaces,
\normalsize{\texttt{\arxivref{hep-th/0206233}{hep-th/0206233}}}\ignorespaces
\bibitem{Exton_1995}
H.~Exton,
\textit{``On the system of partial differential equations associated with
  Appell's function F4''},
\doiref{10.1088/0305-4470/28/3/017}{Journal~of~Physics~A:~Mathematical~and~General
  \textbf{28}, 631 (1995)\ignorespaces}\ignorespaces
\bibitem{Goldberger:2011yp}
W.~D. Goldberger, W.~Skiba \& M.~Son,
\textit{``{Superembedding Methods for 4d N=1 SCFTs}''},
\doiref{10.1103/PhysRevD.86.025019}{Phys.~Rev. \textbf{D86}, 025019
  (2012)\ignorespaces}\ignorespaces,
\normalsize{\texttt{\arxivref{1112.0325}{arXiv:1112.0325}}}\ignorespaces
\bibitem{Goldberger:2012xb}
W.~D. Goldberger, Z.~U. Khandker, D.~Li \& W.~Skiba,
\textit{``{Superembedding Methods for Current Superfields}''},
\doiref{10.1103/PhysRevD.88.125010}{Phys.~Rev. \textbf{D88}, 125010
  (2013)\ignorespaces}\ignorespaces,
\normalsize{\texttt{\arxivref{1211.3713}{arXiv:1211.3713}}}\ignorespaces
\end{thebibliography}

\end{document}